\documentclass{article}

\newcommand{\blind}{1}
\newcommand{\jabes}{1}

\usepackage[english]{babel}
\usepackage{natbib}
\usepackage[letterpaper,top=2cm,bottom=2cm,left=3cm,right=3cm,marginparwidth=1.75cm]{geometry}
\usepackage{amsmath, amsthm, amssymb}
\usepackage{graphicx}
\usepackage[colorlinks=true, allcolors=blue]{hyperref}
\usepackage{subcaption}
\usepackage{mathtools}
\usepackage{wrapfig}
\usepackage{placeins}
\usepackage{float}
\usepackage{multirow}
\usepackage{booktabs}
\graphicspath{{pictures/}}

\usepackage{algorithm}
\usepackage{algpseudocode}

\newcommand{\bm}[1]{\boldsymbol{#1}}

\title{An accuracy-runtime trade-off comparison of scalable Gaussian process approximations for spatial data}
\author{
  Filippo Rambelli\footnotemark[1] \\
  \and
  Fabio Sigrist\footnotemark[1] \footnotemark[3] 
}

\begin{document}
\date{}
\maketitle

\begin{abstract}
Gaussian processes (GPs) are flexible, probabilistic, nonparametric models widely used in fields such as spatial statistics and machine learning. A drawback of Gaussian processes is their computational cost, with $\mathcal{O}(N^3)$ time and $\mathcal{O}(N^2)$ memory complexity, which makes them prohibitive for large data sets. Numerous approximation techniques have been proposed to address this limitation. In this work, we systematically compare the accuracy of different Gaussian process approximations with respect to likelihood evaluation, parameter estimation, and prediction, explicitly accounting for the computational time required. We analyze the trade-off between accuracy and runtime on multiple simulated and large-scale real-world data sets and find that Vecchia approximations consistently provide the best accuracy–runtime trade-off across most settings considered.
\end{abstract}

\noindent%

\if1\blind
{
\footnotetext[1]{Seminar for Statistics, ETH Zurich, Switzerland}
\footnotetext[3]{Corresponding author: fabio.sigrist@stat.math.ethz.ch}
} \fi

\section{Introduction}\label{intro}
Gaussian processes (GPs) are flexible, probabilistic, nonparametric models which are widely used in spatial statistics \citep{cressie2011statistics, banerjee2014hierarchical} and machine learning \citep{williams2006gaussian}. Estimation and prediction with Gaussian processes has $\mathcal{O}(N^3)$ computational complexity and requires $\mathcal{O}(N^2)$ memory. These costs render Gaussian processes computationally prohibitive, even when $N$ is moderately large. To mitigate this problem, various approximations have been proposed; see \citet{a_case_of_study_competition} and \citet{when_GP_meets_BIGDATA} for reviews. The goal of a Gaussian process approximation is to achieve computational efficiency with as little degradation in accuracy as possible compared to exact calculations. Most approximations have tuning parameters which can be adjusted to balance a trade-off between accuracy and computational efficiency. In particular, many approximations recover the exact GP calculations in a limit case for a certain choice of tuning parameters. What matters in practice, though, is which approximation has the highest accuracy for a given computational budget. Analyzing the accuracy of Gaussian process approximations without taking computational cost into account thus provides only limited insights. The goal of this article is to compare various popular Gaussian process approximations regarding this accuracy-runtime trade-off.

We consider the following approximations: Vecchia approximations \citep{vecchia_first,stein2004approximating,datta2016hierarchical,katzfuss2017general}, covariance tapering \citep{tapering_first}, modified predictive process approximations also called fully independent training conditional (FITC) approximations \citep{FITC_ref,ref_spatial_stat,modified_predictive_process}, full-scale approximations (FSA) combining predictive processes with tapering \citep{full_scale_first}, multi-resolution approximations (MRA) \citep{MRA}, the stochastic partial differential equations (SPDE)-based approximation introduced in \citet{SPDE}, fixed rank kriging (FRK) \citep{FRK_first}, and the Fourier transform-based periodic embedding approach of \citet{PeriodicEmbedding}. Note that we are mainly interested in understanding how well the different methods approximate a given exact Gaussian process in a sense specified below. This means that we are not primarily interested in comparing different methods concerning pure prediction accuracy. If this were the goal, we would also have to include methods such as tree-boosting, neural networks, and other machine learning models. Most of the methods we consider (Vecchia, FITC, FSA, SPDE, and MRA) do indeed aim to approximate an exact Gaussian process with a given covariance function. FRK and the periodic embedding approach, however, do not aim to approximate a certain Gaussian process. We nonetheless include them in our comparison since they are frequently used in spatial statistics \citep{a_case_of_study_competition}.

We compare the accuracy of the different approximations concerning log-likelihood evaluation, parameter estimation, and prediction on several simulated and real world data sets. For analyzing predictive distributions, we distinguish between interpolation and extrapolation test sets. For every approximation, we conduct multiple computations with varying complexity by adjusting tuning parameters such as the number of neighbors for Vecchia approximations and the number of inducing points for FITC approximations. This allows us to assess the trade-off between the quality and the computational time of the different approximations. We find that Vecchia approximations consistently provide the best accuracy–runtime trade-off in almost all experiments.

To the best of our knowledge, no comprehensive accuracy-runtime analysis concerning log-likelihood evaluation, parameter estimation, and prediction has been done before. \citet{a_case_of_study_competition} compare scalable methods for modeling spatial data. However, the runtimes of the different methods are vastly different, and the methods are run using different numbers of CPU cores. Similarly, \citet{huang2021competition}, \citet{abdulah2022second}, and \citet{hong2023third} compare various methods for analyzing spatial data but do not consider runtimes in their analysis. \citet{guinness2018permutation} analyzes the prior Kullback–Leibler divergence between exact Gaussian processes and various Vecchia approximations as a function of the runtime and compares this to covariance tapering with three taper ranges and one SPDE approximation. In line with our results, \citet{guinness2018permutation} finds that Vecchia approximations are the closest to an exact Gaussian process prior distribution.

\subsection{Background on Gaussian processes}
A stochastic process $f(\bm{s}), s\in D \subset \mathbb{R}^d,$ is a Gaussian process if $(f(\bm{s}_1),\dots,f(\bm{s}_N))$ follows a multivariate Gaussian distribution for any finite set of input locations $\bm{s}_1,\dots,\bm{s}_N\in D$. A Gaussian process is defined by a mean function \( \mu(\bm{s}) = E[f(\bm{s})] \) and a positive definite covariance function \( \quad k(\bm{s}, \bm{s}') = E\left[(f(\bm{s}) - \mu(\bm{s}))(f(\bm{s}') - \mu(\bm{s}'))\right],
 \quad \bm{s}, \bm{s}' \in D \). In regression settings, one usually has a set of $N$ noisy observations $( y_1, \ldots, y_N)$ at training locations \( \bm{s}_1, \ldots, \bm{s}_N \): $y(\bm{s}_i) = f(\bm{s}_i) + \varepsilon_i, \quad \varepsilon_i \stackrel{iid}{\sim} \mathcal{N}(0, \sigma^2_n), \quad i = 1, \ldots, N.$ Assuming a zero prior mean and a parametric covariance function \( k_\theta(\cdot, \cdot) \), where the covariance parameters denoted by $\theta$ also include the error term variance $\sigma^2_n$, the log-marginal likelihood is given by
\begin{equation*}
    \log p(\mathbf{y} \mid \theta)=-\frac{1}{2} \mathbf{y}^T\left(K_\theta(S,S)+\sigma_n^2 I\right)^{-1} \mathbf{y}-\frac{1}{2} \log \left|K_\theta(S,S)+\sigma_n^2 I\right|-\frac{N}{2} \log (2 \pi),
\end{equation*}
where $\mathbf{y}=(y(\bm{s}_1),\dots,y(\bm{s}_N))^T$, \(K_\theta(S,S)\in\mathbb{R}^{N\times N}\) indicates the covariance matrix evaluated at training locations $S=(s_1,\dots,s_N)$ with entries \(K_\theta(S,S)_{ij} = k_\theta(\bm{s}_i, \bm{s}_j)\). Estimation of the parameters $\theta$ is usually done by maximizing this log-marginal likelihood. Further, probabilistic predictions at a set of locations $S^*=(s_1,\dots,s_{N_p})$ are obtained using the posterior predictive distribution given by:
\begin{equation*}
\begin{split}
\mathbf{f}_* \mid \mathbf{y} &\sim \mathcal{N}\left( K_\theta(S_*, S)\left[K_\theta(S, S)+\sigma_n^2 I\right]^{-1} \mathbf{y},  K_\theta(S_*, S_*)-K_\theta(S_*, S)\left[K_\theta(S, S)+\sigma_n^2 I\right]^{-1} K_\theta(S, S_*) \right).
\end{split}
\end{equation*}

The rest of the article is structured as follows. In Section \ref{setting}, we document the experimental setup including the data sets, the metrics used to evaluate the quality of the approximations, and the hardware and software resources employed. In Section \ref{results} , we present the results from our simulated and real-world data experiments.

\section{Experimental setting}\label{setting}
In the following, we present our experimental setup. The simulated and real-world data sets are described in Sections \ref{sim_data} and \ref{real_world_data}, respectively. In Section \ref{evaluation_criteria}, we present the evaluation criteria for log-likelihood evaluation, parameter estimation, and prediction accuracy. The GP approximations considered in this article and the choice of tuning parameters for the approximations are specified in Section \ref{gp_approx_tune_pars}, and our computational environment is described in Section \ref{soft_hardware}. 

\subsection{Simulated data}\label{sim_data}
\begin{figure}[ht!]
  \centering
    \includegraphics[width=0.65\linewidth]{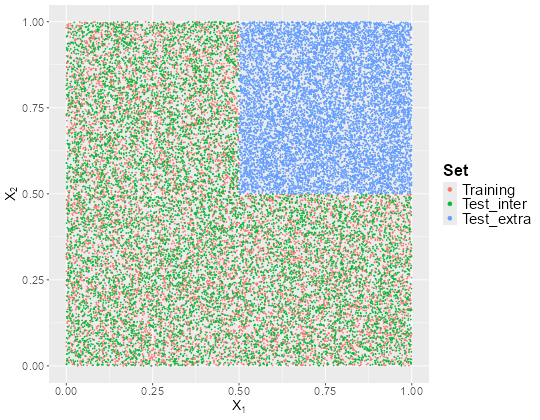} 
  \caption[Illustration of a simulated data set]
  {Illustration of simulated training and test data with $N=10,000$. For the test data, we distinguish between test interpolation (`Test\_inter') and extrapolation (`Test\_extra') data.}
  \label{locations_example}
\end{figure}

We simulate Gaussian process realizations with locations sampled uniformly in the unit square \([0,1]^2\) with two different sample sizes $N=10,000$ and $N=100,000$. Every data set consists of a training set and an “interpolation” test set each containing $N$ locations uniformly sampled from \([0,1]^2\) excluding  \([0.5,1]^2\). Moreover, there is an “extrapolation” test set consisting of additional $N$ locations uniformly distributed in  \([0.5,1]^2\). Figure \ref{locations_example} shows an example of the distribution of such locations. 

Unless stated otherwise, we use a zero-mean Gaussian process with an isotropic Matérn covariance function $ k_{M} \left(\bm{s}, \bm{s}^{\prime}\right)=\sigma^22^{1-\nu}/\Gamma(\nu)(\sqrt{2\nu}\left\|\bm{s}-\bm{s}^{\prime}\right\|/\rho)^\nu K_\nu(\sqrt{2\nu}\left\|\bm{s}-\bm{s}^{\prime}\right\|/\rho),$ where $\sigma^2$, $\rho$, and $\nu$ are the marginal variance, range, and smoothness parameters, respectively. Furthermore, we add an independent Gaussian error term (``nugget effect") with variance $\sigma_n^2$. Unless stated otherwise, we use $\sigma^2 = 1$, $\sigma^2_n=0.5$, $\rho=0.2/2.74$, and $\nu=1.5$ for simulating data. This corresponds to a signal-to-noise ratio of two, and the effective range is $0.2$, i.e., at a distance of $0.2$ the correlation is approximately $0.05$. In addition, we simulate data with smaller and larger range parameters $\rho = 0.05/2.74$ and $\rho=0.5/2.74$, other smoothness parameters $\nu=0.5$ and $\nu=2.5$, a higher signal-to-noise ratio with a smaller error variance of $\sigma^2_n=0.1$, and an anisotropic model. For the latter, we use an anisotropic automatic relevance determination (ARD) Matérn covariance $ k_{M} \left(\bm{s}, \bm{s}^{\prime}\right)=\sigma^22^{1-\nu}/\Gamma(\nu)\left(\sqrt{2\nu}\sqrt{\left((s_x-s_x')/\rho_x\right)^2+\left((s_y-s_y')/\rho_y\right)^2}\right)^\nu K_\nu\left(\sqrt{2\nu}\sqrt{\left((s_x-s_x')/\rho_x\right)^2+\left((s_y-s_y')/\rho_y\right)^2}\right)$ with different range parameters $\rho_x=0.05/2.74$ and $\rho_y=0.2/2.74$ for $s_x$ and $s_y$, $\bm{s}=(s_x,s_y)$. For the smoothness parameters $\nu=0.5$ and $\nu=2.5$, we use $\rho = 0.2/3$ and $\rho = 0.2/2.65$, respectively, which corresponds to an effective range of $0.2$. For every simulation scenario, we repeatedly simulate a certain number of data sets in order to analyze the variability of the evaluation metrics. The specific number of simulation repetitions varies slightly with the sample size and the task; see Section \ref{evaluation_criteria} for more information. 


\subsection{Real-world data sets}\label{real_world_data}
In the following, we present the four large-scale real-world data sets we consider in this article. We use a Matérn $3/2$ covariance function for all real-world data sets and all models where this is possible, i.e., exact GP calculations and all GP approximations, except FRK and periodic embedding. For the MODIS 2016 data set described in Section \ref{modis_2016}, we follow \citet{a_case_of_study_competition} and use an exponential covariance function.

\subsubsection{Lucas County house price data set}\label{house_price_data}
\begin{figure}[ht!]
      \centering     
      \includegraphics[width=\linewidth]{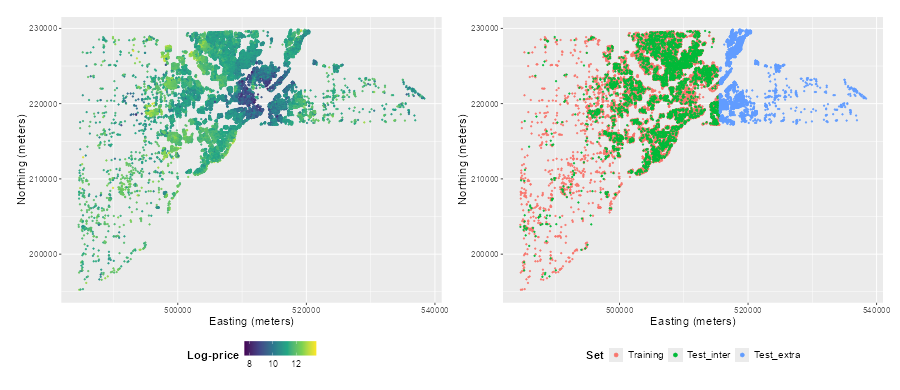} 
  \caption[Illustration of the house price data set.]
  {Logarithmic selling price over space (left) and illustration of training and test locations (right) for the house price data set.}
  \label{house_split}
\end{figure}

The Lucas County house price data set, available from the R package spData, contains information about $25,357$ single family houses sold in Lucas County, Ohio, between 1993 and 1998. The response variable is the logarithmic selling price. We use the spatial coordinates as input locations for the Gaussian processes, and we assume that the prior mean is zero after centering the data using the sample mean of the training set. In addition, we do a second experiment where we use a non-zero linear prior mean with a linear predictor containing the following covariates: $age$, $age^2$, $\log(total\_ living\_area)$, $nb\_rooms$, $\log(lot\_size)$, and $selling\_year$. We do a random $70$/$30$\% training-test split allocating $17,751$ observations to the training set, while the remaining data points are evenly divided between ``interpolation” and ``extrapolation” test sets each comprising $3,803$ data points. An illustration of this split and a spatial visualization of house prices can be found in Figure \ref{house_split}. 

\subsubsection{Laegern canopy height LiDAR data set}
\begin{figure}[ht!]
      \centering     
      \includegraphics[width=\linewidth]{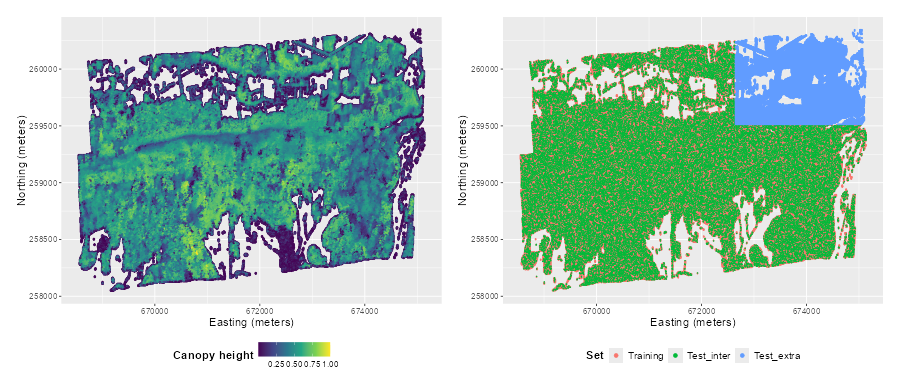} 
  \caption[Illustration of the Laegern data set.]
  {Canopy height over space (left) and illustration of training and test locations (right) for the Laegern data set.}
  \label{laegern_split}
\end{figure}

The Laegern data set \citep{Laegern} includes measurements of plant functional traits of the Laegern temperate mixed forest in Switzerland. We focus on the canopy height which is provided in scaled form between $0$ and $1$. The data is recorded on a $1099\times384$ grid with a $6$ m spatial resolution. After removing missing values and excluding non-forest areas, there are $279,180$ observations left. Similarly to the housing data set, we allocate 70\% of the observations to the training set, 15\% to the ``interpolation” test set, and 15\% to the ``extrapolation” test set. A  graphical illustration can be found in Figure \ref{laegern_split}.

\subsubsection{MODIS 2016 temperature satellite data set}\label{modis_2016}
\begin{figure}[ht!]
      \centering    
      \includegraphics[width=\linewidth]{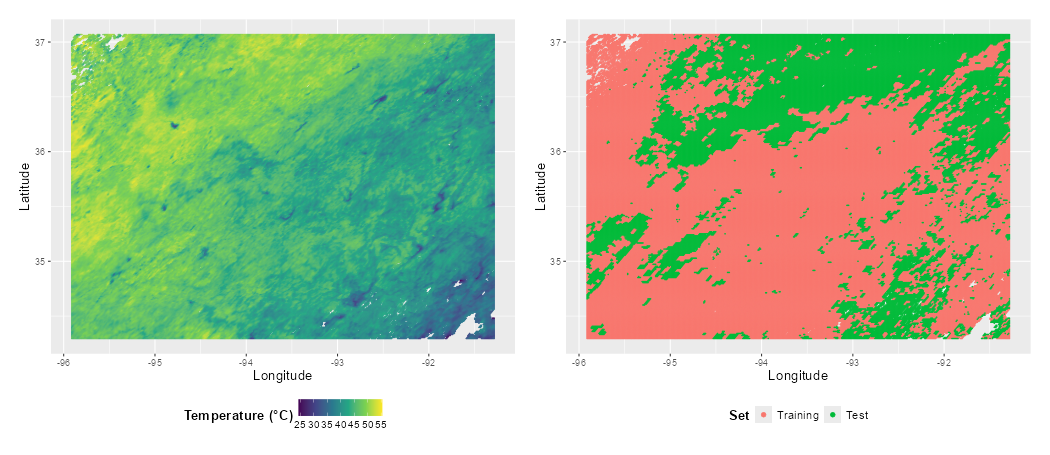} 
  \caption[Illustration of the MODIS 2023 data set.]
  {Land surface temperature over space (left) and illustration of training and test locations (right) for the MODIS 2016 data set.}
  \label{modis16_split}
\end{figure}

Next, we consider land surface temperatures (LST) recorded in degrees Celsius by the MODIS satellite (https://mrtweb.cr.usgs.gov/) on August 4, 2016. Measurements are made on a $500\times300$ spatial grid with a resolution of $0.009273987$ degrees. The same data was used in \citet{a_case_of_study_competition}, and we use the same train-test split as in \citet{a_case_of_study_competition} with $105,569$ observations in the training set and $42,740$ in the test set. Note that a small number of measurements ($1,691$) is missing because of cloud cover. Due to the presence of a clearly visible trend, see Figure \ref{modis16_split}, we include a linear regression term with longitude and latitude coordinates and an intercept term in the GP models. 

\subsubsection{MODIS 2023 temperature satellite data set}
\begin{figure}[ht!]
      \centering    
      \includegraphics[width=\linewidth]{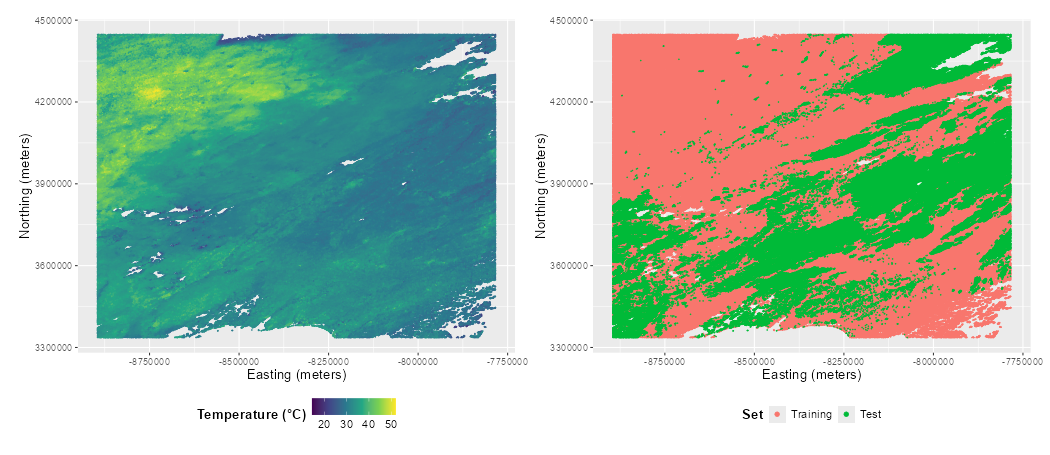} 
  \caption[Illustration of the MODIS 2023 data set.]
  {Land surface temperature over space (left) and illustration of training and test locations (right) for the MODIS 2023 data set.}
  \label{modis23_splitt}
\end{figure}

We additionally use another land surface temperature data set from the MODIS satellite recorded on August 20, 2023. In this case, the spatial grid has dimensions $1200\times1200$ with a resolution of $926.6254$ meters. We use the same train-test split as in \citet{TimFabioReinhard}: out of the $600,000$ available observations, $400,000$ are used for the training set, while the remaining $200,000$ are allocated to the test set. Illustrations of the temperature data and the train-test data split are shown in Figure \ref{modis23_splitt}. We again include an intercept and the coordinates as covariates in a linear predictor. 

\subsection{Evaluation criteria} \label{evaluation_criteria}
The following section introduces the criteria used to assess the quality of the different approximations. We consider multiple accuracy metrics for the three tasks of log-likelihood evaluation, parameter estimation, and prediction. 

\subsubsection{Log-likelihood evaluation}
We first analyze the accuracy of log-likelihood evaluations of the different GP approximations. In experiments with $N=10,000$, we perform $100$ simulation iterations and consider the absolute difference to the exact log-likelihood since exact calculations can still be run in a reasonable amount of time. For the synthetic data sets with $N=100,000$, we inspect the negative log-likelihood value itself by carrying out $20$ repetitions. For the simulated data, the log-likelihood of the training data is evaluated both at the true data-generating parameters and by using another set of parameters obtained by doubling the parameters. For the real-world data sets, we evaluate the log-likelihood for all methods using the set of covariance parameters estimated by the most accurate method that could be implemented within a reasonable amount of time. This corresponds to using estimates of exact Gaussian processes for the housing data, and estimates of Vecchia approximations with the highest number of neighbors for the three other data sets. Additionally, we evaluate the log-likelihood at a second set of parameters obtained again by doubling the former parameters.

\subsubsection{Parameter estimation}\label{estimation}
Next, we measure the accuracy of parameter estimates on the simulated data considering the bias and the mean square error (MSE). We carry out $100$ and $20$ repetitions for the experiment with $N=10,000$ and $N=100,000$, respectively. In general, the parameters that are estimated are $\theta=\left(\sigma_n^2,\sigma^2,\rho\right)$, but for an anisotropic ARD covariance, the parameters are $\theta=\left(\sigma_n^2,\sigma^2,\rho_x,\rho_y\right)$. Since several software implementations used in this article (see Section \ref{soft_hardware}) do not allow for estimating the smoothness parameter, we fix the smoothness parameter at the data-generating values for both estimation and prediction; see Section \ref{sim_data} for the specific values. But we also consider a misspecified scenario where the smoothness parameter used for estimation and prediction is not equal to the data-generating one; see Section \ref{add_exper}.

\subsubsection{Prediction}
We also evaluate the accuracy of predictive distributions. For the synthetic data sets, we use the true data-generating covariance parameters for making predictions. For the real-world data sets, we use each method's estimated parameters. For the simulated data sets, prediction is done for the latent process $f(\cdot)$, and for the real-world data sets, we predict the observable variable $y(\cdot)$. The first metric used is the RMSE between the predicted and true test values given by $\sqrt{\frac{1}{N_s} \sum_{i=1}^{N_s}\left(f_{i}-\mu_{p, i}\right)^2}$ for the simulated data and by $\sqrt{\frac{1}{N_s} \sum_{i=1}^{N_s}\left(y_{i}-\mu_{p, i}\right)^2}$ for the
 real-world data, where $N_s$ is the sample size of the test data, $\mu_{p, i}$ indicates the predictive mean for test sample $i$, $f_i$ is the value of the latent GP at test point $i$, and $y_i$ is the observable test data which equals $f_i$ plus the error term. To assess the accuracy of probabilistic predictions, we consider the univariate log-score defined as the average univariate Gaussian negative log-likelihood (NLL) of the latent and observable processes, 
$$ \frac{1}{N_s} \sum_{i=1}^{N_s} \left[\frac{\left(f_{i}-\mu_{p, i}\right)^2 }{2\sigma_{p, i}^2}+\frac{1}{2} \log \left(2 \pi\sigma_{p, i}^2 \right)\right] ~~\text{and}~~ \frac{1}{N_s} \sum_{i=1}^{N_s} \left[\frac{\left(y_{i}-\mu_{p, i}\right)^2 }{2\left(\sigma_{p, i}^2 + \hat{\sigma}_n^2\right)}+\frac{1}{2} \log \left(2 \pi\left(\sigma_{p, i}^2+ \hat{\sigma}_n^2 \right)\right)\right],$$ 
respectively, where $\sigma_{p, i}^2$ is the predictive variance of test point $i$, and $\hat{\sigma}_n^2$ the estimated noise variance. The former version for the latent process is used for the simulated data, and the second version is used for the real-world data. For the real-world data sets, we additionally use the continuous ranked probability score (CRPS) \citep{CRPS} which, for Gaussian predictive distributions, is given by
$$\frac{1}{N_s} \sum_{i=1}^{N_s} \sqrt{\sigma_{p, i}^2+\hat{\sigma}_n^2}\left(-\frac{1}{\sqrt{\pi}}+2 \phi\left(\frac{y_{i}-\mu_{p, i}}{\sqrt{\sigma_{p, i}^2+\hat{\sigma}_n^2}}\right)+\frac{y_{i}-\mu_{p, i}}{\sqrt{\sigma_{p, i}^2+\hat{\sigma}_n^2}}\left(2 \Phi\left(\frac{y_{i}-\mu_{p, i}}{\sqrt{\sigma_{p, i}^2+\hat{\sigma}_n^2}}\right)-1\right)\right),$$ 
where $\phi$ and $\Phi$ denote the probability density function and the cumulative density function, respectively, of a standard Gaussian distribution.

On data sets with moderate sample sizes, we additionally assess the quality of the approximations by comparing their predictive distributions to the ones obtained using exact calculations. For this, we consider the RMSE of both approximate univariate predictive means and variances compared to the ones of exact GP calculations. Further, we consider the Kullback–Leibler (KL) divergence between the exact and approximate univariate predictive distributions. The latter is given by 
$$ K L(N\left(\mu_q, \sigma_q\right),N\left(\mu_p, \sigma_p\right))=\log \frac{\sigma_p}{\sigma_q}+\frac{\sigma_q^2+\left(\mu_q-\mu_p\right)^2}{2 \sigma_p^2}-\frac{1}{2}.$$ 
The number of simulation runs to evaluate the prediction accuracy for the simulated data is $50$ for \mbox{$N=10,000$} and $10$ for $N=100,000$.

\subsubsection{Runtimes}
For each evaluation metric, we measure the wall-clock time needed to obtain the corresponding quantities using the computing environment described in Section \ref{soft_hardware}. For instance, when evaluating the accuracy of covariance parameter estimates on simulated data, we record the wall-clock time required to fit the models. Furthermore, when analyzing the prediction accuracy on the real-world data sets, we measure the total time including estimation and prediction. For the simulated data sets, on the other hand, we only measure the time to calculate predictions as predictions are done using the data-generating parameters. In the simulation-based experiments, where multiple repetitions are conducted, both times and metrics are averaged over the iterations. 

\subsection{GP approximations and choice of tuning parameters}\label{gp_approx_tune_pars}
We consider the following GP approximations: Vecchia approximations, covariance tapering, modified predictive process aka Fully Independent Training Conditional (FITC) approximations, full-scale approximations (FSA) combining predictive processes with tapering, multi-resolution approximations (MRA), the stochastic partial differential equations (SPDE)-based approximation of \citet{SPDE}, fixed rank kriging (FRK), and the periodic embedding approach of \citet{PeriodicEmbedding}. 

The runtime and accuracy of the different approximation methods are determined by one or several tuning parameters such as the number of neighbors for Vecchia approximations and the number of inducing points for FITC approximations. For every approximation, we use up to five different choices of tuning parameters. The selection of tuning parameters is done such that the different methods have roughly similar run-times. The specific choices of the tuning parameters thus vary across data sets and are reported in the appendix. In the simulated experiments, the choice of the tuning parameters is done based on the time for evaluating the log-likelihood. In real-world applications, we instead use the total time for estimation and prediction on the (“interpolation”) test set. We sometimes deviate from this tuning parameter selection approach for the following reasons. First, Vecchia approximations generally yield very accurate results already for short runtimes, and it would make little sense to consider longer runtimes given that convergence was already achieved with smaller numbers of neighbors. Second, for the SPDE approach, we frequently encounter crashes when setting the tuning parameter (max.edge) to large values, which would yield low computational times, and we therefore cannot explore the full range of runtimes for this method. Furthermore, for some approximations, such as MRA and FRK, runtimes are already long even for very coarse (and hence inaccurate) tuning parameter choices, and we thus cannot analyze these methods for small runtimes. Moreover, we set an upper limit on the runtime for every iteration in the experiments, including likelihood evaluation, estimation, and prediction, and we do not consider tuning parameter choices that exceed this limit. For simulated data sets with sample sizes $N=10,000$ and $N=100,000$, this limit is $10$ and $150$ minutes per repetition, respectively. For the real-world data, we use the following limits: $30$ minutes for the house price data set, two hours for the MODIS 2016 data set, three hours for the Laegern data set, and four hours for the MODIS 2023 data set.

Vecchia approximations are applied to the observable process $y(\cdot)$ and not the latent process $f(\cdot)$, random orderings are used as this has been found to give accurate approximations \citep{guinness2018permutation}, nearest neighbors are determined using the Euclidean distance, except for the anisotropic covariance for which a correlation-distance is used \citep{kang2023correlation}, and for prediction, the observed locations appear first in the ordering and we condition on the observed locations only. For FITC and FSA approximations, the inducing points are determined using the kmeans++ algorithm as this has been reported to give good approximations \citep{TimFabioReinhard}. For covariance tapering, the time for log-likelihood evaluation does not include the time required to perform a symbolic Cholesky decomposition for the sparsity pattern since this needs to be done only once. 

The periodic embedding approach is only applied for the data sets with locations on a lattice, i.e., the Laegern, MODIS 2016, and MODIS 2023 real-world data sets. The SPDE-based approach is not considered for the log-likelihood evaluation as, to the best of our knowledge, the corresponding software implementation does not allow for evaluating the log-likelihood at user-defined covariance parameters. Furthermore, FRK and periodic embedding do not try to approximate a GP with a Matérn covariance function, and it is thus pointless to analyze the accuracy of log-likelihood evaluations. In addition, FRK does not provide parameter estimates for the marginal variance and range parameters. 

\subsection{Software and hardware resources}\label{soft_hardware}
We use the following software implementations for conducting our experiments. For Vecchia approximations, covariance tapering, FITC approximations, and full-scale approximations, we use the GPBoost library version 1.5.1 \citep{sigrist2021gpboost}. All these methods are implemented in C++ in a comparable manner which makes their comparison as fair as possible. For the remaining techniques, we use the following R packages: GPvecchia version 0.1.6 for the multi-resolution approximation, INLA version 24.6.27 and rSPDE version 2.3.3 for the SPDE approach, except for the anisotropic Matérn model for which we use INLA version 24.12.11 and rSPDE version 2.5.1, FRK version 2.3.0 for fixed rank kriging, and npspec version 0.1.0 for periodic embedding (https://github.com/joeguinness/npspec). Parameter estimation is done with the default optimizers in the corresponding software packages. This means that differences in the estimation times of the methods are also influenced by the optimizers used. This is, arguably, not ideal, but it is not possible to use the same optimizers in all packages. However, the runtimes for log-likelihood evaluation for the simulated and real-world data sets and for prediction for the simulated data sets are not affected by this. Furthermore, the four methods (Vecchia, FITC, tapering, full-scale) implemented in the GPBoost library all use the same optimizers.

Our experiments are run on a machine with a 2.25 GHz AMD EPYC 7742 processor with eight cores and 512 GB of random-access memory (RAM). Additionally, to analyze the impact of multi-core parallelization, we repeat the analysis for the simulated data sets with $N=100,000$ and the house price data set using a single processor core; see Section \ref{single_core}. The code to reproduce our experiments and the data sets used in this article can be found in the following GitHub repository: \url{https://github.com/Filippo-Rambelli/accuracy\_runtime\_scalable\_GP\_approx}.

\section{Results}\label{results}

\subsection{Simulated data with $N=10,000$}
We first discuss the results for the simulated data with a sample of size $N=10,000$ and an effective range of $0.2$. We use figures to present the results in the main article, the results in tabular form can be found in the supplementary material. The tuning parameters used for every method are reported in Table \ref{hyp_10k} in Appendix \ref{app_tune_pars}.

In Figure \ref{lik_02}, we show the absolute mean difference between the exact and approximate log-likelihood versus the evaluation time. The left panel shows the results when the evaluation is done at the true data-generating parameters, while the right panel shows the results using ``wrong" parameters obtained by doubling the true parameters. Vecchia approximations clearly achieve the highest accuracy. The difference to the exact log-likelihood decreases quickly with respect to time, approaching small values even for relatively small numbers of neighbors. The second most accurate approximation for a given computational budget is the FITC approximation followed by the MRA and FSA. Tapering is the method with the overall lowest accuracy, achieving a high accuracy only for relatively large taper ranges. The results for larger and smaller effective ranges of $0.5$ and $0.05$ are reported in the supplementary material in Sections \if1\jabes\ref{appendix_05} and \ref{appendix_005}\else B.2 and B.3\fi, respectively. Overall, we observe similar findings. Vecchia approximations are clearly the most accurate approximations. As expected, FITC approximations are more accurate for large range parameters compared to small ones, and vice versa for covariance tapering.  
\begin{figure}[ht!]
      \centering     
      \includegraphics[width=0.7\linewidth]{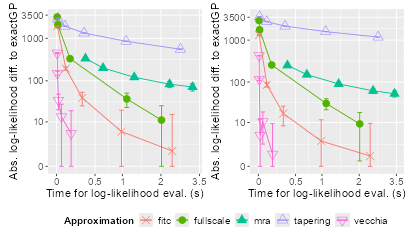} 
  \caption[N=10,000, range 0.2. Log-likelihood evaluation.]
  {Absolute mean difference between approximate and exact log-likelihood on simulated data sets with an effective range of $0.2$ and a sample size $N=10,000$. The true data-generating parameters are used on the left and ``wrong" parameters obtained by doubling the true ones are used on the right.}
  \label{lik_02}
\end{figure}

\begin{figure}[ht!]
      \centering   
      \includegraphics[width=\linewidth]{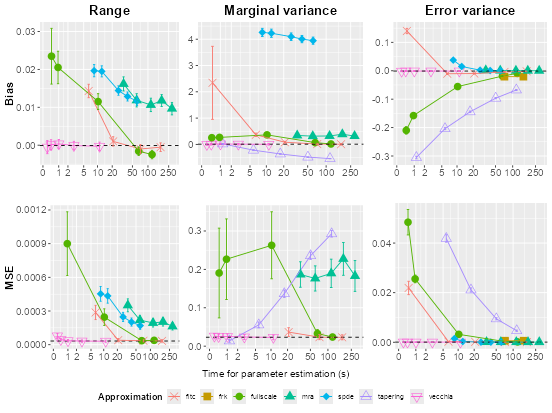} 
  \caption[N=10,000, range 0.2. Parameter estimation accuracy.]
  {Bias and MSE of the estimates for the range, marginal variance, and error term variance on simulated data sets with an effective range of $0.2$ and a sample size $N=10,000$. The dashed lines correspond to the results from the exact calculations.}
  \label{02_parest}
\end{figure}

We next present the results for parameter estimation. In Figure \ref{02_parest}, we show the bias and the MSE for the range, marginal variance, and error variance parameters versus the time required to estimate these parameters. We again find that Vecchia approximations are clearly the most accurate with the lowest bias and MSE for most computational times yielding very small biases and MSEs even for small estimation times. The range and the variance parameters are overestimated by the majority of the other methods. For tapering and FITC, some results with extreme values are excluded from the plots for better visibility of the other results. The full results in tabular form can be found in the supplementary material in Section \if1\jabes\ref{appendix_02}\else B.1\fi. 

\begin{figure}[ht!]
      \centering   
      \includegraphics[width=\linewidth]{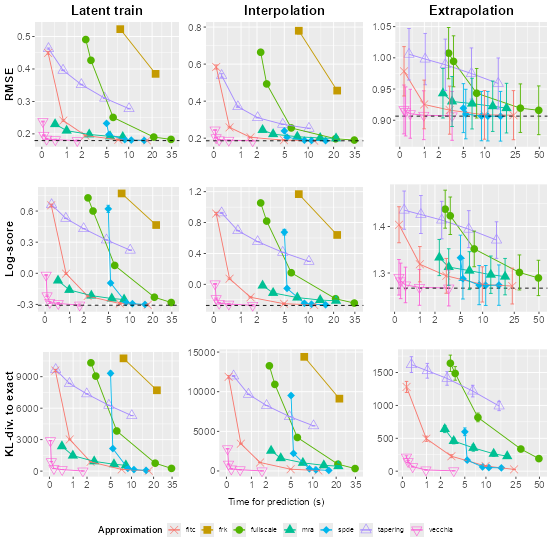} 
  \caption[N=10,000, range 0.2. Prediction accuracy.]
  {Average RMSE, log-score, and KL divergence between exact and approximate predictions on simulated data sets with an effective range of $0.2$ and a sample size $N=10,000$. We distinguish between predicting the latent process at (i) training locations (left), (ii) interpolation test locations (center), and (iii) extrapolation locations (right). The dashed lines correspond to the results from the exact calculations.}
  \label{predacc_02}
\end{figure}

In Figure \ref{predacc_02}, we report the prediction accuracy results including RMSEs, log-scores, and KL divergences to exact calculations for predicting the latent process at the training, “interpolation”, and “extrapolation” test locations. We observe that Vecchia approximations are clearly the most accurate compared to all other methods regarding all evaluation metrics. Overall, the MRA and FITC approximation are the second most accurate methods in this experiment, followed by the SPDE approximation. The FITC, full-scale, MRA, and SPDE methods eventually achieve similar prediction accuracies as Vecchia approximations but for much larger computational times. Tapering and FRK do not attain this level of accuracy for all computational times considered. For the “extrapolation” test set, we observe that the variability is much higher for all methods. Note that for FRK, the “extrapolation” metrics are not included in the figure as they are very large, i.e., very inaccurate. All results in tabular form can be found in the supplementary material in Section \if1\jabes\ref{appendix_02}\else B.1\fi. In the supplementary material in Section \if1\jabes\ref{appendix_02}\else B.1\fi, we additionally report the RMSE comparing the approximate predictive means and variances to the exact values. The results are qualitatively very similar to the ones mentioned above.

The results for experiments conducted on simulated data sets with effective ranges of $0.5$ and $0.05$ are reported in the supplementary material in Sections \if1\jabes\ref{appendix_05} and \ref{appendix_005}\else B.2 and B.3\fi, respectively. In both scenarios, Vecchia approximations yield the most accurate results for a given amount of computational time. As expected, all techniques except tapering generally are more accurate for the larger range parameter. The FITC approximation, in particular, has an accuracy-runtime trade-off almost equal to that of the Vecchia approximation. As expected, covariance tapering is more accurate for small range parameters and less accurate for large ones.

\subsection{Simulated data with $N=100,000$}\label{res_sim_large}
Next, we discuss the results for the simulated data with a sample of size $N=100,000$. The tuning parameters used for every method are reported in Table \ref{par_100k} in Appendix \ref{app_tune_pars}. Figure \ref{lik_02_100k} in Appendix \ref{app_figs_tabs} displays the average approximate negative log-likelihood versus the evaluation time for an effective range of $0.2$. The results are very similar to the ones of the smaller sample size of $N=10,000$. Vecchia approximations again yield the best accuracy-runtime trade-off, with the maximal accuracy being achieved for relatively small runtimes.

Concerning parameter estimation accuracy, Figure \ref{parest_02_100k} in Appendix \ref{app_figs_tabs} reports the bias and MSE versus the time required to estimate these parameters. We again find similar results as for the smaller sample size. Vecchia approximations are clearly the most accurate with the lowest bias and MSE for most computational times and very small biases and MSEs even for small numbers of neighbors. Note that for tapering, some results with extreme values are again excluded from the plots for better visibility of the other results. The full results in tabular form can be found in the supplementary material in Section \if1\jabes\ref{appendix_100000_02}\else B.4\fi.

Figure \ref{predacc_02_100k} in Appendix \ref{app_figs_tabs} shows the RMSE and log-score to analyze the accuracy of predictive distributions. We again find that Vecchia approximations are the most accurate already for small computational times. The SPDE-based approach results in the second most accurate predictions, followed by the MRA and the FITC approximation. FRK and tapering generally generate the least accurate predictions. For FRK, the results have been excluded again from the extrapolation plots due to extreme and very inaccurate values; they can be found in the supplementary material in Section \if1\jabes\ref{appendix_100000_02} \else B.4 \fi in tabular form. The results for the experiments using simulated data with $N=100,000$ and effective ranges of $0.5$ and $0.05$ are reported in the supplementary material in Sections \if1\jabes\ref{appendix_05_100k} and \ref{appendix_005_100k}\else B.5 and B.6\fi, respectively. Overall, the results are qualitatively very similar to the above reported results.

\if1\blind{
\FloatBarrier
}\fi
\subsection{House price data set} \label{house}
\begin{figure}[ht!]
      \centering 
      \includegraphics[width=0.7\linewidth]{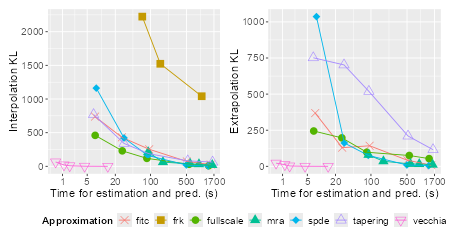} 
  \caption[House price data set. Comparison to exact calculations via KL divergence between predictive distributions.]
  {KL divergence between exact and approximate predictions on the “interpolation” test (left) and “extrapolation” test (right) sets on the house price data set.}
  \label{kl_house}
\end{figure}

In this subsection, we present the results for the house price data set. The tuning parameters used on this data set are reported in Table \ref{hyp_house} in Appendix \ref{app_tune_pars}. In Figure \ref{predacc_house}, we show the prediction accuracy results on the “interpolation” and “extrapolation” test sets when using a zero prior mean function. Specifically, we plot the RMSE, the log-score and the CRPS versus the prediction time. In addition, we also report the results when doing predictions with exact calculations. To further compare the differences between exact and approximate calculations, Figure \ref{kl_house} shows the KL divergence between the exact and approximate predictive distributions versus the runtime on the two test sets. We observe that Vecchia approximations are again the most accurate approximations having almost equal prediction accuracy as exact calculations on both the interpolation and the extrapolation test sets as well as very small KL divergences even for small runtimes. For the interpolation test set, full-scale approximations are the second most accurate. For long runtimes, the prediction accuracy measures are similar across all approximations except for FRK. Note that for the extrapolation test data, the FITC and full-scale approximations with a small number of inducing points lead to more accurate predictions than an exact Gaussian process model. As above, some very high values for FRK were removed from the plots, but can be found in tabular form in the supplementary material in Section \if1\jabes\ref{appendix_house}\else B.12\fi. In Figure \ref{lik_house} in Appendix \ref{app_figs_tabs}, we additionally report the negative log-likelihood values versus the wall-clock time. Except for tapering, all approximations yield a similar log-likelihood for large runtimes. Vecchia approximations followed by full-scale approximations are the most accurate for a given computational budget. Additional comparisons to exact calculations are presented in the supplementary material in Section \if1\jabes\ref{appendix_house}\else B.12\fi, where we report the RMSE between exact and approximate predictive means and variances.
\begin{figure}[ht!]
      \centering  
      \includegraphics[width=\linewidth]{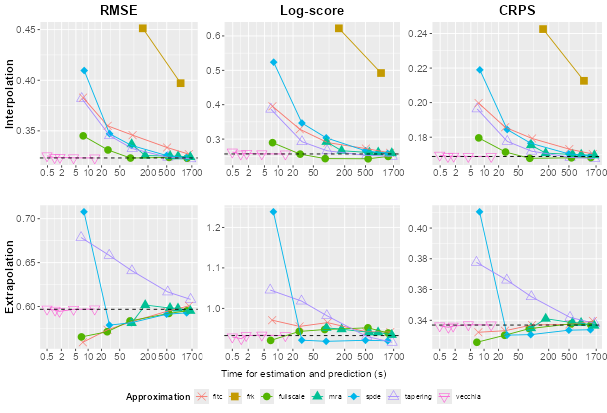} 
  \caption[House price data set. Prediction accuracy.]
  {RMSE, log-score and CRPS on the “interpolation” and “extrapolation” test sets of the house price data set. The dashed lines correspond to the results from the exact calculations.}
  \label{predacc_house}
\end{figure}

The results when including a non-zero prior mean function with the covariates described in Section \ref{house_price_data} are reported in the supplementary material in Section \if1\jabes\ref{appendix_house_with_covs}\else B.13\fi. Overall, we find qualitatively very similar results as when not including a linear predictor. In particular, the Vecchia approximations clearly have the best accuracy-runtime trade-off.

\subsection{Laegern data set}
In the following, we report the results for the Laegern data set. The tuning parameters used are listed in Table \ref{hyp_lae} in Appendix \ref{app_tune_pars}. Figure \ref{predacc_laegern} shows the RMSE, the log-score, and the CRPS computed on the interpolation and extrapolation test sets. We again find that Vecchia approximations are the most accurate even when few neighbors are used. No other approximation reaches the accuracy of Vecchia approximations on the interpolation test set, not even for long runtimes. For the extrapolation test data set, the accuracy of the Vecchia approximation, the full-scale approximation, SPDE and the MRA are similar for large runtimes. Overall, FRK and FITC approximations are the least accurate approximations with poor scores even for large runtimes. Note that some very bad results for FITC and FRK are not shown in Figure \ref{predacc_laegern} for better visibility of the other results, but they can be found in the supplementary material in Section \if1\jabes\ref{appendix_laeggern}\else B.14\fi. Unfortunately, it was not possible to include results from periodic embedding as its run time exceeds our 3-hour limit. In addition, Figure \ref{lae_lik} in Appendix \ref{app_figs_tabs} shows the negative log-likelihood versus the runtime. Vecchia approximations clearly converge the fastest and achieve the lowest negative log-likelihood.
\begin{figure}[ht!]
      \centering  
      \includegraphics[width=\linewidth]{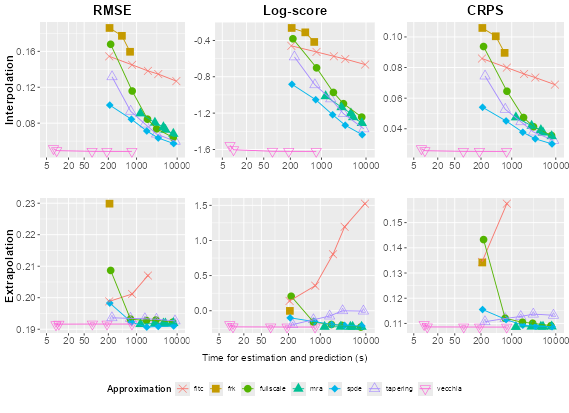} 
  \caption[Laegern data set. Prediction accuracy]
  {RMSE, log-score, and CRPS on the Laegern data set.}
  \label{predacc_laegern}
\end{figure}

\subsection{MODIS 2016 data set}
\begin{figure}[ht!]
      \centering 
      \includegraphics[width=\linewidth]{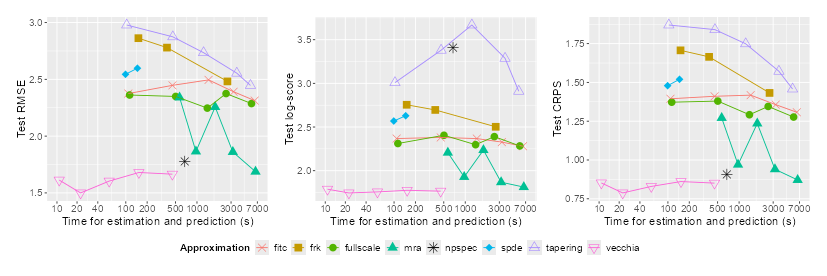} 
  \caption[MODIS 2016 data set. Prediction on the test set.]
  {Test RMSE, log-score, and CRPS on the MODIS 2016 data set.}
  \label{modis16_test}
\end{figure}

\begin{figure}[ht!]
      \centering 
      \includegraphics[width=0.7\linewidth]{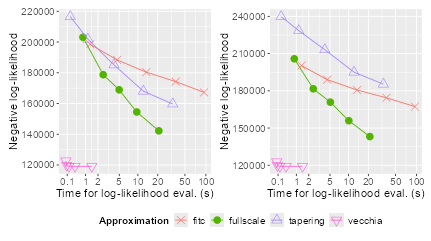} 
  \caption[MODIS 2016 data set. Log-likelihood evaluation.]
  {Negative log-likelihood on the MODIS 2016 data set. The parameters estimated by the Vecchia approximation with the largest considered number of neighbors are used on the left and two times these values are used on the right.}
  \label{lik_modis16}
\end{figure}

Next, we report the results for the MODIS 2016 data set. The choices of the tuning parameters for all methods are reported in Table \ref{modis16_hyp} in Appendix \ref{app_tune_pars}. Note that for the SPDE approach, a limited range of approximation complexities is considered since the runtime exceeds a limit of two hours when the largest triangulation edge is smaller than approximately $0.36$. Figure \ref{modis16_test} presents the prediction accuracy results versus the wall-clock time. Once again, Vecchia approximations emerge as the most accurate, excelling already for short runtimes, followed by the MRA. Periodic embedding also achieves good results in terms of RMSE and CRPS, but shows a higher log-score due to very small estimated variances. In Figure \ref{lik_modis16}, we plot the negative log-likelihood versus the wall-clock time. Vecchia approximations quickly achieve much lower negative log-likelihoods compared to all other methods. Note that the negative log-likelihood values computed with the MRA have not been included in Figure \ref{lik_modis16} as they are very large, but they can be found in the supplementary material in Section \if1\jabes\ref{appendix_modis_2016} \else B.15 \fi in tabular form.

\subsection{MODIS 2023 data set}
\begin{figure}[ht!]
      \centering 
      \includegraphics[width=\linewidth]{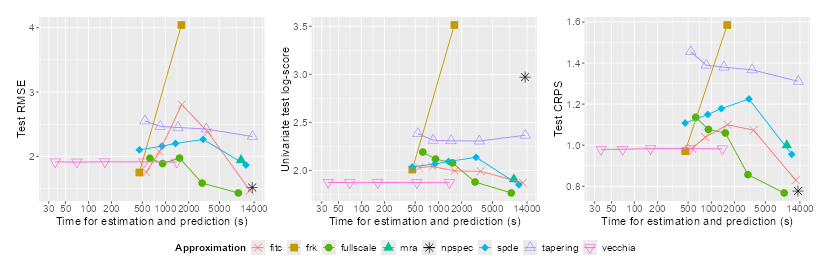} 
  \caption[MODIS 2023 data set. Prediction on the test set.]
  {Test RMSE, log-score, and CRPS on the MODIS 2023 data set.}
  \label{modis23_test}
\end{figure}

\begin{figure}[ht!]
      \centering 
      \includegraphics[width=0.7\linewidth]{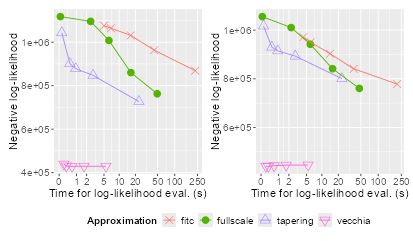} 
  \caption[MODIS 2023 data set. Log-likelihood evaluation.]
  {Negative log-likelihood on the MODIS 2023 data set. The parameters estimated by the Vecchia approximation with 240 neighbors are used on the left, and two times these values are used on the right.
}
  \label{lik_modis23}
\end{figure}

The results for the MODIS 2023 data set are presented in the following. The choices of the tuning parameters for all methods are reported in Table \ref{modis23_hyp} in Appendix \ref{app_tune_pars}. Due to the large data set size, it was only possible to implement one and two levels of complexity for the MRA and FRK, respectively, as otherwise the time limit is exceeded. Figure \ref{modis23_test} reports the prediction accuracy results. Similarly as for other data sets, the accuracy measures for Vecchia approximations quickly converge. The full-scale and FITC approximations attain better prediction accuracy measures for large runtimes compared to Vecchia approximations. This is likely a consequence of the fact that the test data contains a lot of extrapolation areas. It was already observed in Section \ref{house} that low-rank approximations can be more accurate in extrapolation than exact calculations. Figure \ref{lik_modis23} displays the negative log-likelihoods versus time. Vecchia approximations quickly converge to low negative log-likelihoods, whereas all other methods fail to converge. Note that the results of MRA are absent from this plot as they are very large, but they can be found in the supplementary material in Section \if1\jabes\ref{appendix_modis_2023} \else B.16 \fi in tabular form.

\subsection{Additional experiments using simulated data: anisotropy, other error variance and smoothness parameters, and misspecification }\label{add_exper}
We perform further experiments using simulated data with $N=100,000$ and various additional settings as described in Section \ref{sim_data}. First, we consider an anisotropic Matérn covariance function. The results for this are reported in Figures \ref{ard_ll}, \ref{ard_estim}, and \ref{ard_pred} in Appendix \ref{app_figs_tabs}, and in tabular form in the supplementary material in Section \if1\jabes\ref{appendix_100000_02_ard}\else B.7\fi. We find that Vecchia approximations clearly yield the best accuracy-runtime trade-off. For instance, Vecchia approximations have the lowest RMSEs and log-scores by a large margin for relatively short runtimes for both interpolation and extrapolation test data sets; see Figure \ref{ard_pred}. All other methods require much longer runtimes to achieve a similar prediction accuracy. Note that for FRK, it is not possible to estimate both anisotropy parameters $\rho_x$ and $\rho_y$, and we can only estimate a single range parameter and fix the anisotropy ratio at the true data-generating value, which is $4$. In addition, for the SPDE approximation, the function rspde.anisotropic2d() does not allow fixing the off-diagonal anisotropy term $h_{xy}$ exactly to zero (i.e., axis-aligned anisotropy), but we impose an extremely concentrated prior at zero (prior.hxy = list(mean = 0, precision = 1e10)). The choices of the tuning parameters for all methods are reported in Table \ref{aniso_hyp} in Appendix \ref{app_tune_pars} for this experiment.

In the supplementary material in Section \if1\jabes\ref{appendix_100000_02_n01}\else B.8\fi, we report the results when using a smaller error variance of $\sigma^2_n=0.1$, and thus a larger signal-to-noise ratio, for an effective range of $0.2$ and a smoothness of $\nu=1.5$. These results are qualitatively very similar to those reported in Section \ref{res_sim_large}. In the supplementary material in Sections \if1\jabes\ref{appendix_100000_02_s05} and \ref{appendix_100000_02_s25}\else B.9 and B.10\fi, we report the results for different smoothness parameters $\nu=0.5$ and $\nu=2.5$. We find again qualitatively similar results as for the case $\nu=1.5$. The SPDE approach is the most affected by the choice of the smoothness parameter. Specifically, the predictions of the SPDE approach are less accurate for the lower smoothness parameter $\nu=0.5$ and slightly more accurate for the larger smoothness parameter $\nu=2.5$ compared to $\nu=1.5$. Note that the considered software implementation of FRK does not include a Matérn basis function with smoothness of $\nu=2.5$, and we thus cannot include FRK for $\nu=2.5$.

Furthermore, we consider a setting with misspecification. Specifically, we simulate data using a Matérn covariance and a smoothness parameter of $\nu=2.5$ but use $\nu=0.5$ for estimation and prediction. The results of this are reported in the supplementary material in Section \if1\jabes\ref{appendix_100000_02_misspecified}\else B.11\fi. Overall, the results are qualitatively similar as in the correctly specified case. The only exception is the prediction accuracy of the SPDE approach with surprisingly low and high accuracy measures for interpolation and extrapolation, respectively. Furthermore, the RMSE and log-scores are essentially constant across all runtimes for the SPDE approach. The reasons for these findings are unclear to us.

\subsection{Results without multi-core parallelization}\label{single_core}
To analyze the impact of multi-core parallelization, we repeat the analysis for the large simulated data sets with $N=100,000$ and an isotropic Matérn covariance with $\sigma^2 = 1$, $\sigma^2_n=0.5$, $\rho=0.2/2.74$, and $\nu=1.5$ as well as the house price data set with a zero prior mean function using a single processor core. The same processor model, clock speed, amount of RAM, and tuning parameter choices are used as in the multi-core setting. The results for the simulated data sets can be found in the supplementary material in Section \if1\jabes\ref{large_sim_range_2}\else B.17\fi. In the supplementary material in Section \if1\jabes\ref{house_single_core}\else B.18\fi, we report the results for the house price data set. The results are overall similar to the results when using multi-core parallelization with Vecchia approximations yielding the best accuracy-runtime trade-off. Only concerning the prediction accuracy for longer runtimes on the simulated data, the MRA and SPDE approach show similar accuracy-runtime trade-offs. We note that there are differences among the methods regarding the speed-up of multi-core parallelization. However, given that approximations such as Vecchia, FITC, full-scale, and tapering are implemented in a different library compared to the MRA and SPDE approximation, it is difficult to say whether these speed-up differences are method-inherent or due to the specific implementations.

\section{Conclusion}
We conducted an extensive accuracy-runtime trade-off comparison between eight popular Gaussian process approximations on simulated data and multiple large-scale real-world spatial data sets. We analyze the trade-off between accuracy and runtime of the different methods concerning likelihood evaluation, parameter estimation, and prediction. Overall, Vecchia approximations clearly emerge as the most accurate approximation type when accounting for the runtime. 

There are several open questions for future research. First, we only considered the regression case. Future research can analyze the quality of Gaussian process approximations using non-Gaussian likelihoods. For this, a challenge is that there are no closed-form expressions for marginal likelihoods and predictive distributions which means that additional approximations have to be made. Other covariance functions such as non-stationary ones can also be analyzed in future research. Furthermore, while we did consider eight of the most prominent Gaussian process approximations in spatial statistics, we did not include all available approximations. Future research can extend this to additional methods such as hierarchical matrices \citep{abdulah2018parallel, geoga2020scalable, chen2023linear} and those introduced in \citet{majumder2022kryging}. The goal of this article was to compare the quality of different Gaussian process approximations as independently as possible from the specific software implementation. We are aware that this is not fully possible, but at least for four of the methods considered (Vecchia, FITC, covariance tapering, and full-scale approximations), we have used a single software library (GPBoost) where these methods are implemented in a consistent manner. For certain methods such as Vecchia approximations, there exist multiple software packages including spNNGP \citep{datta2016hierarchical}, GPvecchia \citep{katzfuss2017general}, BRISC \citep{saha2018brisc}, GpGp \citep{guinness2018permutation, guinness2019gaussian}, and meshed \citep{peruzzi2022highly}. A comparison of different software libraries for a given method such as Vecchia approximations could also be an interesting goal for future research. Finally, we have done our experiments using a relatively modest computing environment with eight CPU threads that resembles a typical laptop. Future research can analyze and compare different methods and software libraries such as GPyTorch \citep{gardner2018gpytorch}, ExaGeoStat \citep{abdulah2018parallel}, and ParallelVecchiaGP \citep{pan2024gpu} designed for high-performance GPU-based and many-core CPU computing environments.


\section*{Acknowledgments}
This research was partially supported by the Swiss Innovation Agency - Innosuisse (grants number `55463.1 IP-ICT' and `57667.1 IP-ICT').

\bibliographystyle{abbrvnat}
\bibliography{bib_GP_approx.bib}

\appendix

\phantomsection
\section*{Appendix}
\addcontentsline{toc}{section}{Appendix}
\refstepcounter{section}

\subsection{Additional figures}\label{app_figs_tabs}

\begin{figure}[ht!]
      \centering  
      \includegraphics[width=0.7\linewidth]{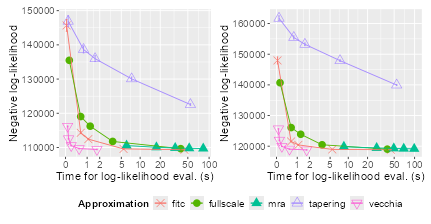} 
  \caption[N=100,000, range 0.2. Log-likelihood evaluation.]
  {Average negative log-likelihood on simulated data sets with a practical range of $0.2$ and a sample size $N=100,000$. The true data-generating parameters are used on the left and ``wrong" parameters obtained by doubling the true ones are used on the right.}
  \label{lik_02_100k}
\end{figure}

\begin{figure}[ht!] 
      \centering   
      \includegraphics[width=\linewidth]{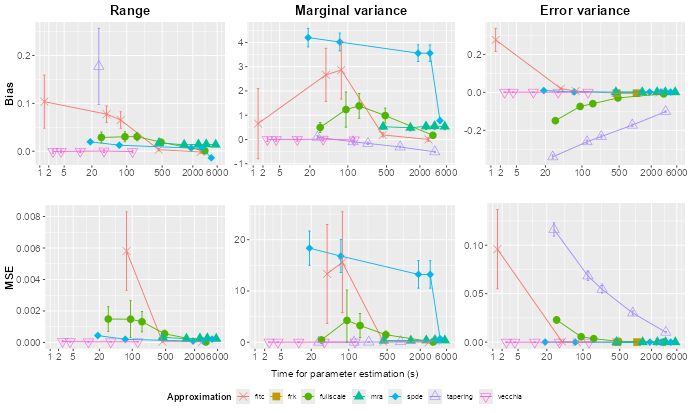} 
  \caption[N=100,000, range 0.2. Parameter estimation accuracy.]
  {Bias and MSE of the estimates for the range, marginal variance, and error term variance on simulated data sets with a practical range of $0.2$ and a sample size N=100,000,}
  \label{parest_02_100k}
\end{figure}

\begin{figure}[ht!]
      \centering   
      \includegraphics[width=\linewidth]{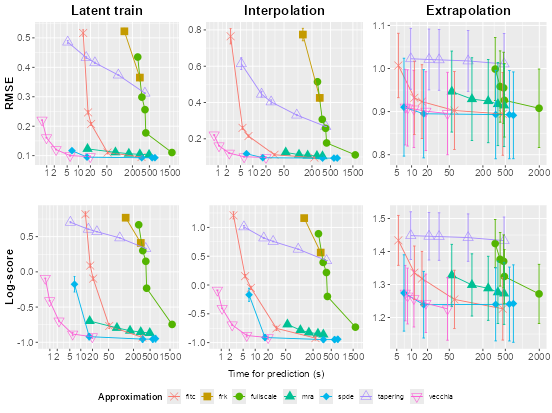} 
  \caption[N=100,000, range 0.2. Prediction accuracy.]
  {Average RMSE and log-score on simulated data sets with
a practical range of $0.2$ and a sample size N=100,000.}
  \label{predacc_02_100k}
\end{figure}

\begin{figure}[ht!]
      \centering  
      \includegraphics[width=0.7\linewidth]{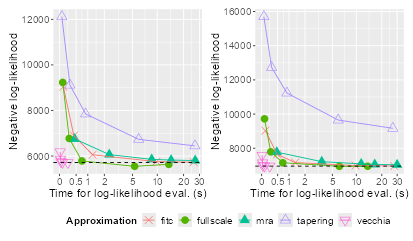} 
  \caption[House price dataset. Log-likelihood evaluation.]
  {Negative log-likelihood on the house price dataset. The parameters estimated via exact calculations are used on the left and
the pointwise doubling of them on the right. The dashed lines correspond to the exact negative log-likelihood.}
  \label{lik_house}
\end{figure}

\begin{figure}[ht!]
      \centering  
      \includegraphics[width=0.7\linewidth]{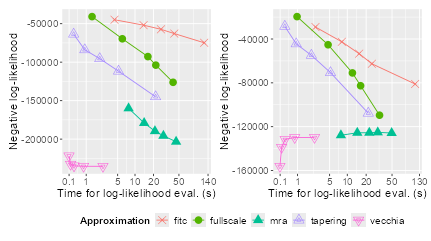} 
  \caption[Laegern data set. Log-likelihood evaluation.]
  {Negative log-likelihood on the Laegern data set. The parameters estimated by the Vecchia approximation with the largest considered number of neighbors are used on the left and two times these values are used on the right.}
  \label{lae_lik}
\end{figure}

\begin{figure}[ht!]
  \centering       
  \includegraphics[width=0.7\linewidth]{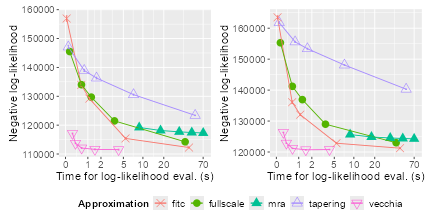} 
  \caption[$N=100,000$, anisotropic Matérn covariance function. Log-likelihood evaluation.]
  {Average negative log-likelihood on
simulated data sets with an anisotropic Matérn covariance function and a sample size of $N=100,000$. The true data-generating parameters are used on the left, and two times these values are used on the right.}
\label{ard_ll}
\end{figure}

\begin{figure}[ht!]
  \centering           
  \includegraphics[width=\linewidth]{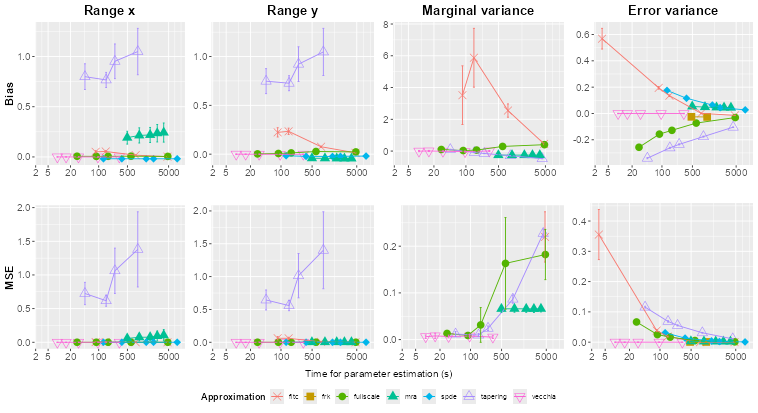} 
  \caption[$N=100,000$, anisotropic Matérn covariance function. Parameter estimation accuracy.]
  {Bias and MSE of the estimates for the ranges, marginal variance, and error term variance on simulated
data sets with an anisotropic Matérn covariance function and a sample size of $N=100,000$.}
\label{ard_estim}
\end{figure}

\begin{figure}[ht!]
  \centering           
  \includegraphics[width=\linewidth]{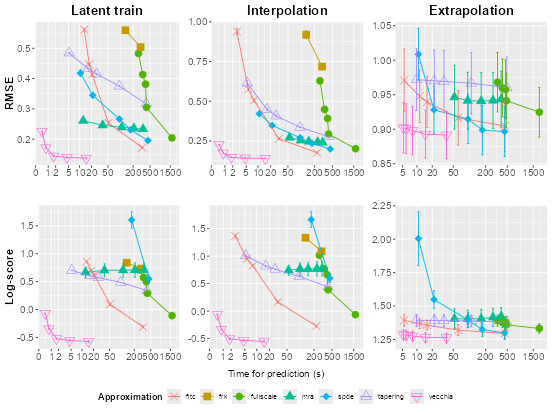} 
  \caption[$N=100,000$, anisotropic Matérn covariance function. Prediction accuracy.]
  {Average RMSE and log-score on simulated data sets with an anisotropic Matérn covariance function and a sample size of $N=100,000$.}
\label{ard_pred}
\end{figure}

\clearpage
\subsection{Tuning parameters}\label{app_tune_pars}

\begin{table}[ht!]
\centering
\resizebox{\textwidth}{!}{%
\begin{tabular}{l|l|ccccc}
  & \textbf{Tuning parameter} & \textbf{} & \textbf{} & \textbf{} & \textbf{} & \textbf{} \\
\midrule
\textbf{Vecchia} & nb. neighbours & 5 & 10 & 20 & 40 & 80\\
\hline
\textbf{Tapering} & avg. nb. non-zeros / row & 11 & 30 & 60 & 130 & 263 \\
\hline
\textbf{FITC} & nb. inducing points & 47 & 254 & 500 & 950 & 1500 \\
\hline
\multirow{2}{*}{\textbf{Full-scale}} & nb. inducing points & 10 & 24 & 120 & 300 & 450 \\
 & avg. nb. non-zeros / row & 5 & 8 & 28 & 100 & 150 \\
\hline
\textbf{FRK} & nb. resolutions & 1  & 2 & exceeds time limit &   &  \\
\hline
\textbf{MRA} & nb. knots per partition & 1 & 2 & 4 & 7 & 9 \\
\hline
\textbf{SPDE} & max.edge & 0.07 & 0.05 & 0.03 & 0.025 & 0.02 \\
\bottomrule
\end{tabular}%
}
\caption{Tuning parameters chosen for the comparison on simulated data sets with an effective range of $0.2$ and a sample size of $N=10,000$.}
\label{hyp_10k}
\end{table}

\begin{table}[ht!]
\centering
\resizebox{\textwidth}{!}{%
\begin{tabular}{l|l|cccccc}
  & \textbf{Tuning parameter} & \textbf{} & \textbf{} & \textbf{} & \textbf{} & \textbf{} & \textbf{} \\
\midrule
\textbf{Vecchia} & nb. neighbours & 5 & 10 & 20 & 40 & 80 \\
\hline
\textbf{Tapering} & avg. nb. non-zeros / row & 8 & 17 & 22 & 40 & 111\\
\hline
\textbf{FITC} & nb. inducing points & 20 & 200 & 275 & 900 & 2000\\
\hline
\multirow{2}{*}{\textbf{Full-scale}} & nb. inducing points & 26 & 91 & 122 & 250 & 650 \\
 & avg. nb. non-zeros / row & 6 & 11 & 15 & 27 & 76 \\
\hline
\textbf{FRK} & nb. resolutions & 1  & 2 & exceeds time limit &   &  &  \\
\hline
\textbf{MRA} & nb. knots per partition & 1 & 2 & 3 & 4 & 5 \\
\hline
\textbf{SPDE} & max.edge & 0.04 & 0.02 & 0.005 & 0.004 & 0.003   \\
\bottomrule
\end{tabular}%
}
\caption{Tuning parameters chosen for the comparison on simulated data sets with an effective range of $0.2$ and a sample size of $N=100,000$.}
\label{par_100k}
\end{table}

\begin{table}[ht!]
\centering
\resizebox{\textwidth}{!}{%
\begin{tabular}{l|l|ccccc}
  & \textbf{Tuning parameter} & \textbf{} & \textbf{} & \textbf{} & \textbf{} & \textbf{} \\
\midrule
\textbf{Vecchia} & nb. neighbours &  5 & 10 & 20 & 40 & 80\\
\hline
\textbf{Tapering} & avg. nb. non-zeros / row & 32 & 100 & 200 & 512 & 739  \\
\hline
\textbf{FITC} & nb. inducing points & 220 & 500 & 1000 & 2200 & 3700 \\
\hline
\multirow{2}{*}{\textbf{Full-scale}} & nb. inducing points & 106 & 252 & 444 & 1050 & 1900 \\
 & avg. nb. non-zeros / row & 17 & 36 & 71 & 256 & 420 \\
\hline
\textbf{FRK} & nb. resolutions & 1  & 2 & 3 & exceeds time limit &  \\
\hline
\textbf{MRA} & nb. knots per partition & 1 & 6 & 15 & 22 & 32 \\
\hline
\textbf{SPDE} & max.edge & 3000 & 1100 & 650 & 298 & 200 \\
\bottomrule
\end{tabular}%
}
\caption{Tuning parameters chosen for the comparison on the house price dataset.}
\label{hyp_house}
\end{table}

\begin{table}[ht!]
\centering
\resizebox{\textwidth}{!}{%
\begin{tabular}{l|l|ccccc}
  & \textbf{Tuning parameter} & \textbf{} & \textbf{} & \textbf{} & \textbf{} & \textbf{} \\
\midrule
\textbf{Vecchia} & nb. neighbours &  5 & 10 & 20 & 40 & 80\\
\hline
\textbf{Tapering} & avg. nb. non-zeros / row & 7 & 10 & 16 & 23 & 53  \\
\hline
\textbf{FITC} & nb. inducing points & 400 & 760 & 1100 & 1400 & 2200 \\
\hline
\multirow{2}{*}{\textbf{Full-scale}} & nb. inducing points & 180 & 450 & 775 & 850 & 1100 \\
 & avg. nb. non-zeros / row & 1 & 7 & 10 & 16 & 29 \\
\hline
\textbf{FRK} & nb. resolutions & 1  & 2 & 3 & exceeds time limit  &  \\
\hline
\textbf{MRA} & nb. knots per partition & 1 & 2  & 3 & 4 & 6 \\
\hline
\textbf{SPDE} & max.edge & 48  & 30 & 20 & 15 & 11 \\
\hline
\textbf{NPSPEC} & m & exceeds time limit & &  &  &  \\
\bottomrule
\end{tabular}%
}
\caption{Tuning parameters chosen for the comparison on the Laegern dataset.}
\label{hyp_lae}
\end{table}

\begin{table}[ht!]
\centering
\resizebox{\textwidth}{!}{%
\begin{tabular}{l|l|ccccc}
  & \textbf{Tuning parameter} & \textbf{} & \textbf{} & \textbf{} & \textbf{} & \textbf{} \\
\midrule
\textbf{Vecchia} & nb. neighbours &  5 & 10 & 20 & 40 & 80\\
\hline
\textbf{Tapering} & avg. nb. non-zeros / row & 8 & 16 & 34 & 64 & 114  \\
\hline
\textbf{FITC} & nb. inducing points & 380 & 800 & 1400 & 2000 & 3000 \\
\hline
\multirow{2}{*}{\textbf{Full-scale}} & nb. inducing points & 250 & 400 & 580 & 800 & 1100 \\
 & avg. nb. non-zeros / row & 2 & 10 & 17 & 25 & 48 \\
\hline
\textbf{FRK} & nb. resolutions & 1  & 2 & 3 & exceeds time limit &  \\
\hline
\textbf{MRA} & nb. knots per partition & 1 & 3 & 5 & 8 & 17 \\
\hline
\textbf{SPDE} & max.edge & 0.042 & 0.036 & exceeds time limit &  &  \\
\hline
\textbf{NPSPEC} & m & 1.2 & &  &  &  \\
\bottomrule
\end{tabular}%
}
\caption{Tuning parameters chosen for the comparison on the MODIS 2016 dataset.}
\label{modis16_hyp}
\end{table}

\begin{table}[ht!]
\centering
\resizebox{\textwidth}{!}{%
\begin{tabular}{l|l|ccccc}
  & \textbf{Tuning parameter} & \textbf{} & \textbf{} & \textbf{} & \textbf{} & \textbf{} \\
\midrule
\textbf{Vecchia} & nb. neighbours &  5 & 10 & 20 & 40 & 80\\
\hline
\textbf{Tapering} & avg. nb. non-zeros / row & 4 & 8 & 10 & 14 & 31 \\
\hline
\textbf{FITC} & nb. inducing points & 338 & 425 & 690 & 1200 & 2200 \\
\hline
\multirow{2}{*}{\textbf{Full-scale}} & nb. inducing points & 5 & 180 & 345 & 625 & 1100 \\
 & avg. nb. non-zeros / row & 1 & 1 & 2 & 8 & 16 \\
\hline
\textbf{FRK} & nb. resolutions & 1  & 2 & exceeds time limit &  &  \\
\hline
\textbf{MRA} & nb. knots per partition & 1 & exceeds time limit &  &  &  \\
\hline
\textbf{SPDE} & max.edge & 0.9  & 0.7 & 0.58 & 0.45 & 0.256 \\
\hline
\textbf{NPSPEC} & m & 1.2 & &  &  &  \\
\bottomrule
\end{tabular}%
}
\caption{Tuning parameters chosen for the comparison on the MODIS 2023 dataset.}
\label{modis23_hyp}
\end{table}

\begin{table}[ht!]
\centering
\resizebox{\textwidth}{!}{%
\begin{tabular}{l|l|cccccc}
  & \textbf{Tuning parameter} & \textbf{} & \textbf{} & \textbf{} & \textbf{} & \textbf{} & \textbf{} \\
\midrule
\textbf{Vecchia} & nb. neighbours & 5 & 10 & 20 & 40 & 80 \\
\hline
\textbf{Tapering} & avg. nb. non-zeros / row & 8 & 17 & 22 & 40 & 111\\
\hline
\textbf{FITC} & nb. inducing points & 20 & 200 & 275 & 900 & 2000\\
\hline
\multirow{2}{*}{\textbf{Full-scale}} & nb. inducing points & 26 & 91 & 122 & 250 & 650 \\
 & avg. nb. non-zeros / row & 6 & 11 & 15 & 27 & 76 \\
\hline
\textbf{FRK} & nb. resolutions & 1  & 2 & exceeds time limit &   &  &  \\
\hline
\textbf{MRA} & nb. knots per partition & 1 & 2 & 3 & 4 & 5 \\
\hline
\textbf{SPDE} & max edge & 0.05 & 0.04 & 0.03 & 0.025 & 0.02   \\
\bottomrule
\end{tabular}%
}
\caption{Tuning parameters chosen for the comparison on simulated data sets with an anisotropic Matérn covariance function and a sample size of N=100,000.}\label{aniso_hyp}
\end{table}


\clearpage

\phantomsection
\section*{Supplementary material}\label{app:complement}
\addcontentsline{toc}{section}{Supplementary material}
\refstepcounter{section}

\subsection{Simulated data with $N=10,000$ and an effective range of $0.2$}\label{appendix_02}

\subsubsection{Log-likelihood evaluation}

\begin{table}[H]
\centering
\resizebox{\textwidth}{!}{%
\begin{tabular}{llcccccccc}
\toprule 
 &  & \textbf{Vecchia} & \textbf{Tapering} & \textbf{FITC} & \textbf{Full-scale} & \textbf{FRK} & \textbf{MRA} & \textbf{SPDE} \\
\midrule
\multirow{2}{*}{\textbf{1}} 
& Value & 450 & 2948 & 1951 & 3170 &  & 337 &  \\
& (SE) & 7.45 & 27 & 19.9 & 30.9 &  & 7.63 &  \\
& Time (s) & 7.99e-03 & 0.0141 & 9.09e-03 & 7.12e-03 &  & 0.356 &  \\
\midrule
\multirow{2}{*}{\textbf{2}} 
& Value & 150 & 1922 & 195 & 2070 &  & 201 &  \\
& (SE) & 7.25 & 14.1 & 8.41 & 19 &  & 7.49 &  \\
& Time (s) & 9.82e-03 & 0.0938 & 0.103 & 0.0143 &  & 0.633 &  \\
\midrule
\multirow{2}{*}{\textbf{3}} 
& Value & 34.3 & 1330 & 39.3 & 333 &  & 121 &  \\
& (SE) & 7.18 & 8.57 & 7.28 & 8.01 &  & 7.4 &  \\
& Time (s) & 0.0207 & 0.337 & 0.311 & 0.153 &  & 1.27 &  \\
\midrule
\multirow{2}{*}{\textbf{4}} 
& Value & 13.5 & 853 & 5.53 & 37 &  & 83.3 &  \\
& (SE) & 7.19 & 5.97 & 7.22 & 7.22 &  & 7.32 &  \\
& Time (s) & 0.0533 & 1.08 & 0.981 & 1.1 &  & 2.29 &  \\
\midrule
\multirow{2}{*}{\textbf{5}} 
& Value & 4.87 & 555 & 1.29 & 11.1 &  & 71.4 &  \\
& (SE) & 7.24 & 5.52 & 7.23 & 7.2 &  & 7.24 &  \\
& Time (s) & 0.166 & 2.68 & 2.38 & 2.02 &  & 3.19 &  \\
\bottomrule
\end{tabular}%
}
\caption{Absolute mean difference between approximate and exact log-likelihood, standard error and time needed for evaluating the approximate log-likelihood on simulated data sets with an effective range of $0.2$ of and a sample size N=10,000. Log-likelihoods are evaluated at the true data-generating parameters. }
\end{table}

\begin{table}[H]
\centering
\resizebox{\textwidth}{!}{%
\begin{tabular}{llcccccccc}
\toprule 
 &  & \textbf{Vecchia} & \textbf{Tapering} & \textbf{FITC} & \textbf{Full-scale} & \textbf{FRK} & \textbf{MRA} & \textbf{SPDE} \\
\midrule
\multirow{2}{*}{\textbf{1}} 
& Value & 428 & 3597 & 1313 & 2927 &  & 251 &  \\
& (SE) & 3.96 & 13.4 & 14.9 & 21.4 &  & 4.83 &  \\
& Time (s) & 7.98e-03 & 0.0126 & 8.58e-03 & 6.92e-03 &  & 0.372 &  \\
\midrule
\multirow{2}{*}{\textbf{2}} 
& Value & 116 & 2744 & 85.6 & 1766 &  & 150 &  \\
& (SE) & 3.87 & 6.92 & 5.11 & 15.6 &  & 4.72 &  \\
& Time (s) & 9.77e-03 & 0.0949 & 0.0978 & 0.0134 &  & 0.696 &  \\
\midrule
\multirow{2}{*}{\textbf{3}} 
& Value & 4.48 & 2155 & 17.1 & 256 &  & 89.9 &  \\
& (SE) & 3.92 & 4.15 & 4.55 & 6.19 &  & 4.73 &  \\
& Time (s) & 0.0189 & 0.337 & 0.311 & 0.156 &  & 1.4 &  \\
\midrule
\multirow{2}{*}{\textbf{4}} 
& Value & 10.5 & 1593 & 2.99 & 30.2 &  & 61.5 &  \\
& (SE) & 4.14 & 2.91 & 4.48 & 4.65 &  & 4.6 &  \\
& Time (s) & 0.0498 & 1.08 & 0.982 & 1.09 &  & 2.49 &  \\
\midrule
\multirow{2}{*}{\textbf{5}} 
& Value & 0.951 & 1181 & 0.763 & 9.18 &  & 52.3 &  \\
& (SE) & 4.29 & 2.74 & 4.47 & 4.51 &  & 4.54 &  \\
& Time (s) & 0.167 & 2.69 & 2.38 & 2.02 &  & 3.41 &  \\
\bottomrule
\end{tabular}%
}
\caption{Absolute mean difference between approximate and exact log-likelihood, standard error and time needed for evaluating the approximate log-likelihood on simulated data sets with an effective range of $0.2$ of and a sample size N=10,000. Two times the true parameters are used.}
\end{table}

\subsubsection{Parameter estimation}
\begin{table}[H]
\centering
\resizebox{\textwidth}{!}{%
\begin{tabular}{llcccccccc}
\toprule 
 &  & \textbf{Vecchia} & \textbf{Tapering} & \textbf{FITC} & \textbf{Full-scale} & \textbf{FRK} & \textbf{MRA} & \textbf{SPDE} \\
\midrule
\multirow{2}{*}{\textbf{1}} 
& Bias & -5.15e-04 & 2.36 & 0.102 & 0.0235 &  & 0.0162 & 0.0197 \\
& (SE 1) & 8.81e-04 & 0.0839 & 0.0109 & 3.75e-03 &  & 9.49e-04 & 8.19e-04 \\
& MSE & 7.79e-05 & 6.27 & 0.0222 & 1.96e-03 &  & 3.51e-04 & 4.53e-04 \\
& (SE 2) & 1.09e-05 & 0.426 & 5.68e-03 & 3.93e-04 &  & 2.94e-05 & 3.36e-05 \\
& Time (s) & 0.213 & 1.18 & 0.494 & 0.458 & 71.7 & 33.6 & 8.47 \\
\midrule
\multirow{2}{*}{\textbf{2}} 
& Bias & 1.12e-04 & 3.55 & 0.0143 & 0.0205 &  & 0.0118 & 0.0195 \\
& (SE 1) & 6.77e-04 & 0.161 & 9.06e-04 & 2.19e-03 &  & 8.92e-04 & 7.43e-04 \\
& MSE & 4.58e-05 & 15.1 & 2.87e-04 & 9e-04 &  & 2.19e-04 & 4.34e-04 \\
& (SE 2) & 5.35e-06 & 1.45 & 3.24e-05 & 1.45e-04 &  & 1.99e-05 & 3.29e-05 \\
& Time (s) & 0.429 & 5.82 & 6.46 & 0.942 & 154 & 60.5 & 12.3 \\
\midrule
\multirow{2}{*}{\textbf{3}} 
& Bias & 3.39e-04 & 5.43 & 9.99e-04 & 0.0115 &  & 0.0106 & 0.0144 \\
& (SE 1) & 5.93e-04 & 0.181 & 6.25e-04 & 1.06e-03 &  & 9.03e-04 & 6.3e-04 \\
& MSE & 3.52e-05 & 32.8 & 4e-05 & 2.45e-04 &  & 1.94e-04 & 2.47e-04 \\
& (SE 2) & 5.62e-06 & 1.96 & 7.27e-06 & 3.78e-05 &  & 1.86e-05 & 2.08e-05 \\
& Time (s) & 1.01 & 17.5 & 20.9 & 10.3 &  & 114 & 26.6 \\
\midrule
\multirow{2}{*}{\textbf{4}} 
& Bias & -9.39e-05 & 5.96 & -9.99e-04 & -1.59e-03 &  & 0.0118 & 0.0129 \\
& (SE 1) & 5.67e-04 & 0.143 & 5.81e-04 & 5.74e-04 &  & 8.07e-04 & 6.07e-04 \\
& MSE & 3.22e-05 & 37.5 & 3.48e-05 & 3.55e-05 &  & 2.03e-04 & 2.02e-04 \\
& (SE 2) & 5.27e-06 & 1.88 & 4.97e-06 & 5.24e-06 &  & 1.88e-05 & 1.83e-05 \\
& Time (s) & 2.9 & 50.4 & 71.7 & 66.2 &  & 183 & 40.2 \\
\midrule
\multirow{2}{*}{\textbf{5}} 
& Bias & -2.84e-04 & 6.34 & -5.23e-04 & -2.47e-03 &  & 9.65e-03 & 0.0116 \\
& (SE 1) & 5.56e-04 & 0.201 & 5.77e-04 & 5.71e-04 &  & 8.39e-04 & 5.94e-04 \\
& MSE & 3.1e-05 & 44.2 & 3.36e-05 & 3.87e-05 &  & 1.63e-04 & 1.7e-04 \\
& (SE 2) & 4.85e-06 & 2.26 & 5.29e-06 & 5.42e-06 &  & 1.64e-05 & 1.68e-05 \\
& Time (s) & 11.1 & 116 & 175 & 120 &  & 289 & 61.3 \\
\bottomrule
\end{tabular}%
}
\caption{Bias and MSE for the GP range, standard errors and time needed for estimating covariance parameters on simulated
data sets with an effective range of $0.2$ and a sample size of N=10,000.
}
\label{table_range_02}
\end{table}

\begin{table}[H]
\centering
\resizebox{\textwidth}{!}{%
\begin{tabular}{llcccccccc}
\toprule 
 &  & \textbf{Vecchia} & \textbf{Tapering} & \textbf{FITC} & \textbf{Full-scale} & \textbf{FRK} & \textbf{MRA} & \textbf{SPDE} \\
\midrule
\multirow{2}{*}{\textbf{1}} 
& Bias & -3.72e-03 & 0.019 & 2.34 & 0.255 &  & 0.34 & 4.26 \\
& (SE 1) & 0.0162 & 0.0116 & 0.708 & 0.0355 &  & 0.0268 & 0.0788 \\
& MSE & 0.0261 & 0.0137 & 55.7 & 0.191 &  & 0.187 & 18.8 \\
& (SE 2) & 3.54e-03 & 1.82e-03 & 28.3 & 0.0597 &  & 0.0169 & 0.717 \\
& Time (s) & 0.213 & 1.18 & 0.494 & 0.458 & 71.7 & 33.6 & 8.47 \\
\midrule
\multirow{2}{*}{\textbf{2}} 
& Bias & 7.04e-03 & -0.219 & 0.365 & 0.266 &  & 0.314 & 4.23 \\
& (SE 1) & 0.0158 & 8.63e-03 & 0.0314 & 0.0395 &  & 0.0281 & 0.0795 \\
& MSE & 0.0252 & 0.0554 & 0.232 & 0.227 &  & 0.176 & 18.5 \\
& (SE 2) & 3.11e-03 & 3.84e-03 & 0.0435 & 0.0535 &  & 0.0167 & 0.725 \\
& Time (s) & 0.429 & 5.82 & 6.46 & 0.942 & 154 & 60.5 & 12.3 \\
\midrule
\multirow{2}{*}{\textbf{3}} 
& Bias & 0.0167 & -0.364 & 0.0833 & 0.363 &  & 0.321 & 4.1 \\
& (SE 1) & 0.0158 & 6.72e-03 & 0.0172 & 0.0362 &  & 0.0296 & 0.0727 \\
& MSE & 0.0252 & 0.137 & 0.0366 & 0.263 &  & 0.19 & 17.3 \\
& (SE 2) & 3.47e-03 & 4.87e-03 & 5.82e-03 & 0.0445 &  & 0.0188 & 0.621 \\
& Time (s) & 1.01 & 17.5 & 20.9 & 10.3 &  & 114 & 26.6 \\
\midrule
\multirow{2}{*}{\textbf{4}} 
& Bias & 8.79e-03 & -0.483 & 0.0118 & 0.068 &  & 0.386 & 4 \\
& (SE 1) & 0.0154 & 5.04e-03 & 0.0153 & 0.0168 &  & 0.0282 & 0.0708 \\
& MSE & 0.0239 & 0.236 & 0.0234 & 0.0328 &  & 0.227 & 16.5 \\
& (SE 2) & 3.28e-03 & 4.86e-03 & 3.34e-03 & 5.25e-03 &  & 0.0218 & 0.598 \\
& Time (s) & 2.9 & 50.4 & 71.7 & 66.2 &  & 183 & 40.2 \\
\midrule
\multirow{2}{*}{\textbf{5}} 
& Bias & 3.63e-03 & -0.54 & 6.7e-03 & 0.0159 &  & 0.318 & 3.94 \\
& (SE 1) & 0.0151 & 3.97e-03 & 0.0152 & 0.0153 &  & 0.0288 & 0.0695 \\
& MSE & 0.0229 & 0.293 & 0.0231 & 0.0238 &  & 0.183 & 16 \\
& (SE 2) & 3.15e-03 & 4.28e-03 & 3.32e-03 & 3.53e-03 &  & 0.0204 & 0.575 \\
& Time (s) & 11.1 & 116 & 175 & 120 &  & 289 & 61.3 \\
\bottomrule
\end{tabular}%
}
\caption{Bias and MSE for the GP marginal variance, standard errors and time needed for estimating covariance parameters on simulated
data sets with an effective range of $0.2$ and a sample size of N=10,000.}
\end{table}

\begin{table}[H]
\centering
\resizebox{\textwidth}{!}{%
\begin{tabular}{llcccccccc}
\toprule 
 &  & \textbf{Vecchia} & \textbf{Tapering} & \textbf{FITC} & \textbf{Full-scale} & \textbf{FRK} & \textbf{MRA} & \textbf{SPDE} \\
\midrule
\multirow{2}{*}{\textbf{1}} 
& Bias & -1.55e-03 & -0.305 & 0.14 & -0.21 & -0.0207 & 1.62e-03 & 0.0375 \\
& (SE 1) & 8.38e-04 & 2.33e-03 & 4.82e-03 & 6.62e-03 & 1.36e-03 & 9.14e-04 & 8.67e-04 \\
& MSE & 7.26e-05 & 0.0937 & 0.0219 & 0.0484 & 6.14e-04 & 8.54e-05 & 1.48e-03 \\
& (SE 2) & 1.05e-05 & 1.43e-03 & 1.38e-03 & 2.61e-03 & 6.44e-05 & 1.28e-05 & 6.66e-05 \\
& Time (s) & 0.213 & 1.18 & 0.494 & 0.458 & 71.7 & 33.6 & 8.47 \\
\midrule
\multirow{2}{*}{\textbf{2}} 
& Bias & -9.44e-04 & -0.204 & -0.0105 & -0.158 & -0.0207 & 5.42e-04 & 0.015 \\
& (SE 1) & 8.13e-04 & 1.83e-03 & 1.11e-03 & 2.51e-03 & 1.36e-03 & 8.38e-04 & 7.58e-04 \\
& MSE & 6.71e-05 & 0.0419 & 2.33e-04 & 0.0255 & 6.14e-04 & 6.98e-05 & 2.83e-04 \\
& (SE 2) & 9.47e-06 & 7.6e-04 & 2.81e-05 & 7.98e-04 & 6.44e-05 & 9.76e-06 & 2.59e-05 \\
& Time (s) & 0.429 & 5.82 & 6.46 & 0.942 & 154 & 60.5 & 12.3 \\
\midrule
\multirow{2}{*}{\textbf{3}} 
& Bias & -9.79e-04 & -0.145 & -9.86e-03 & -0.0557 &  & 2.08e-04 & 2.04e-03 \\
& (SE 1) & 8.06e-04 & 1.42e-03 & 8.33e-04 & 9.42e-04 &  & 8.03e-04 & 7.58e-04 \\
& MSE & 6.59e-05 & 0.0211 & 1.67e-04 & 3.19e-03 &  & 6.39e-05 & 6.1e-05 \\
& (SE 2) & 9.81e-06 & 4.2e-04 & 1.85e-05 & 1.06e-04 &  & 1.03e-05 & 9.84e-06 \\
& Time (s) & 1.01 & 17.5 & 20.9 & 10.3 &  & 114 & 26.6 \\
\midrule
\multirow{2}{*}{\textbf{4}} 
& Bias & -1e-03 & -0.0966 & -4.21e-03 & -0.0186 &  & 1.5e-04 & 7.82e-04 \\
& (SE 1) & 7.84e-04 & 1.1e-03 & 7.79e-04 & 7.94e-04 &  & 7.98e-04 & 7.67e-04 \\
& MSE & 6.24e-05 & 9.45e-03 & 7.84e-05 & 4.1e-04 &  & 6.31e-05 & 5.88e-05 \\
& (SE 2) & 9.41e-06 & 2.18e-04 & 1.08e-05 & 2.97e-05 &  & 1.1e-05 & 9.7e-06 \\
& Time (s) & 2.9 & 50.4 & 71.7 & 66.2 &  & 183 & 40.2 \\
\midrule
\multirow{2}{*}{\textbf{5}} 
& Bias & -9.87e-04 & -0.0676 & -2.04e-03 & -9.22e-03 &  & 1.2e-06 & 2.08e-04 \\
& (SE 1) & 7.71e-04 & 9.32e-04 & 7.77e-04 & 7.85e-04 &  & 7.69e-04 & 7.65e-04 \\
& MSE & 6.05e-05 & 4.66e-03 & 6.45e-05 & 1.47e-04 &  & 5.85e-05 & 5.79e-05 \\
& (SE 2) & 9.18e-06 & 1.29e-04 & 9.79e-06 & 1.61e-05 &  & 1e-05 & 9.31e-06 \\
& Time (s) & 11.1 & 116 & 175 & 120 &  & 289 & 61.3 \\
\bottomrule
\end{tabular}%
}
\caption{Bias and MSE for the error term variance, standard errors and time needed for estimating covariance parameters on simulated
data sets with an effective range of $0.2$ and a sample size of N=10,000.}
\end{table}

\subsubsection{Prediction}
\begin{table}[H]
\centering
\resizebox{\textwidth}{!}{%
\begin{tabular}{llcccccccc}
\toprule 
 &  & \textbf{Vecchia} & \textbf{Tapering} & \textbf{FITC} & \textbf{Full-scale} & \textbf{FRK} & \textbf{MRA} & \textbf{SPDE} \\
\midrule
\multirow{2}{*}{\textbf{1}} 
& RMSE & 0.238 & 0.463 & 0.449 & 0.491 & 0.523 & 0.23 & 0.232 \\
& (SE 1) & 6.42e-04 & 8.91e-04 & 1.9e-03 & 1.4e-03 & 2.63e-03 & 7.84e-04 & 1.15e-03 \\
& Log-score & -0.0181 & 0.662 & 0.656 & 0.727 & 0.77 & -0.0699 & 0.621 \\
& (SE 2) & 2.7e-03 & 1.38e-03 & 5.5e-03 & 3.73e-03 & 5.04e-03 & 3.51e-03 & 0.0178 \\
& Time (s) & 0.0308 & 0.201 & 0.19 & 2.36 & 7.71 & 0.441 & 4.97 \\
\midrule
\multirow{2}{*}{\textbf{2}} 
& RMSE & 0.196 & 0.396 & 0.241 & 0.427 & 0.385 & 0.21 & 0.198 \\
& (SE 1) & 7.56e-04 & 5.54e-04 & 1.1e-03 & 1.08e-03 & 1.96e-03 & 7.38e-04 & 8.68e-04 \\
& Log-score & -0.217 & 0.53 & -3.57e-03 & 0.6 & 0.465 & -0.16 & -0.094 \\
& (SE 2) & 3.91e-03 & 8.83e-04 & 3.96e-03 & 3.08e-03 & 5.03e-03 & 3.73e-03 & 8.04e-03 \\
& Time (s) & 0.0576 & 0.797 & 0.811 & 2.88 & 22.1 & 0.984 & 5.37 \\
\midrule
\multirow{2}{*}{\textbf{3}} 
& RMSE & 0.184 & 0.352 & 0.195 & 0.251 &  & 0.198 & 0.182 \\
& (SE 1) & 8.42e-04 & 4.89e-04 & 8.74e-04 & 8.37e-04 &  & 7.65e-04 & 8.66e-04 \\
& Log-score & -0.282 & 0.43 & -0.22 & 0.0757 &  & -0.215 & -0.282 \\
& (SE 2) & 4.68e-03 & 8.23e-04 & 4.12e-03 & 2.31e-03 &  & 4.01e-03 & 5.4e-03 \\
& Time (s) & 0.141 & 1.98 & 2.33 & 6.16 &  & 2.7 & 8.88 \\
\midrule
\multirow{2}{*}{\textbf{4}} 
& RMSE & 0.182 & 0.31 & 0.182 & 0.19 &  & 0.192 & 0.181 \\
& (SE 1) & 8.59e-04 & 4.92e-04 & 8.92e-04 & 7.44e-04 &  & 7.71e-04 & 8.73e-04 \\
& Log-score & -0.292 & 0.321 & -0.292 & -0.231 &  & -0.244 & -0.293 \\
& (SE 2) & 4.83e-03 & 8.74e-04 & 4.77e-03 & 3.2e-03 &  & 4.21e-03 & 5.26e-03 \\
& Time (s) & 0.451 & 4.67 & 7.27 & 21 &  & 5.59 & 10.8 \\
\midrule
\multirow{2}{*}{\textbf{5}} 
& RMSE & 0.179 & 0.278 & 0.18 & 0.183 &  & 0.19 & 0.18 \\
& (SE 1) & 8.7e-04 & 5.02e-04 & 8.78e-04 & 8.44e-04 &  & 7.88e-04 & 8.67e-04 \\
& Log-score & -0.306 & 0.223 & -0.304 & -0.282 &  & -0.253 & -0.301 \\
& (SE 2) & 4.95e-03 & 9.51e-04 & 4.9e-03 & 4.18e-03 &  & 4.32e-03 & 5.13e-03 \\
& Time (s) & 1.66 & 10.1 & 17.2 & 34.4 &  & 8.34 & 15.9 \\
\bottomrule
\end{tabular}%
}
\caption{Average RMSE and log-score on the training set, standard errors and time needed for making predictions on the training set on simulated data sets with an effective range of $0.2$ and a sample size of N=10,000.}
\end{table}

\begin{table}[H]
\centering
\resizebox{\textwidth}{!}{%
\begin{tabular}{llcccccccc}
\toprule 
 &  & \textbf{Vecchia} & \textbf{Tapering} & \textbf{FITC} & \textbf{Full-scale} & \textbf{FRK} & \textbf{MRA} & \textbf{SPDE} \\
\midrule
\multirow{2}{*}{\textbf{1}} 
& RMSE & 0.244 & 0.539 & 0.584 & 0.664 & 0.78 & 0.244 & 0.239 \\
& (SE 1) & 6.34e-04 & 4.84e-03 & 5.54e-03 & 6.82e-03 & 8.96e-03 & 8.19e-04 & 1.26e-03 \\
& Log-score & 5.68e-03 & 0.925 & 0.916 & 1.05 & 1.17 & -0.0161 & 0.676 \\
& (SE 2) & 2.53e-03 & 3.32e-03 & 8.2e-03 & 6.76e-03 & 0.0113 & 3.26e-03 & 0.0188 \\
& Time (s) & 0.0343 & 0.22 & 0.0639 & 2.29 & 7.67 & 2.46 & 5.01 \\
\midrule
\multirow{2}{*}{\textbf{2}} 
& RMSE & 0.202 & 0.37 & 0.258 & 0.493 & 0.457 & 0.222 & 0.205 \\
& (SE 1) & 8.37e-04 & 1.67e-03 & 1.55e-03 & 4.19e-03 & 3.25e-03 & 7.92e-04 & 9.24e-04 \\
& Log-score & -0.185 & 0.695 & 0.0729 & 0.819 & 0.639 & -0.111 & -0.0526 \\
& (SE 2) & 4.1e-03 & 1.1e-03 & 4.06e-03 & 4.83e-03 & 6.95e-03 & 3.61e-03 & 8.01e-03 \\
& Time (s) & 0.0619 & 0.844 & 0.493 & 2.84 & 22 & 3.52 & 5.43 \\
\midrule
\multirow{2}{*}{\textbf{3}} 
& RMSE & 0.191 & 0.311 & 0.204 & 0.255 &  & 0.208 & 0.189 \\
& (SE 1) & 9.3e-04 & 8.68e-04 & 1.02e-03 & 1.4e-03 &  & 8.29e-04 & 9.41e-04 \\
& Log-score & -0.245 & 0.553 & -0.168 & 0.15 &  & -0.171 & -0.245 \\
& (SE 2) & 4.93e-03 & 6.61e-04 & 4.04e-03 & 2.81e-03 &  & 3.97e-03 & 5.51e-03 \\
& Time (s) & 0.146 & 2.05 & 1.54 & 6.16 &  & 7.35 & 8.99 \\
\midrule
\multirow{2}{*}{\textbf{4}} 
& RMSE & 0.189 & 0.275 & 0.189 & 0.197 &  & 0.201 & 0.187 \\
& (SE 1) & 9.36e-04 & 7.02e-04 & 9.84e-04 & 8.55e-04 &  & 8.44e-04 & 9.36e-04 \\
& Log-score & -0.256 & 0.414 & -0.251 & -0.187 &  & -0.203 & -0.258 \\
& (SE 2) & 5.01e-03 & 6.55e-04 & 4.74e-03 & 3.11e-03 &  & 4.2e-03 & 5.31e-03 \\
& Time (s) & 0.463 & 4.76 & 4.95 & 21.1 &  & 13.9 & 11 \\
\midrule
\multirow{2}{*}{\textbf{5}} 
& RMSE & 0.186 & 0.253 & 0.187 & 0.19 &  & 0.199 & 0.187 \\
& (SE 1) & 9.36e-04 & 6.67e-04 & 9.51e-04 & 9.01e-04 &  & 8.6e-04 & 9.24e-04 \\
& Log-score & -0.271 & 0.298 & -0.267 & -0.243 &  & -0.213 & -0.265 \\
& (SE 2) & 5.08e-03 & 7.49e-04 & 4.92e-03 & 4.07e-03 &  & 4.37e-03 & 5.14e-03 \\
& Time (s) & 1.69 & 10.2 & 11.8 & 34.5 &  & 21.3 & 16.1 \\
\bottomrule
\end{tabular}%
}
\caption{Average RMSE and log-score on the test “interpolation” set, standard errors and time needed for making predictions on the test “interpolation” set on simulated data sets with an effective range of $0.2$ and a sample size of N=10,000.}
\end{table}

\begin{table}[H]
\centering
\resizebox{\textwidth}{!}{%
\begin{tabular}{llcccccccc}
\toprule 
 &  & \textbf{Vecchia} & \textbf{Tapering} & \textbf{FITC} & \textbf{Full-scale} & \textbf{FRK} & \textbf{MRA} & \textbf{SPDE} \\
\midrule
\multirow{2}{*}{\textbf{1}} 
& RMSE & 0.917 & 1.01 & 0.978 & 1.01 & 1.77 & 0.943 & 0.919 \\
& (SE 1) & 0.02 & 0.0211 & 0.0202 & 0.021 & 0.128 & 0.0202 & 0.02 \\
& Log-score & 1.29 & 1.44 & 1.4 & 1.44 & 3.37 & 1.33 & 1.33 \\
& (SE 2) & 0.0195 & 0.021 & 0.0198 & 0.0209 & 0.402 & 0.0199 & 0.0248 \\
& Time (s) & 0.126 & 0.297 & 0.127 & 2.96 & 7.66 & 2.38 & 5.01 \\
\midrule
\multirow{2}{*}{\textbf{2}} 
& RMSE & 0.914 & 0.998 & 0.925 & 0.994 & 10.9 & 0.93 & 0.91 \\
& (SE 1) & 0.02 & 0.021 & 0.0199 & 0.0208 & 0.843 & 0.0204 & 0.0203 \\
& Log-score & 1.28 & 1.43 & 1.32 & 1.42 & 6.45 & 1.31 & 1.29 \\
& (SE 2) & 0.0195 & 0.0208 & 0.0193 & 0.0206 & 0.46 & 0.02 & 0.0223 \\
& Time (s) & 0.179 & 1.05 & 0.988 & 3.55 & 21.9 & 3.37 & 5.43 \\
\midrule
\multirow{2}{*}{\textbf{3}} 
& RMSE & 0.91 & 0.989 & 0.916 & 0.943 &  & 0.927 & 0.907 \\
& (SE 1) & 0.02 & 0.021 & 0.0196 & 0.0203 &  & 0.0207 & 0.0202 \\
& Log-score & 1.27 & 1.41 & 1.29 & 1.35 &  & 1.31 & 1.27 \\
& (SE 2) & 0.0195 & 0.0206 & 0.0191 & 0.0196 &  & 0.0208 & 0.0215 \\
& Time (s) & 0.344 & 2.61 & 3.08 & 7.74 &  & 6.67 & 8.99 \\
\midrule
\multirow{2}{*}{\textbf{4}} 
& RMSE & 0.908 & 0.975 & 0.909 & 0.919 &  & 0.922 & 0.907 \\
& (SE 1) & 0.02 & 0.0209 & 0.0199 & 0.0198 &  & 0.0206 & 0.0203 \\
& Log-score & 1.27 & 1.39 & 1.28 & 1.3 &  & 1.3 & 1.27 \\
& (SE 2) & 0.0196 & 0.0203 & 0.0193 & 0.0192 &  & 0.0205 & 0.0219 \\
& Time (s) & 0.966 & 6.54 & 9.9 & 28.9 &  & 12.7 & 11 \\
\midrule
\multirow{2}{*}{\textbf{5}} 
& RMSE & 0.907 & 0.959 & 0.908 & 0.916 &  & 0.919 & 0.906 \\
& (SE 1) & 0.02 & 0.0209 & 0.0201 & 0.0199 &  & 0.0203 & 0.0203 \\
& Log-score & 1.27 & 1.37 & 1.27 & 1.29 &  & 1.29 & 1.27 \\
& (SE 2) & 0.0197 & 0.02 & 0.0197 & 0.0194 &  & 0.02 & 0.0221 \\
& Time (s) & 3.37 & 15 & 23.6 & 49.3 &  & 19.3 & 16.1 \\
\bottomrule
\end{tabular}%
}
\caption{Average RMSE and log-score on the test “extrapolation” set, standard errors and time needed for making predictions on the test “extrapolation” set on simulated data sets with an effective range of $0.2$ and a sample size of N=10,000.}
\label{table_extrapolation_02}
\end{table}

\subsubsection{Comparison to exact calculations}

\begin{figure}[H]

  \centering            \includegraphics[width=\linewidth]{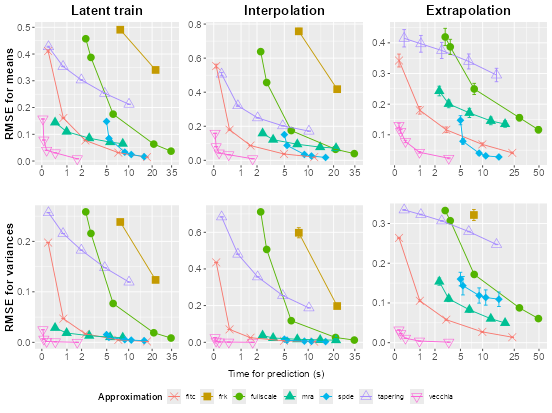} 
  \caption[$N=10,000$, range $0.2$. Comparison to exact calculations.]
  {RMSE between exact and approximate predictive means and variances on
simulated data sets with an effective range of $0.2$ and a sample size N=100,000. Every approximation makes predictions using its parameter estimates.}
  \label{compexact_02}
\end{figure}

\begin{table}[H]
\centering
\resizebox{\textwidth}{!}{%
\begin{tabular}{llcccccccc}
\toprule 
 &  & \textbf{Vecchia} & \textbf{Tapering} & \textbf{FITC} & \textbf{Full-scale} & \textbf{FRK} & \textbf{MRA} & \textbf{SPDE} \\
\midrule
\multirow{2}{*}{\textbf{1}} 
& RMSE means & 0.157 & 0.427 & 0.411 & 0.457 & 0.491 & 0.144 & 0.148 \\
& (SE 1) & 3.2e-04 & 9.19e-04 & 1.88e-03 & 1.39e-03 & 2.68e-03 & 5.6e-04 & 1.17e-03 \\
& RMSE variances & 0.0249 & 0.257 & 0.198 & 0.258 & 0.238 & 0.0286 & 0.0153 \\
& (SE 2) & 7.5e-06 & 4.43e-05 & 9.45e-05 & 6.17e-05 & 2.48e-03 & 3.78e-05 & 4.51e-05 \\
& KL & 2899 & 9683 & 9588 & 10327 & 10749 & 2369 & 9321 \\
& (SE 3) & 8.97 & 13 & 48.8 & 34.3 & 48.2 & 13.7 & 109 \\
& Time (s) & 0.0308 & 0.201 & 0.19 & 2.36 & 7.71 & 0.441 & 4.97 \\
\midrule
\multirow{2}{*}{\textbf{2}} 
& RMSE means & 0.0786 & 0.353 & 0.161 & 0.387 & 0.341 & 0.11 & 0.0845 \\
& (SE 1) & 1.84e-04 & 5.2e-04 & 1.09e-03 & 1.14e-03 & 2.04e-03 & 4.68e-04 & 4.54e-04 \\
& RMSE variances & 6.36e-03 & 0.215 & 0.0469 & 0.216 & 0.124 & 0.0185 & 9.87e-03 \\
& (SE 2) & 5.34e-06 & 3.85e-05 & 1e-04 & 6.93e-05 & 1.56e-03 & 3e-05 & 5.62e-05 \\
& KL & 899 & 8357 & 3017 & 9056 & 7707 & 1479 & 2151 \\
& (SE 3) & 3.9 & 7.16 & 24.1 & 27.4 & 46.6 & 9.74 & 17 \\
& Time (s) & 0.0576 & 0.797 & 0.811 & 2.88 & 22.1 & 0.984 & 5.37 \\
\midrule
\multirow{2}{*}{\textbf{3}} 
& RMSE means & 0.0402 & 0.303 & 0.077 & 0.175 &  & 0.0853 & 0.0331 \\
& (SE 1) & 1.95e-04 & 4.22e-04 & 4.4e-04 & 5.93e-04 &  & 4.03e-04 & 2.12e-04 \\
& RMSE variances & 1.69e-03 & 0.183 & 0.0166 & 0.0769 &  & 0.013 & 5.3e-03 \\
& (SE 2) & 9.66e-07 & 3.24e-05 & 6.05e-05 & 8.42e-05 &  & 2.18e-05 & 4.57e-05 \\
& KL & 245 & 7364 & 883 & 3825 &  & 932 & 262 \\
& (SE 3) & 2.33 & 5.93 & 6.8 & 12 &  & 6.93 & 4.43 \\
& Time (s) & 0.141 & 1.98 & 2.33 & 6.16 &  & 2.7 & 8.88 \\
\midrule
\multirow{2}{*}{\textbf{4}} 
& RMSE means & 0.0306 & 0.253 & 0.0313 & 0.0633 &  & 0.0707 & 0.0244 \\
& (SE 1) & 1.29e-04 & 3.79e-04 & 1.75e-04 & 3.27e-04 &  & 3.86e-04 & 2.49e-04 \\
& RMSE variances & 1.01e-03 & 0.148 & 4.85e-03 & 0.0189 &  & 0.0103 & 4.36e-03 \\
& (SE 2) & 4.45e-07 & 2.62e-05 & 2.96e-05 & 5.04e-05 &  & 2.05e-05 & 5.51e-05 \\
& KL & 146 & 6272 & 159 & 755 &  & 653 & 146 \\
& (SE 3) & 1.25 & 5.42 & 1.36 & 4.39 &  & 5.62 & 4.14 \\
& Time (s) & 0.451 & 4.67 & 7.27 & 21 &  & 5.59 & 10.8 \\
\midrule
\multirow{2}{*}{\textbf{5}} 
& RMSE means & 8.43e-03 & 0.212 & 0.0146 & 0.0369 &  & 0.0647 & 0.0162 \\
& (SE 1) & 2.88e-05 & 3.44e-04 & 7.61e-05 & 2.38e-04 &  & 3.92e-04 & 3.25e-04 \\
& RMSE variances & 8.36e-05 & 0.119 & 1.71e-03 & 8.48e-03 &  & 9.27e-03 & 3.26e-03 \\
& (SE 2) & 2.42e-07 & 2.06e-05 & 1.08e-05 & 3.51e-05 &  & 1.99e-05 & 7.92e-05 \\
& KL & 11.2 & 5290 & 34.6 & 257 &  & 551 & 71 \\
& (SE 3) & 0.0813 & 4.95 & 0.301 & 2.28 &  & 5.2 & 4.2 \\
& Time (s) & 1.66 & 10.1 & 17.2 & 34.4 &  & 8.34 & 15.9 \\
\bottomrule
\end{tabular}%
}
\caption{Average RMSE between means, RMSE between variances, KL divergence between exact and approximate predictions, standard errors and time needed for making predictions on the training set on simulated data sets with an effective range of $0.2$ and a sample size of N=10,000. }
\end{table}

\begin{table}[H]
\centering
\resizebox{\textwidth}{!}{%
\begin{tabular}{llcccccccc}
\toprule 
 &  & \textbf{Vecchia} & \textbf{Tapering} & \textbf{FITC} & \textbf{Full-scale} & \textbf{FRK} & \textbf{MRA} & \textbf{SPDE} \\
\midrule
\multirow{2}{*}{\textbf{1}} 
& RMSE means & 0.159 & 0.506 & 0.554 & 0.638 & 0.757 & 0.158 & 0.151 \\
& (SE 1) & 3.48e-04 & 5.15e-03 & 5.51e-03 & 7.01e-03 & 9.01e-03 & 6.72e-04 & 1.19e-03 \\
& RMSE variances & 0.0254 & 0.684 & 0.436 & 0.711 & 0.597 & 0.0369 & 0.0169 \\
& (SE 2) & 1.14e-05 & 2.33e-04 & 5.53e-04 & 1.98e-04 & 0.0141 & 8.69e-05 & 4.67e-05 \\
& KL & 2797 & 11981 & 11875 & 13267 & 14400 & 2551 & 9521 \\
& (SE 3) & 9.15 & 33.3 & 77.4 & 66.8 & 111 & 14.4 & 106 \\
& Time (s) & 0.0343 & 0.22 & 0.0639 & 2.29 & 7.67 & 2.46 & 5.01 \\
\midrule
\multirow{2}{*}{\textbf{2}} 
& RMSE means & 0.0796 & 0.32 & 0.179 & 0.457 & 0.418 & 0.121 & 0.0872 \\
& (SE 1) & 2.07e-04 & 1.81e-03 & 1.59e-03 & 4.4e-03 & 3.37e-03 & 6.02e-04 & 4.99e-04 \\
& RMSE variances & 6.57e-03 & 0.481 & 0.0703 & 0.506 & 0.198 & 0.0234 & 0.0111 \\
& (SE 2) & 6.62e-06 & 2.36e-04 & 2.38e-04 & 2.64e-04 & 3.71e-03 & 7.57e-05 & 5.89e-05 \\
& KL & 864 & 9672 & 3442 & 10922 & 9122 & 1615 & 2221 \\
& (SE 3) & 3.94 & 10.3 & 28.9 & 47.2 & 66.1 & 11.5 & 17 \\
& Time (s) & 0.0619 & 0.844 & 0.493 & 2.84 & 22 & 3.52 & 5.43 \\
\midrule
\multirow{2}{*}{\textbf{3}} 
& RMSE means & 0.0428 & 0.249 & 0.0864 & 0.175 &  & 0.0941 & 0.0343 \\
& (SE 1) & 2.49e-04 & 8.06e-04 & 6.59e-04 & 1.16e-03 &  & 5.1e-04 & 2.36e-04 \\
& RMSE variances & 1.92e-03 & 0.357 & 0.0249 & 0.118 &  & 0.0165 & 6.1e-03 \\
& (SE 2) & 2.31e-06 & 1.82e-04 & 1.15e-04 & 1.85e-04 &  & 5.43e-05 & 5.14e-05 \\
& KL & 258 & 8256 & 1067 & 4226 &  & 1025 & 273 \\
& (SE 3) & 2.85 & 5.11 & 9.02 & 16.3 &  & 7.97 & 4.48 \\
& Time (s) & 0.146 & 2.05 & 1.54 & 6.16 &  & 7.35 & 8.99 \\
\midrule
\multirow{2}{*}{\textbf{4}} 
& RMSE means & 0.0325 & 0.203 & 0.037 & 0.0645 &  & 0.0773 & 0.0251 \\
& (SE 1) & 1.48e-04 & 4.9e-04 & 2.07e-04 & 4.32e-04 &  & 4.43e-04 & 2.6e-04 \\
& RMSE variances & 1.14e-03 & 0.255 & 8.05e-03 & 0.0261 &  & 0.0128 & 5.02e-03 \\
& (SE 2) & 1.06e-06 & 1.25e-04 & 5.63e-05 & 9.79e-05 &  & 4.94e-05 & 6.19e-05 \\
& KL & 153 & 6865 & 220 & 855 &  & 711 & 152 \\
& (SE 3) & 1.36 & 3.76 & 1.64 & 4.94 &  & 5.9 & 4.16 \\
& Time (s) & 0.463 & 4.76 & 4.95 & 21.1 &  & 13.9 & 11 \\
\midrule
\multirow{2}{*}{\textbf{5}} 
& RMSE means & 8.86e-03 & 0.171 & 0.0188 & 0.0391 &  & 0.0712 & 0.0166 \\
& (SE 1) & 3.13e-05 & 3.93e-04 & 1.18e-04 & 3.07e-04 &  & 4.02e-04 & 3.27e-04 \\
& RMSE variances & 9.24e-05 & 0.188 & 3.3e-03 & 0.012 &  & 0.0116 & 3.76e-03 \\
& (SE 2) & 3.04e-07 & 8.7e-05 & 2.21e-05 & 5.83e-05 &  & 4.85e-05 & 8.91e-05 \\
& KL & 11.6 & 5697 & 56.4 & 304 &  & 607 & 73.7 \\
& (SE 3) & 0.0804 & 3.38 & 0.465 & 2.7 &  & 4.97 & 4.37 \\
& Time (s) & 1.69 & 10.2 & 11.8 & 34.5 &  & 21.3 & 16.1 \\
\bottomrule
\end{tabular}%
}
\caption{Average RMSE between means, RMSE between variances, KL divergence between exact and approximate predictions, standard errors and time needed for making predictions on the test “interpolation” set on simulated data sets with an effective range of $0.2$ and a sample size of N=10,000.}
\end{table}

\begin{table}[H]
\centering
\resizebox{\textwidth}{!}{%
\begin{tabular}{llcccccccc}
\toprule 
 &  & \textbf{Vecchia} & \textbf{Tapering} & \textbf{FITC} & \textbf{Full-scale} & \textbf{FRK} & \textbf{MRA} & \textbf{SPDE} \\
\midrule
\multirow{2}{*}{\textbf{1}} 
& RMSE means & 0.131 & 0.415 & 0.342 & 0.419 & 1.46 & 0.243 & 0.147 \\
& (SE 1) & 3.71e-03 & 0.0144 & 0.0107 & 0.0144 & 0.146 & 8.05e-03 & 7.52e-03 \\
& RMSE variances & 0.0321 & 0.334 & 0.264 & 0.333 & 0.321 & 0.154 & 0.16 \\
& (SE 2) & 1.41e-04 & 5.14e-04 & 7.63e-04 & 5.36e-04 & 7e-03 & 4.64e-03 & 8.73e-03 \\
& KL & 214 & 1620 & 1281 & 1641 & 21613 & 641 & 599 \\
& (SE 3) & 10 & 61.4 & 43.1 & 63.3 & 4039 & 28 & 30.1 \\
& Time (s) & 0.126 & 0.297 & 0.127 & 2.96 & 7.66 & 2.38 & 5.01 \\
\midrule
\multirow{2}{*}{\textbf{2}} 
& RMSE means & 0.111 & 0.397 & 0.18 & 0.387 & 10.9 & 0.201 & 0.0788 \\
& (SE 1) & 3.11e-03 & 0.014 & 6.34e-03 & 0.0126 & 0.849 & 6.4e-03 & 4.88e-03 \\
& RMSE variances & 0.0231 & 0.323 & 0.105 & 0.307 & 25.1 & 0.11 & 0.143 \\
& (SE 2) & 1.47e-04 & 4.54e-04 & 1.14e-03 & 5.16e-04 & 2.62 & 3.63e-03 & 0.0119 \\
& KL & 146 & 1527 & 494 & 1487 & 53044 & 459 & 167 \\
& (SE 3) & 7.14 & 57.3 & 20 & 52.6 & 4664 & 22.5 & 12.1 \\
& Time (s) & 0.179 & 1.05 & 0.988 & 3.55 & 21.9 & 3.37 & 5.43 \\
\midrule
\multirow{2}{*}{\textbf{3}} 
& RMSE means & 0.0782 & 0.375 & 0.116 & 0.25 &  & 0.171 & 0.0401 \\
& (SE 1) & 1.97e-03 & 0.0134 & 4.7e-03 & 9.36e-03 &  & 7.65e-03 & 3.04e-03 \\
& RMSE variances & 0.0113 & 0.307 & 0.0579 & 0.172 &  & 0.0829 & 0.119 \\
& (SE 2) & 1.09e-04 & 4.08e-04 & 8.13e-04 & 1.19e-03 &  & 3.19e-03 & 9.76e-03 \\
& KL & 68.3 & 1410 & 227 & 817 &  & 354 & 64.1 \\
& (SE 3) & 3.21 & 52.7 & 11.5 & 33.9 &  & 25.7 & 7.11 \\
& Time (s) & 0.344 & 2.61 & 3.08 & 7.74 &  & 6.67 & 8.99 \\
\midrule
\multirow{2}{*}{\textbf{4}} 
& RMSE means & 0.043 & 0.339 & 0.0687 & 0.156 &  & 0.145 & 0.0313 \\
& (SE 1) & 1.36e-03 & 0.0125 & 2.43e-03 & 5.36e-03 &  & 6.4e-03 & 2.7e-03 \\
& RMSE variances & 4.15e-03 & 0.28 & 0.0272 & 0.0877 &  & 0.0604 & 0.114 \\
& (SE 2) & 3.91e-05 & 3.5e-04 & 5.81e-04 & 1.01e-03 &  & 2.16e-03 & 9.77e-03 \\
& KL & 18.9 & 1218 & 78.3 & 336 &  & 264 & 54.4 \\
& (SE 3) & 1.22 & 45.2 & 4.44 & 13.8 &  & 19 & 7.15 \\
& Time (s) & 0.966 & 6.54 & 9.9 & 28.9 &  & 12.7 & 11 \\
\midrule
\multirow{2}{*}{\textbf{5}} 
& RMSE means & 0.0234 & 0.295 & 0.0411 & 0.117 &  & 0.135 & 0.0277 \\
& (SE 1) & 8.41e-04 & 0.0112 & 1.25e-03 & 5.18e-03 &  & 5.3e-03 & 2.69e-03 \\
& RMSE variances & 1.4e-03 & 0.247 & 0.0138 & 0.0607 &  & 0.0497 & 0.109 \\
& (SE 2) & 1.42e-05 & 2.9e-04 & 2.76e-04 & 8.7e-04 &  & 1.84e-03 & 9.28e-03 \\
& KL & 5.16 & 991 & 27.1 & 191 &  & 228 & 50.1 \\
& (SE 3) & 0.43 & 36.3 & 1.27 & 11.3 &  & 13.6 & 6.99 \\
& Time (s) & 3.37 & 15 & 23.6 & 49.3 &  & 19.3 & 16.1 \\
\bottomrule
\end{tabular}%
}
\caption{Average RMSE between means, RMSE between variances, KL divergence between exact and approximate predictions, standard errors and time needed for making predictions on the test “extrapolation” set on simulated data sets with an effective range of $0.2$ and a sample size of N=10,000. 
}
\label{table_kl_extra_02}
\end{table}

\newpage
\subsection{Simulated data with $N=10,000$ and an effective range of $0.5$}\label{appendix_05}
\begin{table}[H]
\centering
\resizebox{\textwidth}{!}{%
\begin{tabular}{l|l|ccccc}
  & \textbf{Tuning parameter} & \textbf{} & \textbf{} & \textbf{} & \textbf{} & \textbf{} \\
\midrule
\textbf{Vecchia} & nb. neighbours & 5 & 10 & 20 & 40 & 80 \\
\hline
\textbf{Tapering} & num. non-zero entries & 11 & 30 & 60 & 130 & 263 \\
\hline
\textbf{FITC} & num. inducing points & 47 & 254 & 500 & 950 & 1500 \\
\hline
\multirow{2}{*}{\textbf{Full-scale}} & num. inducing points & 10 & 24 & 120 & 300 & 450 \\
 & num. non-zero entries & 5 & 8 & 28 & 100 & 150 \\
\hline
\textbf{FRK} & num. resolutions & 1  & 2 & exceeds time limit &   &  \\
\hline
\textbf{MRA} & num. knots per partition & 1 & 2 & 4 & 7 & 9 \\
\hline
\textbf{SPDE} & max edge & 0.0625 & 0.05 & 0.03 & 0.025 & 0.02 \\
\bottomrule
\end{tabular}%
}
\caption{Tuning parameters chosen for the comparison on simulated data sets with an effective range of $0.5$ and a sample size of $N=10,000$.}
\end{table}
\subsubsection{Log-likelihood evaluation}
\begin{figure}[H]
  \centering            \includegraphics[width=0.7\linewidth]{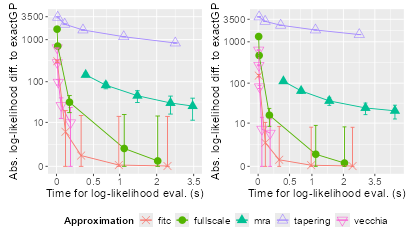} 
  \caption[$N=10,000$, range $0.5$ Log-likelihood evaluation.]
  {Absolute mean difference between approximate and exact log-likelihood on simulated data sets with an effective range $0.5$ of and a sample size $N=10,000$. The true data-generating parameters are used on the left, and two times these values are used on the right.}
\end{figure}

\begin{table}[H]
\centering
\resizebox{\textwidth}{!}{%
\begin{tabular}{llcccccccc}
\toprule 
 &  & \textbf{Vecchia} & \textbf{Tapering} & \textbf{FITC} & \textbf{Full-scale} & \textbf{FRK} & \textbf{MRA} & \textbf{SPDE} \\
\midrule
\multirow{2}{*}{\textbf{1}} 
& Value & 642 & 3402 & 308 & 1773 &  & 146 &  \\
& (SE) & 7.7 & 59.7 & 9.54 & 40.5 &  & 7.74 &  \\
& Time (s) & 8.15e-03 & 0.0142 & 9.09e-03 & 7.05e-03 &  & 0.378 &  \\
\midrule
\multirow{2}{*}{\textbf{2}} 
& Value & 284 & 2327 & 5.58 & 704 &  & 82.6 &  \\
& (SE) & 7.47 & 30.6 & 7.31 & 15.3 &  & 7.65 &  \\
& Time (s) & 9.84e-03 & 0.0948 & 0.104 & 0.0142 &  & 0.719 &  \\
\midrule
\multirow{2}{*}{\textbf{3}} 
& Value & 97.4 & 1697 & 0.799 & 32 &  & 46.1 &  \\
& (SE) & 7.25 & 17.3 & 7.25 & 7.32 &  & 7.47 &  \\
& Time (s) & 0.021 & 0.337 & 0.311 & 0.154 &  & 1.43 &  \\
\midrule
\multirow{2}{*}{\textbf{4}} 
& Value & 26.7 & 1175 & 0.0727 & 1.65 &  & 30.7 &  \\
& (SE) & 7.18 & 9.61 & 7.25 & 7.28 &  & 7.35 &  \\
& Time (s) & 0.053 & 1.09 & 0.982 & 1.1 &  & 2.49 &  \\
\midrule
\multirow{2}{*}{\textbf{5}} 
& Value & 9.72 & 834 & 0.0164 & 0.342 &  & 25.8 &  \\
& (SE) & 7.17 & 6.88 & 7.25 & 7.25 &  & 7.35 &  \\
& Time (s) & 0.166 & 2.68 & 2.38 & 2.03 &  & 3.45 &  \\
\bottomrule
\end{tabular}%
}
\caption{Absolute mean difference between approximate and exact log-likelihood, standard error and time needed for evaluating the approximate log-likelihood on simulated data sets with an effective range of $0.5$ of and a sample size N=10,000. Log-likelihoods are evaluated at the true data-generating parameters.}
\end{table}

\begin{table}[H]
\centering
\resizebox{\textwidth}{!}{%
\begin{tabular}{llcccccccc}
\toprule 
 &  & \textbf{Vecchia} & \textbf{Tapering} & \textbf{FITC} & \textbf{Full-scale} & \textbf{FRK} & \textbf{MRA} & \textbf{SPDE} \\
\midrule
\multirow{2}{*}{\textbf{1}} 
& Value & 633 & 3993 & 153 & 1343 &  & 111 &  \\
& (SE) & 3.94 & 29.7 & 5.28 & 30.9 &  & 4.87 &  \\
& Time (s) & 7.65e-03 & 0.0126 & 8.59e-03 & 6.99e-03 &  & 0.385 &  \\
\midrule
\multirow{2}{*}{\textbf{2}} 
& Value & 271 & 3119 & 2.7 & 469 &  & 64.9 &  \\
& (SE) & 3.79 & 15.2 & 3.95 & 10.7 &  & 4.56 &  \\
& Time (s) & 9.8e-03 & 0.0947 & 0.0973 & 0.0134 &  & 0.74 &  \\
\midrule
\multirow{2}{*}{\textbf{3}} 
& Value & 81.3 & 2513 & 0.436 & 16.1 &  & 36.8 &  \\
& (SE) & 3.7 & 8.54 & 3.94 & 4 &  & 4.21 &  \\
& Time (s) & 0.0186 & 0.337 & 0.312 & 0.156 &  & 1.5 &  \\
\midrule
\multirow{2}{*}{\textbf{4}} 
& Value & 6.86 & 1931 & 0.064 & 0.954 &  & 24.5 &  \\
& (SE) & 3.73 & 4.73 & 3.94 & 3.95 &  & 4.05 &  \\
& Time (s) & 0.0496 & 1.09 & 0.981 & 1.1 &  & 2.99 &  \\
\midrule
\multirow{2}{*}{\textbf{5}} 
& Value & 5.29 & 1496 & 0.0149 & 0.2 &  & 20.8 &  \\
& (SE) & 3.85 & 3.39 & 3.94 & 3.94 &  & 4.06 &  \\
& Time (s) & 0.167 & 2.68 & 2.38 & 2.04 &  & 4.83 &  \\
\midrule
\bottomrule
\end{tabular}%
}
\caption{Absolute mean difference between approximate and exact log-likelihood, standard error and time needed for evaluating the approximate log-likelihood on simulated data sets with an effective range of $0.5$ of and a sample size N=10,000. Log-likelihoods are evaluated at two-times the true data-generating parameters.}
\end{table}

\subsubsection{Parameter estimation}
\begin{figure}[H]
  \centering            \includegraphics[width=\linewidth]{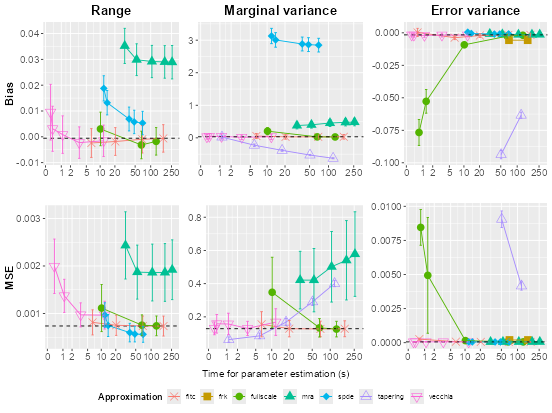} 
  \caption[$N=10,000$, range $0.5$. Parameter estimation accuracy.]
  {Bias and MSE of the estimates for the GP range, GP variance end error term variance on simulated data sets with an effective range of $0.5$ and a sample size N=100,000,}
\end{figure}

\begin{table}[H]
\centering
\resizebox{\textwidth}{!}{%
\begin{tabular}{llcccccccc}
\toprule 
 &  & \textbf{Vecchia} & \textbf{Tapering} & \textbf{FITC} & \textbf{Full-scale} & \textbf{FRK} & \textbf{MRA} & \textbf{SPDE} \\
\midrule
\multirow{2}{*}{\textbf{1}} 
& Bias & 9.54e-03 & 2.11 & 0.182 & 1.08 &  & 0.0352 & 0.0188 \\
& (SE 1) & 5.51e-03 & 0.0785 & 0.0259 & 0.62 &  & 3.47e-03 & 2.49e-03 \\
& MSE & 3.12e-03 & 5.08 & 0.1 & 39.6 &  & 2.43e-03 & 9.68e-04 \\
& (SE 2) & 1.01e-03 & 0.347 & 0.0287 & 33.6 &  & 3.61e-04 & 1.31e-04 \\
& Time (s) & 0.252 & 1.16 & 0.598 & 0.694 & 70.8 & 31.3 & 11.9 \\
\midrule
\multirow{2}{*}{\textbf{2}} 
& Bias & 3.12e-03 & 3.76 & -2.4e-03 & 0.556 &  & 0.0299 & 0.0132 \\
& (SE 1) & 4.45e-03 & 0.153 & 2.84e-03 & 0.41 &  & 3.14e-03 & 2.4e-03 \\
& MSE & 1.99e-03 & 16.5 & 8.13e-04 & 17.1 &  & 1.87e-03 & 7.47e-04 \\
& (SE 2) & 2.92e-04 & 1.37 & 1.35e-04 & 16.7 &  & 2.9e-04 & 1.18e-04 \\
& Time (s) & 0.368 & 5.95 & 6.48 & 1.28 & 156 & 53.5 & 14 \\
\midrule
\multirow{2}{*}{\textbf{3}} 
& Bias & 8.6e-04 & 5.37 & -1.82e-03 & 3.05e-03 &  & 0.0292 & 6.9e-03 \\
& (SE 1) & 3.7e-03 & 0.175 & 2.74e-03 & 3.32e-03 &  & 3.2e-03 & 2.36e-03 \\
& MSE & 1.37e-03 & 31.9 & 7.52e-04 & 1.11e-03 &  & 1.86e-03 & 5.98e-04 \\
& (SE 2) & 1.78e-04 & 2.07 & 1.08e-04 & 2.53e-04 &  & 3.05e-04 & 9.98e-05 \\
& Time (s) & 1.12 & 17.9 & 20.4 & 10.2 &  & 105 & 38.1 \\
\midrule
\multirow{2}{*}{\textbf{4}} 
& Bias & -2.24e-03 & 6.88 & -9.58e-04 & -3.17e-03 &  & 0.029 & 5.76e-03 \\
& (SE 1) & 3.11e-03 & 0.199 & 2.71e-03 & 2.73e-03 &  & 3.21e-03 & 2.32e-03 \\
& MSE & 9.71e-04 & 51.3 & 7.37e-04 & 7.54e-04 &  & 1.86e-03 & 5.67e-04 \\
& (SE 2) & 1.29e-04 & 3.01 & 1.07e-04 & 1.1e-04 &  & 3.1e-04 & 8.43e-05 \\
& Time (s) & 3.4 & 50.7 & 68.8 & 66.2 &  & 186 & 48 \\
\midrule
\multirow{2}{*}{\textbf{5}} 
& Bias & -1.55e-03 & 8.03 & -6.19e-04 & -1.76e-03 &  & 0.0289 & 5.34e-03 \\
& (SE 1) & 3.1e-03 & 0.199 & 2.71e-03 & 2.71e-03 &  & 3.31e-03 & 2.32e-03 \\
& MSE & 9.65e-04 & 68.5 & 7.35e-04 & 7.4e-04 &  & 1.92e-03 & 5.62e-04 \\
& (SE 2) & 1.49e-04 & 3.26 & 1.07e-04 & 1.06e-04 &  & 3.21e-04 & 8.54e-05 \\
& Time (s) & 11.8 & 118 & 174 & 127 &  & 257 & 69.8 \\
\bottomrule
\end{tabular}%
}
\caption{Bias and MSE for the GP range, standard errors and time needed for estimating covariance parameters on simulated data sets with an effective range of $0.5$ and a sample size of N=10,000.
}
\end{table}

\begin{table}[H]
\centering
\resizebox{\textwidth}{!}{%
\begin{tabular}{llcccccccc}
\toprule 
 &  & \textbf{Vecchia} & \textbf{Tapering} & \textbf{FITC} & \textbf{Full-scale} & \textbf{FRK} & \textbf{MRA} & \textbf{SPDE} \\
\midrule
\multirow{2}{*}{\textbf{1}} 
& Bias & 0.0323 & 0.0231 & 7.7 & 29437 &  & 0.379 & 3.13 \\
& (SE 1) & 0.036 & 0.025 & 2.47 & 28222 &  & 0.053 & 0.12 \\
& MSE & 0.131 & 0.0631 & 667 & 8.05e+10 &  & 0.421 & 11.2 \\
& (SE 2) & 0.0197 & 8.19e-03 & 444 & 8.01e+10 &  & 0.0886 & 0.88 \\
& Time (s) & 0.252 & 1.16 & 0.598 & 0.694 & 70.8 & 31.3 & 11.9 \\
\midrule
\multirow{2}{*}{\textbf{2}} 
& Bias & 0.0471 & -0.227 & 0.0556 & 3487 &  & 0.4 & 3.01 \\
& (SE 1) & 0.0395 & 0.0185 & 0.039 & 3340 &  & 0.0513 & 0.114 \\
& MSE & 0.158 & 0.0856 & 0.155 & 1.13e+09 &  & 0.421 & 10.3 \\
& (SE 2) & 0.0285 & 8.62e-03 & 0.0386 & 1.12e+09 &  & 0.0974 & 0.799 \\
& Time (s) & 0.368 & 5.95 & 6.48 & 1.28 & 156 & 53.5 & 14 \\
\midrule
\multirow{2}{*}{\textbf{3}} 
& Bias & 0.0434 & -0.384 & 0.029 & 0.205 &  & 0.454 & 2.88 \\
& (SE 1) & 0.0394 & 0.0144 & 0.0358 & 0.0553 &  & 0.0546 & 0.109 \\
& MSE & 0.157 & 0.168 & 0.129 & 0.348 &  & 0.502 & 9.48 \\
& (SE 2) & 0.0332 & 0.0108 & 0.0251 & 0.108 &  & 0.108 & 0.729 \\
& Time (s) & 1.12 & 17.9 & 20.4 & 10.2 &  & 105 & 38.1 \\
\midrule
\multirow{2}{*}{\textbf{4}} 
& Bias & 0.0162 & -0.529 & 0.0297 & 0.0296 &  & 0.476 & 2.86 \\
& (SE 1) & 0.036 & 0.0105 & 0.0355 & 0.0365 &  & 0.0563 & 0.109 \\
& MSE & 0.13 & 0.291 & 0.127 & 0.134 &  & 0.541 & 9.36 \\
& (SE 2) & 0.022 & 0.011 & 0.0241 & 0.0282 &  & 0.122 & 0.72 \\
& Time (s) & 3.4 & 50.7 & 68.8 & 66.2 &  & 186 & 48 \\
\midrule
\multirow{2}{*}{\textbf{5}} 
& Bias & 0.0336 & -0.628 & 0.0318 & 0.0269 &  & 0.484 & 2.85 \\
& (SE 1) & 0.0409 & 7.86e-03 & 0.0357 & 0.0354 &  & 0.0589 & 0.108 \\
& MSE & 0.169 & 0.401 & 0.128 & 0.126 &  & 0.578 & 9.26 \\
& (SE 2) & 0.0409 & 9.76e-03 & 0.0245 & 0.024 &  & 0.13 & 0.704 \\
& Time (s) & 11.8 & 118 & 174 & 127 &  & 257 & 69.8 \\
\bottomrule
\end{tabular}%
}
\caption{Bias and MSE for the GP marginal variance, standard errors and time needed for estimating covariance parameters on simulated data sets with an effective range of $0.5$ and a sample size of N=10,000.}
\end{table}

\begin{table}[H]
\centering
\resizebox{\textwidth}{!}{%
\begin{tabular}{llcccccccc}
\toprule 
 &  & \textbf{Vecchia} & \textbf{Tapering} & \textbf{FITC} & \textbf{Full-scale} & \textbf{FRK} & \textbf{MRA} & \textbf{SPDE} \\
\midrule
\multirow{2}{*}{\textbf{1}} 
& Bias & -1.69e-03 & -0.305 & 3.04e-04 & -0.0767 & -5.67e-03 & -8.11e-04 & 4.81e-04 \\
& (SE 1) & 7.38e-04 & 4.44e-03 & 1.62e-03 & 5.07e-03 & 1.29e-03 & 7.47e-04 & 7.38e-04 \\
& MSE & 5.74e-05 & 0.0951 & 2.61e-04 & 8.45e-03 & 1.98e-04 & 5.59e-05 & 5.42e-05 \\
& (SE 2) & 8.2e-06 & 2.68e-03 & 3.62e-05 & 6.68e-04 & 2.51e-05 & 8.22e-06 & 8.2e-06 \\
& Time (s) & 0.252 & 1.16 & 0.598 & 0.694 & 70.8 & 31.3 & 11.9 \\
\midrule
\multirow{2}{*}{\textbf{2}} 
& Bias & -1.77e-03 & -0.203 & -3.74e-03 & -0.0527 & -5.67e-03 & -1.18e-03 & -4.86e-04 \\
& (SE 1) & 7.79e-04 & 3.34e-03 & 7.59e-04 & 4.65e-03 & 1.29e-03 & 7.7e-04 & 7.47e-04 \\
& MSE & 6.38e-05 & 0.0423 & 7.16e-05 & 4.94e-03 & 1.98e-04 & 6.01e-05 & 5.54e-05 \\
& (SE 2) & 9.29e-06 & 1.36e-03 & 9.76e-06 & 2.17e-03 & 2.51e-05 & 8.42e-06 & 8.17e-06 \\
& Time (s) & 0.368 & 5.95 & 6.48 & 1.28 & 156 & 53.5 & 14 \\
\midrule
\multirow{2}{*}{\textbf{3}} 
& Bias & -1.95e-03 & -0.143 & -2.07e-03 & -9.22e-03 &  & -1.38e-03 & -9.54e-04 \\
& (SE 1) & 7.58e-04 & 2.43e-03 & 7.37e-04 & 8.12e-04 &  & 7.61e-04 & 7.4e-04 \\
& MSE & 6.12e-05 & 0.021 & 5.86e-05 & 1.51e-04 &  & 5.93e-05 & 5.52e-05 \\
& (SE 2) & 8.85e-06 & 6.99e-04 & 8.39e-06 & 1.8e-05 &  & 8.7e-06 & 8.22e-06 \\
& Time (s) & 1.12 & 17.9 & 20.4 & 10.2 &  & 105 & 38.1 \\
\midrule
\multirow{2}{*}{\textbf{4}} 
& Bias & -1.75e-03 & -0.0937 & -1.68e-03 & -2.59e-03 &  & -1.36e-03 & -8.98e-04 \\
& (SE 1) & 7.49e-04 & 1.64e-03 & 7.37e-04 & 7.46e-04 &  & 7.61e-04 & 7.4e-04 \\
& MSE & 5.92e-05 & 9.05e-03 & 5.71e-05 & 6.23e-05 &  & 5.91e-05 & 5.5e-05 \\
& (SE 2) & 8.73e-06 & 3.11e-04 & 8.3e-06 & 8.85e-06 &  & 8.87e-06 & 8.1e-06 \\
& Time (s) & 3.4 & 50.7 & 68.8 & 66.2 &  & 186 & 48 \\
\midrule
\multirow{2}{*}{\textbf{5}} 
& Bias & -1.72e-03 & -0.0634 & -1.6e-03 & -1.87e-03 &  & -1.33e-03 & -8.34e-04 \\
& (SE 1) & 7.38e-04 & 1.19e-03 & 7.37e-04 & 7.39e-04 &  & 7.77e-04 & 7.39e-04 \\
& MSE & 5.75e-05 & 4.16e-03 & 5.69e-05 & 5.81e-05 &  & 6.15e-05 & 5.48e-05 \\
& (SE 2) & 8.69e-06 & 1.55e-04 & 8.31e-06 & 8.49e-06 &  & 9.62e-06 & 8.05e-06 \\
& Time (s) & 11.8 & 118 & 174 & 127 &  & 257 & 69.8 \\
\bottomrule
\end{tabular}%
}
\caption{Bias and MSE for the error term variance, standard errors and time needed for estimating covariance parameters on simulated data sets with an effective range of $0.5$ and a sample size of N=10,000.}
\end{table}

\subsubsection{Prediction}
\begin{figure}[H]
  \centering            \includegraphics[width=\linewidth]{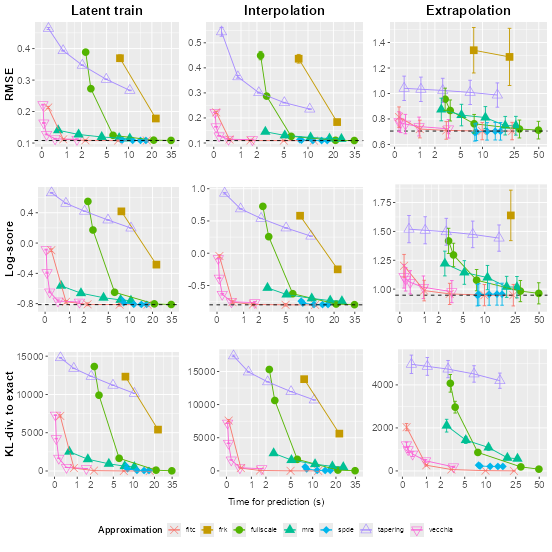} 
  \caption[$N=10,000$, range $0.5$. Prediction accuracy.]
  {Average RMSE and log-score on simulated data sets with
an effective range of $0.5$ and a sample size N=100,000. Predictions are done using the true data-generating parameters.}
\end{figure}

\begin{table}[H]
\centering
\resizebox{\textwidth}{!}{%
\begin{tabular}{llcccccccc}
\toprule 
 &  & \textbf{Vecchia} & \textbf{Tapering} & \textbf{FITC} & \textbf{Full-scale} & \textbf{FRK} & \textbf{MRA} & \textbf{SPDE} \\
\midrule
\multirow{2}{*}{\textbf{1}} 
& RMSE & 0.222 & 0.464 & 0.213 & 0.389 & 0.369 & 0.141 & 0.112 \\
& (SE 1) & 6.39e-04 & 1.62e-03 & 1.84e-03 & 3.37e-03 & 5.43e-03 & 8.5e-04 & 1.02e-03 \\
& Log-score & -0.086 & 0.664 & -0.0948 & 0.55 & 0.419 & -0.564 & -0.757 \\
& (SE 2) & 2.91e-03 & 2.53e-03 & 9.19e-03 & 0.013 & 0.0148 & 6.41e-03 & 0.0115 \\
& Time (s) & 0.0307 & 0.201 & 0.193 & 2.36 & 7.68 & 0.569 & 7.89 \\
\midrule
\multirow{2}{*}{\textbf{2}} 
& RMSE & 0.164 & 0.394 & 0.112 & 0.273 & 0.178 & 0.127 & 0.11 \\
& (SE 1) & 6.02e-04 & 6.8e-04 & 1.09e-03 & 2.03e-03 & 1.82e-03 & 8.44e-04 & 1.01e-03 \\
& Log-score & -0.391 & 0.525 & -0.776 & 0.172 & -0.285 & -0.662 & -0.787 \\
& (SE 2) & 3.69e-03 & 1.09e-03 & 9.2e-03 & 7.33e-03 & 0.0109 & 7.08e-03 & 0.0107 \\
& Time (s) & 0.0575 & 0.803 & 0.811 & 2.88 & 22.4 & 1.77 & 8.25 \\
\midrule
\multirow{2}{*}{\textbf{3}} 
& RMSE & 0.127 & 0.347 & 0.109 & 0.125 &  & 0.119 & 0.108 \\
& (SE 1) & 6.67e-04 & 5.3e-04 & 1.01e-03 & 1.18e-03 &  & 9e-04 & 1e-03 \\
& Log-score & -0.646 & 0.421 & -0.809 & -0.65 &  & -0.722 & -0.809 \\
& (SE 2) & 5.31e-03 & 9.1e-04 & 9.37e-03 & 8.11e-03 &  & 7.96e-03 & 0.0101 \\
& Time (s) & 0.14 & 2.06 & 2.33 & 6.17 &  & 4.24 & 11.3 \\
\midrule
\multirow{2}{*}{\textbf{4}} 
& RMSE & 0.113 & 0.302 & 0.108 & 0.109 &  & 0.116 & 0.108 \\
& (SE 1) & 8.23e-04 & 5.15e-04 & 9.78e-04 & 1.01e-03 &  & 9.28e-04 & 1e-03 \\
& Log-score & -0.765 & 0.305 & -0.813 & -0.803 &  & -0.751 & -0.809 \\
& (SE 2) & 7.48e-03 & 9.28e-04 & 9.3e-03 & 9.22e-03 &  & 8.39e-03 & 0.0101 \\
& Time (s) & 0.442 & 4.94 & 7.27 & 21.1 &  & 7.49 & 16.8 \\
\midrule
\multirow{2}{*}{\textbf{5}} 
& RMSE & 0.112 & 0.267 & 0.108 & 0.108 &  & 0.114 & 0.108 \\
& (SE 1) & 8.57e-04 & 5.14e-04 & 9.73e-04 & 9.88e-04 &  & 9.25e-04 & 1.01e-03 \\
& Log-score & -0.78 & 0.198 & -0.814 & -0.811 &  & -0.761 & -0.81 \\
& (SE 2) & 7.91e-03 & 9.79e-04 & 9.3e-03 & 9.33e-03 &  & 8.42e-03 & 0.0102 \\
& Time (s) & 1.64 & 10.4 & 17.2 & 34.7 &  & 10.4 & 14.3 \\
\bottomrule
\end{tabular}%
}
\caption{Average RMSE and log-score on the training set, standard errors and time needed for making predictions on the training set on simulated data sets with an effective
range of $0.5$ and a sample size of N=10,000.
}
\end{table}

\begin{table}[H]
\centering
\resizebox{\textwidth}{!}{%
\begin{tabular}{llcccccccc}
\toprule 
 &  & \textbf{Vecchia} & \textbf{Tapering} & \textbf{FITC} & \textbf{Full-scale} & \textbf{FRK} & \textbf{MRA} & \textbf{SPDE} \\
\midrule
\multirow{2}{*}{\textbf{1}} 
& RMSE & 0.223 & 0.544 & 0.222 & 0.449 & 0.436 & 0.145 & 0.113 \\
& (SE 1) & 6.13e-04 & 8.9e-03 & 2.38e-03 & 8.58e-03 & 8.86e-03 & 9.15e-04 & 1.05e-03 \\
& Log-score & -0.0834 & 0.93 & -0.0432 & 0.726 & 0.581 & -0.537 & -0.743 \\
& (SE 2) & 2.75e-03 & 6.27e-03 & 9.83e-03 & 0.0176 & 0.0203 & 6.74e-03 & 0.0116 \\
& Time (s) & 0.0342 & 0.219 & 0.0639 & 2.29 & 7.68 & 2.72 & 7.93 \\
\midrule
\multirow{2}{*}{\textbf{2}} 
& RMSE & 0.165 & 0.364 & 0.114 & 0.287 & 0.184 & 0.13 & 0.111 \\
& (SE 1) & 5.98e-04 & 2.68e-03 & 1.15e-03 & 3.33e-03 & 1.97e-03 & 9.14e-04 & 1.04e-03 \\
& Log-score & -0.386 & 0.688 & -0.758 & 0.256 & -0.25 & -0.64 & -0.773 \\
& (SE 2) & 3.62e-03 & 1.72e-03 & 9.31e-03 & 8.63e-03 & 0.0113 & 7.41e-03 & 0.0109 \\
& Time (s) & 0.0615 & 0.846 & 0.494 & 2.84 & 22.1 & 5.09 & 8.31 \\
\midrule
\multirow{2}{*}{\textbf{3}} 
& RMSE & 0.129 & 0.301 & 0.11 & 0.127 &  & 0.122 & 0.11 \\
& (SE 1) & 7.02e-04 & 9.82e-04 & 1.05e-03 & 1.27e-03 &  & 9.69e-04 & 1.03e-03 \\
& Log-score & -0.637 & 0.539 & -0.794 & -0.628 &  & -0.701 & -0.795 \\
& (SE 2) & 5.55e-03 & 7.2e-04 & 9.52e-03 & 8.11e-03 &  & 8.29e-03 & 0.0104 \\
& Time (s) & 0.145 & 2.15 & 1.55 & 6.17 &  & 10.7 & 11.4 \\
\midrule
\multirow{2}{*}{\textbf{4}} 
& RMSE & 0.115 & 0.261 & 0.11 & 0.111 &  & 0.118 & 0.11 \\
& (SE 1) & 8.48e-04 & 6.32e-04 & 1.01e-03 & 1.05e-03 &  & 9.78e-04 & 1.03e-03 \\
& Log-score & -0.752 & 0.392 & -0.8 & -0.788 &  & -0.735 & -0.796 \\
& (SE 2) & 7.67e-03 & 5.82e-04 & 9.51e-03 & 9.39e-03 &  & 8.66e-03 & 0.0104 \\
& Time (s) & 0.454 & 5.03 & 4.95 & 21.2 &  & 18 & 16.9 \\
\midrule
\multirow{2}{*}{\textbf{5}} 
& RMSE & 0.113 & 0.235 & 0.11 & 0.11 &  & 0.116 & 0.11 \\
& (SE 1) & 8.69e-04 & 5.66e-04 & 1e-03 & 1.02e-03 &  & 9.7e-04 & 1.03e-03 \\
& Log-score & -0.766 & 0.265 & -0.8 & -0.797 &  & -0.744 & -0.796 \\
& (SE 2) & 8.01e-03 & 6.3e-04 & 9.52e-03 & 9.51e-03 &  & 8.68e-03 & 0.0105 \\
& Time (s) & 1.67 & 10.5 & 11.8 & 34.7 &  & 24.8 & 14.6 \\
\bottomrule
\end{tabular}%
}
\caption{Average RMSE and log-score on the test “interpolation” set, standard errors and time needed for making predictions on the test “interpolation” set on simulated
data sets with an effective range of $0.5$ and a sample size of N=10,000.}
\end{table}

\begin{table}[H]
\centering
\resizebox{\textwidth}{!}{%
\begin{tabular}{llcccccccc}
\toprule 
 &  & \textbf{Vecchia} & \textbf{Tapering} & \textbf{FITC} & \textbf{Full-scale} & \textbf{FRK} & \textbf{MRA} & \textbf{SPDE} \\
\midrule
\multirow{2}{*}{\textbf{1}} 
& RMSE & 0.778 & 1.04 & 0.815 & 0.955 & 1.34 & 0.874 & 0.694 \\
& (SE 1) & 0.0398 & 0.0488 & 0.0408 & 0.0445 & 0.0911 & 0.0454 & 0.0344 \\
& Log-score & 1.11 & 1.52 & 1.2 & 1.42 & 5.35 & 1.22 & 0.947 \\
& (SE 2) & 0.0519 & 0.0602 & 0.0503 & 0.0559 & 0.688 & 0.0548 & 0.0473 \\
& Time (s) & 0.125 & 0.296 & 0.127 & 2.97 & 7.65 & 2.57 & 7.93 \\
\midrule
\multirow{2}{*}{\textbf{2}} 
& RMSE & 0.77 & 1.03 & 0.721 & 0.867 & 1.29 & 0.831 & 0.699 \\
& (SE 1) & 0.0395 & 0.0488 & 0.0356 & 0.0421 & 0.115 & 0.0432 & 0.035 \\
& Log-score & 1.08 & 1.51 & 0.987 & 1.3 & 1.64 & 1.14 & 0.953 \\
& (SE 2) & 0.0519 & 0.0598 & 0.0456 & 0.0519 & 0.112 & 0.0521 & 0.0479 \\
& Time (s) & 0.179 & 1.05 & 0.988 & 3.55 & 22 & 5.18 & 8.31 \\
\midrule
\multirow{2}{*}{\textbf{3}} 
& RMSE & 0.76 & 1.02 & 0.71 & 0.762 &  & 0.811 & 0.702 \\
& (SE 1) & 0.0389 & 0.0487 & 0.0352 & 0.0375 &  & 0.0437 & 0.0349 \\
& Log-score & 1.06 & 1.5 & 0.959 & 1.08 &  & 1.1 & 0.955 \\
& (SE 2) & 0.0514 & 0.0595 & 0.0469 & 0.047 &  & 0.0538 & 0.0484 \\
& Time (s) & 0.342 & 2.71 & 3.09 & 7.75 &  & 10.9 & 11.4 \\
\midrule
\multirow{2}{*}{\textbf{4}} 
& RMSE & 0.74 & 1.01 & 0.705 & 0.721 &  & 0.747 & 0.703 \\
& (SE 1) & 0.0379 & 0.0487 & 0.0349 & 0.0359 &  & 0.0372 & 0.0352 \\
& Log-score & 1.02 & 1.47 & 0.947 & 0.983 &  & 1.02 & 0.956 \\
& (SE 2) & 0.05 & 0.0588 & 0.0469 & 0.0469 &  & 0.0465 & 0.0487 \\
& Time (s) & 0.948 & 6.84 & 9.9 & 29.1 &  & 19.4 & 16.9 \\
\midrule
\multirow{2}{*}{\textbf{5}} 
& RMSE & 0.722 & 0.988 & 0.705 & 0.712 &  & 0.747 & 0.704 \\
& (SE 1) & 0.0369 & 0.0486 & 0.035 & 0.0354 &  & 0.0376 & 0.0352 \\
& Log-score & 0.978 & 1.44 & 0.947 & 0.963 &  & 1.02 & 0.956 \\
& (SE 2) & 0.0489 & 0.058 & 0.0476 & 0.0467 &  & 0.0473 & 0.0487 \\
& Time (s) & 3.34 & 15.5 & 23.6 & 49.5 &  & 26.3 & 14.6 \\
\bottomrule
\end{tabular}%
}
\caption{Average RMSE and log-score on the test “extrapolation” set, standard errors and time needed for making predictions on the test “extrapolation” set on simulated
data sets with an effective range of $0.5$ and a sample size of N=10,000.
}
\end{table}

\subsubsection{Comparison to exact calculations}

\begin{figure}[H]
  \centering            \includegraphics[width=\linewidth]{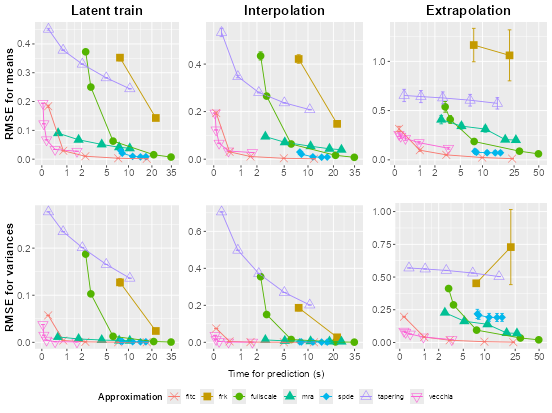} 
  \caption[$N=10,000$, range $0.5$. Comparison to exact calculations.]
  {RMSE between exact and approximate predictive means and variances on
simulated data sets with an effective range of $0.5$ and a sample size N=100,000. Every approximation makes predictions using its parameter estimates.}
\end{figure}

\begin{table}[H]
\centering
\resizebox{\textwidth}{!}{%
\begin{tabular}{llcccccccc}
\toprule 
 &  & \textbf{Vecchia} & \textbf{Tapering} & \textbf{FITC} & \textbf{Full-scale} & \textbf{FRK} & \textbf{MRA} & \textbf{SPDE} \\
\midrule
\multirow{2}{*}{\textbf{1}} 
& RMSE means & 0.194 & 0.451 & 0.184 & 0.373 & 0.352 & 0.0902 & 0.0295 \\
& (SE 1) & 5.44e-04 & 1.6e-03 & 1.8e-03 & 3.42e-03 & 5.61e-03 & 5.69e-04 & 2.17e-04 \\
& RMSE variances & 0.0376 & 0.277 & 0.0572 & 0.187 & 0.127 & 0.0114 & 2.39e-03 \\
& (SE 2) & 2.23e-06 & 5.01e-05 & 1.63e-04 & 8.82e-05 & 4.02e-03 & 2.56e-05 & 2.01e-05 \\
& KL & 7298 & 14796 & 7194 & 13648 & 12327 & 2523 & 609 \\
& (SE 3) & 21.8 & 24.1 & 74 & 133 & 146 & 22 & 7.66 \\
& Time (s) & 0.0307 & 0.201 & 0.193 & 2.36 & 7.68 & 0.569 & 7.89 \\
\midrule
\multirow{2}{*}{\textbf{2}} 
& RMSE means & 0.123 & 0.378 & 0.0292 & 0.25 & 0.143 & 0.0673 & 0.0206 \\
& (SE 1) & 4.08e-04 & 6.58e-04 & 2.94e-04 & 2.13e-03 & 1.86e-03 & 4.51e-04 & 1.94e-04 \\
& RMSE variances & 0.0151 & 0.235 & 3.86e-03 & 0.103 & 0.024 & 7.1e-03 & 1.96e-03 \\
& (SE 2) & 3.16e-06 & 4.3e-05 & 3.51e-05 & 1.56e-04 & 1.49e-03 & 1.54e-05 & 2.53e-05 \\
& KL & 4228 & 13409 & 384 & 9900 & 5406 & 1537 & 294 \\
& (SE 3) & 19.3 & 10.1 & 5.48 & 67.8 & 88.6 & 15.2 & 7.49 \\
& Time (s) & 0.0575 & 0.803 & 0.811 & 2.88 & 22.4 & 1.77 & 8.25 \\
\midrule
\multirow{2}{*}{\textbf{3}} 
& RMSE means & 0.067 & 0.33 & 0.0101 & 0.0626 &  & 0.0511 & 9.68e-03 \\
& (SE 1) & 2.59e-04 & 5.18e-04 & 1.08e-04 & 6.85e-04 &  & 4.21e-04 & 3.63e-04 \\
& RMSE variances & 4.53e-03 & 0.201 & 9.16e-04 & 0.0127 &  & 4.84e-03 & 1.19e-03 \\
& (SE 2) & 2.32e-06 & 3.55e-05 & 1.17e-05 & 6.59e-05 &  & 1.07e-05 & 5.44e-05 \\
& KL & 1673 & 12367 & 46.1 & 1658 &  & 939 & 81.2 \\
& (SE 3) & 11.1 & 8.44 & 0.745 & 21.2 &  & 11.6 & 8.62 \\
& Time (s) & 0.14 & 2.06 & 2.33 & 6.17 &  & 4.24 & 11.3 \\
\midrule
\multirow{2}{*}{\textbf{4}} 
& RMSE means & 0.0337 & 0.282 & 3.11e-03 & 0.0153 &  & 0.0413 & 8.55e-03 \\
& (SE 1) & 1.51e-04 & 4.98e-04 & 2.37e-05 & 1.69e-04 &  & 3.82e-04 & 3.95e-04 \\
& RMSE variances & 1.16e-03 & 0.165 & 1.9e-04 & 1.61e-03 &  & 3.72e-03 & 1.09e-03 \\
& (SE 2) & 7.99e-07 & 2.83e-05 & 2.86e-06 & 1.71e-05 &  & 1.18e-05 & 6.29e-05 \\
& KL & 468 & 11206 & 4.2 & 109 &  & 631 & 68.4 \\
& (SE 3) & 4.05 & 8.45 & 0.0541 & 1.97 &  & 9.18 & 8.87 \\
& Time (s) & 0.442 & 4.94 & 7.27 & 21.1 &  & 7.49 & 16.8 \\
\midrule
\multirow{2}{*}{\textbf{5}} 
& RMSE means & 0.0274 & 0.244 & 1.26e-03 & 7.2e-03 &  & 0.0376 & 7.86e-03 \\
& (SE 1) & 1.77e-04 & 4.93e-04 & 1.22e-05 & 8.74e-05 &  & 4.47e-04 & 4.37e-04 \\
& RMSE variances & 7.63e-04 & 0.136 & 5.93e-05 & 5.23e-04 &  & 3.34e-03 & 1.02e-03 \\
& (SE 2) & 1.29e-07 & 2.22e-05 & 1.31e-06 & 8.11e-06 &  & 1.16e-05 & 7.21e-05 \\
& KL & 324 & 10139 & 0.669 & 22.6 &  & 529 & 62.9 \\
& (SE 3) & 4.24 & 8.64 & 0.0108 & 0.5 &  & 9.92 & 9.51 \\
& Time (s) & 1.64 & 10.4 & 17.2 & 34.7 &  & 10.4 & 14.3 \\
\bottomrule
\end{tabular}%
}
\caption{Average RMSE between means, RMSE between variances, KL divergence between exact and approximate predictions, standard errors and time needed for making predictions on the training set on simulated data sets with an effective range of $0.5$ and a sample size
of N=10,000.
}
\end{table}

\begin{table}[H]
\centering
\resizebox{\textwidth}{!}{%
\begin{tabular}{llcccccccc}
\toprule 
 &  & \textbf{Vecchia} & \textbf{Tapering} & \textbf{FITC} & \textbf{Full-scale} & \textbf{FRK} & \textbf{MRA} & \textbf{SPDE} \\
\midrule
\multirow{2}{*}{\textbf{1}} 
& RMSE means & 0.194 & 0.533 & 0.194 & 0.434 & 0.421 & 0.095 & 0.0299 \\
& (SE 1) & 5.21e-04 & 8.94e-03 & 2.41e-03 & 8.77e-03 & 9.09e-03 & 6.05e-04 & 2.26e-04 \\
& RMSE variances & 0.0377 & 0.705 & 0.075 & 0.355 & 0.186 & 0.0128 & 2.5e-03 \\
& (SE 2) & 3.92e-06 & 3.02e-04 & 3e-04 & 2.72e-04 & 8.43e-03 & 3.64e-05 & 2.02e-05 \\
& KL & 7206 & 17344 & 7587 & 15296 & 13834 & 2664 & 618 \\
& (SE 3) & 20.4 & 61.6 & 81.3 & 183 & 201 & 23.5 & 7.7 \\
& Time (s) & 0.0342 & 0.219 & 0.0639 & 2.29 & 7.68 & 2.72 & 7.93 \\
\midrule
\multirow{2}{*}{\textbf{2}} 
& RMSE means & 0.123 & 0.347 & 0.0309 & 0.265 & 0.149 & 0.071 & 0.0208 \\
& (SE 1) & 3.77e-04 & 2.71e-03 & 3.26e-04 & 3.54e-03 & 2.01e-03 & 5.28e-04 & 1.99e-04 \\
& RMSE variances & 0.0151 & 0.498 & 4.72e-03 & 0.15 & 0.0269 & 7.96e-03 & 2.06e-03 \\
& (SE 2) & 3.75e-06 & 2.86e-04 & 4.57e-05 & 3.72e-04 & 1.74e-03 & 1.86e-05 & 2.68e-05 \\
& KL & 4171 & 14917 & 429 & 10620 & 5622 & 1633 & 298 \\
& (SE 3) & 17.4 & 16.5 & 5.87 & 80.7 & 92.7 & 17.1 & 7.38 \\
& Time (s) & 0.0615 & 0.846 & 0.494 & 2.84 & 22.1 & 5.09 & 8.31 \\
\midrule
\multirow{2}{*}{\textbf{3}} 
& RMSE means & 0.0672 & 0.28 & 0.0111 & 0.0638 &  & 0.0545 & 9.82e-03 \\
& (SE 1) & 2.39e-04 & 9.79e-04 & 1.27e-04 & 7.89e-04 &  & 4.86e-04 & 3.67e-04 \\
& RMSE variances & 4.55e-03 & 0.372 & 1.21e-03 & 0.015 &  & 5.52e-03 & 1.25e-03 \\
& (SE 2) & 3.06e-06 & 2.13e-04 & 1.6e-05 & 8.76e-05 &  & 1.75e-05 & 5.68e-05 \\
& KL & 1647 & 13433 & 56 & 1746 &  & 1014 & 82.6 \\
& (SE 3) & 9.91 & 6.74 & 0.884 & 22.2 &  & 13.4 & 8.79 \\
& Time (s) & 0.145 & 2.15 & 1.55 & 6.17 &  & 10.7 & 11.4 \\
\midrule
\multirow{2}{*}{\textbf{4}} 
& RMSE means & 0.0342 & 0.237 & 3.68e-03 & 0.0162 &  & 0.0437 & 8.67e-03 \\
& (SE 1) & 1.7e-04 & 6.11e-04 & 3.06e-05 & 1.86e-04 &  & 4.37e-04 & 3.99e-04 \\
& RMSE variances & 1.21e-03 & 0.269 & 2.93e-04 & 1.97e-03 &  & 4.15e-03 & 1.14e-03 \\
& (SE 2) & 1.75e-06 & 1.42e-04 & 5.1e-06 & 2.01e-05 &  & 1.46e-05 & 6.55e-05 \\
& KL & 471 & 11957 & 5.94 & 122 &  & 677 & 69.6 \\
& (SE 3) & 4.45 & 5.28 & 0.0749 & 2.14 &  & 10 & 9.06 \\
& Time (s) & 0.454 & 5.03 & 4.95 & 21.2 &  & 18 & 16.9 \\
\midrule
\multirow{2}{*}{\textbf{5}} 
& RMSE means & 0.028 & 0.208 & 1.62e-03 & 7.75e-03 &  & 0.0398 & 7.95e-03 \\
& (SE 1) & 1.84e-04 & 5.6e-04 & 1.55e-05 & 9.65e-05 &  & 4.61e-04 & 4.4e-04 \\
& RMSE variances & 7.97e-04 & 0.202 & 1.08e-04 & 6.79e-04 &  & 3.72e-03 & 1.08e-03 \\
& (SE 2) & 5.01e-07 & 9.74e-05 & 2.37e-06 & 1.09e-05 &  & 1.59e-05 & 7.5e-05 \\
& KL & 330 & 10693 & 1.1 & 26.1 &  & 568 & 63.8 \\
& (SE 3) & 4.36 & 5.55 & 0.016 & 0.564 &  & 9.97 & 9.66 \\
& Time (s) & 1.67 & 10.5 & 11.8 & 34.7 &  & 24.8 & 14.6 \\
\bottomrule
\end{tabular}%
}
\caption{Average RMSE between means, RMSE between variances, KL divergence between exact and approximate predictions, standard errors and time needed for making predictions
on the test “interpolation” set on simulated data sets with an effective range of $0.5$ and a
sample size of N=10,000.
}
\end{table}

\begin{table}[H]
\centering
\resizebox{\textwidth}{!}{%
\begin{tabular}{llcccccccc}
\toprule 
 &  & \textbf{Vecchia} & \textbf{Tapering} & \textbf{FITC} & \textbf{Full-scale} & \textbf{FRK} & \textbf{MRA} & \textbf{SPDE} \\
\midrule
\multirow{2}{*}{\textbf{1}} 
& RMSE means & 0.259 & 0.654 & 0.314 & 0.538 & 1.17 & 0.408 & 0.0871 \\
& (SE 1) & 0.0107 & 0.0319 & 0.0134 & 0.0278 & 0.0866 & 0.0219 & 0.0114 \\
& RMSE variances & 0.0864 & 0.569 & 0.197 & 0.412 & 0.452 & 0.228 & 0.218 \\
& (SE 2) & 1.83e-04 & 4.17e-04 & 1.28e-03 & 6.7e-04 & 5.52e-03 & 5.45e-03 & 0.0178 \\
& KL & 1214 & 4960 & 2035 & 4074 & 44168 & 2105 & 269 \\
& (SE 3) & 64.4 & 218 & 84.9 & 209 & 5540 & 151 & 33.5 \\
& Time (s) & 0.125 & 0.296 & 0.127 & 2.97 & 7.65 & 2.57 & 7.93 \\
\midrule
\multirow{2}{*}{\textbf{2}} 
& RMSE means & 0.24 & 0.642 & 0.0979 & 0.41 & 1.06 & 0.343 & 0.0766 \\
& (SE 1) & 0.0108 & 0.0315 & 5.42e-03 & 0.0206 & 0.132 & 0.0155 & 0.0107 \\
& RMSE variances & 0.077 & 0.562 & 0.0416 & 0.287 & 0.729 & 0.163 & 0.209 \\
& (SE 2) & 1.94e-04 & 4.03e-04 & 7.59e-04 & 1.32e-03 & 0.146 & 1.13e-03 & 0.0162 \\
& KL & 988 & 4862 & 265 & 2965 & 8106 & 1434 & 227 \\
& (SE 3) & 66 & 212 & 17.1 & 139 & 1177 & 78.2 & 34 \\
& Time (s) & 0.179 & 1.05 & 0.988 & 3.55 & 22 & 5.18 & 8.31 \\
\midrule
\multirow{2}{*}{\textbf{3}} 
& RMSE means & 0.217 & 0.629 & 0.0509 & 0.186 &  & 0.312 & 0.0724 \\
& (SE 1) & 9.51e-03 & 0.0311 & 2.49e-03 & 9.32e-03 &  & 0.0158 & 9.57e-03 \\
& RMSE variances & 0.0642 & 0.551 & 0.017 & 0.0944 &  & 0.139 & 0.192 \\
& (SE 2) & 2.43e-04 & 3.86e-04 & 4.22e-04 & 1.23e-03 &  & 1.53e-03 & 0.0153 \\
& KL & 784 & 4735 & 65.4 & 860 &  & 1092 & 211 \\
& (SE 3) & 55.5 & 205 & 4.31 & 42.9 &  & 67.7 & 31.4 \\
& Time (s) & 0.342 & 2.71 & 3.09 & 7.75 &  & 10.9 & 11.4 \\
\midrule
\multirow{2}{*}{\textbf{4}} 
& RMSE means & 0.174 & 0.606 & 0.0234 & 0.0883 &  & 0.205 & 0.072 \\
& (SE 1) & 6.86e-03 & 0.0303 & 1.6e-03 & 4.32e-03 &  & 0.0103 & 9.53e-03 \\
& RMSE variances & 0.0418 & 0.531 & 6.06e-03 & 0.0343 &  & 0.0727 & 0.192 \\
& (SE 2) & 2.2e-04 & 3.61e-04 & 1.81e-04 & 6.27e-04 &  & 8.77e-04 & 0.0154 \\
& KL & 471 & 4510 & 13.3 & 189 &  & 609 & 209 \\
& (SE 3) & 33.6 & 195 & 1.28 & 11.5 &  & 42.6 & 30.6 \\
& Time (s) & 0.948 & 6.84 & 9.9 & 29.1 &  & 19.4 & 16.9 \\
\midrule
\multirow{2}{*}{\textbf{5}} 
& RMSE means & 0.118 & 0.574 & 0.0124 & 0.061 &  & 0.201 & 0.0715 \\
& (SE 1) & 4.88e-03 & 0.0294 & 8.04e-04 & 3.23e-03 &  & 0.0104 & 9.64e-03 \\
& RMSE variances & 0.0221 & 0.504 & 2.62e-03 & 0.0204 &  & 0.0696 & 0.191 \\
& (SE 2) & 1.29e-04 & 3.38e-04 & 8.53e-05 & 5.19e-04 &  & 8.24e-04 & 0.0157 \\
& KL & 191 & 4196 & 3.29 & 82.6 &  & 567 & 209 \\
& (SE 3) & 15.4 & 181 & 0.332 & 6.18 &  & 39.7 & 31.1 \\
& Time (s) & 3.34 & 15.5 & 23.6 & 49.5 &  & 26.3 & 14.6 \\
\bottomrule
\end{tabular}%
}
\caption{Average RMSE between means, RMSE between variances, KL divergence between exact and approximate predictions, standard errors and time needed for making predictions
on the test “extrapolation” set on simulated data sets with an effective range of $0.5$ and a
sample size of N=10,000.
}
\end{table}

\newpage
\subsection{Simulated data with $N=10,000$ and an effective range of $0.05$}\label{appendix_005}

\begin{table}[H]
\centering
\resizebox{\textwidth}{!}{%
\begin{tabular}{l|l|ccccc}
  & \textbf{Tuning parameter} & \textbf{} & \textbf{} & \textbf{} & \textbf{} & \textbf{} \\
\midrule
\textbf{Vecchia} & nb. neighbours & 5 & 10 & 20 & 40 & 80 \\
\hline
\textbf{Tapering} & num. non-zero entries & 11 & 30 & 60 & 130 & 263 \\
\hline
\textbf{FITC} & num. inducing points & 47 & 254 & 500 & 950 & 1500 \\
\hline
\multirow{2}{*}{\textbf{Full-scale}} & num. inducing points & 10 & 24 & 120 & 300 & 450 \\
 & num. non-zero entries & 5 & 8 & 28 & 100 & 150 \\
\hline
\textbf{FRK} & num. resolutions & 1  & 2 & exceeds time limit &   &  \\
\hline
\textbf{MRA} & num. knots per partition & 1 & 2 & 4 & 7 & 9 \\
\hline
\textbf{SPDE} & max edge & 0.07 & $0.05$ & 0.03 & 0.025 & 0.022 \\
\bottomrule
\end{tabular}%
}
\caption{Tuning parameters chosen for the comparison on simulated data sets with an effective range of $0.05$ and a sample size of $N=10,000$.}
\end{table}

\subsubsection{Log-likelihood evaluation}

\begin{figure}[H]
  \centering            
  \includegraphics[width=0.7\linewidth]{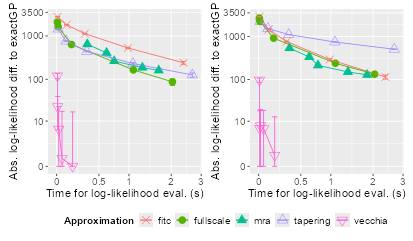} 
  \caption[$N=10,000$, range $0.05$ Log-likelihood evaluation.]
  {Absolute mean difference between approximate and exact log-likelihood on simulated data sets with an effective range of $0.05$ of and a sample size $N=10,000$. The true data-generating parameters are used on the left, and two times these values are used on the right.}
\end{figure}

\begin{table}[H]
\centering
\resizebox{\textwidth}{!}{%
\begin{tabular}{llcccccccc}
\toprule 
 &  & \textbf{Vecchia} & \textbf{Tapering} & \textbf{FITC} & \textbf{Full-scale} & \textbf{FRK} & \textbf{MRA} & \textbf{SPDE} \\
\midrule
\multirow{2}{*}{\textbf{1}} 
& Value & 119 & 1444 & 2757 & 2065 &  & 646 &  \\
& (SE) & 8.59 & 9.73 & 15.4 & 12.3 &  & 9.08 &  \\
& Time (s) & 7.99e-03 & 0.0141 & 8.78e-03 & 7.11e-03 &  & 0.341 &  \\
\midrule
\multirow{2}{*}{\textbf{2}} 
& Value & 22.6 & 763 & 1799 & 1626 &  & 408 &  \\
& (SE) & 8.53 & 7.33 & 12.1 & 10.2 &  & 8.91 &  \\
& Time (s) & 9.65e-03 & 0.0943 & 0.101 & 0.0141 &  & 0.614 &  \\
\midrule
\multirow{2}{*}{\textbf{3}} 
& Value & 6.17 & 441 & 1128 & 635 &  & 264 &  \\
& (SE) & 8.53 & 6.73 & 11.1 & 7.12 &  & 8.93 &  \\
& Time (s) & 0.0205 & 0.336 & 0.31 & 0.153 &  & 0.736 &  \\
\midrule
\multirow{2}{*}{\textbf{4}} 
& Value & 0.523 & 231 & 528 & 165 &  & 186 &  \\
& (SE) & 8.63 & 6.77 & 9.63 & 7.23 &  & 8.88 &  \\
& Time (s) & 0.0525 & 1.08 & 0.981 & 1.09 &  & 1.27 &  \\
\midrule
\multirow{2}{*}{\textbf{5}} 
& Value & 4.55e-03 & 126 & 240 & 87 &  & 160 &  \\
& (SE) & 8.61 & 7.07 & 9.16 & 7.57 &  & 8.94 &  \\
& Time (s) & 0.166 & 2.67 & 2.38 & 2.04 &  & 1.66 &  \\
\bottomrule
\end{tabular}%
}
\caption{Absolute mean difference between approximate and exact log-likelihood, standard error and time needed for evaluating the approximate log-likelihood on simulated
data sets with an effective range of $0.05$ of and a sample size N=10,000. The true data-
generating parameters are used}
\end{table}

\begin{table}[H]
\centering
\resizebox{\textwidth}{!}{%
\begin{tabular}{llcccccccc}
\toprule 
 &  & \textbf{Vecchia} & \textbf{Tapering} & \textbf{FITC} & \textbf{Full-scale} & \textbf{FRK} & \textbf{MRA} & \textbf{SPDE} \\
\midrule
\multirow{2}{*}{\textbf{1}} 
& Value & 97.1 & 2202 & 2852 & 2639 &  & 537 &  \\
& (SE) & 5.63 & 4.72 & 8.24 & 6.3 &  & 6.3 &  \\
& Time (s) & 7.8e-03 & 0.0125 & 7.9e-03 & 6.94e-03 &  & 0.345 &  \\
\midrule
\multirow{2}{*}{\textbf{2}} 
& Value & 6.5 & 1522 & 1443 & 2213 &  & 335 &  \\
& (SE) & 5.95 & 3.55 & 8.5 & 5.5 &  & 6.35 &  \\
& Time (s) & 9.79e-03 & 0.0942 & 0.101 & 0.0133 &  & 0.626 &  \\
\midrule
\multirow{2}{*}{\textbf{3}} 
& Value & 7.64 & 1101 & 761 & 911 &  & 214 &  \\
& (SE) & 6.1 & 3.43 & 8.1 & 4.42 &  & 6.39 &  \\
& Time (s) & 0.0188 & 0.336 & 0.309 & 0.155 &  & 0.768 &  \\
\midrule
\multirow{2}{*}{\textbf{4}} 
& Value & 6.44 & 743 & 292 & 240 &  & 150 &  \\
& (SE) & 6.2 & 3.69 & 6.91 & 5.25 &  & 6.33 &  \\
& Time (s) & 0.0493 & 1.08 & 0.981 & 1.09 &  & 1.36 &  \\
\midrule
\multirow{2}{*}{\textbf{5}} 
& Value & 0.791 & 506 & 115 & 133 &  & 129 &  \\
& (SE) & 6.19 & 4.08 & 6.54 & 5.76 &  & 6.37 &  \\
& Time (s) & 0.167 & 2.68 & 2.38 & 2.04 &  & 1.83 &  \\
\midrule
\bottomrule
\end{tabular}%
}
\caption{Absolute mean difference between approximate and exact log-likelihood, standard error and time needed for evaluating the approximate log-likelihood on simulated
data sets with an effective range of $0.05$ of and a sample size N=10,000. Log-likelihoods are evaluated at two-times the true data-generating parameters.}
\end{table}

\subsubsection{Parameter estimation}

\begin{figure}[H]
  \centering            \includegraphics[width=\linewidth]{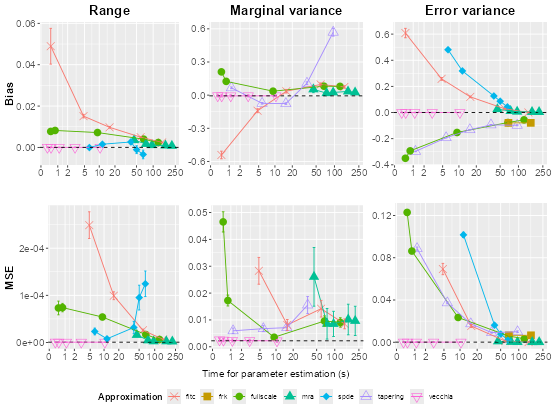} 
  \caption[$N=10,000$, range $0.05$. Parameter estimation accuracy.]
  {Bias and MSE of the estimates for the GP range, GP variance end error term variance on simulated data sets with an effective range of $0.05$ and a sample size N=100,000,}
\end{figure}

\begin{table}[H]
\centering
\resizebox{\textwidth}{!}{%
\begin{tabular}{llcccccccc}
\toprule 
 &  & \textbf{Vecchia} & \textbf{Tapering} & \textbf{FITC} & \textbf{Full-scale} & \textbf{FRK} & \textbf{MRA} & \textbf{SPDE} \\
\midrule
\multirow{2}{*}{\textbf{1}} 
& Bias & 6.95e-05 & 2.25 & 0.0489 & 7.73e-03 &  & 3.51e-03 & -6.35e-05 \\
& (SE 1) & 8.66e-05 & 0.0628 & 4.41e-03 & 3.66e-04 &  & 1.94e-04 & 4.85e-04 \\
& MSE & 7.55e-07 & 5.46 & 4.34e-03 & 7.32e-05 &  & 1.61e-05 & 2.33e-05 \\
& (SE 2) & 1e-07 & 0.285 & 8.57e-04 & 7.44e-06 &  & 1.79e-06 & 3.68e-06 \\
& Time (s) & 0.305 & 1.17 & 0.496 & 0.496 & 68.6 & 45.1 & 6.33 \\
\midrule
\multirow{2}{*}{\textbf{2}} 
& Bias & 3.52e-05 & 3.08 & 0.0151 & 8.21e-03 &  & 1.69e-03 & 1.54e-03 \\
& (SE 1) & 7.8e-05 & 0.11 & 4.44e-04 & 2.58e-04 &  & 1.29e-04 & 2.32e-04 \\
& MSE & 6.09e-07 & 10.7 & 2.49e-04 & 7.4e-05 &  & 4.49e-06 & 7.73e-06 \\
& (SE 2) & 8.03e-08 & 0.739 & 1.45e-05 & 4.57e-06 &  & 9.63e-07 & 1.18e-06 \\
& Time (s) & 0.513 & 5.82 & 4.76 & 0.788 & 159 & 74.5 & 11.5 \\
\midrule
\multirow{2}{*}{\textbf{3}} 
& Bias & 1.78e-05 & 1.97 & 9.72e-03 & 7.2e-03 &  & 1.11e-03 & 2.65e-03 \\
& (SE 1) & 7.71e-05 & 0.0836 & 2.15e-04 & 1.47e-04 &  & 1.09e-04 & 5e-04 \\
& MSE & 5.94e-07 & 4.59 & 9.91e-05 & 5.4e-05 &  & 2.41e-06 & 3.17e-05 \\
& (SE 2) & 8.85e-08 & 0.395 & 4.27e-06 & 2.32e-06 &  & 4.2e-07 & 6.69e-06 \\
& Time (s) & 1.12 & 15.5 & 15.8 & 9.25 &  & 95.4 & 39.4 \\
\midrule
\multirow{2}{*}{\textbf{4}} 
& Bias & 3.86e-05 & 0.198 & 4.96e-03 & 3.99e-03 &  & 9.34e-04 & -1.09e-03 \\
& (SE 1) & 7.75e-05 & 0.0162 & 1.17e-04 & 9.39e-05 &  & 1.26e-04 & 9.75e-04 \\
& MSE & 6.02e-07 & 0.0654 & 2.6e-05 & 1.68e-05 &  & 2.44e-06 & 9.52e-05 \\
& (SE 2) & 8.26e-08 & 0.014 & 1.18e-06 & 7.93e-07 &  & 6.43e-07 & 1.34e-05 \\
& Time (s) & 3.07 & 35.3 & 59.4 & 67.4 &  & 167 & 50.2 \\
\midrule
\multirow{2}{*}{\textbf{5}} 
& Bias & 3.63e-05 & 1.08 & 2.4e-03 & 2.36e-03 &  & 7.74e-04 & -3.4e-03 \\
& (SE 1) & 7.69e-05 & 0.311 & 8.25e-05 & 8.43e-05 &  & 1.21e-04 & 1.07e-03 \\
& MSE & 5.93e-07 & 10.8 & 6.45e-06 & 6.3e-06 &  & 2.06e-06 & 1.24e-04 \\
& (SE 2) & 8.61e-08 & 5.99 & 4.07e-07 & 4.45e-07 &  & 5.22e-07 & 1.38e-05 \\
& Time (s) & 10.4 & 95.5 & 145 & 124 &  & 217 & 66 \\
\bottomrule
\end{tabular}%
}
\caption{Bias and MSE for the GP range, standard errors and time needed for estimating covariance parameters on simulated data sets with an effective range of $0.05$ and a sample size of N=10,000.}
\end{table}

\begin{table}[H]
\centering
\resizebox{\textwidth}{!}{%
\begin{tabular}{llcccccccc}
\toprule 
 &  & \textbf{Vecchia} & \textbf{Tapering} & \textbf{FITC} & \textbf{Full-scale} & \textbf{FRK} & \textbf{MRA} & \textbf{SPDE} \\
\midrule
\multirow{2}{*}{\textbf{1}} 
& Bias & -5.7e-03 & 0.0662 & -0.542 & 0.211 &  & 0.0523 & 9.02 \\
& (SE 1) & 5.03e-03 & 3.96e-03 & 0.0176 & 4.62e-03 &  & 0.0153 & 0.317 \\
& MSE & 2.56e-03 & 5.95e-03 & 0.324 & 0.0465 &  & 0.0261 & 91.2 \\
& (SE 2) & 3.32e-04 & 5.23e-04 & 0.0172 & 1.92e-03 &  & 5.58e-03 & 7.83 \\
& Time (s) & 0.305 & 1.17 & 0.496 & 0.496 & 68.6 & 45.1 & 6.33 \\
\midrule
\multirow{2}{*}{\textbf{2}} 
& Bias & -6.6e-03 & -0.0741 & -0.141 & 0.125 &  & 0.018 & 6.31 \\
& (SE 1) & 4.94e-03 & 3.56e-03 & 9.15e-03 & 4.12e-03 &  & 9.13e-03 & 0.0758 \\
& MSE & 2.48e-03 & 6.76e-03 & 0.0283 & 0.0172 &  & 8.57e-03 & 40.4 \\
& (SE 2) & 3.12e-04 & 5.67e-04 & 2.55e-03 & 1.04e-03 &  & 2.91e-03 & 1 \\
& Time (s) & 0.513 & 5.82 & 4.76 & 0.788 & 159 & 74.5 & 11.5 \\
\midrule
\multirow{2}{*}{\textbf{3}} 
& Bias & -6.63e-03 & -0.0766 & 0.0357 & 0.0358 &  & 0.0245 & 22502 \\
& (SE 1) & 4.81e-03 & 3.55e-03 & 8.29e-03 & 4.76e-03 &  & 8.95e-03 & 22432 \\
& MSE & 2.36e-03 & 7.13e-03 & 8.14e-03 & 3.55e-03 &  & 8.53e-03 & 5.03e+10 \\
& (SE 2) & 2.79e-04 & 5.74e-04 & 1.04e-03 & 4.44e-04 &  & 2.39e-03 & 5.03e+10 \\
& Time (s) & 1.12 & 15.5 & 15.8 & 9.25 &  & 95.4 & 39.4 \\
\midrule
\multirow{2}{*}{\textbf{4}} 
& Bias & -6.18e-03 & 0.108 & 0.0967 & 0.0808 &  & 0.0335 & 6003347 \\
& (SE 1) & 4.74e-03 & 6.28e-03 & 6.95e-03 & 5.54e-03 &  & 9.56e-03 & 1384950 \\
& MSE & 2.28e-03 & 0.0155 & 0.0142 & 9.6e-03 &  & 0.0102 & 2.26e+14 \\
& (SE 2) & 2.79e-04 & 1.63e-03 & 1.59e-03 & 9.75e-04 &  & 2.91e-03 & 5.82e+13 \\
& Time (s) & 3.07 & 35.3 & 59.4 & 67.4 &  & 167 & 50.2 \\
\midrule
\multirow{2}{*}{\textbf{5}} 
& Bias & -6.36e-03 & 0.569 & 0.074 & 0.0792 &  & 0.0246 & 107647 \\
& (SE 1) & 4.66e-03 & 0.018 & 5.53e-03 & 5.29e-03 &  & 9.56e-03 & 39612 \\
& MSE & 2.21e-03 & 0.357 & 8.53e-03 & 9.08e-03 &  & 9.65e-03 & 1.67e+11 \\
& (SE 2) & 2.69e-04 & 0.0183 & 9.36e-04 & 8.85e-04 &  & 2.78e-03 & 8.61e+10 \\
& Time (s) & 10.4 & 95.5 & 145 & 124 &  & 217 & 66 \\
\bottomrule
\end{tabular}%
}
\caption{Bias and MSE for the GP marginal variance, standard errors and time needed for estimating covariance parameters on simulated data sets with an effective range of $0.05$ and a sample size of N=10,000.}
\end{table}

\begin{table}[H]
\centering
\resizebox{\textwidth}{!}{%
\begin{tabular}{llcccccccc}
\toprule 
 &  & \textbf{Vecchia} & \textbf{Tapering} & \textbf{FITC} & \textbf{Full-scale} & \textbf{FRK} & \textbf{MRA} & \textbf{SPDE} \\
\midrule
\multirow{2}{*}{\textbf{1}} 
& Bias & 4.13e-04 & -0.297 & 0.608 & -0.35 & -0.0791 & 0.0255 & 0.48 \\
& (SE 1) & 1.2e-03 & 1.29e-03 & 0.0185 & 2.03e-03 & 1.81e-03 & 1.47e-03 & 2.44e-03 \\
& MSE & 1.45e-04 & 0.0887 & 0.404 & 0.123 & 6.58e-03 & 8.66e-04 & 0.231 \\
& (SE 2) & 1.87e-05 & 7.61e-04 & 0.0203 & 1.4e-03 & 2.8e-04 & 8.33e-05 & 2.35e-03 \\
& Time (s) & 0.305 & 1.17 & 0.496 & 0.496 & 68.6 & 45.1 & 6.33 \\
\midrule
\multirow{2}{*}{\textbf{2}} 
& Bias & 1.39e-04 & -0.193 & 0.257 & -0.293 & -0.0791 & 0.012 & 0.318 \\
& (SE 1) & 1.12e-03 & 1.14e-03 & 5.5e-03 & 1.59e-03 & 1.81e-03 & 1.36e-03 & 1.98e-03 \\
& MSE & 1.24e-04 & 0.0373 & 0.0693 & 0.0863 & 6.58e-03 & 3.27e-04 & 0.102 \\
& (SE 2) & 1.66e-05 & 4.37e-04 & 2.74e-03 & 9.27e-04 & 2.8e-04 & 4.25e-05 & 1.26e-03 \\
& Time (s) & 0.513 & 5.82 & 4.76 & 0.788 & 159 & 74.5 & 11.5 \\
\midrule
\multirow{2}{*}{\textbf{3}} 
& Bias & 5.25e-05 & -0.132 & 0.12 & -0.153 &  & 6.53e-03 & 0.126 \\
& (SE 1) & 1.1e-03 & 1.06e-03 & 3.19e-03 & 1.18e-03 &  & 1.25e-03 & 1.31e-03 \\
& MSE & 1.22e-04 & 0.0174 & 0.0154 & 0.0235 &  & 1.98e-04 & 0.0162 \\
& (SE 2) & 1.63e-05 & 2.78e-04 & 7.61e-04 & 3.55e-04 &  & 2.91e-05 & 3.36e-04 \\
& Time (s) & 1.12 & 15.5 & 15.8 & 9.25 &  & 95.4 & 39.4 \\
\midrule
\multirow{2}{*}{\textbf{4}} 
& Bias & 1.8e-04 & -0.0947 & 0.0287 & -0.0764 &  & 4.46e-03 & 0.0866 \\
& (SE 1) & 1.1e-03 & 1.38e-03 & 1.82e-03 & 1.09e-03 &  & 1.22e-03 & 1.45e-03 \\
& MSE & 1.22e-04 & 9.15e-03 & 1.16e-03 & 5.95e-03 &  & 1.66e-04 & 7.71e-03 \\
& (SE 2) & 1.61e-05 & 2.63e-04 & 1.04e-04 & 1.63e-04 &  & 2.31e-05 & 2.45e-04 \\
& Time (s) & 3.07 & 35.3 & 59.4 & 67.4 &  & 167 & 50.2 \\
\midrule
\multirow{2}{*}{\textbf{5}} 
& Bias & 1.73e-04 & -0.0975 & 7.77e-04 & -0.0563 &  & 4.03e-03 & 0.0429 \\
& (SE 1) & 1.11e-03 & 2.06e-03 & 1.47e-03 & 1.1e-03 &  & 1.14e-03 & 1.72e-03 \\
& MSE & 1.23e-04 & 9.93e-03 & 2.16e-04 & 3.28e-03 &  & 1.45e-04 & 2.13e-03 \\
& (SE 2) & 1.6e-05 & 4.11e-04 & 2.57e-05 & 1.21e-04 &  & 2.1e-05 & 1.37e-04 \\
& Time (s) & 10.4 & 95.5 & 145 & 124 &  & 217 & 66 \\
\bottomrule
\end{tabular}%
}
\caption{Bias and MSE for the error term variance, standard errors and time needed for estimating covariance parameters on simulated data sets with an effective range of $0.05$ and a sample size of N=10,000.}
\end{table}

\subsubsection{Prediction}
\begin{figure}[H]

  \centering            \includegraphics[width=\linewidth]{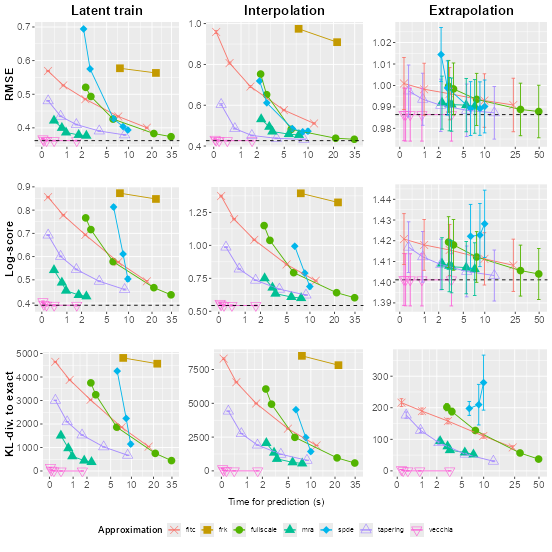} 
  \caption[$N=10,000$, range $0.05$. Prediction accuracy.]
  {Average RMSE and log-score on simulated data sets with
an effective range of $0.05$ and a sample size N=100,000. Predictions are done using the true data-generating parameters.}

\end{figure}

\begin{table}[H]
\centering
\resizebox{\textwidth}{!}{%
\begin{tabular}{llcccccccc}
\toprule 
 &  & \textbf{Vecchia} & \textbf{Tapering} & \textbf{FITC} & \textbf{Full-scale} & \textbf{FRK} & \textbf{MRA} & \textbf{SPDE} \\
\midrule
\multirow{2}{*}{\textbf{1}} 
& RMSE & 0.367 & 0.48 & 0.569 & 0.521 & 0.577 & 0.422 & 0.694 \\
& (SE 1) & 6.3e-04 & 5.53e-04 & 7.93e-04 & 5.91e-04 & 8.57e-04 & 6.53e-04 & 2.22e-03 \\
& Log-score & 0.406 & 0.693 & 0.856 & 0.767 & 0.873 & 0.542 & 7.32 \\
& (SE 2) & 1.69e-03 & 8.89e-04 & 1.39e-03 & 9.85e-04 & 1.69e-03 & 1.64e-03 & 0.0366 \\
& Time (s) & 0.0309 & 0.191 & 0.188 & 2.36 & 7.66 & 0.399 & 2.17 \\
\midrule
\multirow{2}{*}{\textbf{2}} 
& RMSE & 0.363 & 0.434 & 0.526 & 0.493 & 0.563 & 0.4 & 0.575 \\
& (SE 1) & 6.49e-04 & 5.42e-04 & 7.4e-04 & 5.47e-04 & 1.03e-03 & 6.74e-04 & 1.51e-03 \\
& Log-score & 0.395 & 0.601 & 0.778 & 0.716 & 0.848 & 0.488 & 3.06 \\
& (SE 2) & 1.76e-03 & 8.8e-04 & 1.45e-03 & 8.97e-04 & 1.98e-03 & 1.71e-03 & 0.0154 \\
& Time (s) & 0.0573 & 0.681 & 0.81 & 2.88 & 22.4 & 0.74 & 2.8 \\
\midrule
\multirow{2}{*}{\textbf{3}} 
& RMSE & 0.362 & 0.41 & 0.485 & 0.426 &  & 0.386 & 0.427 \\
& (SE 1) & 6.55e-04 & 5.52e-04 & 7.59e-04 & 5.61e-04 &  & 6.49e-04 & 9.57e-04 \\
& Log-score & 0.391 & 0.545 & 0.694 & 0.578 &  & 0.453 & 0.813 \\
& (SE 2) & 1.78e-03 & 9.29e-04 & 1.56e-03 & 9.69e-04 &  & 1.73e-03 & 4.43e-03 \\
& Time (s) & 0.139 & 1.6 & 2.33 & 6.17 &  & 0.952 & 6.26 \\
\midrule
\multirow{2}{*}{\textbf{4}} 
& RMSE & 0.362 & 0.391 & 0.434 & 0.383 &  & 0.379 & 0.404 \\
& (SE 1) & 6.55e-04 & 5.68e-04 & 7.34e-04 & 5.78e-04 &  & 6.37e-04 & 1.23e-03 \\
& Log-score & 0.391 & 0.494 & 0.576 & 0.466 &  & 0.435 & 0.611 \\
& (SE 2) & 1.78e-03 & 1.02e-03 & 1.52e-03 & 1.14e-03 &  & 1.72e-03 & 3.2e-03 \\
& Time (s) & 0.404 & 3.89 & 7.26 & 21.1 &  & 1.77 & 8.44 \\
\midrule
\multirow{2}{*}{\textbf{5}} 
& RMSE & 0.362 & 0.379 & 0.4 & 0.374 &  & 0.377 & 0.394 \\
& (SE 1) & 6.54e-04 & 5.84e-04 & 7.63e-04 & 6.09e-04 &  & 6.26e-04 & 2.67e-03 \\
& Log-score & 0.391 & 0.458 & 0.494 & 0.436 &  & 0.429 & 0.504 \\
& (SE 2) & 1.78e-03 & 1.11e-03 & 1.77e-03 & 1.27e-03 &  & 1.71e-03 & 5.31e-03 \\
& Time (s) & 1.62 & 8.85 & 17.2 & 34.5 &  & 2.42 & 9.78 \\
\bottomrule
\end{tabular}%
}
\caption{Average RMSE and log-score on the training set, standard errors and time
needed for making predictions on the training set on simulated data sets with an effective
range of $0.05$ and a sample size of N=10,000.
}
\end{table}

\begin{table}[H]
\centering
\resizebox{\textwidth}{!}{%
\begin{tabular}{llcccccccc}
\toprule 
 &  & \textbf{Vecchia} & \textbf{Tapering} & \textbf{FITC} & \textbf{Full-scale} & \textbf{FRK} & \textbf{MRA} & \textbf{SPDE} \\
\midrule
\multirow{2}{*}{\textbf{1}} 
& RMSE & 0.433 & 0.604 & 0.959 & 0.753 & 0.974 & 0.532 & 0.719 \\
& (SE 1) & 1.01e-03 & 1.84e-03 & 3.34e-03 & 2.44e-03 & 3.34e-03 & 1.24e-03 & 2.34e-03 \\
& Log-score & 0.561 & 0.987 & 1.37 & 1.15 & 1.39 & 0.75 & 7.58 \\
& (SE 2) & 2.19e-03 & 1.35e-03 & 3.29e-03 & 2.02e-03 & 3.41e-03 & 2.07e-03 & 0.0398 \\
& Time (s) & 0.0343 & 0.206 & 0.063 & 2.29 & 7.65 & 2.34 & 2.21 \\
\midrule
\multirow{2}{*}{\textbf{2}} 
& RMSE & 0.428 & 0.486 & 0.807 & 0.652 & 0.909 & 0.494 & 0.612 \\
& (SE 1) & 1.02e-03 & 1.18e-03 & 2.43e-03 & 2.01e-03 & 3.66e-03 & 1.13e-03 & 1.7e-03 \\
& Log-score & 0.549 & 0.82 & 1.2 & 1.04 & 1.33 & 0.679 & 3.27 \\
& (SE 2) & 2.25e-03 & 9.18e-04 & 2.57e-03 & 1.56e-03 & 4.02e-03 & 2.05e-03 & 0.017 \\
& Time (s) & 0.062 & 0.71 & 0.494 & 2.84 & 22 & 3.11 & 2.85 \\
\midrule
\multirow{2}{*}{\textbf{3}} 
& RMSE & 0.427 & 0.452 & 0.691 & 0.482 &  & 0.471 & 0.485 \\
& (SE 1) & 1.04e-03 & 9.92e-04 & 2.16e-03 & 1.15e-03 &  & 1.07e-03 & 1.4e-03 \\
& Log-score & 0.545 & 0.736 & 1.04 & 0.793 &  & 0.633 & 0.993 \\
& (SE 2) & 2.3e-03 & 9.13e-04 & 2.42e-03 & 1.02e-03 &  & 2.14e-03 & 5.63e-03 \\
& Time (s) & 0.145 & 1.63 & 1.54 & 6.17 &  & 3.44 & 6.38 \\
\midrule
\multirow{2}{*}{\textbf{4}} 
& RMSE & 0.426 & 0.438 & 0.576 & 0.438 &  & 0.458 & 0.47 \\
& (SE 1) & 1.04e-03 & 9.67e-04 & 1.64e-03 & 9.51e-04 &  & 1.04e-03 & 2.42e-03 \\
& Log-score & 0.544 & 0.668 & 0.855 & 0.641 &  & 0.608 & 0.789 \\
& (SE 2) & 2.32e-03 & 1.06e-03 & 2.11e-03 & 1.2e-03 &  & 2.18e-03 & 4.22e-03 \\
& Time (s) & 0.412 & 3.92 & 4.95 & 21.2 &  & 5.75 & 8.6 \\
\midrule
\multirow{2}{*}{\textbf{5}} 
& RMSE & 0.426 & 0.432 & 0.511 & 0.433 &  & 0.454 & 0.473 \\
& (SE 1) & 1.04e-03 & 9.8e-04 & 1.47e-03 & 9.78e-04 &  & 1.1e-03 & 5.37e-03 \\
& Log-score & 0.544 & 0.623 & 0.732 & 0.603 &  & 0.6 & 0.688 \\
& (SE 2) & 2.32e-03 & 1.25e-03 & 2.17e-03 & 1.41e-03 &  & 2.32e-03 & 8.88e-03 \\
& Time (s) & 1.65 & 8.91 & 11.8 & 34.7 &  & 7.72 & 9.97 \\
\bottomrule
\end{tabular}%
}
\caption{Average RMSE and log-score on the test “interpolation” set, standard errors and time needed for making predictions on the test “interpolation” set on simulated
data sets with an effective range of $0.05$ and a sample size of N=10,000.
}
\end{table}

\begin{table}[H]
\centering
\resizebox{\textwidth}{!}{%
\begin{tabular}{llcccccccc}
\toprule 
 &  & \textbf{Vecchia} & \textbf{Tapering} & \textbf{FITC} & \textbf{Full-scale} & \textbf{FRK} & \textbf{MRA} & \textbf{SPDE} \\
\midrule
\multirow{2}{*}{\textbf{1}} 
& RMSE & 0.987 & 0.997 & 1 & 1 & 1.17 & 0.992 & 1.01 \\
& (SE 1) & 6.28e-03 & 6.18e-03 & 6.18e-03 & 6.17e-03 & 0.034 & 6.31e-03 & 6.44e-03 \\
& Log-score & 1.4 & 1.42 & 1.42 & 1.42 & 1.62 & 1.41 & 1.79 \\
& (SE 2) & 6.39e-03 & 6.24e-03 & 6.26e-03 & 6.25e-03 & 0.0452 & 6.38e-03 & 0.0172 \\
& Time (s) & 0.125 & 0.282 & 0.125 & 2.96 & 7.63 & 2.29 & 2.21 \\
\midrule
\multirow{2}{*}{\textbf{2}} 
& RMSE & 0.986 & 0.994 & 0.998 & 0.998 & 3.82 & 0.991 & 0.999 \\
& (SE 1) & 6.29e-03 & 6.19e-03 & 6.19e-03 & 6.17e-03 & 0.658 & 6.34e-03 & 6.45e-03 \\
& Log-score & 1.4 & 1.41 & 1.42 & 1.42 & 2.89 & 1.41 & 1.52 \\
& (SE 2) & 6.41e-03 & 6.23e-03 & 6.25e-03 & 6.24e-03 & 0.266 & 6.4e-03 & 0.0108 \\
& Time (s) & 0.179 & 0.91 & 0.988 & 3.55 & 21.9 & 3.04 & 2.85 \\
\midrule
\multirow{2}{*}{\textbf{3}} 
& RMSE & 0.986 & 0.991 & 0.996 & 0.993 &  & 0.991 & 0.989 \\
& (SE 1) & 6.29e-03 & 6.22e-03 & 6.22e-03 & 6.2e-03 &  & 6.33e-03 & 6.41e-03 \\
& Log-score & 1.4 & 1.41 & 1.42 & 1.41 &  & 1.41 & 1.42 \\
& (SE 2) & 6.42e-03 & 6.24e-03 & 6.27e-03 & 6.23e-03 &  & 6.41e-03 & 7.85e-03 \\
& Time (s) & 0.342 & 2.18 & 3.08 & 7.74 &  & 3.31 & 6.37 \\
\midrule
\multirow{2}{*}{\textbf{4}} 
& RMSE & 0.986 & 0.988 & 0.993 & 0.989 &  & 0.991 & 0.989 \\
& (SE 1) & 6.29e-03 & 6.24e-03 & 6.36e-03 & 6.26e-03 &  & 6.34e-03 & 6.36e-03 \\
& Log-score & 1.4 & 1.41 & 1.41 & 1.41 &  & 1.41 & 1.42 \\
& (SE 2) & 6.42e-03 & 6.26e-03 & 6.44e-03 & 6.27e-03 &  & 6.45e-03 & 7.77e-03 \\
& Time (s) & 0.866 & 5.6 & 9.89 & 29.1 &  & 5.52 & 8.6 \\
\midrule
\multirow{2}{*}{\textbf{5}} 
& RMSE & 0.986 & 0.987 & 0.991 & 0.988 &  & 0.99 & 0.99 \\
& (SE 1) & 6.29e-03 & 6.25e-03 & 6.37e-03 & 6.26e-03 &  & 6.38e-03 & 6.29e-03 \\
& Log-score & 1.4 & 1.4 & 1.41 & 1.4 &  & 1.41 & 1.43 \\
& (SE 2) & 6.42e-03 & 6.28e-03 & 6.43e-03 & 6.28e-03 &  & 6.49e-03 & 8.29e-03 \\
& Time (s) & 3.29 & 13.4 & 23.6 & 49.5 &  & 7.29 & 9.97 \\
\bottomrule
\end{tabular}%
}
\caption{Average RMSE and log-score on the test “extrapolation” set, standard errors and time needed for making predictions on the test “extrapolation” set on simulated
data sets with an effective range of $0.05$ and a sample size of N=10,000.
}
\end{table}

\subsubsection{Comparison to exact calculations}

\begin{figure}[H]

  \centering            \includegraphics[width=\linewidth]{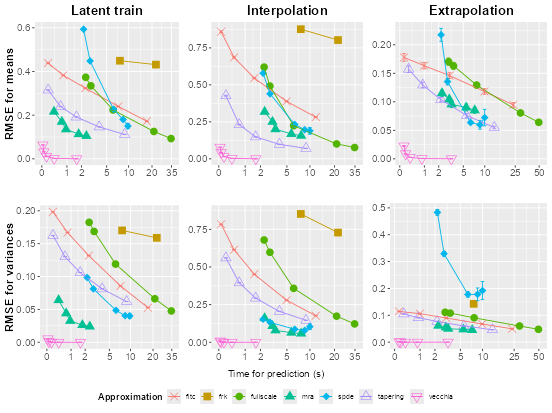} 
  \caption[$N=10,000$, range $0.05$. Comparison to exact calculations.]
  {RMSE between exact and approximate predictive means and variances on
simulated data sets with an effective range of $0.05$ and a sample size N=100,000. Every approximation makes predictions using its parameter estimates.}
\end{figure}

\begin{table}[H]
\centering
\resizebox{\textwidth}{!}{%
\begin{tabular}{llcccccccc}
\toprule 
 &  & \textbf{Vecchia} & \textbf{Tapering} & \textbf{FITC} & \textbf{Full-scale} & \textbf{FRK} & \textbf{MRA} & \textbf{SPDE} \\
\midrule
\multirow{2}{*}{\textbf{1}} 
& RMSE means & 0.0634 & 0.315 & 0.439 & 0.373 & 0.449 & 0.216 & 0.593 \\
& (SE 1) & 1.63e-04 & 4.72e-04 & 6.76e-04 & 5.08e-04 & 7.95e-04 & 3.93e-04 & 2.12e-03 \\
& RMSE variances & 5.23e-03 & 0.163 & 0.198 & 0.182 & 0.17 & 0.0642 & 0.0986 \\
& (SE 2) & 9.84e-06 & 3.48e-05 & 4.3e-05 & 4.51e-05 & 1.04e-03 & 8.26e-05 & 1.27e-04 \\
& KL & 153 & 3009 & 4645 & 3750 & 4812 & 1507 & 69453 \\
& (SE 3) & 0.84 & 4.99 & 9.42 & 5.94 & 12 & 4.55 & 267 \\
& Time (s) & 0.0309 & 0.191 & 0.188 & 2.36 & 7.66 & 0.399 & 2.17 \\
\midrule
\multirow{2}{*}{\textbf{2}} 
& RMSE means & 0.032 & 0.239 & 0.382 & 0.334 & 0.431 & 0.17 & 0.448 \\
& (SE 1) & 7.78e-05 & 3.78e-04 & 6.77e-04 & 4.92e-04 & 1.03e-03 & 3.58e-04 & 1.54e-03 \\
& RMSE variances & 1.21e-03 & 0.13 & 0.167 & 0.169 & 0.159 & 0.0441 & 0.0812 \\
& (SE 2) & 1.17e-06 & 2.78e-05 & 6.45e-05 & 4.64e-05 & 1.06e-03 & 6.74e-05 & 1.57e-04 \\
& KL & 38.4 & 2094 & 3878 & 3245 & 4566 & 969 & 26771 \\
& (SE 3) & 0.189 & 3.45 & 9.08 & 5.38 & 14.7 & 3.45 & 110 \\
& Time (s) & 0.0573 & 0.681 & 0.81 & 2.88 & 22.4 & 0.74 & 2.8 \\
\midrule
\multirow{2}{*}{\textbf{3}} 
& RMSE means & 0.0114 & 0.191 & 0.324 & 0.224 &  & 0.135 & 0.227 \\
& (SE 1) & 3.12e-05 & 3.14e-04 & 6.29e-04 & 3.65e-04 &  & 3.91e-04 & 8.11e-04 \\
& RMSE variances & 1.76e-04 & 0.106 & 0.132 & 0.119 &  & 0.0326 & 0.0486 \\
& (SE 2) & 4.12e-07 & 2.28e-05 & 7.52e-05 & 3.55e-05 &  & 6.91e-05 & 1.84e-04 \\
& KL & 4.85 & 1532 & 3028 & 1866 &  & 626 & 4253 \\
& (SE 3) & 0.0281 & 2.55 & 8.23 & 3.23 &  & 3.24 & 19.8 \\
& Time (s) & 0.139 & 1.6 & 2.33 & 6.17 &  & 0.952 & 6.26 \\
\midrule
\multirow{2}{*}{\textbf{4}} 
& RMSE means & 2.03e-03 & 0.146 & 0.24 & 0.125 &  & 0.113 & 0.18 \\
& (SE 1) & 5.75e-06 & 2.51e-04 & 4.95e-04 & 2.45e-04 &  & 4.03e-04 & 1.88e-03 \\
& RMSE variances & 5.51e-06 & 0.0813 & 0.0857 & 0.0662 &  & 0.0262 & 0.0405 \\
& (SE 2) & 1.68e-08 & 1.76e-05 & 1.01e-04 & 2.49e-05 &  & 6.37e-05 & 3.63e-04 \\
& KL & 0.152 & 1026 & 1861 & 751 &  & 442 & 2237 \\
& (SE 3) & 9.04e-04 & 1.76 & 5.85 & 1.56 &  & 2.76 & 10.8 \\
& Time (s) & 0.404 & 3.89 & 7.26 & 21.1 &  & 1.77 & 8.44 \\
\midrule
\multirow{2}{*}{\textbf{5}} 
& RMSE means & 1.87e-04 & 0.112 & 0.172 & 0.0928 &  & 0.105 & 0.15 \\
& (SE 1) & 7.26e-07 & 1.98e-04 & 3.91e-04 & 1.76e-04 &  & 4.22e-04 & 6.08e-03 \\
& RMSE variances & 5.73e-08 & 0.0619 & 0.0526 & 0.0477 &  & 0.024 & 0.0401 \\
& (SE 2) & 4.72e-10 & 1.34e-05 & 9.22e-05 & 2.53e-05 &  & 5.87e-05 & 2.07e-03 \\
& KL & 1.27e-03 & 671 & 1037 & 442 &  & 381 & 1146 \\
& (SE 3) & 1e-05 & 1.18 & 3.69 & 0.877 &  & 2.7 & 45.2 \\
& Time (s) & 1.62 & 8.85 & 17.2 & 34.5 &  & 2.42 & 9.78 \\
\bottomrule
\end{tabular}%
}
\caption{Average RMSE between means, RMSE between variances, KL divergence between exact and approximate predictions, standard errors and time needed for making predictions
on the training set on simulated data sets with an effective range of $0.05$ and a sample size
of N=10,000.
}
\end{table}

\begin{table}[H]
\centering
\resizebox{\textwidth}{!}{%
\begin{tabular}{llcccccccc}
\toprule 
 &  & \textbf{Vecchia} & \textbf{Tapering} & \textbf{FITC} & \textbf{Full-scale} & \textbf{FRK} & \textbf{MRA} & \textbf{SPDE} \\
\midrule
\multirow{2}{*}{\textbf{1}} 
& RMSE means & 0.0784 & 0.427 & 0.859 & 0.62 & 0.876 & 0.318 & 0.579 \\
& (SE 1) & 1.79e-04 & 1.58e-03 & 2.97e-03 & 2.07e-03 & 3.02e-03 & 9.52e-04 & 2.12e-03 \\
& RMSE variances & 8.5e-03 & 0.561 & 0.784 & 0.679 & 0.852 & 0.16 & 0.154 \\
& (SE 2) & 2.47e-05 & 1.83e-04 & 1.45e-04 & 1.94e-04 & 7.28e-03 & 3.3e-04 & 2.04e-04 \\
& KL & 163 & 4421 & 8301 & 6057 & 8509 & 2062 & 70297 \\
& (SE 3) & 0.706 & 8.11 & 26.7 & 14.2 & 28.5 & 7.76 & 231 \\
& Time (s) & 0.0343 & 0.206 & 0.063 & 2.29 & 7.65 & 2.34 & 2.21 \\
\midrule
\multirow{2}{*}{\textbf{2}} 
& RMSE means & 0.0428 & 0.231 & 0.686 & 0.492 & 0.803 & 0.25 & 0.44 \\
& (SE 1) & 1.13e-04 & 8.69e-04 & 2.2e-03 & 1.69e-03 & 3.61e-03 & 8.11e-04 & 1.58e-03 \\
& RMSE variances & 2.32e-03 & 0.395 & 0.615 & 0.598 & 0.729 & 0.108 & 0.133 \\
& (SE 2) & 5.25e-06 & 1.6e-04 & 2.52e-04 & 1.88e-04 & 7.53e-03 & 2.31e-04 & 2.19e-04 \\
& KL & 48.3 & 2758 & 6558 & 4934 & 7819 & 1347 & 27328 \\
& (SE 3) & 0.237 & 2.99 & 19.8 & 9.94 & 35.7 & 5.58 & 104 \\
& Time (s) & 0.062 & 0.71 & 0.494 & 2.84 & 22 & 3.11 & 2.85 \\
\midrule
\multirow{2}{*}{\textbf{3}} 
& RMSE means & 0.0151 & 0.148 & 0.545 & 0.224 &  & 0.2 & 0.232 \\
& (SE 1) & 4.33e-05 & 4.93e-04 & 1.94e-03 & 8.93e-04 &  & 7.87e-04 & 1.11e-03 \\
& RMSE variances & 3.38e-04 & 0.293 & 0.452 & 0.358 &  & 0.0791 & 0.0863 \\
& (SE 2) & 1.42e-06 & 1.33e-04 & 2.98e-04 & 1.77e-04 &  & 1.94e-04 & 2.72e-04 \\
& KL & 5.89 & 1910 & 4988 & 2482 &  & 886 & 4526 \\
& (SE 3) & 0.031 & 1.39 & 16.5 & 3.33 &  & 4.53 & 17.9 \\
& Time (s) & 0.145 & 1.63 & 1.54 & 6.17 &  & 3.44 & 6.38 \\
\midrule
\multirow{2}{*}{\textbf{4}} 
& RMSE means & 2.77e-03 & 0.0982 & 0.389 & 0.101 &  & 0.168 & 0.197 \\
& (SE 1) & 9.5e-06 & 2.71e-04 & 1.47e-03 & 3.11e-04 &  & 7.88e-04 & 4.62e-03 \\
& RMSE variances & 1.13e-05 & 0.205 & 0.28 & 0.173 &  & 0.0637 & 0.0788 \\
& (SE 2) & 8.49e-08 & 1e-04 & 3.33e-04 & 1.18e-04 &  & 1.97e-04 & 2.15e-03 \\
& KL & 0.198 & 1230 & 3122 & 961 &  & 633 & 2467 \\
& (SE 3) & 1.26e-03 & 0.7 & 11.8 & 0.907 &  & 4.26 & 17.6 \\
& Time (s) & 0.412 & 3.92 & 4.95 & 21.2 &  & 5.75 & 8.6 \\
\midrule
\multirow{2}{*}{\textbf{5}} 
& RMSE means & 2.58e-04 & 0.0707 & 0.282 & 0.0766 &  & 0.156 & 0.19 \\
& (SE 1) & 1.18e-06 & 1.78e-04 & 9.42e-04 & 2.17e-04 &  & 7.54e-04 & 0.0117 \\
& RMSE variances & 1.22e-07 & 0.146 & 0.176 & 0.122 &  & 0.0581 & 0.104 \\
& (SE 2) & 1.7e-09 & 7.38e-05 & 3.11e-04 & 9.58e-05 &  & 1.81e-04 & 9.14e-03 \\
& KL & 1.68e-03 & 785 & 1880 & 582 &  & 549 & 1431 \\
& (SE 3) & 1.34e-05 & 0.424 & 6.95 & 0.65 &  & 3.95 & 79.1 \\
& Time (s) & 1.65 & 8.91 & 11.8 & 34.7 &  & 7.72 & 9.97 \\
\bottomrule
\end{tabular}%
}
\caption{Average RMSE between means, RMSE between variances, KL divergence between exact and approximate predictions, standard errors and time needed for making predictions
on the test “interpolation” set on simulated data sets with an effective range of $0.05$ and a
sample size of N=10,000.}
\end{table}

\begin{table}[H]
\centering
\resizebox{\textwidth}{!}{%
\begin{tabular}{llcccccccc}
\toprule 
 &  & \textbf{Vecchia} & \textbf{Tapering} & \textbf{FITC} & \textbf{Full-scale} & \textbf{FRK} & \textbf{MRA} & \textbf{SPDE} \\
\midrule
\multirow{2}{*}{\textbf{1}} 
& RMSE means & 0.0225 & 0.157 & 0.178 & 0.17 & 0.569 & 0.115 & 0.218 \\
& (SE 1) & 4.57e-04 & 3e-03 & 3.41e-03 & 3.24e-03 & 0.0522 & 2.81e-03 & 5.81e-03 \\
& RMSE variances & 1.82e-03 & 0.105 & 0.114 & 0.111 & 0.143 & 0.0612 & 0.483 \\
& (SE 2) & 2.29e-05 & 4.79e-04 & 5.85e-04 & 5.43e-04 & 4.22e-03 & 1.16e-03 & 5.45e-03 \\
& KL & 4 & 177 & 216 & 202 & 2230 & 94.3 & 3714 \\
& (SE 3) & 0.157 & 5.02 & 6.36 & 5.81 & 429 & 3.87 & 68.3 \\
& Time (s) & 0.125 & 0.282 & 0.125 & 2.96 & 7.63 & 2.29 & 2.21 \\
\midrule
\multirow{2}{*}{\textbf{2}} 
& RMSE means & 8.78e-03 & 0.129 & 0.163 & 0.163 & 3.57 & 0.104 & 0.135 \\
& (SE 1) & 1.58e-04 & 2.52e-03 & 2.85e-03 & 3.11e-03 & 0.672 & 2.69e-03 & 2.95e-03 \\
& RMSE variances & 3.32e-04 & 0.091 & 0.106 & 0.108 & 6.94 & 0.0549 & 0.329 \\
& (SE 2) & 7.15e-06 & 3.69e-04 & 5.68e-04 & 5.06e-04 & 1.47 & 1.2e-03 & 4.23e-03 \\
& KL & 0.584 & 128 & 189 & 188 & 14949 & 77.9 & 1086 \\
& (SE 3) & 0.0213 & 3.58 & 5.01 & 5.37 & 2651 & 3.45 & 25.9 \\
& Time (s) & 0.179 & 0.91 & 0.988 & 3.55 & 21.9 & 3.04 & 2.85 \\
\midrule
\multirow{2}{*}{\textbf{3}} 
& RMSE means & 2.47e-03 & 0.104 & 0.146 & 0.129 &  & 0.0955 & 0.0637 \\
& (SE 1) & 4.95e-05 & 2.09e-03 & 2.65e-03 & 2.51e-03 &  & 2.96e-03 & 1.25e-03 \\
& RMSE variances & 3.29e-05 & 0.077 & 0.09 & 0.0906 &  & 0.0501 & 0.178 \\
& (SE 2) & 1.42e-06 & 2.87e-04 & 6e-04 & 3.76e-04 &  & 1.24e-03 & 4.82e-03 \\
& KL & 0.0447 & 89.2 & 158 & 128 &  & 65.7 & 198 \\
& (SE 3) & 1.82e-03 & 2.44 & 4.44 & 3.58 &  & 3.53 & 11.1 \\
& Time (s) & 0.342 & 2.18 & 3.08 & 7.74 &  & 3.31 & 6.37 \\
\midrule
\multirow{2}{*}{\textbf{4}} 
& RMSE means & 4.76e-04 & 0.076 & 0.118 & 0.0805 &  & 0.09 & 0.0598 \\
& (SE 1) & 1.6e-05 & 1.58e-03 & 2.48e-03 & 1.59e-03 &  & 3.05e-03 & 3.94e-03 \\
& RMSE variances & 1.73e-06 & 0.0604 & 0.0675 & 0.0602 &  & 0.0475 & 0.178 \\
& (SE 2) & 1.29e-07 & 2.08e-04 & 5.5e-04 & 2.52e-04 &  & 1.16e-03 & 0.0121 \\
& KL & 1.67e-03 & 52.9 & 110 & 56.7 &  & 58.3 & 210 \\
& (SE 3) & 1.33e-04 & 1.4 & 3.5 & 1.5 &  & 3.51 & 32.8 \\
& Time (s) & 0.866 & 5.6 & 9.89 & 29.1 &  & 5.52 & 8.6 \\
\midrule
\multirow{2}{*}{\textbf{5}} 
& RMSE means & 5.91e-05 & 0.055 & 0.0946 & 0.064 &  & 0.0851 & 0.0724 \\
& (SE 1) & 2.58e-06 & 1.16e-03 & 2.05e-03 & 1.38e-03 &  & 2.69e-03 & 7.39e-03 \\
& RMSE variances & 2.72e-08 & 0.0463 & 0.0496 & 0.0478 &  & 0.0453 & 0.192 \\
& (SE 2) & 1.83e-09 & 1.52e-04 & 5.26e-04 & 2.46e-04 &  & 1.04e-03 & 0.0171 \\
& KL & 2.57e-05 & 30.6 & 74.4 & 37.4 &  & 52.1 & 280 \\
& (SE 3) & 3.08e-06 & 0.767 & 2.56 & 1.11 &  & 2.87 & 44.5 \\
& Time (s) & 3.29 & 13.4 & 23.6 & 49.5 &  & 7.29 & 9.97 \\
\bottomrule
\end{tabular}%
}
\caption{Average RMSE between means, RMSE between variances, KL divergence between exact and approximate predictions, standard errors and time needed for making predictions
on the test “extrapolation” set on simulated data sets with an effective range of $0.05$ and a
sample size of N=10,000.
}
\end{table}

\newpage
\subsection{Simulated data with $N=100,000$ and an effective range of $0.2$}\label{appendix_100000_02}

\subsubsection{Log-likelihood evaluation}

\begin{table}[H]
\centering
\resizebox{\textwidth}{!}{%
\begin{tabular}{llcccccccc}
\toprule
 &  & \textbf{Vecchia} & \textbf{Tapering} & \textbf{FITC} & \textbf{Full-scale} & \textbf{FRK} & \textbf{MRA} & \textbf{SPDE} \\
\midrule
\multirow{2}{*}{\textbf{1}}
& Value & 116185 & 146919 & 145685 & 135483 &  & 110554 &  \\
& (SE) & 62.2 & 800 & 870 & 377 &  & 57.1 &  \\
& Time (s) & 0.0685 & 0.0868 & 0.0337 & 0.127 &  & 6.11 &  \\
\midrule
\multirow{2}{*}{\textbf{2}}
& Value & 112531 & 138655 & 114284 & 119040 &  & 110044 &  \\
& (SE) & 51.3 & 521 & 127 & 200 &  & 53.6 &  \\
& Time (s) & 0.102 & 0.782 & 0.629 & 0.626 &  & 17.6 &  \\
\midrule
\multirow{2}{*}{\textbf{3}}
& Value & 110554 & 135966 & 112433 & 116220 &  & 109859 &  \\
& (SE) & 47.9 & 437 & 92.2 & 130 &  & 53.4 &  \\
& Time (s) & 0.2 & 1.58 & 1.08 & 1.21 &  & 33 &  \\
\midrule
\multirow{2}{*}{\textbf{4}}
& Value & 109667 & 130065 & 109574 & 111761 &  & 109755 &  \\
& (SE) & 49.1 & 274 & 56.8 & 75.6 &  & 53.1 &  \\
& Time (s) & 0.546 & 7.36 & 5.45 & 3.55 &  & 52.3 &  \\
\midrule
\multirow{2}{*}{\textbf{5}}
& Value & 109411 & 122548 & 109329 & 109645 &  & 109688 &  \\
& (SE) & 49.6 & 119 & 51.6 & 53.9 &  & 53 &  \\
& Time (s) & 1.74 & 54.4 & 44.6 & 39.5 &  & 82.7 &  \\
\bottomrule
\end{tabular}
}
\caption{Mean approximate negative log-likelihood, standard error and time needed for evaluating the approximate log-likelihood on simulated data sets with an effective range of $0.2$ of and a sample size N=100,000. Log-likelihoods are evaluated at the true data-generating parameters.}
\end{table}

\begin{table}[H]
\centering
\resizebox{\textwidth}{!}{%
\begin{tabular}{llcccccccc}
\toprule
 &  & \textbf{Vecchia} & \textbf{Tapering} & \textbf{FITC} & \textbf{Full-scale} & \textbf{FRK} & \textbf{MRA} & \textbf{SPDE} \\
\midrule
\multirow{2}{*}{\textbf{1}}
& Value & 125645 & 161733 & 147970 & 140699 &  & 119828 &  \\
& (SE) & 31.9 & 400 & 587 & 357 &  & 40.8 &  \\
& Time (s) & 0.0693 & 0.0884 & 0.0352 & 0.126 &  & 8.51 &  \\
\midrule
\multirow{2}{*}{\textbf{2}}
& Value & 121954 & 155465 & 121541 & 126066 &  & 119432 &  \\
& (SE) & 27 & 260 & 72.1 & 156 &  & 35.9 &  \\
& Time (s) & 0.103 & 0.784 & 0.629 & 0.633 &  & 27.4 &  \\
\midrule
\multirow{2}{*}{\textbf{3}}
& Value & 119948 & 153238 & 120418 & 123889 &  & 119284 &  \\
& (SE) & 25.7 & 219 & 50.8 & 96 &  & 35.2 &  \\
& Time (s) & 0.198 & 1.59 & 1.08 & 1.25 &  & 49.4 &  \\
\midrule
\multirow{2}{*}{\textbf{4}}
& Value & 119036 & 147907 & 118948 & 120464 &  & 119198 &  \\
& (SE) & 27.2 & 137 & 34.6 & 53 &  & 34.9 &  \\
& Time (s) & 0.545 & 7.36 & 5.45 & 3.59 &  & 69.3 &  \\
\midrule
\multirow{2}{*}{\textbf{5}}
& Value & 118786 & 139948 & 118847 & 119033 &  & 119146 &  \\
& (SE) & 29.2 & 59.4 & 32.9 & 35.1 &  & 34.4 &  \\
& Time (s) & 1.74 & 54.4 & 44.7 & 39.6 &  & 98.6 &  \\
\bottomrule
\end{tabular}
}
\caption{Mean approximate negative log-likelihood, standard error and time needed for evaluating the approximate log-likelihood on simulated data sets with an effective range of $0.2$ of and a sample size N=100,000. Log-likelihoods are evaluated at two-times the true data-generating parameters.}
\end{table}

\subsubsection{Parameter estimation}

\begin{table}[H]
\centering
\resizebox{\textwidth}{!}{%
\begin{tabular}{llcccccccc}
\toprule
 &  & \textbf{Vecchia} & \textbf{Tapering} & \textbf{FITC} & \textbf{Full-scale} & \textbf{FRK} & \textbf{MRA} & \textbf{SPDE} \\
\midrule
\multirow{2}{*}{\textbf{1}}
& Bias & -5.45e-04 & 0.177 & 0.104 & 0.0288 &  & 0.0173 & 0.0196 \\
& (SE 1) & 1.87e-03 & 0.0407 & 0.0283 & 5.74e-03 &  & 1.15e-03 & 1.68e-03 \\
& MSE & 7.04e-05 & 0.0645 & 0.0268 & 1.49e-03 &  & 3.24e-04 & 4.39e-04 \\
& (SE 2) & 2.06e-05 & 0.0247 & 0.013 & 4e-04 &  & 4e-05 & 6.9e-05 \\
& Time (s) & 2.61 & 27.7 & 1.45 & 31.2 & 452 & 435 & 18.5 \\
\midrule
\multirow{2}{*}{\textbf{2}}
& Bias & -3.94e-04 & 0.728 & 0.0775 & 0.0301 &  & 0.0135 & 0.0125 \\
& (SE 1) & 1.56e-03 & 0.117 & 8.78e-03 & 5.34e-03 &  & 1.76e-03 & 1.51e-03 \\
\bottomrule
\end{tabular}
}
\caption{Bias and MSE for the GP range, standard errors and time needed for estimating covariance parameters on simulated data sets with an effective range of $0.2$ and a sample size of N=100,000.
}
\label{table_GPrange_02_100k}
\end{table}

\begin{table}[H]
\centering
\resizebox{\textwidth}{!}{%
\begin{tabular}{llcccccccc}
\toprule
 &  & \textbf{Vecchia} & \textbf{Tapering} & \textbf{FITC} & \textbf{Full-scale} & \textbf{FRK} & \textbf{MRA} & \textbf{SPDE} \\
\midrule
\multirow{2}{*}{\textbf{1}}
& Bias & -7.63e-03 & 0.0922 & 0.654 & 0.496 &  & 0.525 & 4.2 \\
& (SE 1) & 0.039 & 0.0311 & 0.74 & 0.102 &  & 0.0487 & 0.195 \\
& MSE & 0.0305 & 0.0279 & 11.4 & 0.453 &  & 0.321 & 18.4 \\
& (SE 2) & 7.95e-03 & 8.54e-03 & 10.9 & 0.134 &  & 0.0487 & 1.71 \\
& Time (s) & 2.61 & 27.7 & 1.45 & 31.2 & 452 & 435 & 18.5 \\
\midrule
\multirow{2}{*}{\textbf{2}}
& Bias & -3.56e-03 & -0.0968 & 2.66 & 1.23 &  & 0.483 & 4.02 \\
& (SE 1) & 0.0367 & 0.0254 & 0.558 & 0.37 &  & 0.0673 & 0.186 \\
& MSE & 0.0269 & 0.0223 & 13.3 & 4.26 &  & 0.32 & 16.8 \\
& (SE 2) & 6.64e-03 & 5.09e-03 & 4.91 & 3.01 &  & 0.0619 & 1.64 \\
& Time (s) & 4.21 & 125 & 39.8 & 92.5 & 1061 & 1352 & 71.5 \\
\midrule
\multirow{2}{*}{\textbf{3}}
& Bias & 1.7e-04 & -0.162 & 2.86 & 1.38 &  & 0.54 & 3.56 \\
& (SE 1) & 0.035 & 0.0234 & 0.609 & 0.26 &  & 0.0672 & 0.174 \\
& MSE & 0.0245 & 0.0372 & 15.6 & 3.27 &  & 0.378 & 13.2 \\
& (SE 2) & 6.5e-03 & 7.36e-03 & 5.02 & 1.2 &  & 0.0664 & 1.37 \\
& Time (s) & 11.5 & 231 & 76.1 & 161 &  & 2578 & 1856 \\
\midrule
\multirow{2}{*}{\textbf{4}}
& Bias & 0.0125 & -0.31 & 0.189 & 0.984 &  & 0.553 & 3.56 \\
& (SE 1) & 0.0401 & 0.0187 & 0.0454 & 0.157 &  & 0.0694 & 0.173 \\
& MSE & 0.0323 & 0.103 & 0.0769 & 1.46 &  & 0.397 & 13.2 \\
& (SE 2) & 9.95e-03 & 0.0111 & 0.0218 & 0.373 &  & 0.0684 & 1.36 \\
& Time (s) & 35.3 & 881 & 429 & 474 &  & 3676 & 3034 \\
\midrule
\multirow{2}{*}{\textbf{5}}
& Bias & -0.0205 & -0.51 & -9.38e-04 & 0.172 &  & 0.534 & 0.777 \\
& (SE 1) & 0.0455 & 0.0128 & 0.0372 & 0.0396 &  & 0.0658 & 0.0601 \\
& MSE & 0.0419 & 0.263 & 0.0276 & 0.061 &  & 0.367 & 0.673 \\
& (SE 2) & 0.0168 & 0.0128 & 9.28e-03 & 0.0226 &  & 0.0634 & 0.109 \\
& Time (s) & 131 & 3712 & 2856 & 3432 &  & 5605 & 4599 \\
\bottomrule
\end{tabular}
}
\caption{Bias and MSE for the GP marginal variance, standard errors and time needed for estimating covariance parameters on simulated data sets with an effective range of $0.2$ and a sample size of N=100,000.
}
\label{table_GPvar_02_100k}
\end{table}

\begin{table}[H]
\centering
\resizebox{\textwidth}{!}{%
\begin{tabular}{llcccccccc}
\toprule
 &  & \textbf{Vecchia} & \textbf{Tapering} & \textbf{FITC} & \textbf{Full-scale} & \textbf{FRK} & \textbf{MRA} & \textbf{SPDE} \\
\midrule
\multirow{2}{*}{\textbf{1}}
& Bias & 7.48e-04 & -0.34 & 0.277 & -0.149 & -4.45e-03 & 1.13e-03 & 9.6e-03 \\
& (SE 1) & 4.65e-04 & 5.15e-03 & 0.0311 & 5.71e-03 & 1.18e-03 & 4.84e-04 & 5.26e-04 \\
& MSE & 4.88e-06 & 0.116 & 0.0959 & 0.0229 & 4.65e-05 & 5.72e-06 & 9.73e-05 \\
& (SE 2) & 1.42e-06 & 3.52e-03 & 0.0209 & 1.76e-03 & 1.31e-05 & 1.75e-06 & 1.07e-05 \\
& Time (s) & 2.61 & 27.7 & 1.45 & 31.2 & 452 & 435 & 18.5 \\
\midrule
\multirow{2}{*}{\textbf{2}}
& Bias & 9.48e-04 & -0.26 & 0.0198 & -0.0743 & -4.18e-03 & 1.01e-03 & 2.09e-03 \\
& (SE 1) & 4.51e-04 & 4.58e-03 & 1.68e-03 & 1.5e-03 & 1.45e-03 & 4.82e-04 & 4.7e-04 \\
\bottomrule
\end{tabular}
}
\caption{Bias and MSE for the error term variance, standard errors and time needed for estimating covariance parameters on simulated data sets with an effective range of $0.2$ and a sample size of N=100,000.
}
\end{table}

\subsubsection{Prediction}

\begin{table}[H]
\centering
\resizebox{\textwidth}{!}{%
\begin{tabular}{llcccccccc}
\toprule
 &  & \textbf{Vecchia} & \textbf{Tapering} & \textbf{FITC} & \textbf{Full-scale} & \textbf{FRK} & \textbf{MRA} & \textbf{SPDE} \\
\midrule
\multirow{2}{*}{\textbf{1}}
& RMSE & 0.221 & 0.486 & 0.517 & 0.435 & 0.522 & 0.123 & 0.117 \\
& (SE 1) & 5.18e-04 & 2.91e-03 & 5.5e-03 & 2.61e-03 & 5.18e-03 & 1.25e-03 & 2.26e-03 \\
& Log-score & -0.0918 & 0.704 & 0.818 & 0.668 & 0.769 & -0.698 & -0.176 \\
& (SE 2) & 2.37e-03 & 4.59e-03 & 0.0177 & 0.0129 & 9.84e-03 & 0.0101 & 0.0561 \\
& Time (s) & 0.553 & 5.19 & 13.2 & 267 & 130 & 16.9 & 6.77 \\
\midrule
\multirow{2}{*}{\textbf{2}}
& RMSE & 0.161 & 0.433 & 0.247 & 0.299 & 0.365 & 0.111 & 0.0949 \\
& (SE 1) & 3.67e-04 & 1.45e-03 & 2.84e-03 & 2.28e-03 & 5.61e-03 & 1.37e-03 & 1.79e-03 \\
\bottomrule
\end{tabular}
}
\caption{Average RMSE and log-score on the training set, standard errors and time
needed for making predictions on the training set on simulated data sets with an effective
range of $0.2$ and a sample size of N=100,000.}
\end{table}

\begin{table}[H]
\centering
\resizebox{\textwidth}{!}{%
\begin{tabular}{llcccccccc}
\toprule
 &  & \textbf{Vecchia} & \textbf{Tapering} & \textbf{FITC} & \textbf{Full-scale} & \textbf{FRK} & \textbf{MRA} & \textbf{SPDE} \\
\midrule
\multirow{2}{*}{\textbf{1}}
& RMSE & 0.221 & 0.614 & 0.765 & 0.514 & 0.774 & 0.125 & 0.118 \\
& (SE 1) & 4.11e-04 & 0.0165 & 0.0214 & 8.47e-03 & 0.0175 & 1.3e-03 & 2.33e-03 \\
& Log-score & -0.0894 & 1.01 & 1.21 & 0.89 & 1.16 & -0.684 & -0.169 \\
& (SE 2) & 1.87e-03 & 0.0121 & 0.0315 & 0.0156 & 0.0222 & 0.0105 & 0.0579 \\
& Time (s) & 0.649 & 5.25 & 2.7 & 263 & 128 & 54.5 & 7.07 \\
\midrule
\multirow{2}{*}{\textbf{2}}
& RMSE & 0.161 & 0.446 & 0.26 & 0.307 & 0.426 & 0.113 & 0.0957 \\
& (SE 1) & 4.6e-04 & 8.06e-03 & 4.09e-03 & 4.46e-03 & 9.39e-03 & 1.47e-03 & 1.88e-03 \\
\bottomrule
\end{tabular}
}
\caption{Average RMSE and log-score on the test “interpolation” set, standard errors and time needed for making predictions on the test “interpolation” set on simulated
data sets with an effective range of $0.2$ and a sample size of N=100,000.
}
\end{table}

\begin{table}[H]
\centering
\resizebox{\textwidth}{!}{%
\begin{tabular}{llcccccccc}
\toprule
 &  & \textbf{Vecchia} & \textbf{Tapering} & \textbf{FITC} & \textbf{Full-scale} & \textbf{FRK} & \textbf{MRA} & \textbf{SPDE} \\
\midrule
\multirow{2}{*}{\textbf{1}}
& RMSE & 0.911 & 1.02 & 1.01 & 0.999 & 2.04 & 0.946 & 0.91 \\
& (SE 1) & 0.0448 & 0.036 & 0.0381 & 0.0373 & 0.465 & 0.0479 & 0.0579 \\
& Log-score & 1.27 & 1.45 & 1.43 & 1.42 & 5.49 & 1.33 & 1.27 \\
& (SE 2) & 0.0429 & 0.0368 & 0.0386 & 0.0372 & 2.29 & 0.0479 & 0.059 \\
& Time (s) & 7.47 & 9.62 & 5.4 & 331 & 127 & 55.5 & 7.06 \\
\midrule
\multirow{2}{*}{\textbf{2}}
& RMSE & 0.909 & 1.02 & 0.932 & 0.958 & 22.8 & 0.929 & 0.895 \\
& (SE 1) & 0.0451 & 0.036 & 0.0434 & 0.0422 & 7.23 & 0.0489 & 0.0522 \\
& Log-score & 1.27 & 1.45 & 1.34 & 1.38 & 49.1 & 1.3 & 1.24 \\
& (SE 2) & 0.0432 & 0.0367 & 0.0412 & 0.0414 & 16.3 & 0.0481 & 0.0515 \\
& Time (s) & 8.33 & 21.3 & 11.3 & 404 & 293 & 128 & 17 \\
\midrule
\multirow{2}{*}{\textbf{3}}
& RMSE & 0.906 & 1.02 & 0.923 & 0.955 &  & 0.924 & 0.893 \\
& (SE 1) & 0.0456 & 0.036 & 0.0437 & 0.0422 &  & 0.0493 & 0.0524 \\
& Log-score & 1.26 & 1.45 & 1.32 & 1.37 &  & 1.29 & 1.24 \\
& (SE 2) & 0.0439 & 0.0367 & 0.042 & 0.0408 &  & 0.0495 & 0.0564 \\
& Time (s) & 11.1 & 32.7 & 15.8 & 468 &  & 248 & 337 \\
\midrule
\multirow{2}{*}{\textbf{4}}
& RMSE & 0.901 & 1.02 & 0.903 & 0.926 &  & 0.917 & 0.893 \\
& (SE 1) & 0.0467 & 0.036 & 0.0462 & 0.0435 &  & 0.0491 & 0.0523 \\
& Log-score & 1.24 & 1.44 & 1.25 & 1.32 &  & 1.28 & 1.24 \\
& (SE 2) & 0.0451 & 0.0366 & 0.0451 & 0.0415 &  & 0.0483 & 0.0571 \\
& Time (s) & 18.7 & 109 & 61.9 & 481 &  & 370 & 591 \\
\midrule
\multirow{2}{*}{\textbf{5}}
& RMSE & 0.896 & 1.01 & 0.895 & 0.908 &  & 0.914 & 0.891 \\
& (SE 1) & 0.0488 & 0.036 & 0.0489 & 0.0464 &  & 0.0487 & 0.0514 \\
& Log-score & 1.23 & 1.43 & 1.23 & 1.27 &  & 1.27 & 1.24 \\
& (SE 2) & 0.0485 & 0.0364 & 0.0484 & 0.0457 &  & 0.0478 & 0.0598 \\
& Time (s) & 46.5 & 475 & 451 & 1981 &  & 475 & 695 \\
\bottomrule
\end{tabular}
}
\caption{Average RMSE and log-score on the test “extrapolation” set, standard errors and time needed for making predictions on the test “extrapolation” set on simulated
data sets with an effective range of $0.2$ and a sample size of N=100,000.
}
\label{table_extrapolation_02_100k}
\end{table}

\newpage
\subsection{Simulated data with $N=100,000$ and an effective range of $0.5$}\label{appendix_05_100k}

\begin{table}[H]
\centering
\resizebox{\textwidth}{!}{%
\begin{tabular}{l|l|cccccc}
  & \textbf{Tuning parameter} & \textbf{} & \textbf{} & \textbf{} & \textbf{} & \textbf{} & \textbf{} \\
\midrule
\textbf{Vecchia} & nb. neighbours & 5 & 10 & 20 & 40 & 80 \\
\hline
\textbf{Tapering} & num. non-zero entries & 8 & 17 & 22 & 40 & 111\\
\hline
\textbf{FITC} & num. inducing points & 20 & 200 & 275 & 900 & 2000\\
\hline
\multirow{2}{*}{\textbf{Full-scale}} & num. inducing points & 26 & 91 & 122 & 250 & 330 \\
 & num. non-zero entries & 6 & 11 & 15 & 27 & 73 \\
\hline
\textbf{FRK} & num. resolutions & 1  & 2 & exceeds time limit &   &  &  \\
\hline
\textbf{MRA} & num. knots per partition & 1 & 2 & 3 & 4 & 5 \\
\hline
\textbf{SPDE} & max edge & 0.005 & 0.0045 & 0.00425 & 0.004  & 0.003  \\
\bottomrule
\end{tabular}%
}
\caption{Tuning parameters chosen for the comparison on simulated data sets with an effective range of $0.5$ and a sample size of $N=100,000$.}
\end{table}

\subsubsection{Log-likelihood evaluation}
\begin{figure}[H]

  \centering            \includegraphics[width=0.7\linewidth]{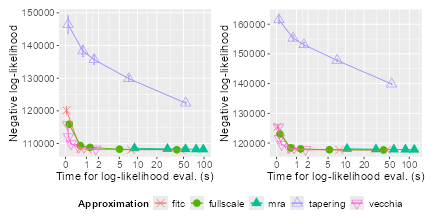} 
  \caption[$N=100,000$, range $0.5$. Log-likelihood evaluation.]
  {Average negative log-likelihood on
simulated data sets with an effective range $0.5$ of and a sample size N=100,000. The true data-generating parameters are used on the left, and two times these values are used on the right.}
\end{figure}

\begin{table}[H]
\centering
\resizebox{\textwidth}{!}{%
\begin{tabular}{llcccccccc}
\toprule
 &  & \textbf{Vecchia} & \textbf{Tapering} & \textbf{FITC} & \textbf{Full-scale} & \textbf{FRK} & \textbf{MRA} & \textbf{SPDE} \\
\midrule
\multirow{2}{*}{\textbf{1}}
& Value & 115800 & 146444 & 120165 & 115905 &  & 108560 &  \\
& (SE) & 82 & 1349 & 579 & 400 &  & 51.9 &  \\
& Time (s) & 0.0692 & 0.0869 & 0.0339 & 0.127 &  & 8.93 &  \\
\midrule
\multirow{2}{*}{\textbf{2}}
& Value & 112035 & 138333 & 108396 & 109383 &  & 108359 &  \\
& (SE) & 56.1 & 875 & 48.1 & 61.3 &  & 49.5 &  \\
& Time (s) & 0.102 & 0.779 & 0.63 & 0.639 &  & 28.8 &  \\
\midrule
\multirow{2}{*}{\textbf{3}}
& Value & 109918 & 135685 & 108240 & 108821 &  & 108285 &  \\
& (SE) & 51.7 & 732 & 47.3 & 50.6 &  & 49 &  \\
& Time (s) & 0.198 & 1.59 & 1.08 & 1.26 &  & 53.5 &  \\
\midrule
\multirow{2}{*}{\textbf{4}}
& Value & 108824 & 129857 & 108089 & 108255 &  & 108239 &  \\
& (SE) & 49.4 & 451 & 48.5 & 46.6 &  & 49 &  \\
& Time (s) & 0.547 & 7.23 & 7.9 & 5.03 &  & 76.6 &  \\
\midrule
\multirow{2}{*}{\textbf{5}}
& Value & 108330 & 122392 & 108084 & 108095 &  & 108215 &  \\
& (SE) & 49.5 & 183 & 48.2 & 48.3 &  & 49 &  \\
& Time (s) & 1.74 & 53.7 & 45.1 & 39.7 &  & 97.4 &  \\
\bottomrule
\end{tabular}
}
\caption{Mean approximate negative log-likelihood, standard error and time needed
for evaluating the approximate log-likelihood on simulated data sets with an effective range of $0.5$ of and a sample size N=100,000. Log-likelihoods are evaluated at the true data-generating parameters.}
\end{table}

\begin{table}[H]
\centering
\resizebox{\textwidth}{!}{%
\begin{tabular}{llcccccccc}
\toprule
 &  & \textbf{Vecchia} & \textbf{Tapering} & \textbf{FITC} & \textbf{Full-scale} & \textbf{FRK} & \textbf{MRA} & \textbf{SPDE} \\
\midrule
\multirow{2}{*}{\textbf{1}}
& Value & 125373 & 161496 & 125476 & 123036 &  & 118070 &  \\
& (SE) & 41.3 & 675 & 342 & 268 &  & 32.8 &  \\
& Time (s) & 0.0699 & 0.088 & 0.0343 & 0.126 &  & 10.6 &  \\
\midrule
\multirow{2}{*}{\textbf{2}}
& Value & 121596 & 155308 & 117807 & 118380 &  & 117909 &  \\
& (SE) & 28.3 & 438 & 25 & 40 &  & 27.3 &  \\
& Time (s) & 0.103 & 0.78 & 0.63 & 0.643 &  & 30.7 &  \\
\midrule
\multirow{2}{*}{\textbf{3}}
& Value & 119469 & 153102 & 117738 & 118052 &  & 117847 &  \\
& (SE) & 26.3 & 366 & 24.2 & 29.1 &  & 26.1 &  \\
& Time (s) & 0.197 & 1.6 & 1.08 & 1.31 &  & 58 &  \\
\midrule
\multirow{2}{*}{\textbf{4}}
& Value & 118365 & 147811 & 117678 & 117750 &  & 117808 &  \\
& (SE) & 25.4 & 226 & 23.7 & 24 &  & 25.6 &  \\
& Time (s) & 0.546 & 7.23 & 7.9 & 5.28 &  & 87.5 &  \\
\midrule
\multirow{2}{*}{\textbf{5}}
& Value & 117856 & 139886 & 117675 & 117681 &  & 117788 &  \\
& (SE) & 25.7 & 91.7 & 23.7 & 23.7 &  & 25.2 &  \\
& Time (s) & 1.74 & 53.7 & 45.1 & 40.4 &  & 119 &  \\
\bottomrule
\end{tabular}
}
\caption{Mean approximate negative log-likelihood, standard error and time needed
for evaluating the approximate log-likelihood on simulated data sets with an effective range of $0.5$ of and a sample size N=100,000. Log-likelihoods are evaluated at two-times the true data-generating parameters.}
\end{table}

\subsubsection{Parameter estimation}

\begin{figure}[H]
  \centering            \includegraphics[width=\linewidth]{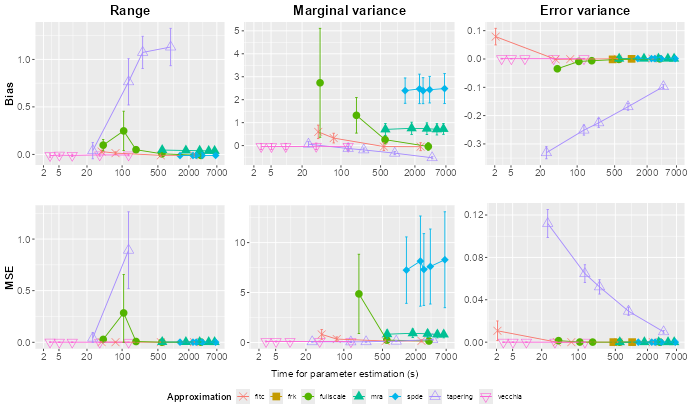} 
  \caption[$N=100,000$, range $0.5$. Parameter estimation accuracy.]
  {Bias and MSE of the estimates for the GP range, GP variance end error term variance on simulated
data sets with an effective range of $0.5$ and a sample size N=100,000,}

\end{figure}

\begin{table}[H]
\centering
\resizebox{\textwidth}{!}{%
\begin{tabular}{llcccccccc}
\toprule
 &  & \textbf{Vecchia} & \textbf{Tapering} & \textbf{FITC} & \textbf{Full-scale} & \textbf{FRK} & \textbf{MRA} & \textbf{SPDE} \\
\midrule
\multirow{2}{*}{\textbf{1}}
& Bias & -0.0122 & 0.0396 & 4.39 & 0.0974 &  & 0.0457 & -0.0121 \\
& (SE 1) & 9.05e-03 & 0.0439 & 2.47 & 0.0303 &  & 6.54e-03 & 6.61e-03 \\
& MSE & 1.79e-03 & 0.04 & 142 & 0.0278 &  & 2.9e-03 & 9.77e-04 \\
& (SE 2) & 5.39e-04 & 0.0271 & 99.9 & 0.0126 &  & 6.44e-04 & 1.98e-04 \\
& Time (s) & 3.02 & 26.2 & 2.12 & 42.4 & 459 & 607 & 1347 \\
\midrule
\multirow{2}{*}{\textbf{2}}
& Bias & -7.7e-03 & 0.764 & 0.0299 & 0.247 &  & 0.0409 & -0.0118 \\
& (SE 1) & 6.74e-03 & 0.124 & 7.69e-03 & 0.106 &  & 6.54e-03 & 6.56e-03 \\
\bottomrule
\end{tabular}
}
\caption{Bias and MSE for the GP range, standard errors and time needed for estimating covariance parameters on simulated data sets with an effective range of $0.5$ and a sample size of N=100,000.
}
\end{table}

\begin{table}[H]
\centering
\resizebox{\textwidth}{!}{%
\begin{tabular}{llcccccccc}
\toprule
 &  & \textbf{Vecchia} & \textbf{Tapering} & \textbf{FITC} & \textbf{Full-scale} & \textbf{FRK} & \textbf{MRA} & \textbf{SPDE} \\
\midrule
\multirow{2}{*}{\textbf{1}}
& Bias & -0.0384 & 0.0636 & 52108 & 2.74 &  & 0.707 & 2.4 \\
& (SE 1) & 0.0689 & 0.0561 & 37935 & 1.21 &  & 0.132 & 0.281 \\
& MSE & 0.0963 & 0.067 & 3.15e+10 & 36.8 &  & 0.829 & 7.25 \\
& (SE 2) & 0.0253 & 0.0152 & 2.63e+10 & 27.4 &  & 0.21 & 1.7 \\
& Time (s) & 3.02 & 26.2 & 2.12 & 42.4 & 459 & 607 & 1347 \\
\midrule
\multirow{2}{*}{\textbf{2}}
& Bias & -0.0397 & -0.125 & 0.599 & 54.6 &  & 0.754 & 2.47 \\
& (SE 1) & 0.0687 & 0.0455 & 0.15 & 39.4 &  & 0.136 & 0.328 \\
& MSE & 0.096 & 0.0572 & 0.808 & 33985 &  & 0.919 & 8.15 \\
& (SE 2) & 0.0235 & 0.019 & 0.251 & 31568 &  & 0.218 & 2.3 \\
& Time (s) & 5.1 & 134 & 39.3 & 107 & 1032 & 1752 & 2408 \\
\midrule
\multirow{2}{*}{\textbf{3}}
& Bias & -0.0387 & -0.187 & 0.329 & 1.32 &  & 0.739 & 2.39 \\
& (SE 1) & 0.07 & 0.0419 & 0.108 & 0.395 &  & 0.13 & 0.288 \\
& MSE & 0.0996 & 0.0701 & 0.34 & 4.87 &  & 0.865 & 7.31 \\
& (SE 2) & 0.0221 & 0.0214 & 0.11 & 2.03 &  & 0.19 & 1.83 \\
& Time (s) & 9.86 & 252 & 74.3 & 188 &  & 3245 & 2780 \\
\midrule
\multirow{2}{*}{\textbf{4}}
& Bias & -0.0287 & -0.328 & -0.0406 & 0.26 &  & 0.711 & 2.44 \\
& (SE 1) & 0.0711 & 0.0342 & 0.0847 & 0.0916 &  & 0.125 & 0.296 \\
& MSE & 0.102 & 0.131 & 0.145 & 0.236 &  & 0.803 & 7.61 \\
& (SE 2) & 0.0221 & 0.0257 & 0.0565 & 0.0754 &  & 0.174 & 1.92 \\
& Time (s) & 36.1 & 885 & 570 & 599 &  & 4862 & 3632 \\
\midrule
\multirow{2}{*}{\textbf{5}}
& Bias & -0.0312 & -0.528 & -0.0467 & -0.0374 &  & 0.727 & 2.48 \\
& (SE 1) & 0.0654 & 0.0231 & 0.0884 & 0.0845 &  & 0.12 & 0.333 \\
& MSE & 0.0865 & 0.289 & 0.158 & 0.144 &  & 0.803 & 8.28 \\
& (SE 2) & 0.025 & 0.0257 & 0.0545 & 0.057 &  & 0.172 & 2.45 \\
& Time (s) & 134 & 4008 & 2517 & 3435 &  & 6333 & 6631 \\
\bottomrule
\end{tabular}
}
\caption{Bias and MSE for the GP marginal variance, standard errors and time needed for estimating covariance parameters on simulated data sets with an effective range of $0.5$ and a sample size of N=100,000.
}
\end{table}

\begin{table}[H]
\centering
\resizebox{\textwidth}{!}{%
\begin{tabular}{llcccccccc}
\toprule
 &  & \textbf{Vecchia} & \textbf{Tapering} & \textbf{FITC} & \textbf{Full-scale} & \textbf{FRK} & \textbf{MRA} & \textbf{SPDE} \\
\midrule
\multirow{2}{*}{\textbf{1}}
& Bias & 1.12e-03 & -0.331 & 0.079 & -0.0345 & -1.39e-03 & 1.15e-03 & 1.45e-03 \\
& (SE 1) & 6.56e-04 & 0.0109 & 0.015 & 3.41e-03 & 1.46e-03 & 5.24e-04 & 5.68e-04 \\
& MSE & 9.86e-06 & 0.112 & 0.0107 & 1.42e-03 & 4.23e-05 & 6.55e-06 & 8.25e-06 \\
& (SE 2) & 2.41e-06 & 6.72e-03 & 4.74e-03 & 2.44e-04 & 1.86e-05 & 1.46e-06 & 1.97e-06 \\
& Time (s) & 3.02 & 26.2 & 2.12 & 42.4 & 459 & 607 & 1347 \\
\midrule
\multirow{2}{*}{\textbf{2}}
& Bias & 1.13e-03 & -0.251 & -1.31e-03 & -8.63e-03 & 4.93e-04 & 1.22e-03 & 1.41e-03 \\
& (SE 1) & 5.41e-04 & 9.02e-03 & 6.22e-04 & 1.26e-03 & 1.38e-03 & 5.56e-04 & 5.28e-04 \\
\bottomrule
\end{tabular}
}
\caption{Bias and MSE for the error term variance, standard errors and time needed for estimating covariance parameters on simulated data sets with an effective range of $0.5$ and a sample size of N=100,000.
}
\end{table}

\subsubsection{Prediction}

\begin{figure}[H]

  \centering            \includegraphics[width=\linewidth]{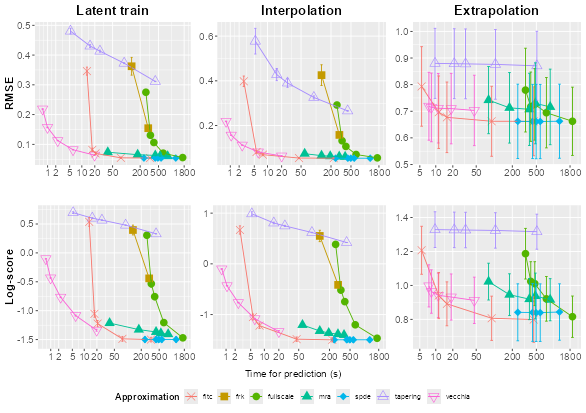} 
  \caption[$N=100,000$, range $0.5$. Prediction accuracy.]
  {Average RMSE and log-score on simulated data sets with
an effective range of $0.5$ and a sample size N=100,000. Predictions are done using the true data-generating parameters.}
\end{figure}

\begin{table}[H]
\centering
\resizebox{\textwidth}{!}{%
\begin{tabular}{llcccccccc}
\toprule
 &  & \textbf{Vecchia} & \textbf{Tapering} & \textbf{FITC} & \textbf{Full-scale} & \textbf{FRK} & \textbf{MRA} & \textbf{SPDE} \\
\midrule
\multirow{2}{*}{\textbf{1}}
& RMSE & 0.219 & 0.481 & 0.346 & 0.275 & 0.362 & 0.0735 & 0.054 \\
& (SE 1) & 6.35e-04 & 4.79e-03 & 6.87e-03 & 3.44e-03 & 0.0152 & 8.58e-04 & 1.28e-03 \\
& Log-score & -0.101 & 0.696 & 0.537 & 0.306 & 0.396 & -1.21 & -1.5 \\
& (SE 2) & 2.91e-03 & 7.54e-03 & 0.039 & 0.0216 & 0.0434 & 0.0129 & 0.024 \\
& Time (s) & 0.56 & 5.25 & 13.2 & 267 & 131 & 39.1 & 242 \\
\midrule
\multirow{2}{*}{\textbf{2}}
& RMSE & 0.157 & 0.43 & 0.0815 & 0.131 & 0.154 & 0.0654 & 0.054 \\
& (SE 1) & 4.14e-04 & 2.34e-03 & 2.92e-03 & 3.39e-03 & 3.37e-03 & 9.11e-04 & 1.28e-03 \\
\bottomrule
\end{tabular}
}
\caption{Average RMSE and log-score on the training set, standard errors and time
needed for making predictions on the training set on simulated data sets with an effective
range of $0.5$ and a sample size of N=100,000.
}
\end{table}

\begin{table}[H]
\centering
\resizebox{\textwidth}{!}{%
\begin{tabular}{llcccccccc}
\toprule
 &  & \textbf{Vecchia} & \textbf{Tapering} & \textbf{FITC} & \textbf{Full-scale} & \textbf{FRK} & \textbf{MRA} & \textbf{SPDE} \\
\midrule
\multirow{2}{*}{\textbf{1}}
& RMSE & 0.219 & 0.577 & 0.398 & 0.292 & 0.425 & 0.0742 & 0.0541 \\
& (SE 1) & 6.48e-04 & 0.0293 & 0.012 & 5.26e-03 & 0.0236 & 8.43e-04 & 1.31e-03 \\
& Log-score & -0.1 & 0.986 & 0.668 & 0.384 & 0.55 & -1.2 & -1.5 \\
& (SE 2) & 2.97e-03 & 0.0203 & 0.0413 & 0.0225 & 0.0574 & 0.0129 & 0.0246 \\
& Time (s) & 0.65 & 5.3 & 2.71 & 263 & 129 & 59.2 & 247 \\
\midrule
\multirow{2}{*}{\textbf{2}}
& RMSE & 0.157 & 0.428 & 0.0819 & 0.132 & 0.158 & 0.0656 & 0.0541 \\
& (SE 1) & 3.61e-04 & 0.0139 & 3.06e-03 & 3.77e-03 & 3.68e-03 & 9.34e-04 & 1.31e-03 \\
\bottomrule
\end{tabular}
}
\caption{Average RMSE and log-score on the test “interpolation” set, standard errors and time needed for making predictions on the test “interpolation” set on simulated
data sets with an effective range of $0.5$ and a sample size of N=100,000.
}
\end{table}

\begin{table}[H]
\centering
\resizebox{\textwidth}{!}{%
\begin{tabular}{llcccccccc}
\toprule
 &  & \textbf{Vecchia} & \textbf{Tapering} & \textbf{FITC} & \textbf{Full-scale} & \textbf{FRK} & \textbf{MRA} & \textbf{SPDE} \\
\midrule
\multirow{2}{*}{\textbf{1}}
& RMSE & 0.719 & 0.88 & 0.793 & 0.779 & 1.47 & 0.742 & 0.662 \\
& (SE 1) & 0.0651 & 0.0671 & 0.0762 & 0.0803 & 0.218 & 0.0644 & 0.0712 \\
& Log-score & 0.997 & 1.33 & 1.21 & 1.19 & 6.75 & 1.02 & 0.841 \\
& (SE 2) & 0.0633 & 0.0536 & 0.0721 & 0.0761 & 1.81 & 0.0553 & 0.0856 \\
& Time (s) & 7.47 & 9.68 & 5.42 & 331 & 128 & 80.7 & 247 \\
\midrule
\multirow{2}{*}{\textbf{2}}
& RMSE & 0.714 & 0.879 & 0.696 & 0.717 & 9.41 & 0.713 & 0.662 \\
& (SE 1) & 0.0658 & 0.067 & 0.0694 & 0.0664 & 2.02 & 0.0675 & 0.0712 \\
& Log-score & 0.964 & 1.33 & 0.94 & 1.02 & 17.4 & 0.945 & 0.841 \\
& (SE 2) & 0.0652 & 0.0535 & 0.0694 & 0.0591 & 4.47 & 0.0628 & 0.0867 \\
& Time (s) & 8.3 & 21.2 & 11.3 & 405 & 294 & 179 & 433 \\
\midrule
\multirow{2}{*}{\textbf{3}}
& RMSE & 0.711 & 0.878 & 0.677 & 0.726 &  & 0.707 & 0.661 \\
& (SE 1) & 0.0665 & 0.0669 & 0.0677 & 0.0685 &  & 0.0692 & 0.0714 \\
& Log-score & 0.941 & 1.33 & 0.891 & 1.01 &  & 0.919 & 0.839 \\
& (SE 2) & 0.0664 & 0.0534 & 0.066 & 0.066 &  & 0.0617 & 0.0861 \\
& Time (s) & 11.1 & 33.1 & 15.8 & 471 &  & 384 & 496 \\
\midrule
\multirow{2}{*}{\textbf{4}}
& RMSE & 0.71 & 0.876 & 0.662 & 0.694 &  & 0.73 & 0.661 \\
& (SE 1) & 0.0676 & 0.0667 & 0.0671 & 0.0674 &  & 0.0776 & 0.0712 \\
& Log-score & 0.932 & 1.32 & 0.806 & 0.921 &  & 0.937 & 0.84 \\
& (SE 2) & 0.0693 & 0.0532 & 0.0664 & 0.0671 &  & 0.0689 & 0.0864 \\
& Time (s) & 18.7 & 105 & 90.8 & 733 &  & 482 & 579 \\
\midrule
\multirow{2}{*}{\textbf{5}}
& RMSE & 0.703 & 0.871 & 0.662 & 0.662 &  & 0.716 & 0.663 \\
& (SE 1) & 0.0675 & 0.0663 & 0.066 & 0.0654 &  & 0.0725 & 0.071 \\
& Log-score & 0.912 & 1.32 & 0.799 & 0.816 &  & 0.916 & 0.843 \\
& (SE 2) & 0.0698 & 0.0527 & 0.0683 & 0.0621 &  & 0.0633 & 0.0847 \\
& Time (s) & 46.4 & 510 & 453 & 1972 &  & 846 & 1207 \\
\bottomrule
\end{tabular}
}
\caption{Average RMSE and log-score on the test “extrapolation” set, standard errors and time needed for making predictions on the test “extrapolation” set on simulated
data sets with an effective range of $0.5$ and a sample size of N=100,000.
}
\end{table}

\newpage
\subsection{Simulated data with $N=100,000$ and an effective range of $0.05$}\label{appendix_005_100k}

\begin{table}[H]
\centering
\resizebox{\textwidth}{!}{%
\begin{tabular}{l|l|cccccc}
  & \textbf{Tuning parameter} & \textbf{} & \textbf{} & \textbf{} & \textbf{} & \textbf{} & \textbf{} \\
\midrule
\textbf{Vecchia} & nb. neighbours & 5 & 10 & 20 & 40 & 80 \\
\hline
\textbf{Tapering} & num. non-zero entries & 8 & 17 & 22 & 40 & 111\\
\hline
\textbf{FITC} & num. inducing points & 20 & 200 & 275 & 900 & 2000\\
\hline
\multirow{2}{*}{\textbf{Full-scale}} & num. inducing points & 26 & 91 & 122 & 250 & 330 \\
 & num. non-zero entries & 6 & 11 & 15 & 27 & 73 \\
\hline
\textbf{FRK} & num. resolutions & 1  & 2 & exceeds time limit &   &  &  \\
\hline
\textbf{MRA} & num. knots per partition & 1 & 2 & 3 & 4 & 5 \\
\hline
\textbf{SPDE} & max edge & 0.035 & 0.02 & 0.005 & 0.004  & 0.003  \\
\bottomrule
\end{tabular}%
}
\caption{Tuning parameters chosen for the comparison on simulated data sets with an effective range of $0.05$ and a sample size of $N=100,000$.}
\end{table}

\subsubsection{Log-likelihood evaluation}
\begin{figure}[H]

  \centering            \includegraphics[width=0.7\linewidth]{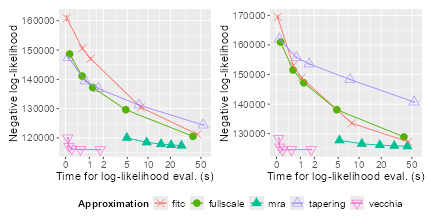} 
  \caption[$N=100,000$, range $0.05$. Log-likelihood evaluation.]
  {Average negative log-likelihood on
simulated data sets with an effective range $0.05$ of and a sample size N=100,000. The true data-generating parameters are used on the left, and two times these values are used on the right.}
\end{figure}

\begin{table}[H]
\centering
\resizebox{\textwidth}{!}{%
\begin{tabular}{llcccccccc}
\toprule
 &  & \textbf{Vecchia} & \textbf{Tapering} & \textbf{FITC} & \textbf{Full-scale} & \textbf{FRK} & \textbf{MRA} & \textbf{SPDE} \\
\midrule
\multirow{2}{*}{\textbf{1}}
& Value & 119974 & 147484 & 160877 & 148634 &  & 120030 &  \\
& (SE) & 54.9 & 225 & 399 & 252 &  & 73.2 &  \\
& Time (s) & 0.0694 & 0.0863 & 0.0356 & 0.127 &  & 4.95 &  \\
\midrule
\multirow{2}{*}{\textbf{2}}
& Value & 117142 & 139428 & 150590 & 141100 &  & 118479 &  \\
& (SE) & 58.3 & 148 & 210 & 188 &  & 72.5 &  \\
& Time (s) & 0.102 & 0.778 & 0.629 & 0.62 &  & 9.56 &  \\
\midrule
\multirow{2}{*}{\textbf{3}}
& Value & 116207 & 136833 & 147030 & 137217 &  & 117896 &  \\
& (SE) & 63 & 125 & 198 & 154 &  & 71.1 &  \\
& Time (s) & 0.202 & 1.59 & 1.08 & 1.2 &  & 15 &  \\
\midrule
\multirow{2}{*}{\textbf{4}}
& Value & 116057 & 131215 & 130417 & 129609 &  & 117572 &  \\
& (SE) & 62 & 82.5 & 163 & 71 &  & 68.7 &  \\
& Time (s) & 0.545 & 7.43 & 7.91 & 4.77 &  & 20.5 &  \\
\midrule
\multirow{2}{*}{\textbf{5}}
& Value & 115963 & 124391 & 121202 & 120531 &  & 117360 &  \\
& (SE) & 62.4 & 46.7 & 79.9 & 55.6 &  & 68.5 &  \\
& Time (s) & 1.74 & 53.4 & 45 & 39.5 &  & 28 &  \\
\bottomrule
\end{tabular}
}
\caption{Mean approximate negative log-likelihood, standard error and time needed
for evaluating the approximate log-likelihood on simulated data sets with an effective range
of $0.05$ of and a sample size N=100,000. Log-likelihoods are evaluated at the true data-generating parameters.}
\end{table}

\begin{table}[H]
\centering
\resizebox{\textwidth}{!}{%
\begin{tabular}{llcccccccc}
\toprule
 &  & \textbf{Vecchia} & \textbf{Tapering} & \textbf{FITC} & \textbf{Full-scale} & \textbf{FRK} & \textbf{MRA} & \textbf{SPDE} \\
\midrule
\multirow{2}{*}{\textbf{1}}
& Value & 128442 & 161978 & 169443 & 160964 &  & 127756 &  \\
& (SE) & 31.5 & 111 & 208 & 132 &  & 55.1 &  \\
& Time (s) & 0.0696 & 0.0877 & 0.034 & 0.126 &  & 5.23 &  \\
\midrule
\multirow{2}{*}{\textbf{2}}
& Value & 125493 & 155774 & 153045 & 151532 &  & 126588 &  \\
& (SE) & 34.5 & 72.6 & 147 & 112 &  & 54.9 &  \\
& Time (s) & 0.103 & 0.781 & 0.629 & 0.619 &  & 10.9 &  \\
\midrule
\multirow{2}{*}{\textbf{3}}
& Value & 124535 & 153578 & 148892 & 147230 &  & 126145 &  \\
& (SE) & 39.9 & 61.1 & 147 & 91.4 &  & 53.2 &  \\
& Time (s) & 0.198 & 1.59 & 1.08 & 1.21 &  & 19.1 &  \\
\midrule
\multirow{2}{*}{\textbf{4}}
& Value & 124604 & 148359 & 133536 & 138116 &  & 125904 &  \\
& (SE) & 43.2 & 39.6 & 113 & 63.3 &  & 51.1 &  \\
& Time (s) & 0.544 & 7.43 & 7.91 & 4.77 &  & 29.6 &  \\
\midrule
\multirow{2}{*}{\textbf{5}}
& Value & 124631 & 140718 & 127253 & 128908 &  & 125747 &  \\
& (SE) & 44.6 & 22.3 & 54.4 & 52.8 &  & 50.9 &  \\
& Time (s) & 1.74 & 53.4 & 45 & 39.5 &  & 43.8 &  \\
\bottomrule
\end{tabular}
}
\caption{Mean approximate negative log-likelihood, standard error and time needed
for evaluating the approximate log-likelihood on simulated data sets with an effective range of $0.05$ of and a sample size N=100,000. Two times the true parameters are used.}
\end{table}

\subsubsection{Parameter estimation}
\begin{figure}[H]
  \centering            \includegraphics[width=\linewidth]{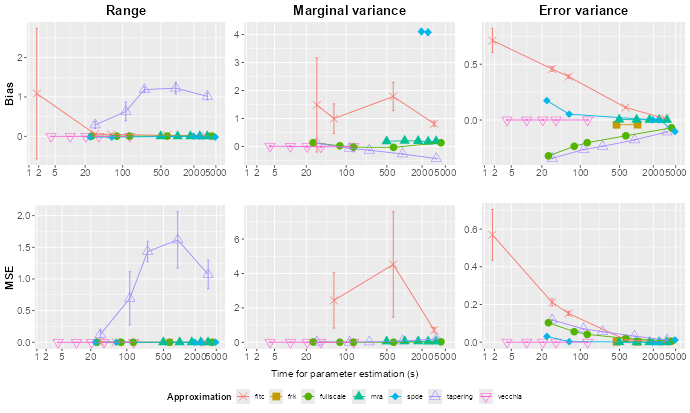} 
  \caption[$N=100,000$, range $0.05$. Parameter estimation accuracy.]
  {Bias and MSE of the estimates for the GP range, GP variance end error term variance on simulated
data sets with an effective range of $0.05$ and a sample size N=100,000,}

\end{figure}

\begin{table}[H]
\centering
\resizebox{\textwidth}{!}{%
\begin{tabular}{llcccccccc}
\toprule
 &  & \textbf{Vecchia} & \textbf{Tapering} & \textbf{FITC} & \textbf{Full-scale} & \textbf{FRK} & \textbf{MRA} & \textbf{SPDE} \\
\midrule
\multirow{2}{*}{\textbf{1}}
& Bias & 1.69e-04 & 0.29 & 1.09 & 3.75e-03 &  & 3.16e-03 & -0.0171 \\
& (SE 1) & 1.36e-04 & 0.0442 & 0.846 & 5.97e-04 &  & 2.64e-04 & 6.16e-05 \\
& MSE & 3.96e-07 & 0.123 & 15.5 & 2.12e-05 &  & 1.13e-05 & 2.94e-04 \\
& (SE 2) & 1.99e-07 & 0.0303 & 14.9 & 5.81e-06 &  & 1.57e-06 & 2.17e-06 \\
& Time (s) & 4.05 & 30.9 & 1.78 & 26.5 & 444 & 495 & 24.7 \\
\midrule
\multirow{2}{*}{\textbf{2}}
& Bias & 5.57e-05 & 0.633 & 0.0563 & 4.7e-03 &  & 2.47e-03 & -0.0182 \\
& (SE 1) & 1.45e-04 & 0.121 & 0.0101 & 4.24e-04 &  & 2.51e-04 & 4.22e-09 \\
\bottomrule
\end{tabular}
}
\caption{Bias and MSE for the GP range, standard errors and time needed for estimating covariance parameters on simulated data sets with an effective range of $0.05$ and a sample size of N=100,000.
}
\end{table}

\begin{table}[H]
\centering
\resizebox{\textwidth}{!}{%
\begin{tabular}{llcccccccc}
\toprule
 &  & \textbf{Vecchia} & \textbf{Tapering} & \textbf{FITC} & \textbf{Full-scale} & \textbf{FRK} & \textbf{MRA} & \textbf{SPDE} \\
\midrule
\multirow{2}{*}{\textbf{1}}
& Bias & 6.68e-03 & 0.112 & 804 & 0.132 &  & 0.182 & 1.34e+09 \\
& (SE 1) & 0.0117 & 9.29e-03 & 783 & 9.31e-03 &  & 0.0292 & 1.34e+09 \\
& MSE & 2.8e-03 & 0.0142 & 1.29e+07 & 0.0192 &  & 0.0492 & 3.57e+19 \\
& (SE 2) & 7.64e-04 & 2.1e-03 & 1.26e+07 & 2.59e-03 &  & 9.51e-03 & 3.57e+19 \\
& Time (s) & 4.05 & 30.9 & 1.78 & 26.5 & 444 & 495 & 24.7 \\
\midrule
\multirow{2}{*}{\textbf{2}}
& Bias & 3.28e-03 & -0.0715 & 1.48 & 0.0218 &  & 0.203 & 9.78e+10 \\
& (SE 1) & 0.0122 & 6.87e-03 & 0.859 & 9.88e-03 &  & 0.0326 & 1.3e+10 \\
\bottomrule
\end{tabular}
}
\caption{Bias and MSE for the GP marginal variance, standard errors and time needed for estimating covariance parameters on simulated data sets with an effective range of $0.05$ and a sample size of N=100,000.
}
\end{table}

\begin{table}[H]
\centering
\resizebox{\textwidth}{!}{%
\begin{tabular}{llcccccccc}
\toprule
 &  & \textbf{Vecchia} & \textbf{Tapering} & \textbf{FITC} & \textbf{Full-scale} & \textbf{FRK} & \textbf{MRA} & \textbf{SPDE} \\
\midrule
\multirow{2}{*}{\textbf{1}}
& Bias & 3.81e-04 & -0.343 & 0.713 & -0.321 & -0.0428 & 4.01e-03 & 0.174 \\
& (SE 1) & 6.13e-04 & 2.08e-03 & 0.0554 & 2.47e-03 & 1.32e-03 & 6.26e-04 & 1.54e-03 \\
& MSE & 7.65e-06 & 0.118 & 0.569 & 0.103 & 1.86e-03 & 2.36e-05 & 0.0303 \\
& (SE 2) & 1.84e-06 & 1.43e-03 & 0.0686 & 1.58e-03 & 1.17e-04 & 6.51e-06 & 5.42e-04 \\
& Time (s) & 4.05 & 30.9 & 1.78 & 26.5 & 444 & 495 & 24.7 \\
\midrule
\multirow{2}{*}{\textbf{2}}
& Bias & 3.12e-05 & -0.265 & 0.457 & -0.236 & -0.0416 & 2.56e-03 & 0.0522 \\
& (SE 1) & 5.91e-04 & 1.36e-03 & 0.0111 & 2.22e-03 & 1.45e-03 & 6.18e-04 & 8.72e-04 \\
\bottomrule
\end{tabular}
}
\caption{Bias and MSE for the error term variance, standard errors and time needed for estimating covariance parameters on simulated data sets with an effective range of $0.05$ and a sample size of N=100,000.
}
\end{table}

\subsubsection{Prediction}
\begin{figure}[H]

  \centering            \includegraphics[width=\linewidth]{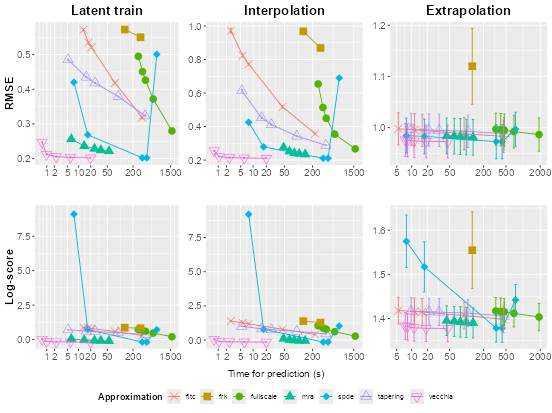} 
  \caption[$N=100,000$, range $0.05$. Prediction accuracy.]
  {Average RMSE and log-score on simulated data sets with
an effective range of $0.05$ and a sample size N=100,000. Predictions are done using the true data-generating parameters.}
\end{figure}

\begin{table}[H]
\centering
\resizebox{\textwidth}{!}{%
\begin{tabular}{llcccccccc}
\toprule
 &  & \textbf{Vecchia} & \textbf{Tapering} & \textbf{FITC} & \textbf{Full-scale} & \textbf{FRK} & \textbf{MRA} & \textbf{SPDE} \\
\midrule
\multirow{2}{*}{\textbf{1}}
& RMSE & 0.247 & 0.485 & 0.572 & 0.495 & 0.572 & 0.256 & 0.42 \\
& (SE 1) & 7.11e-04 & 8.65e-04 & 1.64e-03 & 9.96e-04 & 1.75e-03 & 9.54e-04 & 2.04e-03 \\
& Log-score & 0.0179 & 0.704 & 0.864 & 0.723 & 0.862 & 0.0391 & 9.1 \\
& (SE 2) & 2.81e-03 & 1.35e-03 & 2.8e-03 & 1.89e-03 & 3.11e-03 & 3.88e-03 & 0.0894 \\
& Time (s) & 0.552 & 5.22 & 13.2 & 267 & 128 & 6.42 & 7.53 \\
\midrule
\multirow{2}{*}{\textbf{2}}
& RMSE & 0.212 & 0.435 & 0.535 & 0.451 & 0.55 & 0.236 & 0.268 \\
& (SE 1) & 1.03e-03 & 6.19e-04 & 1.12e-03 & 7.89e-04 & 1.82e-03 & 9.66e-04 & 1.36e-03 \\
\bottomrule
\end{tabular}
}
\caption{Average RMSE and log-score on the training set, standard errors and time
needed for making predictions on the training set on simulated data sets with an effective
range of $0.05$ and a sample size of N=100,000.
}
\end{table}

\begin{table}[H]
\centering
\resizebox{\textwidth}{!}{%
\begin{tabular}{llcccccccc}
\toprule
 &  & \textbf{Vecchia} & \textbf{Tapering} & \textbf{FITC} & \textbf{Full-scale} & \textbf{FRK} & \textbf{MRA} & \textbf{SPDE} \\
\midrule
\multirow{2}{*}{\textbf{1}}
& RMSE & 0.255 & 0.614 & 0.972 & 0.654 & 0.968 & 0.276 & 0.426 \\
& (SE 1) & 1.05e-03 & 4.06e-03 & 7.17e-03 & 4.32e-03 & 7.29e-03 & 1.27e-03 & 2.14e-03 \\
& Log-score & 0.0492 & 1.01 & 1.39 & 1.05 & 1.39 & 0.103 & 9.2 \\
& (SE 2) & 3.97e-03 & 2.99e-03 & 7.04e-03 & 3.75e-03 & 7.58e-03 & 4.63e-03 & 0.0962 \\
& Time (s) & 0.645 & 5.33 & 2.7 & 263 & 127 & 47.8 & 7.86 \\
\midrule
\multirow{2}{*}{\textbf{2}}
& RMSE & 0.221 & 0.453 & 0.824 & 0.513 & 0.869 & 0.252 & 0.279 \\
& (SE 1) & 1.35e-03 & 2.44e-03 & 4.49e-03 & 3.13e-03 & 5.99e-03 & 1.3e-03 & 1.45e-03 \\
\bottomrule
\end{tabular}
}
\caption{Average RMSE and log-score on the test “interpolation” set, standard errors and time needed for making predictions on the test “interpolation” set on simulated
data sets with an effective range of $0.05$ and a sample size of N=100,000.
}
\end{table}

\begin{table}[H]
\centering
\resizebox{\textwidth}{!}{%
\begin{tabular}{llcccccccc}
\toprule
 &  & \textbf{Vecchia} & \textbf{Tapering} & \textbf{FITC} & \textbf{Full-scale} & \textbf{FRK} & \textbf{MRA} & \textbf{SPDE} \\
\midrule
\multirow{2}{*}{\textbf{1}}
& RMSE & 0.976 & 0.997 & 0.998 & 0.997 & 1.12 & 0.984 & 0.984 \\
& (SE 1) & 0.0164 & 0.016 & 0.0159 & 0.016 & 0.0378 & 0.0178 & 0.0173 \\
& Log-score & 1.38 & 1.42 & 1.42 & 1.42 & 1.55 & 1.4 & 1.57 \\
& (SE 2) & 0.0154 & 0.0155 & 0.0154 & 0.0155 & 0.0444 & 0.0171 & 0.0303 \\
& Time (s) & 7.46 & 9.75 & 5.4 & 331 & 127 & 44.4 & 7.85 \\
\midrule
\multirow{2}{*}{\textbf{2}}
& RMSE & 0.975 & 0.995 & 0.996 & 0.996 & 26.3 & 0.983 & 0.983 \\
& (SE 1) & 0.0164 & 0.016 & 0.0161 & 0.0161 & 3.19 & 0.0178 & 0.0173 \\
& Log-score & 1.38 & 1.42 & 1.42 & 1.42 & 33.2 & 1.39 & 1.52 \\
& (SE 2) & 0.0154 & 0.0155 & 0.0156 & 0.0155 & 6.39 & 0.0173 & 0.0289 \\
& Time (s) & 8.31 & 21.1 & 11.3 & 404 & 294 & 60.4 & 17.4 \\
\midrule
\multirow{2}{*}{\textbf{3}}
& RMSE & 0.974 & 0.995 & 0.995 & 0.995 &  & 0.983 & 0.973 \\
& (SE 1) & 0.0164 & 0.0161 & 0.0161 & 0.016 &  & 0.018 & 0.0171 \\
& Log-score & 1.38 & 1.41 & 1.41 & 1.42 &  & 1.39 & 1.38 \\
& (SE 2) & 0.0154 & 0.0155 & 0.0155 & 0.0155 &  & 0.0175 & 0.0162 \\
& Time (s) & 11.1 & 32.4 & 15.8 & 465 &  & 77.7 & 336 \\
\midrule
\multirow{2}{*}{\textbf{4}}
& RMSE & 0.973 & 0.993 & 0.988 & 0.992 &  & 0.981 & 0.973 \\
& (SE 1) & 0.0163 & 0.0161 & 0.0161 & 0.0162 &  & 0.0178 & 0.0171 \\
& Log-score & 1.38 & 1.41 & 1.41 & 1.41 &  & 1.39 & 1.38 \\
& (SE 2) & 0.0153 & 0.0156 & 0.0154 & 0.0156 &  & 0.0175 & 0.0164 \\
& Time (s) & 18.7 & 106 & 90.8 & 690 &  & 99 & 427 \\
\midrule
\multirow{2}{*}{\textbf{5}}
& RMSE & 0.973 & 0.988 & 0.983 & 0.986 &  & 0.981 & 0.997 \\
& (SE 1) & 0.0163 & 0.0163 & 0.0163 & 0.0165 &  & 0.0177 & 0.0168 \\
& Log-score & 1.38 & 1.41 & 1.4 & 1.4 &  & 1.39 & 1.44 \\
& (SE 2) & 0.0153 & 0.0156 & 0.0155 & 0.0158 &  & 0.0174 & 0.0178 \\
& Time (s) & 46.4 & 463 & 452 & 1908 &  & 131 & 737 \\
\bottomrule
\end{tabular}
}
\caption{Average RMSE and log-score on the test “extrapolation” set, standard errors and time needed for making predictions on the test “extrapolation” set on simulated
data sets with an effective range of $0.05$ and a sample size of N=100,000.
}
\end{table}

\newpage
\subsection{Simulated data with $N=100,000$ and an anisotropic Matérn covariance function}\label{appendix_100000_02_ard}

\subsubsection{Log-likelihood evaluation}

\begin{table}[H]
\centering
\resizebox{\textwidth}{!}{%
\begin{tabular}{llcccccccc}
\toprule
 &  & \textbf{Vecchia} & \textbf{Tapering} & \textbf{FITC} & \textbf{Full-scale} & \textbf{FRK} & \textbf{MRA} & \textbf{SPDE} \\
\midrule
\multirow{2}{*}{\textbf{1}}
& Value & 117135 & 147078 & 156948 & 145501 &  & 119206 &  \\
& (SE) & 70.8 & 301 & 470 & 279 &  & 170 &  \\
& Time (s) & 0.241 & 0.0906 & 0.0359 & 0.138 &  & 8.81 &  \\
\midrule
\multirow{2}{*}{\textbf{2}}
& Value & 113699 & 138954 & 133931 & 134010 &  & 118143 &  \\
& (SE) & 73.8 & 190 & 276 & 202 &  & 165 &  \\
& Time (s) & 0.364 & 0.787 & 0.631 & 0.641 &  & 18 &  \\
\midrule
\multirow{2}{*}{\textbf{3}}
& Value & 112072 & 136320 & 129197 & 129703 &  & 117684 &  \\
& (SE) & 79.2 & 157 & 187 & 146 &  & 166 &  \\
& Time (s) & 0.648 & 1.61 & 1.1 & 1.22 &  & 32.9 &  \\
\midrule
\multirow{2}{*}{\textbf{4}}
& Value & 111606 & 130570 & 115399 & 121472 &  & 117442 &  \\
& (SE) & 82.2 & 93.8 & 113 & 103 &  & 167 &  \\
& Time (s) & 1.49 & 7.23 & 5.42 & 3.56 &  & 49 &  \\
\midrule
\multirow{2}{*}{\textbf{5}}
& Value & 111518 & 123381 & 112241 & 114232 &  & 117292 &  \\
& (SE) & 82.4 & 47.4 & 93.7 & 80 &  & 168 &  \\
& Time (s) & 4.18 & 54.5 & 44.9 & 39.7 &  & 70.2 &  \\
\bottomrule
\end{tabular}
}
\caption{Mean approximate negative log-likelihood, standard error and time needed
for evaluating the approximate log-likelihood on simulated data sets with an anisotropic Matérn covariance function and a sample size N=100,000. Log-likelihoods are evaluated at the true data-generating parameters.}
\end{table}

\begin{table}[H]
\centering
\resizebox{\textwidth}{!}{%
\begin{tabular}{llcccccccc}
\toprule
 &  & \textbf{Vecchia} & \textbf{Tapering} & \textbf{FITC} & \textbf{Full-scale} & \textbf{FRK} & \textbf{MRA} & \textbf{SPDE} \\
\midrule
\multirow{2}{*}{\textbf{1}}
& Value & 126348 & 161795 & 163510 & 155295 &  & 125643 &  \\
& (SE) & 36.3 & 150 & 245 & 190 &  & 105 &  \\
& Time (s) & 0.243 & 0.0931 & 0.0386 & 0.133 &  & 8.8 &  \\
\midrule
\multirow{2}{*}{\textbf{2}}
& Value & 122851 & 155576 & 136085 & 141218 &  & 124856 &  \\
& (SE) & 38.6 & 94.4 & 221 & 189 &  & 100 &  \\
& Time (s) & 0.365 & 0.787 & 0.632 & 0.64 &  & 18 &  \\
\midrule
\multirow{2}{*}{\textbf{3}}
& Value & 121171 & 153369 & 132056 & 136918 &  & 124533 &  \\
& (SE) & 42.6 & 77.8 & 141 & 169 &  & 101 &  \\
& Time (s) & 0.65 & 1.62 & 1.1 & 1.23 &  & 32.9 &  \\
\midrule
\multirow{2}{*}{\textbf{4}}
& Value & 120711 & 148098 & 122834 & 128984 &  & 124365 &  \\
& (SE) & 47.2 & 45.9 & 70 & 111 &  & 102 &  \\
& Time (s) & 1.49 & 7.23 & 5.42 & 3.57 &  & 49 &  \\
\midrule
\multirow{2}{*}{\textbf{5}}
& Value & 120793 & 140292 & 121236 & 122993 &  & 124260 &  \\
& (SE) & 49.2 & 23.6 & 55.9 & 66.6 &  & 102 &  \\
& Time (s) & 4.18 & 54.5 & 44.9 & 39.7 &  & 70 &  \\
\bottomrule
\end{tabular}
}
\caption{Mean approximate negative log-likelihood, standard error and time needed
for evaluating the approximate log-likelihood on simulated data sets with an anisotropic Matérn covariance function and a sample size N=100,000. Log-likelihoods are evaluated at two-times the true data-generating parameters.}
\end{table}

\subsubsection{Parameter estimation}

\begin{table}[H]
\centering
\resizebox{\textwidth}{!}{%
\begin{tabular}{llcccccccc}
\toprule
 &  & \textbf{Vecchia} & \textbf{Tapering} & \textbf{FITC} & \textbf{Full-scale} & \textbf{FRK} & \textbf{MRA} & \textbf{SPDE} \\
\midrule
\multirow{2}{*}{\textbf{1}}
& Bias & 4.98e-05 & 0.799 & 2.47 & 6.93e-03 &  & 0.192 & -0.0181 \\
& (SE 1) & 3.02e-04 & 0.0654 & 1.46 & 2.23e-03 &  & 0.0327 & 1.52e-04 \\
& MSE & 1.82e-06 & 0.724 & 48.7 & 1.48e-04 &  & 0.0571 & 3.28e-04 \\
& (SE 2) & 4.58e-07 & 0.0848 & 38.3 & 6.18e-05 &  & 0.0215 & 5.08e-06 \\
& Time (s) & 9.06 & 45.9 & 3.33 & 29.2 & 475 & 496 & 131 \\
\midrule
\multirow{2}{*}{\textbf{2}}
& Bias & 1.84e-04 & 0.767 & 0.0508 & 5.57e-03 &  & 0.214 & -0.0182 \\
& (SE 1) & 2.42e-04 & 0.0384 & 5.69e-03 & 8.97e-04 &  & 0.0385 & 1.94e-05 \\
\bottomrule
\end{tabular}
}
\caption{Bias and MSE for the x-axis GP range, standard errors and time needed for estimating covariance parameters on simulated data sets with an anisotropic Matérn covariance function and a sample size N=100,000.
}
\end{table}

\begin{table}[H]
\centering
\resizebox{\textwidth}{!}{%
\begin{tabular}{llcccccccc}
\toprule
 &  & \textbf{Vecchia} & \textbf{Tapering} & \textbf{FITC} & \textbf{Full-scale} & \textbf{FRK} & \textbf{MRA} & \textbf{SPDE} \\
\midrule
\multirow{2}{*}{\textbf{1}}
& Bias & -1.21e-04 & 0.748 & 2.67 & 2.75e-03 &  & -0.0419 & -0.0163 \\
& (SE 1) & 1.08e-03 & 0.0652 & 1.61 & 3.94e-03 &  & 1.6e-03 & 9.45e-04 \\
& MSE & 2.34e-05 & 0.644 & 59 & 3.18e-04 &  & 1.8e-03 & 2.83e-04 \\
& (SE 2) & 8.09e-06 & 0.0777 & 50.4 & 1.8e-04 &  & 1.29e-04 & 3.39e-05 \\
& Time (s) & 9.06 & 45.9 & 3.33 & 29.2 & 475 & 496 & 131 \\
\midrule
\multirow{2}{*}{\textbf{2}}
& Bias & -2.6e-04 & 0.726 & 0.223 & 6.83e-03 &  & -0.0427 & -0.0225 \\
& (SE 1) & 9.57e-04 & 0.0392 & 0.0243 & 3.09e-03 &  & 1.69e-03 & 6.02e-04 \\
\bottomrule
\end{tabular}
}
\caption{Bias and MSE for the y-axis GP range, standard errors and time needed for estimating covariance parameters on simulated data sets with an anisotropic Matérn covariance function and a sample size N=100,000.
}
\end{table}

\begin{table}[H]
\centering
\resizebox{\textwidth}{!}{%
\begin{tabular}{llcccccccc}
\toprule
 &  & \textbf{Vecchia} & \textbf{Tapering} & \textbf{FITC} & \textbf{Full-scale} & \textbf{FRK} & \textbf{MRA} & \textbf{SPDE} \\
\midrule
\multirow{2}{*}{\textbf{1}}
& Bias & -4.71e-03 & 0.0957 & 20928 & 0.0986 &  & -0.255 & 8.08e+07 \\
& (SE 1) & 0.0177 & 0.0118 & 19819 & 0.0132 &  & 7.46e-03 & 2.94e+07 \\
& MSE & 6.27e-03 & 0.0119 & 8.29e+09 & 0.0132 &  & 0.0659 & 2.3e+16 \\
& (SE 2) & 1.94e-03 & 2.69e-03 & 8.08e+09 & 3.09e-03 &  & 3.79e-03 & 1.28e+16 \\
& Time (s) & 9.06 & 45.9 & 3.33 & 29.2 & 475 & 496 & 131 \\
\midrule
\multirow{2}{*}{\textbf{2}}
& Bias & 5.8e-03 & -0.089 & 3.52 & 0.0336 &  & -0.255 & 2.67e+08 \\
& (SE 1) & 0.0199 & 9.77e-03 & 0.943 & 0.0196 &  & 7.28e-03 & 2.62e+08 \\
\bottomrule
\end{tabular}
}
\caption{Bias and MSE for the GP marginal variance, standard errors and time needed for estimating covariance parameters on simulated data sets an anisotropic Matérn covariance function and a sample size N=100,000.
}
\end{table}

\begin{table}[H]
\centering
\resizebox{\textwidth}{!}{%
\begin{tabular}{llcccccccc}
\toprule
 &  & \textbf{Vecchia} & \textbf{Tapering} & \textbf{FITC} & \textbf{Full-scale} & \textbf{FRK} & \textbf{MRA} & \textbf{SPDE} \\
\midrule
\multirow{2}{*}{\textbf{1}}
& Bias & -3.24e-04 & -0.34 & 0.568 & -0.257 & -0.0255 & 0.0531 & 0.175 \\
& (SE 1) & 7.32e-04 & 2.23e-03 & 0.0402 & 5.96e-03 & 1.49e-03 & 1.5e-03 & 3.62e-03 \\
& MSE & 1.08e-05 & 0.116 & 0.355 & 0.0667 & 6.93e-04 & 2.86e-03 & 0.0309 \\
& (SE 2) & 4.21e-06 & 1.52e-03 & 0.0422 & 3.03e-03 & 7.5e-05 & 1.56e-04 & 1.28e-03 \\
& Time (s) & 9.06 & 45.9 & 3.33 & 29.2 & 475 & 496 & 131 \\
\midrule
\multirow{2}{*}{\textbf{2}}
& Bias & -4.89e-04 & -0.262 & 0.194 & -0.157 & -0.0265 & 0.0483 & 0.116 \\
& (SE 1) & 6.94e-04 & 2.26e-03 & 4.95e-03 & 3.81e-03 & 1.61e-03 & 1.51e-03 & 2.27e-03 \\
\bottomrule
\end{tabular}
}
\caption{Bias and MSE for the error term variance, standard errors and time needed for estimating covariance parameters on simulated data sets with an anisotropic Matérn covariance function and a sample size N=100,000.
}
\end{table}

\subsubsection{Prediction}

\begin{table}[H]
\centering
\resizebox{\textwidth}{!}{%
\begin{tabular}{llcccccccc}
\toprule
 &  & \textbf{Vecchia} & \textbf{Tapering} & \textbf{FITC} & \textbf{Full-scale} & \textbf{FRK} & \textbf{MRA} & \textbf{SPDE} \\
\midrule
\multirow{2}{*}{\textbf{1}}
& RMSE & 0.226 & 0.484 & 0.561 & 0.483 & 0.56 & 0.261 & 0.418 \\
& (SE 1) & 3.91e-04 & 1.19e-03 & 2.25e-03 & 1.34e-03 & 2.77e-03 & 2.85e-03 & 5.71e-03 \\
& Log-score & -0.0691 & 0.702 & 0.86 & 0.709 & 0.838 & 0.668 & 19.8 \\
& (SE 2) & 1.71e-03 & 1.87e-03 & 4.83e-03 & 3.03e-03 & 4.95e-03 & 0.0468 & 0.47 \\
& Time (s) & 0.432 & 5.22 & 13.2 & 267 & 131 & 12.4 & 10.5 \\
\midrule
\multirow{2}{*}{\textbf{2}}
& RMSE & 0.171 & 0.433 & 0.447 & 0.413 & 0.504 & 0.246 & 0.345 \\
& (SE 1) & 4.61e-04 & 6.81e-04 & 1.57e-03 & 8.31e-04 & 2.12e-03 & 3.37e-03 & 3.98e-03 \\
\bottomrule
\end{tabular}
}
\caption{Average RMSE and log-score on the training set, standard errors and time
needed for making predictions on the training set on simulated data sets with an anisotropic Matérn covariance function and a sample size N=100,000.
}
\end{table}

\begin{table}[H]
\centering
\resizebox{\textwidth}{!}{%
\begin{tabular}{llcccccccc}
\toprule
 &  & \textbf{Vecchia} & \textbf{Tapering} & \textbf{FITC} & \textbf{Full-scale} & \textbf{FRK} & \textbf{MRA} & \textbf{SPDE} \\
\midrule
\multirow{2}{*}{\textbf{1}}
& RMSE & 0.228 & 0.615 & 0.938 & 0.629 & 0.919 & 0.272 & 0.422 \\
& (SE 1) & 5.67e-04 & 7.19e-03 & 0.0105 & 6.13e-03 & 0.0129 & 2.75e-03 & 5.67e-03 \\
& Log-score & -0.0612 & 1.01 & 1.37 & 1.02 & 1.33 & 0.741 & 19.9 \\
& (SE 2) & 2.43e-03 & 5.26e-03 & 0.0117 & 5.69e-03 & 0.0142 & 0.0467 & 0.461 \\
& Time (s) & 0.492 & 5.27 & 2.7 & 263 & 129 & 54.4 & 10.7 \\
\midrule
\multirow{2}{*}{\textbf{2}}
& RMSE & 0.174 & 0.449 & 0.576 & 0.449 & 0.718 & 0.254 & 0.348 \\
& (SE 1) & 6.23e-04 & 3.64e-03 & 4.45e-03 & 3.42e-03 & 6.03e-03 & 3.17e-03 & 3.95e-03 \\
\bottomrule
\end{tabular}
}
\caption{Average RMSE and log-score on the test “interpolation” set, standard errors and time needed for making predictions on the test “interpolation” set on simulated
data sets with an anisotropic Matérn covariance function and a sample size N=100,000.
}
\end{table}

\begin{table}[H]
\centering
\resizebox{\textwidth}{!}{%
\begin{tabular}{llcccccccc}
\toprule
 &  & \textbf{Vecchia} & \textbf{Tapering} & \textbf{FITC} & \textbf{Full-scale} & \textbf{FRK} & \textbf{MRA} & \textbf{SPDE} \\
\midrule
\multirow{2}{*}{\textbf{1}}
& RMSE & 0.902 & 0.972 & 0.971 & 0.968 & 8 & 0.946 & 1.01 \\
& (SE 1) & 0.0186 & 0.0228 & 0.0232 & 0.022 & 6.58 & 0.0237 & 0.0192 \\
& Log-score & 1.29 & 1.39 & 1.39 & 1.39 & 6.13 & 1.4 & 2 \\
& (SE 2) & 0.0184 & 0.022 & 0.0225 & 0.0212 & 4.05 & 0.0378 & 0.102 \\
& Time (s) & 5.2 & 9.77 & 5.4 & 331 & 129 & 52.8 & 10.7 \\
\midrule
\multirow{2}{*}{\textbf{2}}
& RMSE & 0.901 & 0.97 & 0.948 & 0.96 & 482 & 0.941 & 0.928 \\
& (SE 1) & 0.0184 & 0.0228 & 0.0194 & 0.0214 & 446 & 0.021 & 0.0177 \\
& Log-score & 1.28 & 1.39 & 1.37 & 1.38 & 23.9 & 1.4 & 1.55 \\
& (SE 2) & 0.0183 & 0.022 & 0.0188 & 0.0207 & 6.8 & 0.0336 & 0.0331 \\
& Time (s) & 5.96 & 21.8 & 11.4 & 407 & 292 & 93.3 & 21.6 \\
\midrule
\multirow{2}{*}{\textbf{3}}
& RMSE & 0.898 & 0.969 & 0.939 & 0.958 &  & 0.941 & 0.915 \\
& (SE 1) & 0.018 & 0.0228 & 0.019 & 0.0217 &  & 0.021 & 0.0181 \\
& Log-score & 1.28 & 1.39 & 1.36 & 1.38 &  & 1.41 & 1.4 \\
& (SE 2) & 0.0178 & 0.022 & 0.0185 & 0.0212 &  & 0.0343 & 0.0281 \\
& Time (s) & 8.14 & 33.5 & 15.9 & 465 &  & 165 & 94.4 \\
\midrule
\multirow{2}{*}{\textbf{4}}
& RMSE & 0.893 & 0.967 & 0.916 & 0.941 &  & 0.942 & 0.899 \\
& (SE 1) & 0.0176 & 0.0227 & 0.0202 & 0.0199 &  & 0.0208 & 0.0183 \\
& Log-score & 1.27 & 1.39 & 1.32 & 1.36 &  & 1.42 & 1.32 \\
& (SE 2) & 0.0176 & 0.0219 & 0.0203 & 0.0196 &  & 0.0347 & 0.0261 \\
& Time (s) & 14.7 & 108 & 61.9 & 480 &  & 273 & 174 \\
\midrule
\multirow{2}{*}{\textbf{5}}
& RMSE & 0.892 & 0.96 & 0.905 & 0.925 &  & 0.944 & 0.897 \\
& (SE 1) & 0.0176 & 0.0225 & 0.019 & 0.0184 &  & 0.0207 & 0.0185 \\
& Log-score & 1.26 & 1.38 & 1.29 & 1.33 &  & 1.42 & 1.3 \\
& (SE 2) & 0.0177 & 0.0216 & 0.0199 & 0.0179 &  & 0.0341 & 0.0243 \\
& Time (s) & 37.5 & 491 & 451 & 1942 &  & 386 & 448 \\
\bottomrule
\end{tabular}
}
\caption{Average RMSE and log-score on the test “extrapolation” set, standard errors and time needed for making predictions on the test “extrapolation” set on simulated
data sets with an anisotropic Matérn covariance function and a sample size N=100,000.
}
\end{table}

\newpage
\subsection{Simulated data with $N=100,000$, effective range of $0.2$ and noise variance of $0.1$}\label{appendix_100000_02_n01}

\begin{table}[ht!]
\centering
\resizebox{\textwidth}{!}{%
\begin{tabular}{l|l|cccccc}
  & \textbf{Tuning parameter} & \textbf{} & \textbf{} & \textbf{} & \textbf{} & \textbf{} & \textbf{} \\
\midrule
\textbf{Vecchia} & nb. neighbours & 5 & 10 & 20 & 40 & 80 \\
\hline
\textbf{Tapering} & avg. nb. non-zeros / row & 8 & 17 & 22 & 40 & 111\\
\hline
\textbf{FITC} & nb. inducing points & 20 & 200 & 275 & 900 & 2000\\
\hline
\multirow{2}{*}{\textbf{Full-scale}} & nb. inducing points & 26 & 91 & 122 & 250 & 650 \\
 & avg. nb. non-zeros / row & 6 & 11 & 15 & 27 & 76 \\
\hline
\textbf{FRK} & nb. resolutions & 1  & 2 & exceeds time limit &   &  &  \\
\hline
\textbf{MRA} & nb. knots per partition & 1 & 2 & 3 & 4 & 5 \\
\hline
\textbf{SPDE} & max edge & 0.04 & 0.02 & 0.005 & 0.004 & 0.003   \\
\bottomrule
\end{tabular}%
}
\caption{Tuning parameters chosen for the comparison on simulated data sets with an effective range of $0.2$, noise variance of $0.1$ and a sample size of $N=100,000$.}
\end{table}

\subsubsection{Log-likelihood evaluation}
\begin{figure}[H]

  \centering            \includegraphics[width=0.7\linewidth]{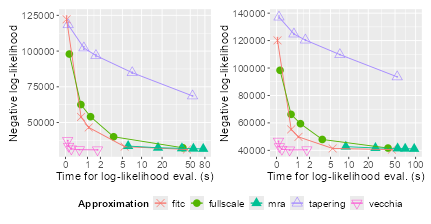} 
  \caption[$N=100,000$, range $0.2$, noise variance $0.1$. Log-likelihood evaluation.]
  {Average negative log-likelihood on
simulated data sets with an effective range of $0.2$, error variance of $0.1$ and a sample size N=100,000. The true data-generating parameters are used on the left, and two times these values are used on the right.}
\end{figure}

\begin{table}[H]
\centering
\resizebox{\textwidth}{!}{%
\begin{tabular}{llcccccccc}
\toprule
 &  & \textbf{Vecchia} & \textbf{Tapering} & \textbf{FITC} & \textbf{Full-scale} & \textbf{FRK} & \textbf{MRA} & \textbf{SPDE} \\
\midrule
\multirow{2}{*}{\textbf{1}}
& Value & 37499 & 118410 & 122210 & 97933 &  & 33593 &  \\
& (SE) & 52.7 & 897 & 1239 & 596 &  & 84 &  \\
& Time (s) & 0.0684 & 0.0861 & 0.0339 & 0.126 &  & 6.2 &  \\
\midrule
\multirow{2}{*}{\textbf{2}}
& Value & 33452 & 102201 & 53932 & 62587 &  & 32401 &  \\
& (SE) & 56.6 & 542 & 280 & 300 &  & 86.7 &  \\
& Time (s) & 0.102 & 0.782 & 0.63 & 0.622 &  & 17.5 &  \\
\midrule
\multirow{2}{*}{\textbf{3}}
& Value & 31590 & 96866 & 46607 & 54096 &  & 31986 &  \\
& (SE) & 65.1 & 444 & 302 & 232 &  & 87.9 &  \\
& Time (s) & 0.201 & 1.61 & 1.08 & 1.2 &  & 38.3 &  \\
\midrule
\multirow{2}{*}{\textbf{4}}
& Value & 31038 & 84893 & 33163 & 40076 &  & 31758 &  \\
& (SE) & 77 & 261 & 167 & 128 &  & 85.6 &  \\
& Time (s) & 0.546 & 7.18 & 5.44 & 3.55 &  & 55.4 &  \\
\midrule
\multirow{2}{*}{\textbf{5}}
& Value & 30910 & 68639 & 31190 & 32370 &  & 31616 &  \\
& (SE) & 79.8 & 99.1 & 107 & 91.3 &  & 86.2 &  \\
& Time (s) & 1.74 & 53.8 & 45 & 39.4 &  & 75.8 &  \\
\bottomrule
\end{tabular}
}
\caption{Mean approximate negative log-likelihood, standard error and time needed
for evaluating the approximate log-likelihood on simulated data sets with an effective range of $0.2$, error variance of $0.1$ and a sample size N=100,000. Log-likelihoods are evaluated at the true data-generating parameters.}
\end{table}

\begin{table}[H]
\centering
\resizebox{\textwidth}{!}{%
\begin{tabular}{llcccccccc}
\toprule
 &  & \textbf{Vecchia} & \textbf{Tapering} & \textbf{FITC} & \textbf{Full-scale} & \textbf{FRK} & \textbf{MRA} & \textbf{SPDE} \\
\midrule
\multirow{2}{*}{\textbf{1}}
& Value & 46690 & 137164 & 120154 & 98351 &  & 42693 &  \\
& (SE) & 30.1 & 448 & 1143 & 705 &  & 78.5 &  \\
& Time (s) & 0.0693 & 0.0867 & 0.0353 & 0.126 &  & 8.97 &  \\
\midrule
\multirow{2}{*}{\textbf{2}}
& Value & 42668 & 124798 & 55259 & 66225 &  & 41732 &  \\
& (SE) & 33.2 & 270 & 224 & 446 &  & 80.9 &  \\
& Time (s) & 0.103 & 0.784 & 0.629 & 0.623 &  & 25.6 &  \\
\midrule
\multirow{2}{*}{\textbf{3}}
& Value & 40797 & 120395 & 49796 & 59423 &  & 41380 &  \\
& (SE) & 41.3 & 221 & 226 & 347 &  & 82 &  \\
& Time (s) & 0.198 & 1.61 & 1.08 & 1.22 &  & 54.6 &  \\
\midrule
\multirow{2}{*}{\textbf{4}}
& Value & 40280 & 109795 & 41377 & 47921 &  & 41191 &  \\
& (SE) & 55.1 & 129 & 105 & 166 &  & 78.9 &  \\
& Time (s) & 0.544 & 7.18 & 5.44 & 3.57 &  & 70.4 &  \\
\midrule
\multirow{2}{*}{\textbf{5}}
& Value & 40366 & 93557 & 40508 & 41666 &  & 41067 &  \\
& (SE) & 65.5 & 48.7 & 81.7 & 97.9 &  & 78.6 &  \\
& Time (s) & 1.74 & 53.8 & 45 & 39.4 &  & 95.3 &  \\
\bottomrule
\end{tabular}
}
\caption{Mean approximate negative log-likelihood, standard error and time needed
for evaluating the approximate log-likelihood on simulated data sets with an effective range of $0.2$, error variance of $0.1$ and a sample size N=100,000. Log-likelihoods are evaluated at two-times the true data-generating parameters.}
\end{table}

\subsubsection{Parameter estimation}

\begin{figure}[H]
  \centering            \includegraphics[width=\linewidth]{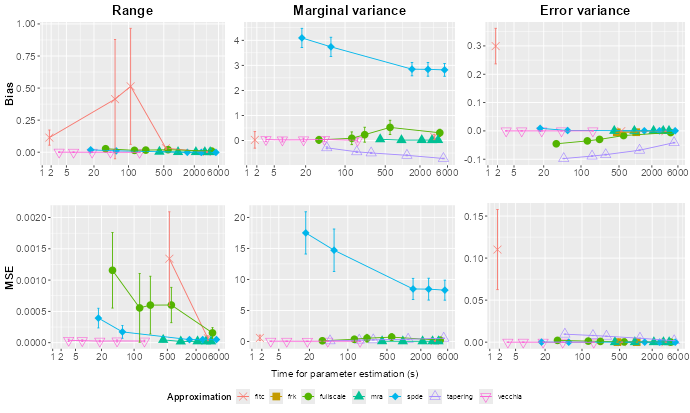} 
  \caption[$N=100,000$, range $0.2$, noise variance $0.1$. Parameter estimation accuracy.]
  {Bias and MSE of the estimates for the GP range, GP variance end error term variance on simulated
data sets with an effective range of $0.2$, error variance of $0.1$ and a sample size N=100,000.}

\end{figure}

\begin{table}[H]
\centering
\resizebox{\textwidth}{!}{%
\begin{tabular}{llcccccccc}
\toprule
 &  & \textbf{Vecchia} & \textbf{Tapering} & \textbf{FITC} & \textbf{Full-scale} & \textbf{FRK} & \textbf{MRA} & \textbf{SPDE} \\
\midrule
\multirow{2}{*}{\textbf{1}}
& Bias & 5.4e-04 & 2.35 & 0.114 & 0.0269 &  & 3.57e-03 & 0.0183 \\
& (SE 1) & 1.37e-03 & 0.311 & 0.0302 & 4.65e-03 &  & 1.3e-03 & 1.75e-03 \\
& MSE & 3.81e-05 & 7.48 & 0.0311 & 1.16e-03 &  & 4.48e-05 & 3.93e-04 \\
& (SE 2) & 9.84e-06 & 1.86 & 0.0153 & 3.08e-04 &  & 1.59e-05 & 8.02e-05 \\
& Time (s) & 3.31 & 47.8 & 1.77 & 34.6 & 456 & 409 & 17.1 \\
\midrule
\multirow{2}{*}{\textbf{2}}
& Bias & -2.04e-05 & 1.81 & 0.414 & 0.0145 &  & 1.04e-03 & 0.0103 \\
& (SE 1) & 1.27e-03 & 0.134 & 0.237 & 4.16e-03 &  & 9.81e-04 & 1.87e-03 \\
\bottomrule
\end{tabular}
}
\caption{Bias and MSE for the GP range, standard errors and time needed for estimating covariance parameters on simulated data sets with an effective range of $0.2$, error variance of $0.1$ and a sample size N=100,000.
}
\end{table}

\begin{table}[H]
\centering
\resizebox{\textwidth}{!}{%
\begin{tabular}{llcccccccc}
\toprule
 &  & \textbf{Vecchia} & \textbf{Tapering} & \textbf{FITC} & \textbf{Full-scale} & \textbf{FRK} & \textbf{MRA} & \textbf{SPDE} \\
\midrule
\multirow{2}{*}{\textbf{1}}
& Bias & 0.0377 & -0.292 & 0.025 & 0.0253 &  & 0.0487 & 4.1 \\
& (SE 1) & 0.0426 & 0.0197 & 0.169 & 0.0722 &  & 0.0378 & 0.196 \\
& MSE & 0.0377 & 0.0928 & 0.569 & 0.105 &  & 0.0296 & 17.5 \\
& (SE 2) & 7.22e-03 & 0.0113 & 0.229 & 0.0388 &  & 9.02e-03 & 1.74 \\
& Time (s) & 3.31 & 47.8 & 1.77 & 34.6 & 456 & 409 & 17.1 \\
\midrule
\multirow{2}{*}{\textbf{2}}
& Bias & 0.0233 & -0.452 & -4.93e+26 & 0.0886 &  & 0.0152 & 3.74 \\
& (SE 1) & 0.042 & 0.0134 & 4.81e+26 & 0.13 &  & 0.031 & 0.195 \\
& MSE & 0.0358 & 0.208 & 4.87e+54 & 0.344 &  & 0.0184 & 14.7 \\
& (SE 2) & 7.06e-03 & 0.012 & 4.74e+54 & 0.245 &  & 4.16e-03 & 1.74 \\
& Time (s) & 7.3 & 160 & 53.7 & 131 & 1034 & 956 & 56.5 \\
\midrule
\multirow{2}{*}{\textbf{3}}
& Bias & 0.0248 & -0.497 & -3.16e+17 & 0.232 &  & 0.0169 & 2.85 \\
& (SE 1) & 0.036 & 0.0119 & 3.08e+17 & 0.152 &  & 0.031 & 0.135 \\
& MSE & 0.0266 & 0.249 & 2e+36 & 0.516 &  & 0.0185 & 8.46 \\
& (SE 2) & 0.0102 & 0.0118 & 1.95e+36 & 0.32 &  & 4.19e-03 & 0.872 \\
& Time (s) & 18.3 & 285 & 110 & 220 &  & 2149 & 1455 \\
\midrule
\multirow{2}{*}{\textbf{4}}
& Bias & 0.0415 & -0.591 & 1.38 & 0.523 &  & 0.0146 & 2.84 \\
& (SE 1) & 0.0457 & 9.18e-03 & 0.316 & 0.145 &  & 0.0313 & 0.139 \\
& MSE & 0.0435 & 0.351 & 3.91 & 0.695 &  & 0.0189 & 8.43 \\
& (SE 2) & 0.0244 & 0.0108 & 1.87 & 0.282 &  & 4.21e-03 & 0.899 \\
& Time (s) & 43.5 & 1189 & 551 & 607 &  & 3232 & 2754 \\
\midrule
\multirow{2}{*}{\textbf{5}}
& Bias & 7.9e-03 & -0.718 & 0.141 & 0.308 &  & 0.0182 & 2.82 \\
& (SE 1) & 0.034 & 6.14e-03 & 0.0548 & 0.0725 &  & 0.0312 & 0.13 \\
& MSE & 0.0231 & 0.516 & 0.08 & 0.2 &  & 0.0188 & 8.27 \\
& (SE 2) & 8.32e-03 & 8.8e-03 & 0.0344 & 0.0615 &  & 4.28e-03 & 0.823 \\
& Time (s) & 165 & 5099 & 3534 & 4412 &  & 4119 & 5338 \\
\bottomrule
\end{tabular}
}
\caption{Bias and MSE for the GP marginal variance, standard errors and time needed for estimating covariance parameters on simulated data sets with an effective range of $0.2$, error variance of $0.1$ and a sample size N=100,000.
}
\end{table}

\begin{table}[H]
\centering
\resizebox{\textwidth}{!}{%
\begin{tabular}{llcccccccc}
\toprule
 &  & \textbf{Vecchia} & \textbf{Tapering} & \textbf{FITC} & \textbf{Full-scale} & \textbf{FRK} & \textbf{MRA} & \textbf{SPDE} \\
\midrule
\multirow{2}{*}{\textbf{1}}
& Bias & -2.3e-05 & -0.0976 & 0.299 & -0.0455 & -5.05e-03 & 1.5e-04 & 9.09e-03 \\
& (SE 1) & 8.53e-05 & 4.35e-04 & 0.032 & 1.32e-03 & 2.65e-04 & 1.19e-04 & 3.2e-04 \\
& MSE & 1.46e-07 & 9.53e-03 & 0.11 & 2.11e-03 & 2.68e-05 & 2.9e-07 & 8.46e-05 \\
& (SE 2) & 3.4e-08 & 8.44e-05 & 0.0243 & 1.22e-04 & 2.75e-06 & 7.32e-08 & 5.98e-06 \\
& Time (s) & 3.31 & 47.8 & 1.77 & 34.6 & 456 & 409 & 17.1 \\
\midrule
\multirow{2}{*}{\textbf{2}}
& Bias & 6.04e-06 & -0.0889 & -8.11e+20 & -0.0353 & -4.64e-03 & 1.16e-04 & 1.42e-03 \\
& (SE 1) & 9.82e-05 & 7.05e-04 & 7.91e+20 & 6.09e-04 & 2.3e-04 & 1.29e-04 & 1.52e-04 \\
\bottomrule
\end{tabular}
}
\caption{Bias and MSE for the error term variance, standard errors and time needed for estimating covariance parameters on simulated data sets with an effective range of $0.2$, error variance of $0.1$ and a sample size N=100,000.
}
\end{table}

\subsubsection{Prediction}

\begin{figure}[H]

  \centering            \includegraphics[width=\linewidth]{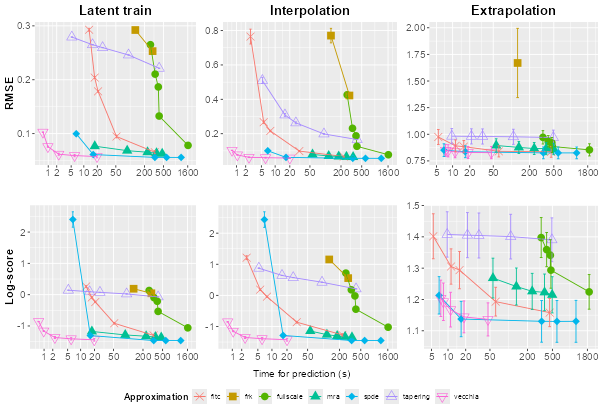} 
  \caption[$N=100,000$, range $0.2$, noise variance $0.1$. Prediction accuracy.]
  {Average RMSE and log-score on simulated data sets with
an effective range of $0.2$, error variance of $0.1$ and a sample size N=100,000. Predictions are done using the true data-generating parameters.}
\end{figure}

\begin{table}[H]
\centering
\resizebox{\textwidth}{!}{%
\begin{tabular}{llcccccccc}
\toprule
 &  & \textbf{Vecchia} & \textbf{Tapering} & \textbf{FITC} & \textbf{Full-scale} & \textbf{FRK} & \textbf{MRA} & \textbf{SPDE} \\
\midrule
\multirow{2}{*}{\textbf{1}}
& RMSE & 0.103 & 0.279 & 0.293 & 0.265 & 0.292 & 0.0767 & 0.0997 \\
& (SE 1) & 2.95e-04 & 1.89e-04 & 1.37e-03 & 4.96e-04 & 1.3e-03 & 8.42e-04 & 1.46e-03 \\
& Log-score & -0.858 & 0.143 & 0.262 & 0.133 & 0.189 & -1.17 & 2.4 \\
& (SE 2) & 2.85e-03 & 6.22e-04 & 0.0126 & 6.19e-03 & 4.49e-03 & 0.0118 & 0.125 \\
& Time (s) & 0.551 & 5.19 & 13.2 & 267 & 129 & 17.6 & 6.65 \\
\midrule
\multirow{2}{*}{\textbf{2}}
& RMSE & 0.0758 & 0.265 & 0.204 & 0.21 & 0.253 & 0.0684 & 0.0612 \\
& (SE 1) & 5.42e-04 & 1.48e-04 & 1.03e-03 & 6.04e-04 & 1.88e-03 & 9.53e-04 & 1.27e-03 \\
\bottomrule
\end{tabular}
}
\caption{Average RMSE and log-score on the training set, standard errors and time
needed for making predictions on the training set on simulated data sets with an effective range of $0.2$, error variance of $0.1$ and a sample size N=100,000.
}
\end{table}

\begin{table}[H]
\centering
\resizebox{\textwidth}{!}{%
\begin{tabular}{llcccccccc}
\toprule
 &  & \textbf{Vecchia} & \textbf{Tapering} & \textbf{FITC} & \textbf{Full-scale} & \textbf{FRK} & \textbf{MRA} & \textbf{SPDE} \\
\midrule
\multirow{2}{*}{\textbf{1}}
& RMSE & 0.103 & 0.51 & 0.765 & 0.426 & 0.771 & 0.0794 & 0.101 \\
& (SE 1) & 3.46e-04 & 9.16e-03 & 0.0221 & 8.06e-03 & 0.0216 & 9.58e-04 & 1.51e-03 \\
& Log-score & -0.851 & 0.874 & 1.21 & 0.716 & 1.15 & -1.15 & 2.44 \\
& (SE 2) & 3.28e-03 & 5.56e-03 & 0.0316 & 0.0108 & 0.0287 & 0.0126 & 0.128 \\
& Time (s) & 0.643 & 5.26 & 2.71 & 263 & 128 & 56.2 & 6.95 \\
\midrule
\multirow{2}{*}{\textbf{2}}
& RMSE & 0.0765 & 0.309 & 0.267 & 0.231 & 0.422 & 0.0703 & 0.0621 \\
& (SE 1) & 6.15e-04 & 4.1e-03 & 3.43e-03 & 3.24e-03 & 8.99e-03 & 1.05e-03 & 1.33e-03 \\
\bottomrule
\end{tabular}
}
\caption{Average RMSE and log-score on the test “interpolation” set, standard errors and time needed for making predictions on the test “interpolation” set on simulated
data sets with an effective range of $0.2$, error variance of $0.1$ and a sample size N=100,000.
}
\end{table}

\begin{table}[H]
\centering
\resizebox{\textwidth}{!}{%
\begin{tabular}{llcccccccc}
\toprule
 &  & \textbf{Vecchia} & \textbf{Tapering} & \textbf{FITC} & \textbf{Full-scale} & \textbf{FRK} & \textbf{MRA} & \textbf{SPDE} \\
\midrule
\multirow{2}{*}{\textbf{1}}
& RMSE & 0.85 & 0.982 & 0.975 & 0.971 & 1.67 & 0.895 & 0.851 \\
& (SE 1) & 0.0283 & 0.0364 & 0.0356 & 0.0319 & 0.166 & 0.032 & 0.0304 \\
& Log-score & 1.2 & 1.41 & 1.4 & 1.4 & 3.2 & 1.27 & 1.21 \\
& (SE 2) & 0.0297 & 0.0369 & 0.0367 & 0.0326 & 0.543 & 0.0324 & 0.0304 \\
& Time (s) & 7.45 & 9.67 & 5.41 & 331 & 127 & 55.9 & 6.94 \\
\midrule
\multirow{2}{*}{\textbf{2}}
& RMSE & 0.845 & 0.98 & 0.891 & 0.934 & 36.5 & 0.877 & 0.836 \\
& (SE 1) & 0.0292 & 0.0363 & 0.0286 & 0.0271 & 6.58 & 0.0288 & 0.0285 \\
& Log-score & 1.19 & 1.41 & 1.31 & 1.36 & 149 & 1.24 & 1.14 \\
& (SE 2) & 0.0305 & 0.0367 & 0.0287 & 0.0278 & 37.6 & 0.0302 & 0.0288 \\
& Time (s) & 8.28 & 21.3 & 11.3 & 405 & 290 & 133 & 16.8 \\
\midrule
\multirow{2}{*}{\textbf{3}}
& RMSE & 0.836 & 0.979 & 0.884 & 0.92 &  & 0.866 & 0.825 \\
& (SE 1) & 0.029 & 0.0362 & 0.0295 & 0.0254 &  & 0.0285 & 0.0282 \\
& Log-score & 1.17 & 1.4 & 1.29 & 1.34 &  & 1.23 & 1.13 \\
& (SE 2) & 0.0283 & 0.0367 & 0.0304 & 0.0253 &  & 0.0293 & 0.034 \\
& Time (s) & 11.1 & 33 & 15.8 & 464 &  & 238 & 337 \\
\midrule
\multirow{2}{*}{\textbf{4}}
& RMSE & 0.829 & 0.977 & 0.84 & 0.885 &  & 0.865 & 0.825 \\
& (SE 1) & 0.0285 & 0.0361 & 0.0268 & 0.0287 &  & 0.0289 & 0.0282 \\
& Log-score & 1.14 & 1.4 & 1.19 & 1.29 &  & 1.22 & 1.13 \\
& (SE 2) & 0.026 & 0.0364 & 0.0234 & 0.0294 &  & 0.0299 & 0.034 \\
& Time (s) & 18.7 & 108 & 61.9 & 473 &  & 375 & 590 \\
\midrule
\multirow{2}{*}{\textbf{5}}
& RMSE & 0.827 & 0.969 & 0.836 & 0.854 &  & 0.862 & 0.825 \\
& (SE 1) & 0.0291 & 0.0356 & 0.0261 & 0.0291 &  & 0.0273 & 0.0282 \\
& Log-score & 1.14 & 1.39 & 1.16 & 1.22 &  & 1.21 & 1.13 \\
& (SE 2) & 0.027 & 0.0357 & 0.0229 & 0.0282 &  & 0.0295 & 0.0342 \\
& Time (s) & 46.5 & 491 & 452 & 1919 &  & 496 & 1186 \\
\bottomrule
\end{tabular}
}
\caption{Average RMSE and log-score on the test “extrapolation” set, standard errors and time needed for making predictions on the test “extrapolation” set on simulated
data sets with an effective range of $0.2$, error variance of $0.1$ and a sample size N=100,000.
}
\end{table}

\newpage
\subsection{Simulated data with $N=100,000$, effective range of $0.2$ and Matérn smoothness of $0.5$}\label{appendix_100000_02_s05}

\begin{table}[ht!]
\centering
\resizebox{\textwidth}{!}{%
\begin{tabular}{l|l|cccccc}
  & \textbf{Tuning parameter} & \textbf{} & \textbf{} & \textbf{} & \textbf{} & \textbf{} & \textbf{} \\
\midrule
\textbf{Vecchia} & nb. neighbours & 5 & 10 & 20 & 40 & 80 \\
\hline
\textbf{Tapering} & avg. nb. non-zeros / row & 8 & 17 & 22 & 40 & 111\\
\hline
\textbf{FITC} & nb. inducing points & 20 & 200 & 275 & 900 & 2000\\
\hline
\multirow{2}{*}{\textbf{Full-scale}} & nb. inducing points & 26 & 91 & 122 & 250 & 650 \\
 & avg. nb. non-zeros / row & 6 & 11 & 15 & 27 & 76 \\
\hline
\textbf{FRK} & nb. resolutions & 1  & 2 & exceeds time limit &   &  &  \\
\hline
\textbf{MRA} & nb. knots per partition & 1 & 2 & 3 & 4 & 5 \\
\hline
\textbf{SPDE} & max edge & 0.04 & 0.02 & 0.005 & 0.004 & 0.003   \\
\bottomrule
\end{tabular}%
}
\caption{Tuning parameters chosen for the comparison on simulated data sets with an effective range of $0.2$, Matérn smoothness of $0.5$ and a sample size of $N=100,000$.}
\end{table}

\subsubsection{Log-likelihood evaluation}
\begin{figure}[H]

  \centering            \includegraphics[width=0.7\linewidth]{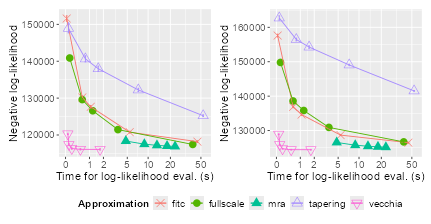} 
  \caption[$N=100,000$, range $0.2$, Matérn smoothness $0.5$. Log-likelihood evaluation.]
  {Average negative log-likelihood on
simulated data sets with an effective range of $0.2$, Matérn smoothness of $0.5$ and a sample size N=100,000. The true data-generating parameters are used on the left, and two times these values are used on the right.}
\end{figure}

\begin{table}[H]
\centering
\resizebox{\textwidth}{!}{%
\begin{tabular}{llcccccccc}
\toprule
 &  & \textbf{Vecchia} & \textbf{Tapering} & \textbf{FITC} & \textbf{Full-scale} & \textbf{FRK} & \textbf{MRA} & \textbf{SPDE} \\
\midrule
\multirow{2}{*}{\textbf{1}}
& Value & 120303 & 148921 & 151578 & 140874 &  & 118362 &  \\
& (SE) & 92.8 & 631 & 371 & 346 &  & 102 &  \\
& Time (s) & 0.0679 & 0.0863 & 0.0326 & 0.127 &  & 4.76 &  \\
\midrule
\multirow{2}{*}{\textbf{2}}
& Value & 117430 & 140638 & 130186 & 129595 &  & 117516 &  \\
& (SE) & 91.5 & 420 & 207 & 159 &  & 99.7 &  \\
& Time (s) & 0.1 & 0.779 & 0.627 & 0.622 &  & 8.86 &  \\
\midrule
\multirow{2}{*}{\textbf{3}}
& Value & 116329 & 137975 & 127645 & 126569 &  & 117195 &  \\
& (SE) & 92.5 & 357 & 179 & 116 &  & 98.2 &  \\
& Time (s) & 0.198 & 1.58 & 1.08 & 1.2 &  & 13.2 &  \\
\midrule
\multirow{2}{*}{\textbf{4}}
& Value & 116110 & 132220 & 120768 & 121459 &  & 117013 &  \\
& (SE) & 95.5 & 234 & 134 & 103 &  & 97.9 &  \\
& Time (s) & 0.539 & 7.27 & 5.44 & 3.58 &  & 18.1 &  \\
\midrule
\multirow{2}{*}{\textbf{5}}
& Value & 116093 & 125227 & 118198 & 117435 &  & 116892 &  \\
& (SE) & 95.6 & 119 & 111 & 87.9 &  & 97.5 &  \\
& Time (s) & 1.72 & 53.4 & 45.1 & 39.2 &  & 23.1 &  \\
\bottomrule
\end{tabular}
}
\caption{Mean approximate negative log-likelihood, standard error and time needed
for evaluating the approximate log-likelihood on simulated data sets with an effective range of $0.2$, Matérn smoothness of $0.5$ and a sample size N=100,000. Log-likelihoods are evaluated at the true data-generating parameters.}
\end{table}

\begin{table}[H]
\centering
\resizebox{\textwidth}{!}{%
\begin{tabular}{llcccccccc}
\toprule
 &  & \textbf{Vecchia} & \textbf{Tapering} & \textbf{FITC} & \textbf{Full-scale} & \textbf{FRK} & \textbf{MRA} & \textbf{SPDE} \\
\midrule
\multirow{2}{*}{\textbf{1}}
& Value & 128952 & 162716 & 157608 & 149798 &  & 126554 &  \\
& (SE) & 48.7 & 314 & 212 & 197 &  & 59.6 &  \\
& Time (s) & 0.0689 & 0.0869 & 0.0359 & 0.126 &  & 4.76 &  \\
\midrule
\multirow{2}{*}{\textbf{2}}
& Value & 125941 & 156460 & 136893 & 138594 &  & 125754 &  \\
& (SE) & 49.7 & 208 & 127 & 106 &  & 57.9 &  \\
& Time (s) & 0.101 & 0.782 & 0.627 & 0.621 &  & 8.86 &  \\
\midrule
\multirow{2}{*}{\textbf{3}}
& Value & 124711 & 154263 & 134649 & 135831 &  & 125451 &  \\
& (SE) & 51.4 & 176 & 110 & 83 &  & 57 &  \\
& Time (s) & 0.196 & 1.58 & 1.08 & 1.21 &  & 13.3 &  \\
\midrule
\multirow{2}{*}{\textbf{4}}
& Value & 124445 & 149068 & 128683 & 130920 &  & 125282 &  \\
& (SE) & 55 & 115 & 78.1 & 71.3 &  & 56.8 &  \\
& Time (s) & 0.537 & 7.27 & 5.44 & 3.59 &  & 18.1 &  \\
\midrule
\multirow{2}{*}{\textbf{5}}
& Value & 124438 & 141524 & 126489 & 126710 &  & 125171 &  \\
& (SE) & 55.7 & 58 & 63.7 & 55.3 &  & 56.8 &  \\
& Time (s) & 1.72 & 53.4 & 45.1 & 39.2 &  & 23 &  \\
\bottomrule
\end{tabular}
}
\caption{Mean approximate negative log-likelihood, standard error and time needed
for evaluating the approximate log-likelihood on simulated data sets with an effective range of $0.2$, Matérn smoothness of $0.5$ and a sample size N=100,000. Log-likelihoods are evaluated at two-times the true data-generating parameters.}
\end{table}

\subsubsection{Parameter estimation}

\begin{figure}[H]
  \centering            \includegraphics[width=\linewidth]{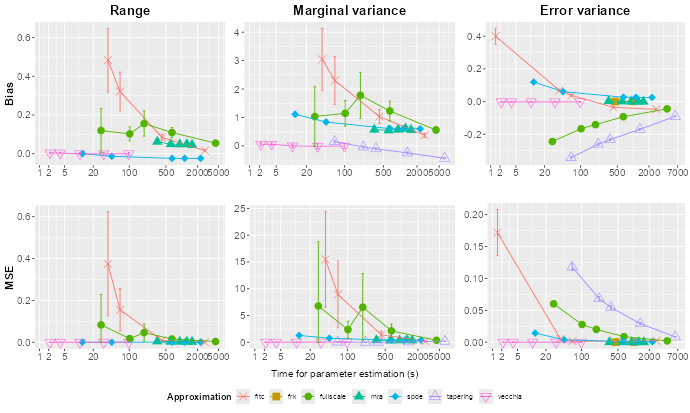} 
  \caption[$N=100,000$, range $0.2$, Matérn smoothness $0.5$. Parameter estimation accuracy.]
  {Bias and MSE of the estimates for the GP range, GP variance end error term variance on simulated
data sets with an effective range of $0.2$, Matérn smoothness of $0.5$ and a sample size N=100,000.}

\end{figure}

\begin{table}[H]
\centering
\resizebox{\textwidth}{!}{%
\begin{tabular}{llcccccccc}
\toprule
 &  & \textbf{Vecchia} & \textbf{Tapering} & \textbf{FITC} & \textbf{Full-scale} & \textbf{FRK} & \textbf{MRA} & \textbf{SPDE} \\
\midrule
\multirow{2}{*}{\textbf{1}}
& Bias & 3.73e-03 & 134 & 20.9 & 0.119 &  & 0.0615 & -4.53e-04 \\
& (SE 1) & 2.2e-03 & 8.12 & 8.54 & 0.0587 &  & 5.4e-03 & 1.16e-03 \\
& MSE & 1.11e-04 & 19168 & 1895 & 0.0831 &  & 4.34e-03 & 2.56e-05 \\
& (SE 2) & 2.67e-05 & 1921 & 834 & 0.0742 &  & 8.66e-04 & 8.65e-06 \\
& Time (s) & 2.17 & 65.8 & 1.48 & 28.5 & 457 & 343 & 12.1 \\
\midrule
\multirow{2}{*}{\textbf{2}}
& Bias & 2.82e-03 & 173 & 0.483 & 0.102 &  & 0.0493 & -0.0138 \\
& (SE 1) & 2.15e-03 & 15.6 & 0.084 & 0.0187 &  & 3.72e-03 & 1.23e-03 \\
\bottomrule
\end{tabular}
}
\caption{Bias and MSE for the GP range, standard errors and time needed for estimating covariance parameters on simulated data sets with an effective range of $0.2$, Matérn smoothness of $0.5$ and a sample size N=100,000.
}
\end{table}

\begin{table}[H]
\centering
\resizebox{\textwidth}{!}{%
\input{tables/variance_100k_s05}
}
\caption{Bias and MSE for the GP marginal variance, standard errors and time needed for estimating covariance parameters on simulated data sets with an effective range of $0.2$, Matérn smoothness of $0.5$ and a sample size N=100,000.
}
\end{table}

\begin{table}[H]
\centering
\resizebox{\textwidth}{!}{%
\begin{tabular}{llcccccccc}
\toprule
 &  & \textbf{Vecchia} & \textbf{Tapering} & \textbf{FITC} & \textbf{Full-scale} & \textbf{FRK} & \textbf{MRA} & \textbf{SPDE} \\
\midrule
\multirow{2}{*}{\textbf{1}}
& Bias & -1.65e-03 & -0.343 & 0.399 & -0.245 & 3.65e-04 & 2.83e-03 & 0.119 \\
& (SE 1) & 9.36e-04 & 3.73e-03 & 0.0249 & 3.54e-03 & 1.85e-03 & 9.69e-04 & 1.71e-03 \\
& MSE & 2.02e-05 & 0.118 & 0.172 & 0.0601 & 6.5e-05 & 2.59e-05 & 0.0142 \\
& (SE 2) & 5.02e-06 & 2.56e-03 & 0.0184 & 1.74e-03 & 2.64e-05 & 6.12e-06 & 4.02e-04 \\
& Time (s) & 2.17 & 65.8 & 1.48 & 28.5 & 457 & 343 & 12.1 \\
\midrule
\multirow{2}{*}{\textbf{2}}
& Bias & -2.15e-03 & -0.261 & 0.0684 & -0.166 & 2.04e-03 & 8.57e-04 & 0.0604 \\
& (SE 1) & 8.67e-04 & 3.28e-03 & 3.82e-03 & 1.62e-03 & 1.7e-03 & 9.85e-04 & 1.2e-03 \\
\bottomrule
\end{tabular}
}
\caption{Bias and MSE for the error term variance, standard errors and time needed for estimating covariance parameters on simulated data sets with an effective range of $0.2$, Matérn smoothness of $0.5$ and a sample size N=100,000.
}
\end{table}

\subsubsection{Prediction}

\begin{figure}[H]

  \centering            \includegraphics[width=\linewidth]{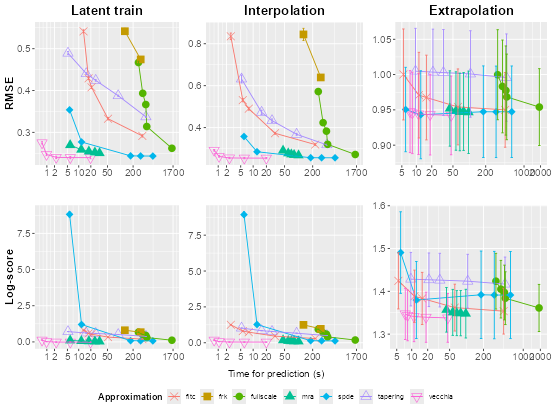} 
  \caption[$N=100,000$, range $0.2$, Matérn smoothness $0.5$. Prediction accuracy.]
  {Average RMSE and log-score on simulated data sets with
an effective range of $0.2$, Matérn smoothness of $0.5$ and a sample size N=100,000. Predictions are done using the true data-generating parameters.}
\end{figure}

\begin{table}[H]
\centering
\resizebox{\textwidth}{!}{%
\input{tables/train_pred_100k_s05}
}
\caption{Average RMSE and log-score on the training set, standard errors and time
needed for making predictions on the training set on simulated data sets with an effective range of $0.2$, Matérn smoothness of $0.5$ and a sample size N=100,000.
}
\end{table}

\begin{table}[H]
\centering
\resizebox{\textwidth}{!}{%
\input{tables/inter_pred_100k_s05}
}
\caption{Average RMSE and log-score on the test “interpolation” set, standard errors and time needed for making predictions on the test “interpolation” set on simulated
data sets with an effective range of $0.2$, Matérn smoothness of $0.5$ and a sample size N=100,000.
}
\end{table}

\begin{table}[H]
\centering
\resizebox{\textwidth}{!}{%
\input{tables/extra_pred_100k_s05}
}
\caption{Average RMSE and log-score on the test “extrapolation” set, standard errors and time needed for making predictions on the test “extrapolation” set on simulated
data sets with an effective range of $0.2$, Matérn smoothness of $0.5$ and a sample size N=100,000.
}
\end{table}

\newpage
\subsection{Simulated data with $N=100,000$, effective range of $0.2$ and Matérn smoothness of $2.5$}\label{appendix_100000_02_s25}

The FRK approximation has not been considered in this comparison as the function auto\_basis() does not include a Matérn basis function type with smoothness of $2.5$.

\begin{table}[ht!]
\centering
\resizebox{\textwidth}{!}{%
\begin{tabular}{l|l|cccccc}
  & \textbf{Tuning parameter} & \textbf{} & \textbf{} & \textbf{} & \textbf{} & \textbf{} & \textbf{} \\
\midrule
\textbf{Vecchia} & nb. neighbours & 5 & 10 & 20 & 40 & 80 \\
\hline
\textbf{Tapering} & avg. nb. non-zeros / row & 8 & 17 & 22 & 40 & 111\\
\hline
\textbf{FITC} & nb. inducing points & 20 & 200 & 275 & 900 & 2000\\
\hline
\multirow{2}{*}{\textbf{Full-scale}} & nb. inducing points & 26 & 91 & 122 & 250 & 650 \\
 & avg. nb. non-zeros / row & 6 & 11 & 15 & 27 & 76 \\
\hline
\textbf{FRK} & nb. resolutions &   &  &  &   &  &  \\
\hline
\textbf{MRA} & nb. knots per partition & 1 & 2 & 3 & 4 & 5 \\
\hline
\textbf{SPDE} & max edge & 0.04 & 0.02 & 0.005 & 0.004 & 0.003   \\
\bottomrule
\end{tabular}%
}
\caption{Tuning parameters chosen for the comparison on simulated data sets with an effective range of $0.2$, Matérn smoothness of $2.5$ and a sample size of $N=100,000$.}
\end{table}

\subsubsection{Log-likelihood evaluation}
\begin{figure}[H]

  \centering            \includegraphics[width=0.7\linewidth]{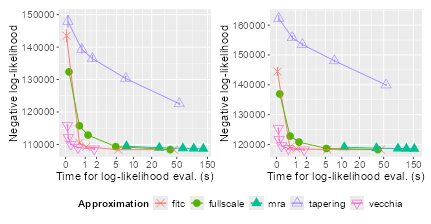} 
  \caption[$N=100,000$, range $0.2$, Matérn smoothness $2.5$. Log-likelihood evaluation.]
  {Average negative log-likelihood on
simulated data sets with an effective range of $0.2$, Matérn smoothness of $2.5$ and a sample size N=100,000. The true data-generating parameters are used on the left, and two times these values are used on the right.}
\end{figure}

\begin{table}[H]
\centering
\resizebox{\textwidth}{!}{%
\input{tables/true_lik_100k_s25}
}
\caption{Mean approximate negative log-likelihood, standard error and time needed
for evaluating the approximate log-likelihood on simulated data sets with an effective range of $0.2$, Matérn smoothness of $2.5$ and a sample size N=100,000. Log-likelihoods are evaluated at the true data-generating parameters.}
\end{table}

\begin{table}[H]
\centering
\resizebox{\textwidth}{!}{%
\input{tables/wrong_lik_100k_s25}
}
\caption{Mean approximate negative log-likelihood, standard error and time needed
for evaluating the approximate log-likelihood on simulated data sets with an effective range of $0.2$, Matérn smoothness of $2.5$ and a sample size N=100,000. Log-likelihoods are evaluated at two-times the true data-generating parameters.}
\end{table}

\subsubsection{Parameter estimation}

\begin{figure}[H]%
  \centering            \includegraphics[width=\linewidth]{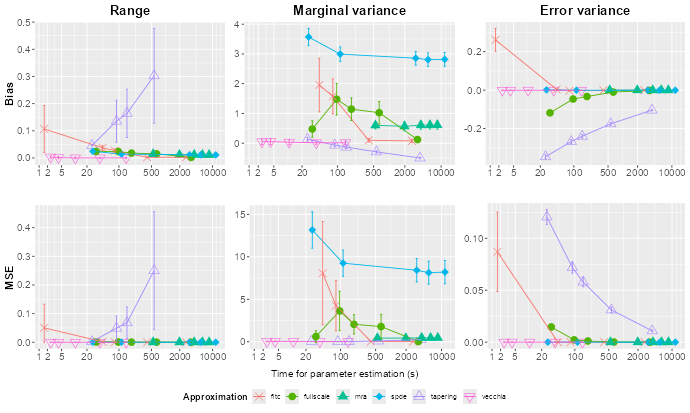} 
  \caption[$N=100,000$, range $0.2$, Matérn smoothness $2.5$. Parameter estimation accuracy.]
  {Bias and MSE of the estimates for the GP range, GP variance end error term variance on simulated
data sets with an effective range of $0.2$, Matérn smoothness of $2.5$ and a sample size N=100,000.}

\end{figure}

\begin{table}[H]
\centering
\resizebox{\textwidth}{!}{%
\input{tables/range_100k_s25}
}
\caption{Bias and MSE for the GP range, standard errors and time needed for estimating covariance parameters on simulated data sets with an effective range of $0.2$, Matérn smoothness of $2.5$ and a sample size N=100,000.
}
\end{table}

\begin{table}[H]
\centering
\resizebox{\textwidth}{!}{%
\input{tables/variance_100k_s25}
}
\caption{Bias and MSE for the GP marginal variance, standard errors and time needed for estimating covariance parameters on simulated data sets with an effective range of $0.2$, Matérn smoothness of $2.5$ and a sample size N=100,000.
}
\end{table}

\begin{table}[H]
\centering
\resizebox{\textwidth}{!}{%
\input{tables/nugget_100k_s25}
}
\caption{Bias and MSE for the error term variance, standard errors and time needed for estimating covariance parameters on simulated data sets with an effective range of $0.2$, Matérn smoothness of $2.5$ and a sample size N=100,000.
}
\end{table}

\subsubsection{Prediction}

\begin{figure}[H]

  \centering            \includegraphics[width=\linewidth]{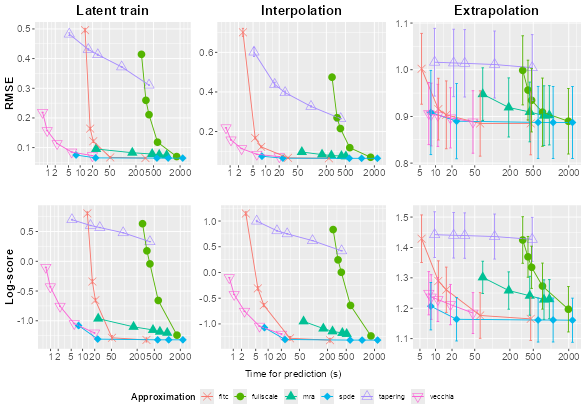} 
  \caption[$N=100,000$, range $0.2$, Matérn smoothness $2.5$. Prediction accuracy.]
  {Average RMSE and log-score on simulated data sets with
an effective range of $0.2$, Matérn smoothness of $2.5$ and a sample size N=100,000. Predictions are done using the true data-generating parameters.}
\end{figure}

\begin{table}[H]
\centering
\resizebox{\textwidth}{!}{%
\input{tables/train_pred_100k_s25}
}
\caption{Average RMSE and log-score on the training set, standard errors and time
needed for making predictions on the training set on simulated data sets with an effective range of $0.2$, Matérn smoothness of $2.5$ and a sample size N=100,000.
}
\end{table}

\begin{table}[H]
\centering
\resizebox{\textwidth}{!}{%
\input{tables/inter_pred_100k_s25}
}
\caption{Average RMSE and log-score on the test “interpolation” set, standard errors and time needed for making predictions on the test “interpolation” set on simulated
data sets with an effective range of $0.2$, Matérn smoothness of $2.5$ and a sample size N=100,000.
}
\end{table}

\begin{table}[H]
\centering
\resizebox{\textwidth}{!}{%
\input{tables/extra_pred_100k_s25}
}
\caption{Average RMSE and log-score on the test “extrapolation” set, standard errors and time needed for making predictions on the test “extrapolation” set on simulated
data sets with an effective range of $0.2$, Matérn smoothness of $2.5$ and a sample size N=100,000.
}
\end{table}

\newpage
\subsection{Simulated data with $N=100,000$, effective range of $0.2$ and misspecified Matérn smoothness}\label{appendix_100000_02_misspecified}

\begin{table}[ht!]
\centering
\resizebox{\textwidth}{!}{%
\begin{tabular}{l|l|cccccc}
  & \textbf{Tuning parameter} & \textbf{} & \textbf{} & \textbf{} & \textbf{} & \textbf{} & \textbf{} \\
\midrule
\textbf{Vecchia} & nb. neighbours & 5 & 10 & 20 & 40 & 80 \\
\hline
\textbf{Tapering} & avg. nb. non-zeros / row & 8 & 17 & 22 & 40 & 111\\
\hline
\textbf{FITC} & nb. inducing points & 20 & 200 & 275 & 900 & 2000\\
\hline
\multirow{2}{*}{\textbf{Full-scale}} & nb. inducing points & 26 & 91 & 122 & 250 & 650 \\
 & avg. nb. non-zeros / row & 6 & 11 & 15 & 27 & 76 \\
\hline
\textbf{FRK} & nb. resolutions & 1  & 2 & exceeds time limit &   &  &  \\
\hline
\textbf{MRA} & nb. knots per partition & 1 & 2 & 3 & 4 & 5 \\
\hline
\textbf{SPDE} & max edge & 0.0275 & 0.02 & 0.005 & 0.004 & 0.003   \\
\bottomrule
\end{tabular}%
}
\caption{Tuning parameters chosen for the comparison on simulated data sets with an effective range of $0.2$, misspecified Matérn smoothness and a sample size of $N=100,000$.}
\end{table}

\subsubsection{Log-likelihood evaluation}
\begin{figure}[H]

  \centering            \includegraphics[width=0.7\linewidth]{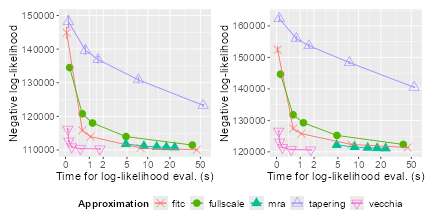} 
  \caption[$N=100,000$, range $0.2$, misspecified Matérn smoothness. Log-likelihood evaluation.]
  {Average negative log-likelihood on
simulated data sets with an effective range of $0.2$, misspecified Matérn smoothness and a sample size N=100,000. The true data-generating parameters are used on the left, and two times these values are used on the right.}
\end{figure}

\begin{table}[H]
\centering
\resizebox{\textwidth}{!}{%
\input{tables/true_lik_100k_miss}
}
\caption{Mean approximate negative log-likelihood, standard error and time needed
for evaluating the approximate log-likelihood on simulated data sets with an effective range of $0.2$, misspecified Matérn smoothness and a sample size N=100,000. Log-likelihoods are evaluated at the true data-generating parameters.}
\end{table}

\begin{table}[H]
\centering
\resizebox{\textwidth}{!}{%
\input{tables/wrong_lik_100k_miss}
}
\caption{Mean approximate negative log-likelihood, standard error and time needed
for evaluating the approximate log-likelihood on simulated data sets with an effective range of $0.2$, misspecified Matérn smoothness and a sample size N=100,000. Log-likelihoods are evaluated at two-times the true data-generating parameters.}
\end{table}

\subsubsection{Parameter estimation}

\begin{figure}[H]
  \centering            \includegraphics[width=\linewidth]{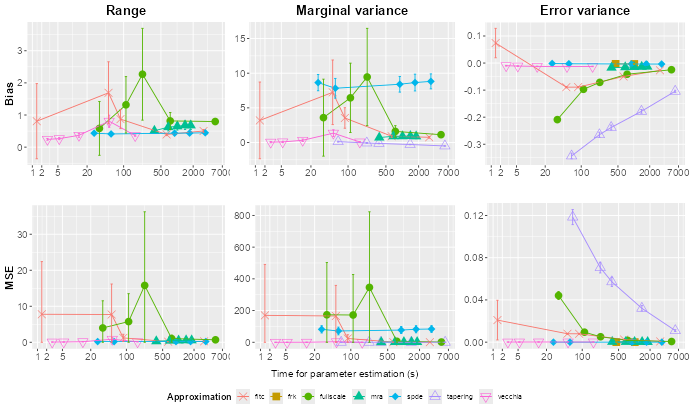} 
  \caption[$N=100,000$, range $0.2$, misspecified Matérn smoothness. Parameter estimation accuracy.]
  {Bias and MSE of the estimates for the GP range, GP variance end error term variance on simulated
data sets with an effective range of $0.2$, misspecified Matérn smoothness and a sample size N=100,000.}

\end{figure}

\begin{table}[H]
\centering
\resizebox{\textwidth}{!}{%
\input{tables/range_100k_miss}
}
\caption{Bias and MSE for the GP range, standard errors and time needed for estimating covariance parameters on simulated data sets with an effective range of $0.2$, misspecified Matérn smoothness and a sample size N=100,000.
}
\end{table}

\begin{table}[H]
\centering
\resizebox{\textwidth}{!}{%
\input{tables/variance_100k_miss}
}
\caption{Bias and MSE for the GP marginal variance, standard errors and time needed for estimating covariance parameters on simulated data sets with an effective range of $0.2$, misspecified Matérn smoothness and a sample size N=100,000.
}
\end{table}

\begin{table}[H]
\centering
\resizebox{\textwidth}{!}{%
\input{tables/nugget_100k_miss}
}
\caption{Bias and MSE for the error term variance, standard errors and time needed for estimating covariance parameters on simulated data sets with an effective range of $0.2$, misspecified Matérn smoothness and a sample size N=100,000.
}
\end{table}

\subsubsection{Prediction}

\begin{figure}[H]

  \centering            \includegraphics[width=\linewidth]{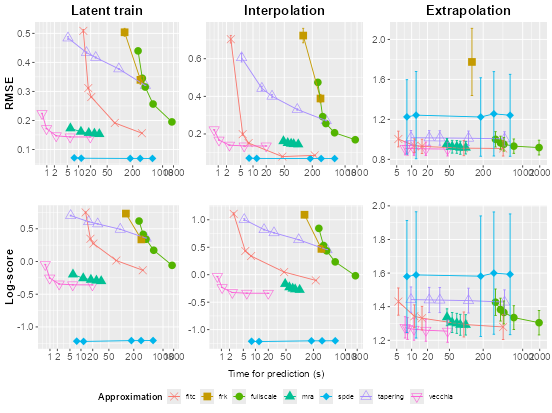} 
  \caption[$N=100,000$, range $0.2$, misspecified Matérn smoothness. Prediction accuracy.]
  {Average RMSE and log-score on simulated data sets with
an effective range of $0.2$, misspecified Matérn smoothness and a sample size N=100,000. Predictions are done using the true data-generating parameters.}
\end{figure}

\begin{table}[H]
\centering
\resizebox{\textwidth}{!}{%
\input{tables/train_pred_100k_miss}
}
\caption{Average RMSE and log-score on the training set, standard errors and time
needed for making predictions on the training set on simulated data sets with an effective range of $0.2$, misspecified Matérn smoothness and a sample size N=100,000.
}
\end{table}

\begin{table}[H]
\centering
\resizebox{\textwidth}{!}{%
\input{tables/inter_pred_100k_miss}
}
\caption{Average RMSE and log-score on the test “interpolation” set, standard errors and time needed for making predictions on the test “interpolation” set on simulated
data sets with an effective range of $0.2$, misspecified Matérn smoothness and a sample size N=100,000.
}
\end{table}

\begin{table}[H]
\centering
\resizebox{\textwidth}{!}{%
\input{tables/extra_pred_100k_miss}
}
\caption{Average RMSE and log-score on the test “extrapolation” set, standard errors and time needed for making predictions on the test “extrapolation” set on simulated
data sets with an effective range of $0.2$, misspecified Matérn smoothness and a sample size N=100,000.
}
\end{table}

\newpage
\subsection{House price data set}\label{appendix_house}
\subsubsection{Prediction}
\begin{table}[H]
\centering
\resizebox{\textwidth}{!}{%
\begin{tabular}{llcccccccc}
\toprule 
 &  & \textbf{Vecchia} & \textbf{Tapering} & \textbf{FITC} & \textbf{Full-scale} & \textbf{FRK} & \textbf{MRA} & \textbf{SPDE} \\
\midrule
\multirow{4}{*}{\textbf{1}} 
& RMSE & 0.325 & 0.382 & 0.383 & 0.345 & 0.544 & 0.337 & 0.41 \\
& Log-score & 0.262 & 0.387 & 0.395 & 0.29 & 0.81 & 0.292 & 0.524 \\
& CRPS & 0.17 & 0.196 & 0.2 & 0.18 & 0.298 & 0.175 & 0.219 \\
& Time (s) & 0.471 & 6.97 & 7.51 & 7.56 & 68.8 & 89.7 & 7.99 \\
\midrule
\multirow{4}{*}{\textbf{2}} 
& RMSE & 0.324 & 0.345 & 0.355 & 0.331 & 0.451 & 0.326 & 0.347 \\
& Log-score & 0.257 & 0.292 & 0.329 & 0.256 & 0.622 & 0.266 & 0.347 \\
& CRPS & 0.169 & 0.178 & 0.186 & 0.171 & 0.243 & 0.171 & 0.184 \\
& Time (s) & 1.13 & 30 & 27.2 & 27.6 & 155 & 174 & 29.9 \\
\midrule
\multirow{4}{*}{\textbf{3}} 
& RMSE & 0.324 & 0.332 & 0.346 & 0.323 & 0.397 & 0.325 & 0.336 \\
& Log-score & 0.258 & 0.266 & 0.29 & 0.243 & 0.492 & 0.261 & 0.304 \\
& CRPS & 0.169 & 0.172 & 0.179 & 0.168 & 0.213 & 0.17 & 0.176 \\
& Time (s) & 1.76 & 92.2 & 94 & 84.7 & 985 & 590 & 88.6 \\
\midrule
\multirow{4}{*}{\textbf{4}} 
& RMSE & 0.323 & 0.324 & 0.334 & 0.324 &  & 0.325 & 0.326 \\
& Log-score & 0.257 & 0.252 & 0.272 & 0.243 &  & 0.261 & 0.266 \\
& CRPS & 0.169 & 0.168 & 0.173 & 0.168 &  & 0.17 & 0.17 \\
& Time (s) & 4.35 & 528 & 508 & 560 &  & 871 & 497 \\
\midrule
\multirow{4}{*}{\textbf{5}} 
& RMSE & 0.323 & 0.322 & 0.327 & 0.323 &  & 0.324 & 0.324 \\
& Log-score & 0.257 & 0.25 & 0.26 & 0.25 &  & 0.259 & 0.258 \\
& CRPS & 0.169 & 0.168 & 0.17 & 0.168 &  & 0.17 & 0.169 \\
& Time (s) & 14 & 1599 & 1467 & 1347 &  & 1576 & 1312 \\
\bottomrule
\end{tabular}%
}
\caption{Test “interpolation” RMSE, log-score, CRPS and time needed for making predictions on the “interpolation” test set on the house price data set. Every approximation makes predictions using its parameter
estimates.
}
\end{table}

\begin{table}[H] 
\centering
\resizebox{\textwidth}{!}{%
\begin{tabular}{llcccccccc}
\toprule 
 &  & \textbf{Vecchia} & \textbf{Tapering} & \textbf{FITC} & \textbf{Full-scale} & \textbf{FRK} & \textbf{MRA} & \textbf{SPDE} \\
\midrule
\multirow{4}{*}{\textbf{1}} 
& RMSE & 0.597 & 0.679 & 0.559 & 0.565 & 1.45 & 0.581 & 0.708 \\
& Log-score & 0.929 & 1.05 & 0.971 & 0.921 & 3.75 & 0.952 & 1.24 \\
& CRPS & 0.336 & 0.378 & 0.332 & 0.326 & 0.976 & 0.335 & 0.411 \\
& Time (s) & 0.498 & 6.82 & 7.51 & 6.97 & 69 & 89.7 & 7.99 \\
\midrule
\multirow{4}{*}{\textbf{2}} 
& RMSE & 0.596 & 0.658 & 0.574 & 0.571 & 10.5 & 0.602 & 0.579 \\
& Log-score & 0.925 & 1.02 & 0.955 & 0.943 & 23.7 & 0.948 & 0.921 \\
& CRPS & 0.335 & 0.366 & 0.333 & 0.33 & 6.35 & 0.341 & 0.33 \\
& Time (s) & 1.21 & 29.6 & 27.2 & 26.8 & 154 & 175 & 29.9 \\
\midrule
\multirow{4}{*}{\textbf{3}} 
& RMSE & 0.595 & 0.641 & 0.583 & 0.584 & 2.76 & 0.598 & 0.582 \\
& Log-score & 0.933 & 0.982 & 0.965 & 0.948 & 2.59 & 0.941 & 0.919 \\
& CRPS & 0.336 & 0.355 & 0.337 & 0.335 & 1.5 & 0.338 & 0.331 \\
& Time (s) & 1.74 & 91.2 & 94.2 & 83 & 979 & 584 & 88.6 \\
\midrule
\multirow{4}{*}{\textbf{4}} 
& RMSE & 0.597 & 0.617 & 0.595 & 0.592 &  & 0.598 & 0.591 \\
& Log-score & 0.934 & 0.931 & 0.942 & 0.952 &  & 0.941 & 0.921 \\
& CRPS & 0.337 & 0.342 & 0.337 & 0.338 &  & 0.338 & 0.334 \\
& Time (s) & 4.42 & 523 & 508 & 552 &  & 861 & 497 \\
\midrule
\multirow{4}{*}{\textbf{5}} 
& RMSE & 0.597 & 0.608 & 0.601 & 0.595 &  & 0.595 & 0.593 \\
& Log-score & 0.933 & 0.916 & 0.941 & 0.939 &  & 0.935 & 0.92 \\
& CRPS & 0.337 & 0.338 & 0.34 & 0.337 &  & 0.336 & 0.334 \\
& Time (s) & 14.1 & 1587 & 1467 & 1334 &  & 1563 & 1312 \\
\bottomrule
\end{tabular}%
}
\caption{Test “extrapolation” RMSE, log-score, CRPS and time needed for making predictions on the “extrapolation” test set on the house price data set. Every approximation makes predictions using its parameter
estimates.}
\label{house_extra_tab}
\end{table}

\subsubsection{Log-likelihood evaluation}
\begin{table}[H]
\centering
\resizebox{\textwidth}{!}{%
\begin{tabular}{llcccccccc}
\toprule 
 &  & \textbf{Vecchia} & \textbf{Tapering} & \textbf{FITC} & \textbf{Full-scale} & \textbf{FRK} & \textbf{MRA} & \textbf{SPDE} \\
\midrule
\multirow{2}{*}{\textbf{1}} 
& Value & 6184 & 12120 & 9047 & 9230 &  & 6750 &  \\
& Time (s) & 0.0132 & 0.057 & 0.0969 & 0.0806 &  & 0.451 &  \\
\midrule
\multirow{2}{*}{\textbf{2}} 
& Value & 5862 & 9114 & 6893 & 6770 &  & 6063 &  \\
& Time (s) & 0.0233 & 0.29 & 0.42 & 0.268 &  & 2.39 &  \\
\midrule
\multirow{2}{*}{\textbf{3}} 
& Value & 5751 & 7847 & 6055 & 5783 &  & 5853 &  \\
& Time (s) & 0.0351 & 0.897 & 1.25 & 0.732 &  & 8.6 &  \\
\midrule
\multirow{2}{*}{\textbf{4}} 
& Value & 5721 & 6732 & 5753 & 5548 &  & 5822 &  \\
& Time (s) & 0.0731 & 5.97 & 9.77 & 5.31 &  & 14.5 &  \\
\midrule
\multirow{2}{*}{\textbf{5}} 
& Value & 5714 & 6446 & 5731 & 5633 &  & 5802 &  \\
& Time (s) & 0.226 & 26.7 & 25.9 & 13.6 &  & 26.9 &  \\
\bottomrule
\end{tabular}%
}
\caption{Approximate negative log-likelihood and time needed for its evaluation on the house price data set. The covariance parameters estimated via exact calculations are used.}
\end{table}

\begin{table}[H]
\centering
\resizebox{\textwidth}{!}{%
\begin{tabular}{llcccccccc}
\toprule 
 &  & \textbf{Vecchia} & \textbf{Tapering} & \textbf{FITC} & \textbf{Full-scale} & \textbf{FRK} & \textbf{MRA} & \textbf{SPDE} \\
\midrule
\multirow{2}{*}{\textbf{1}} 
& Value & 7587 & 15704 & 9027 & 9720 &  & 7773 &  \\
& Time (s) & 9.12e-03 & 0.0565 & 0.0989 & 0.0745 &  & 0.467 &  \\
\midrule
\multirow{2}{*}{\textbf{2}} 
& Value & 7142 & 12729 & 7642 & 7787 &  & 7213 &  \\
& Time (s) & 0.0232 & 0.288 & 0.481 & 0.259 &  & 3.59 &  \\
\midrule
\multirow{2}{*}{\textbf{3}} 
& Value & 6992 & 11229 & 7160 & 7150 &  & 7085 &  \\
& Time (s) & 0.0279 & 0.898 & 1.34 & 0.708 &  & 11.5 &  \\
\midrule
\multirow{2}{*}{\textbf{4}} 
& Value & 6962 & 9649 & 6984 & 6941 &  & 7049 &  \\
& Time (s) & 0.0745 & 5.96 & 9.79 & 6.17 &  & 16.6 &  \\
\midrule
\multirow{2}{*}{\textbf{5}} 
& Value & 6955 & 9166 & 6960 & 6944 &  & 7031 &  \\
& Time (s) & 0.23 & 26.8 & 27.8 & 13.7 &  & 30 &  \\
\bottomrule
\end{tabular}%
}
\caption{Approximate negative log-likelihood and time needed for its evaluation on the house price data set. Two times the values of the covariance parameters estimated via exact calculations are used.}
\end{table}

\subsubsection{Comparison to exact calculations}

\begin{figure}[H]

  \centering            \includegraphics[width=0.66\linewidth]{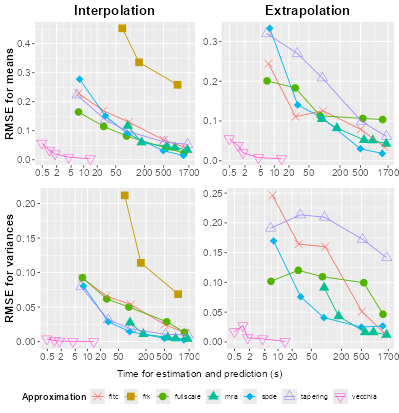} 
  \caption[House price data set. Comparison to exact calculations.]
  {RMSE between exact and approximate predictive means and variances on the house price data set. Every approximation makes predictions using its parameter estimates.}
  \label{compexact_house}
\end{figure}

\begin{table}[H]
\centering
\resizebox{\textwidth}{!}{%
\begin{tabular}{llcccccccc}
\toprule 
 &  & \textbf{Vecchia} & \textbf{Tapering} & \textbf{FITC} & \textbf{Full-scale} & \textbf{FRK} & \textbf{MRA} & \textbf{SPDE} \\
\midrule
\multirow{4}{*}{\textbf{1}} 
& RMSE means & 0.0571 & 0.223 & 0.229 & 0.165 & 0.453 & 0.117 & 0.278 \\
& RMSE vars & 3.78e-03 & 0.0793 & 0.0921 & 0.0925 & 0.212 & 0.0276 & 0.0804 \\
& KL-divergence & 65 & 769 & 739 & 460 & 2224 & 211 & 1162 \\
& Time (s) & 0.471 & 6.97 & 7.51 & 7.56 & 68.8 & 89.7 & 7.99 \\
\midrule
\multirow{4}{*}{\textbf{2}} 
& RMSE means & 0.033 & 0.137 & 0.169 & 0.115 & 0.336 & 0.0598 & 0.151 \\
& RMSE vars & 1.69e-03 & 0.0314 & 0.0657 & 0.0618 & 0.114 & 0.0108 & 0.0286 \\
& KL-divergence & 20.9 & 340 & 431 & 230 & 1524 & 62.4 & 421 \\
& Time (s) & 1.13 & 30 & 27.2 & 27.6 & 155 & 174 & 29.9 \\
\midrule
\multirow{4}{*}{\textbf{3}} 
& RMSE means & 0.0203 & 0.0976 & 0.13 & 0.0818 & 0.258 & 0.0455 & 0.0901 \\
& RMSE vars & 6.33e-04 & 0.0186 & 0.0537 & 0.05 & 0.0686 & 5.88e-03 & 0.0143 \\
& KL-divergence & 8.02 & 188 & 254 & 119 & 1041 & 36.4 & 161 \\
& Time (s) & 1.76 & 92.2 & 94 & 84.7 & 985 & 590 & 88.6 \\
\midrule
\multirow{4}{*}{\textbf{4}} 
& RMSE means & 8.84e-03 & 0.0608 & 0.068 & 0.0406 &  & 0.0421 & 0.031 \\
& RMSE vars & 2.77e-04 & 0.0104 & 0.0245 & 0.0286 &  & 4.55e-03 & 4.69e-03 \\
& KL-divergence & 1.57 & 81.3 & 72.8 & 30.3 &  & 32 & 20.4 \\
& Time (s) & 4.35 & 528 & 508 & 560 &  & 871 & 497 \\
\midrule
\multirow{4}{*}{\textbf{5}} 
& RMSE means & 3.29e-03 & 0.053 & 0.0425 & 0.0232 &  & 0.034 & 0.0146 \\
& RMSE vars & 4.67e-05 & 9.55e-03 & 0.0124 & 0.0133 &  & 3.85e-03 & 2.35e-03 \\
& KL-divergence & 0.205 & 63.3 & 25.7 & 9.22 &  & 20.2 & 4.66 \\
& Time (s) & 14 & 1599 & 1467 & 1347 &  & 1576 & 1312 \\
\bottomrule
\end{tabular}%
}
\caption{RMSE between means, RMSE between variances, KL divergence between exact and approximate predictive distributions and time needed for making predictions on the “interpolation” test set of the house price data set.}
\end{table}

\begin{table}[H]
\centering
\resizebox{\textwidth}{!}{%
\begin{tabular}{llcccccccc}
\toprule 
 &  & \textbf{Vecchia} & \textbf{Tapering} & \textbf{FITC} & \textbf{Full-scale} & \textbf{FRK} & \textbf{MRA} & \textbf{SPDE} \\
\midrule
\multirow{4}{*}{\textbf{1}} 
& RMSE means & 0.0553 & 0.32 & 0.244 & 0.201 & 1.31 & 0.105 & 0.333 \\
& RMSE vars & 0.0172 & 0.191 & 0.246 & 0.102 & 0.245 & 0.0912 & 0.17 \\
& KL-divergence & 22.2 & 752 & 369 & 245 & 11057 & 79.5 & 1036 \\
& Time (s) & 0.498 & 6.82 & 7.51 & 6.97 & 69 & 89.7 & 7.99 \\
\midrule
\multirow{4}{*}{\textbf{2}} 
& RMSE means & 0.038 & 0.27 & 0.111 & 0.183 & 10.5 & 0.082 & 0.14 \\
& RMSE vars & 0.0274 & 0.214 & 0.164 & 0.12 & 3.84 & 0.0432 & 0.0757 \\
& KL-divergence & 11.7 & 702 & 129 & 197 & 89167 & 35.1 & 162 \\
& Time (s) & 1.21 & 29.6 & 27.2 & 26.8 & 154 & 175 & 29.9 \\
\midrule
\multirow{4}{*}{\textbf{3}} 
& RMSE means & 0.0199 & 0.208 & 0.126 & 0.113 & 2.57 & 0.0522 & 0.109 \\
& RMSE vars & 6.98e-03 & 0.21 & 0.16 & 0.11 & 7.41 & 0.0163 & 0.0408 \\
& KL-divergence & 2.38 & 519 & 142 & 97.6 & 6900 & 16.8 & 75.9 \\
& Time (s) & 1.74 & 91.2 & 94.2 & 83 & 979 & 584 & 88.6 \\
\midrule
\multirow{4}{*}{\textbf{4}} 
& RMSE means & 7.85e-03 & 0.099 & 0.0788 & 0.106 &  & 0.0514 & 0.03 \\
& RMSE vars & 5.11e-03 & 0.172 & 0.0511 & 0.0994 &  & 0.0163 & 0.0248 \\
& KL-divergence & 0.458 & 210 & 42.2 & 75 &  & 15.8 & 9.35 \\
& Time (s) & 4.42 & 523 & 508 & 552 &  & 861 & 497 \\
\midrule
\multirow{4}{*}{\textbf{5}} 
& RMSE means & 4.88e-03 & 0.0609 & 0.0342 & 0.104 &  & 0.0426 & 0.0182 \\
& RMSE vars & 7.05e-04 & 0.141 & 0.0131 & 0.0465 &  & 0.012 & 0.0267 \\
& KL-divergence & 0.114 & 115 & 8.11 & 53.4 &  & 11.3 & 4.5 \\
& Time (s) & 14.1 & 1587 & 1467 & 1334 &  & 1563 & 1312 \\
\bottomrule
\end{tabular}%
}
\caption{RMSE between means, RMSE between variances, KL divergence between exact and approximate predictive distributions and time needed for making predictions on the “extrapolation” test set of the house price data set.}
\label{house_kl_extra_tab}
\end{table}

\newpage
\subsection{House price data set including covariates}\label{appendix_house_with_covs}

In this setting, the MRA encountered numerical errors during both estimation and prediction. As a result, only a limited number of resolutions could be considered, and some results on the extrapolation set are missing.

\begin{table}[ht!]
\centering
\resizebox{\textwidth}{!}{%
\begin{tabular}{l|l|ccccc}
  & \textbf{Tuning parameter} & \textbf{} & \textbf{} & \textbf{} & \textbf{} & \textbf{} \\
\midrule
\textbf{Vecchia} & nb. neighbours &  5 & 10 & 20 & 40 & 80\\
\hline
\textbf{Tapering} & avg. nb. non-zeros / row & 32 & 100 & 200 & 512 & 739  \\
\hline
\textbf{FITC} & nb. inducing points & 220 & 500 & 1000 & 2200 & 3700 \\
\hline
\multirow{2}{*}{\textbf{Full-scale}} & nb. inducing points & 106 & 252 & 444 & 1050 & 1900 \\
 & avg. nb. non-zeros / row & 17 & 36 & 71 & 256 & 420 \\
\hline
\textbf{FRK} & nb. resolutions & 1  & 2 & 3 & exceeds time limit &  \\
\hline
\textbf{MRA} & nb. knots per partition & 1 & 2 & 3 & 4 & 6 \\
\hline
\textbf{SPDE} & max edge & 0.3 & 0.11 & 0.065 & 0.0298 & 0.025  \\
\bottomrule
\end{tabular}%
}
\caption{Tuning parameters chosen for the comparison on the house price data set including covariates. Note that coordinates have been scaled by a factor $10^5$ to avoid numerical instabilities.}
\end{table}

\subsubsection{Prediction}

\begin{figure}[ht!]
      \centering  
      \includegraphics[width=\linewidth]{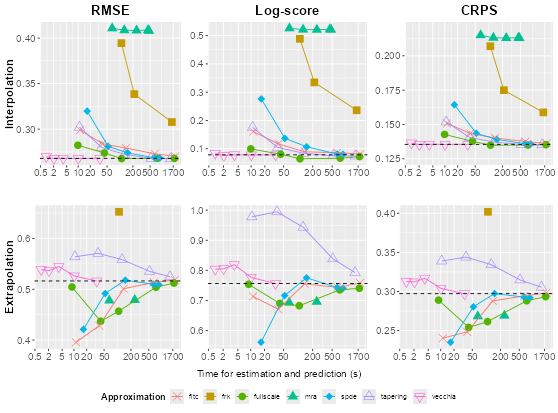} 
  \caption[House price data set including covariates. Prediction accuracy.]
  {RMSE, log-score and CRPS on the “interpolation” and “extrapolation” test sets of the house price data set including covariates. Every approximation makes predictions using its parameter estimates. The dashed lines correspond to the results from the exact calculations.}
\end{figure}

\begin{table}[H]
\centering
\resizebox{\textwidth}{!}{%
\begin{tabular}{llcccccccc}
\toprule 
 &  & \textbf{Vecchia} & \textbf{Tapering} & \textbf{FITC} & \textbf{Full-scale} & \textbf{FRK} & \textbf{MRA} & \textbf{SPDE} \\
\midrule
\multirow{4}{*}{\textbf{1}} 
& RMSE & 0.27 & 0.302 & 0.299 & 0.282 & 0.394 & 0.411 & 0.32 \\
& Log-score & 0.0838 & 0.176 & 0.161 & 0.0993 & 0.489 & 0.526 & 0.276 \\
& CRPS & 0.137 & 0.152 & 0.151 & 0.143 & 0.207 & 0.215 & 0.164 \\
& Time (s) & 1.03 & 10.8 & 11.2 & 9.84 & 108 & 65.9 & 16.8 \\
\midrule
\multirow{4}{*}{\textbf{2}} 
& RMSE & 0.268 & 0.28 & 0.284 & 0.274 & 0.338 & 0.409 & 0.281 \\
& Log-score & 0.0757 & 0.105 & 0.117 & 0.0803 & 0.335 & 0.521 & 0.137 \\
& CRPS & 0.135 & 0.141 & 0.143 & 0.138 & 0.175 & 0.213 & 0.143 \\
& Time (s) & 2.04 & 37.2 & 38.8 & 43.3 & 215 & 126 & 52.9 \\
\midrule
\multirow{4}{*}{\textbf{3}} 
& RMSE & 0.268 & 0.272 & 0.279 & 0.268 & 0.308 & 0.408 & 0.274 \\
& Log-score & 0.0784 & 0.0841 & 0.0889 & 0.0643 & 0.236 & 0.521 & 0.107 \\
& CRPS & 0.136 & 0.137 & 0.14 & 0.135 & 0.159 & 0.213 & 0.139 \\
& Time (s) & 4 & 128 & 137 & 109 & 1589 & 243 & 150 \\
\midrule
\multirow{4}{*}{\textbf{4}} 
& RMSE & 0.268 & 0.268 & 0.273 & 0.268 &  & 0.409 & 0.268 \\
& Log-score & 0.0779 & 0.0748 & 0.086 & 0.0665 &  & 0.522 & 0.0811 \\
& CRPS & 0.136 & 0.135 & 0.138 & 0.135 &  & 0.213 & 0.136 \\
& Time (s) & 10.1 & 522 & 640 & 731 &  & 448 & 634 \\
\midrule
\multirow{4}{*}{\textbf{5}} 
& RMSE & 0.268 & 0.267 & 0.27 & 0.268 &  & 0.408 & 0.268 \\
& Log-score & 0.0782 & 0.0741 & 0.0809 & 0.0721 &  & 0.521 & 0.0803 \\
& CRPS & 0.136 & 0.135 & 0.136 & 0.135 &  & 0.213 & 0.136 \\
& Time (s) & 33.9 & 1485 & 1874 & 1816 &  & 474 & 835 \\
\bottomrule
\end{tabular}%
}
\caption{Test “interpolation” RMSE, log-score, CRPS and time needed for making predictions on the “interpolation” test set on the house price data set including covariates. Every approximation makes predictions using its parameter estimates.
}
\end{table}

\begin{table}[H] 
\centering
\resizebox{\textwidth}{!}{%
\begin{tabular}{llcccccccc}
\toprule 
 &  & \textbf{Vecchia} & \textbf{Tapering} & \textbf{FITC} & \textbf{Full-scale} & \textbf{FRK} & \textbf{MRA} & \textbf{SPDE} \\
\midrule
\multirow{4}{*}{\textbf{1}} 
& RMSE & 0.539 & 0.564 & 0.396 & 0.505 & 0.653 & 0.478 & 0.421 \\
& Log-score & 0.804 & 0.978 & 0.713 & 0.754 & 1.38 & 0.693 & 0.561 \\
& CRPS & 0.313 & 0.339 & 0.24 & 0.289 & 0.402 & 0.268 & 0.235 \\
& Time (s) & 1.07 & 10.6 & 11.2 & 8.92 & 108 & 65.8 & 16.8 \\
\midrule
\multirow{4}{*}{\textbf{2}} 
& RMSE & 0.537 & 0.571 & 0.428 & 0.437 & 8.33 &  & 0.492 \\
& Log-score & 0.805 & 0.995 & 0.671 & 0.69 & 27.7 &  & 0.716 \\
& CRPS & 0.312 & 0.344 & 0.248 & 0.254 & 3.55 &  & 0.28 \\
& Time (s) & 2.08 & 36.6 & 38.8 & 42 & 215 & 124 & 52.9 \\
\midrule
\multirow{4}{*}{\textbf{3}} 
& RMSE & 0.545 & 0.558 & 0.501 & 0.457 & 1.62 & 0.479 & 0.518 \\
& Log-score & 0.819 & 0.943 & 0.755 & 0.682 & 2.24 & 0.696 & 0.775 \\
& CRPS & 0.317 & 0.334 & 0.288 & 0.261 & 0.888 & 0.269 & 0.297 \\
& Time (s) & 4.06 & 127 & 137 & 106 & 1589 & 243 & 150 \\
\midrule
\multirow{4}{*}{\textbf{4}} 
& RMSE & 0.526 & 0.535 & 0.513 & 0.504 &  &  & 0.51 \\
& Log-score & 0.776 & 0.838 & 0.743 & 0.735 &  &  & 0.743 \\
& CRPS & 0.304 & 0.315 & 0.294 & 0.288 &  &  & 0.293 \\
& Time (s) & 10.2 & 517 & 641 & 719 &  & 439 & 634 \\
\midrule
\multirow{4}{*}{\textbf{5}} 
& RMSE & 0.516 & 0.525 & 0.518 & 0.512 &  &  & 0.509 \\
& Log-score & 0.756 & 0.793 & 0.758 & 0.74 &  &  & 0.74 \\
& CRPS & 0.297 & 0.305 & 0.298 & 0.293 &  &  & 0.292 \\
& Time (s) & 34 & 1473 & 1874 & 1799 &  & 460 & 835 \\
\bottomrule
\end{tabular}%
}
\caption{Test “extrapolation” RMSE, log-score, CRPS and time needed for making predictions on the “extrapolation” test set on the house price data set including covariates. Every approximation makes predictions using its parameter
estimates.}
\end{table}

\subsubsection{Log-likelihood evaluation}

\begin{figure}[ht!]
      \centering  
      \includegraphics[width=0.7\linewidth]{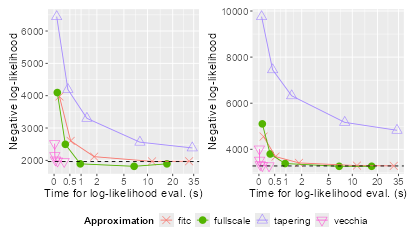} 
  \caption[House price data set including covariates. Log-likelihood evaluation.]
  {Negative log-likelihood on the house price data set including covariates. The parameters estimated via exact calculations are used on the left and
the pointwise doubling of them on the right. The dashed lines correspond to the exact negative log-likelihood.}
\end{figure}

\begin{table}[H]
\centering
\resizebox{\textwidth}{!}{%
\begin{tabular}{llcccccccc}
\toprule 
 &  & \textbf{Vecchia} & \textbf{Tapering} & \textbf{FITC} & \textbf{Full-scale} & \textbf{FRK} & \textbf{MRA} & \textbf{SPDE} \\
\midrule
\multirow{2}{*}{\textbf{1}} 
& Value & 2518 & 6445 & 3975 & 4096 &  & 18324 &  \\
& Time (s) & 0.0156 & 0.0698 & 0.147 & 0.0868 &  & 0.754 &  \\
\midrule
\multirow{2}{*}{\textbf{2}} 
& Value & 2148 & 4193 & 2601 & 2496 &  & 18723 &  \\
& Time (s) & 0.02 & 0.425 & 0.531 & 0.336 &  & 1.48 &  \\
\midrule
\multirow{2}{*}{\textbf{3}} 
& Value & 2012 & 3300 & 2114 & 1892 &  & 18960 &  \\
& Time (s) & 0.0367 & 1.32 & 1.8 & 0.96 &  & 2.43 &  \\
\midrule
\multirow{2}{*}{\textbf{4}} 
& Value & 1980 & 2562 & 1968 & 1815 &  & 19032 &  \\
& Time (s) & 0.0926 & 8.05 & 11.4 & 6.8 &  & 3.77 &  \\
\midrule
\multirow{2}{*}{\textbf{5}} 
& Value & 1965 & 2384 & 1972 & 1892 &  & 19304 &  \\
& Time (s) & 0.296 & 33.7 & 31 & 17.1 &  & 5.17 &  \\
\bottomrule
\end{tabular}%
}
\caption{Approximate negative log-likelihood and time needed for its evaluation on the house price data set including covariates. The covariance parameters estimated via exact calculations are used.
}
\end{table}

\begin{table}[H]
\centering
\resizebox{\textwidth}{!}{%
\begin{tabular}{llcccccccc}
\toprule 
 &  & \textbf{Vecchia} & \textbf{Tapering} & \textbf{FITC} & \textbf{Full-scale} & \textbf{FRK} & \textbf{MRA} & \textbf{SPDE} \\
\midrule
\multirow{2}{*}{\textbf{1}} 
& Value & 4013 & 9756 & 4552 & 5103 &  & 36408 &  \\
& Time (s) & 0.0119 & 0.0708 & 0.127 & 0.0911 &  & 0.772 &  \\
\midrule
\multirow{2}{*}{\textbf{2}} 
& Value & 3529 & 7449 & 3683 & 3796 &  & 28902 &  \\
& Time (s) & 0.0177 & 0.422 & 0.531 & 0.336 &  & 1.73 &  \\
\midrule
\multirow{2}{*}{\textbf{3}} 
& Value & 3333 & 6321 & 3409 & 3382 &  & 26738 &  \\
& Time (s) & 0.0341 & 1.32 & 1.8 & 0.962 &  & 2.63 &  \\
\midrule
\multirow{2}{*}{\textbf{4}} 
& Value & 3294 & 5168 & 3291 & 3261 &  & 25989 &  \\
& Time (s) & 0.0926 & 8.06 & 11.4 & 6.85 &  & 3.75 &  \\
\midrule
\multirow{2}{*}{\textbf{5}} 
& Value & 3283 & 4824 & 3280 & 3263 &  & 24923 &  \\
& Time (s) & 0.297 & 33.7 & 30.9 & 17.1 &  & 5.43 &  \\
\bottomrule
\end{tabular}%
}
\caption{Approximate negative log-likelihood and time needed for its evaluation on the house price data set including covariates. Two times the covariance parameters estimated via exact calculations are used.}
\end{table}

\subsubsection{Comparison to exact calculations}

\begin{figure}[H]
      \centering 
      \includegraphics[width=0.7\linewidth]{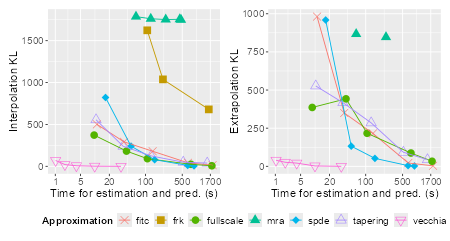} 
  \caption[House price data set including covariates. Comparison to exact calculations via KL divergence between predictive distributions.]
  {KL divergence between exact and approximate predictions on the “interpolation” test (left) and “extrapolation” test (right) sets on the house price data set including covariates. Every approximation makes predictions using its parameter estimates.}
\end{figure}

\begin{figure}[H]
  \centering            \includegraphics[width=0.66\linewidth]{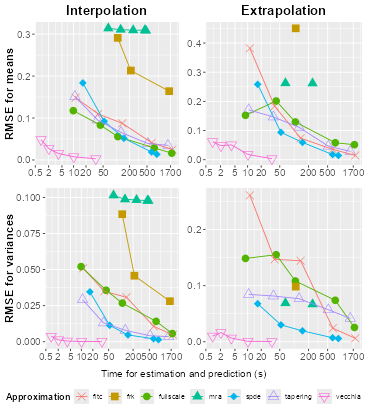} 
  \caption[House price data set, single-core setting. Comparison to exact calculations.]
  {RMSE between exact and approximate predictive means and variances on the house price data set including covariates. Every approximation makes predictions using its parameter estimates.}
\end{figure}

\begin{table}[H]
\centering
\resizebox{\textwidth}{!}{%
\begin{tabular}{llcccccccc}
\toprule 
 &  & \textbf{Vecchia} & \textbf{Tapering} & \textbf{FITC} & \textbf{Full-scale} & \textbf{FRK} & \textbf{MRA} & \textbf{SPDE} \\
\midrule
\multirow{4}{*}{\textbf{1}} 
& RMSE means & 0.0486 & 0.152 & 0.148 & 0.118 & 0.291 & 0.314 & 0.184 \\
& RMSE vars & 3.74e-03 & 0.0291 & 0.0509 & 0.0521 & 0.0884 & 0.101 & 0.0346 \\
& KL-divergence & 70.1 & 556 & 501 & 373 & 1623 & 1787 & 822 \\
& Time (s) & 1.03 & 10.8 & 11.2 & 9.84 & 108 & 65.9 & 16.8 \\
\midrule
\multirow{4}{*}{\textbf{2}} 
& RMSE means & 0.0271 & 0.0954 & 0.11 & 0.083 & 0.213 & 0.311 & 0.0923 \\
& RMSE vars & 1.31e-03 & 0.0131 & 0.035 & 0.0356 & 0.0457 & 0.0987 & 0.0113 \\
& KL-divergence & 21.1 & 240 & 292 & 182 & 1038 & 1761 & 241 \\
& Time (s) & 2.04 & 37.2 & 38.8 & 43.3 & 215 & 126 & 52.9 \\
\midrule
\multirow{4}{*}{\textbf{3}} 
& RMSE means & 0.0154 & 0.0657 & 0.0874 & 0.0557 & 0.164 & 0.309 & 0.0516 \\
& RMSE vars & 4.36e-04 & 7.9e-03 & 0.0308 & 0.0267 & 0.0281 & 0.0982 & 4.7e-03 \\
& KL-divergence & 6.91 & 124 & 180 & 91.4 & 680 & 1752 & 79.2 \\
& Time (s) & 4 & 128 & 137 & 109 & 1589 & 243 & 150 \\
\midrule
\multirow{4}{*}{\textbf{4}} 
& RMSE means & 7.54e-03 & 0.0393 & 0.041 & 0.0287 &  & 0.31 & 0.019 \\
& RMSE vars & 1.65e-04 & 4.12e-03 & 0.0106 & 0.0141 &  & 0.0977 & 2.08e-03 \\
& KL-divergence & 1.71 & 49.6 & 42.2 & 24.1 &  & 1755 & 11.3 \\
& Time (s) & 10.1 & 522 & 640 & 731 &  & 448 & 634 \\
\midrule
\multirow{4}{*}{\textbf{5}} 
& RMSE means & 2.81e-03 & 0.0348 & 0.0249 & 0.0162 &  & 0.308 & 0.0135 \\
& RMSE vars & 1.89e-04 & 3.55e-03 & 4.8e-03 & 5.64e-03 &  & 0.0977 & 1.42e-03 \\
& KL-divergence & 0.232 & 38.7 & 14 & 6.59 &  & 1749 & 5.75 \\
& Time (s) & 33.9 & 1485 & 1874 & 1816 &  & 474 & 835 \\
\bottomrule
\end{tabular}%
}
\caption{RMSE between means, RMSE between variances, KL divergence between exact and approximate predictive distributions and time needed for making predictions on the “interpolation” test set of the house price data set including covariates.}
\end{table}

\begin{table}[H]
\centering
\resizebox{\textwidth}{!}{%
\begin{tabular}{llcccccccc}
\toprule 
 &  & \textbf{Vecchia} & \textbf{Tapering} & \textbf{FITC} & \textbf{Full-scale} & \textbf{FRK} & \textbf{MRA} & \textbf{SPDE} \\
\midrule
\multirow{4}{*}{\textbf{1}} 
& RMSE means & 0.062 & 0.171 & 0.382 & 0.152 & 0.451 & 0.262 & 0.258 \\
& RMSE vars & 0.0101 & 0.0845 & 0.261 & 0.149 & 0.0983 & 0.0689 & 0.0678 \\
& KL-divergence & 39.9 & 527 & 981 & 386 & 2831 & 867 & 959 \\
& Time (s) & 1.07 & 10.6 & 11.2 & 8.92 & 108 & 65.8 & 16.8 \\
\midrule
\multirow{4}{*}{\textbf{2}} 
& RMSE means & 0.048 & 0.147 & 0.184 & 0.201 & 8.21 &  & 0.0938 \\
& RMSE vars & 0.0164 & 0.0816 & 0.147 & 0.155 & 1.21 &  & 0.0301 \\
& KL-divergence & 26.7 & 417 & 350 & 442 & 99884 &  & 132 \\
& Time (s) & 2.08 & 36.6 & 38.8 & 42 & 215 & 124 & 52.9 \\
\midrule
\multirow{4}{*}{\textbf{3}} 
& RMSE means & 0.0508 & 0.111 & 0.0737 & 0.129 & 1.35 & 0.262 & 0.0593 \\
& RMSE vars & 6.55e-03 & 0.0768 & 0.145 & 0.108 & 0.829 & 0.0669 & 0.0195 \\
& KL-divergence & 21.2 & 285 & 215 & 217 & 4944 & 847 & 52.2 \\
& Time (s) & 4.06 & 127 & 137 & 106 & 1589 & 243 & 150 \\
\midrule
\multirow{4}{*}{\textbf{4}} 
& RMSE means & 0.018 & 0.0463 & 0.0354 & 0.0582 &  &  & 0.0176 \\
& RMSE vars & 8.51e-04 & 0.0571 & 0.0242 & 0.0739 &  &  & 7.13e-03 \\
& KL-divergence & 2.53 & 92.2 & 21.8 & 87.2 &  &  & 4.63 \\
& Time (s) & 10.2 & 517 & 641 & 719 &  & 439 & 634 \\
\midrule
\multirow{4}{*}{\textbf{5}} 
& RMSE means & 3.13e-03 & 0.0287 & 0.0146 & 0.0514 &  &  & 0.0144 \\
& RMSE vars & 5.13e-04 & 0.0417 & 6.27e-03 & 0.0255 &  &  & 5.76e-03 \\
& KL-divergence & 0.102 & 43.6 & 3.31 & 33.2 &  &  & 2.39 \\
& Time (s) & 34 & 1473 & 1874 & 1799 &  & 460 & 835 \\
\bottomrule
\end{tabular}%
}
\caption{RMSE between means, RMSE between variances, KL divergence between exact and approximate predictive distributions and time needed for making predictions on the “extrapolation” test set of the house price data set including covariates.}
\end{table}

\newpage
\subsection{Laegern data set}\label{appendix_laeggern}
\subsubsection{Prediction}
\begin{table}[H]
\centering
\resizebox{\textwidth}{!}{%
\begin{tabular}{llcccccccc}
\toprule 
 &  & \textbf{Vecchia} & \textbf{Tapering} & \textbf{FITC} & \textbf{Full-scale} & \textbf{FRK} & \textbf{MRA} & \textbf{SPDE} \\
\midrule
\multirow{4}{*}{\textbf{1}} 
& RMSE & 0.0516 & 0.132 & 0.154 & 0.168 & 0.186 & 0.0909 & 0.1 \\
& Log-score & -1.56 & -0.581 & -0.46 & -0.382 & -0.263 & -1.01 & -0.883 \\
& CRPS & 0.027 & 0.0745 & 0.0859 & 0.0937 & 0.106 & 0.0475 & 0.054 \\
& Time (s) & 7.94 & 247 & 211 & 229 & 215 & 1243 & 217 \\
\midrule
\multirow{4}{*}{\textbf{2}} 
& RMSE & 0.0494 & 0.0932 & 0.145 & 0.116 & 0.177 & 0.0805 & 0.0843 \\
& Log-score & -1.6 & -0.888 & -0.527 & -0.702 & -0.311 & -1.13 & -1.05 \\
& CRPS & 0.0257 & 0.0527 & 0.08 & 0.0645 & 0.1 & 0.042 & 0.0451 \\
& Time (s) & 9.71 & 713 & 775 & 779 & 434 & 2829 & 745 \\
\midrule
\multirow{4}{*}{\textbf{3}} 
& RMSE & 0.0486 & 0.079 & 0.138 & 0.0844 & 0.16 & 0.0755 & 0.0713 \\
& Log-score & -1.62 & -1.04 & -0.575 & -0.972 & -0.417 & -1.2 & -1.22 \\
& CRPS & 0.0253 & 0.0446 & 0.0758 & 0.0474 & 0.0896 & 0.0392 & 0.0377 \\
& Time (s) & 78.9 & 1619 & 1904 & 1856 & 689 & 4588 & 1783 \\
\midrule
\multirow{4}{*}{\textbf{4}} 
& RMSE & 0.0486 & 0.0678 & 0.135 & 0.0736 &  & 0.0726 & 0.0634 \\
& Log-score & -1.62 & -1.21 & -0.605 & -1.1 &  & -1.24 & -1.33 \\
& CRPS & 0.0253 & 0.0379 & 0.0735 & 0.0413 &  & 0.0376 & 0.0334 \\
& Time (s) & 184 & 3121 & 3425 & 3122 &  & 5130 & 3422 \\
\midrule
\multirow{4}{*}{\textbf{5}} 
& RMSE & 0.0486 & 0.0598 & 0.127 & 0.064 &  & 0.0681 & 0.0573 \\
& Log-score & -1.62 & -1.37 & -0.665 & -1.24 &  & -1.31 & -1.44 \\
& CRPS & 0.0252 & 0.0325 & 0.0688 & 0.0356 &  & 0.0352 & 0.0301 \\
& Time (s) & 764 & 8940 & 9555 & 8017 &  & 8058 & 8212 \\
\bottomrule
\end{tabular}%
}
\caption{Test “interpolation” RMSE, log-score, CRPS and time needed for making predictions on the “interpolation” test set on the Laegern data set. Every approximation makes predictions using its parameter estimates.}
\label{inter_lae_tab}
\end{table}

\begin{table}[H]
\centering
\resizebox{\textwidth}{!}{%
\begin{tabular}{llcccccccc}
\toprule 
 &  & \textbf{Vecchia} & \textbf{Tapering} & \textbf{FITC} & \textbf{Full-scale} & \textbf{FRK} & \textbf{MRA} & \textbf{SPDE} \\
\midrule
\multirow{4}{*}{\textbf{1}} 
& RMSE & 0.191 & 0.194 & 0.199 & 0.209 & 0.23 & 0.192 & 0.198 \\
& Log-score & -0.198 & -0.197 & 0.139 & 0.206 & -2.21e-03 & -0.229 & -0.101 \\
& CRPS & 0.11 & 0.111 & 0.134 & 0.143 & 0.134 & 0.109 & 0.116 \\
& Time (s) & 9.31 & 246 & 211 & 229 & 214 & 1244 & 217 \\
\midrule
\multirow{4}{*}{\textbf{2}} 
& RMSE & 0.192 & 0.193 & 0.201 & 0.193 & 1.45 & 0.192 & 0.192 \\
& Log-score & -0.229 & -0.128 & 0.356 & -0.161 & 13.4 & -0.227 & -0.154 \\
& CRPS & 0.109 & 0.112 & 0.158 & 0.112 & 0.752 & 0.109 & 0.112 \\
& Time (s) & 11.1 & 707 & 775 & 708 & 434 & 2827 & 745 \\
\midrule
\multirow{4}{*}{\textbf{3}} 
& RMSE & 0.192 & 0.193 & 0.207 & 0.193 & 2247 & 0.192 & 0.191 \\
& Log-score & -0.231 & -0.0753 & 0.804 & -0.196 & 11.4 & -0.229 & -0.196 \\
& CRPS & 0.109 & 0.113 & 0.242 & 0.11 & 704 & 0.109 & 0.109 \\
& Time (s) & 80.5 & 1603 & 1904 & 1755 & 690 & 4588 & 1783 \\
\midrule
\multirow{4}{*}{\textbf{4}} 
& RMSE & 0.192 & 0.193 & 0.258 & 0.193 &  & 0.192 & 0.191 \\
& Log-score & -0.232 & -3.3e-04 & 1.19 & -0.208 &  & -0.229 & -0.208 \\
& CRPS & 0.109 & 0.114 & 0.375 & 0.11 &  & 0.109 & 0.109 \\
& Time (s) & 186 & 3090 & 3425 & 2986 &  & 5129 & 3422 \\
\midrule
\multirow{4}{*}{\textbf{5}} 
& RMSE & 0.192 & 0.193 & 0.418 & 0.192 &  & 0.192 & 0.191 \\
& Log-score & -0.232 & -4.48e-03 & 1.52 & -0.232 &  & -0.229 & -0.218 \\
& CRPS & 0.109 & 0.113 & 0.57 & 0.109 &  & 0.109 & 0.109 \\
& Time (s) & 767 & 8853 & 9561 & 7758 &  & 8059 & 8212 \\
\bottomrule
\end{tabular}%
}
\caption{Test “extrapolation” RMSE, log-score, CRPS and time needed for making predictions on the “extrapolation” test set on the Laegern data set. Every approximation makes predictions using its parameter estimates.}
\end{table}

\subsubsection{Log-likelihood evaluation}
\begin{table}[H]
\centering
\resizebox{\textwidth}{!}{%
\begin{tabular}{llcccccccc}
\toprule 
 &  & \textbf{Vecchia} & \textbf{Tapering} & \textbf{FITC} & \textbf{Full-scale} & \textbf{FRK} & \textbf{MRA} & \textbf{SPDE} \\
\midrule
\multirow{2}{*}{\textbf{1}} 
& Value & -221776 & -63605 & -44870 & -40836 &  & -159974 &  \\
& Time (s) & 0.0968 & 0.29 & 4.38 & 1.45 &  & 7.79 &  \\
\midrule
\multirow{2}{*}{\textbf{2}} 
& Value & -232645 & -83954 & -51950 & -69641 &  & -179025 &  \\
& Time (s) & 0.158 & 0.903 & 13.8 & 6.06 &  & 14.1 &  \\
\midrule
\multirow{2}{*}{\textbf{3}} 
& Value & -234768 & -95433 & -57332 & -92766 &  & -189639 &  \\
& Time (s) & 0.307 & 2.21 & 26.2 & 16.3 &  & 21.1 &  \\
\midrule
\multirow{2}{*}{\textbf{4}} 
& Value & -235276 & -111701 & -62877 & -103885 &  & -195678 &  \\
& Time (s) & 0.824 & 5.19 & 42.4 & 21.7 &  & 28.7 &  \\
\midrule
\multirow{2}{*}{\textbf{5}} 
& Value & -235332 & -144967 & -74816 & -126180 &  & -203352 &  \\
& Time (s) & 2.59 & 21.6 & 122 & 40.6 &  & 45.2 &  \\
\bottomrule
\end{tabular}%
}
\caption{Approximate negative log-likelihood and time needed for its evaluation on the Laegern data set. The covariance parameters estimated by a Vecchia approximation with 240 neighbors are used.}
\end{table}

\begin{table}[H]
\centering
\resizebox{\textwidth}{!}{%
\begin{tabular}{llcccccccc}
\toprule 
 &  & \textbf{Vecchia} & \textbf{Tapering} & \textbf{FITC} & \textbf{Full-scale} & \textbf{FRK} & \textbf{MRA} & \textbf{SPDE} \\
\midrule
\multirow{2}{*}{\textbf{1}} 
& Value & -156187 & -28386 & -29081 & -19570 &  & -127829 &  \\
& Time (s) & 0.0925 & 0.291 & 2.69 & 0.965 &  & 7.84 &  \\
\midrule
\multirow{2}{*}{\textbf{2}} 
& Value & -138867 & -44428 & -42569 & -45256 &  & -125582 &  \\
& Time (s) & 0.157 & 0.909 & 8.12 & 4.69 &  & 14.5 &  \\
\midrule
\multirow{2}{*}{\textbf{3}} 
& Value & -131620 & -54872 & -53401 & -70953 &  & -125547 &  \\
& Time (s) & 0.304 & 2.21 & 15.5 & 12.1 &  & 22.4 &  \\
\midrule
\multirow{2}{*}{\textbf{4}} 
& Value & -129886 & -70668 & -62599 & -82656 &  & -125221 &  \\
& Time (s) & 0.812 & 5.18 & 24.6 & 16.4 &  & 30.3 &  \\
\midrule
\multirow{2}{*}{\textbf{5}} 
& Value & -129805 & -107711 & -81065 & -109549 &  & -125789 &  \\
& Time (s) & 2.58 & 21.6 & 112 & 32.2 &  & 49.4 &  \\
\bottomrule
\end{tabular}%
}
\caption{Approximate negative log-likelihood and time needed for its evaluation on the Laegern data set. Two times the covariance parameters estimated by a Vecchia approximation with the largest considered number of neighbors are used.}
\end{table}

\newpage
\subsection{MODIS 2016 data set}\label{appendix_modis_2016}
\subsubsection{Prediction}
\begin{table}[H]
\centering
\resizebox{\textwidth}{!}{%
\begin{tabular}{llcccccccc}
\toprule 
 &  & \textbf{Vecchia} & \textbf{Tapering} & \textbf{FITC} & \textbf{Full-scale}  & \textbf{FRK} & \textbf{MRA} & \textbf{SPDE} & \textbf{npspec} \\
\midrule
\multirow{2}{*}{\textbf{1}} 
& RMSE & 1.61 & 2.98 & 2.37 & 2.36 & 2.86 & 2.34 & 2.54 & 1.78 \\
& Log-score & 1.79 & 3.01 & 2.37 & 2.31 & 2.75 & 2.21 & 2.57 & 3.41 \\
& CRPS & 0.852 & 1.87 & 1.39 & 1.37 & 1.71 & 1.27 & 1.48 & 0.907 \\
& Time (s) & 10.8 & 103 & 109 & 113 & 151 & 570 & 99.1 & 674 \\
\midrule
\multirow{2}{*}{\textbf{2}} 
& RMSE & 1.5 & 2.87 & 2.45 & 2.35 & 2.78 & 1.87 & 2.6 &  \\
& Log-score & 1.75 & 3.38 & 2.38 & 2.41 & 2.7 & 1.93 & 2.63 &  \\
& CRPS & 0.788 & 1.84 & 1.41 & 1.38 & 1.66 & 0.97 & 1.52 &  \\
& Time (s) & 22.5 & 459 & 453 & 503 & 380 & 967 & 146 &  \\
\midrule
\multirow{2}{*}{\textbf{3}} 
& RMSE & 1.6 & 2.73 & 2.5 & 2.25 & 2.48 & 2.26 &  &  \\
& Log-score & 1.76 & 3.67 & 2.37 & 2.3 & 2.5 & 2.24 &  &  \\
& CRPS & 0.83 & 1.75 & 1.42 & 1.29 & 1.43 & 1.24 &  &  \\
& Time (s) & 58.2 & 1237 & 1453 & 1396 & 2671 & 1797 &  &  \\
\midrule
\multirow{2}{*}{\textbf{4}} 
& RMSE & 1.68 & 2.55 & 2.39 & 2.37 &  & 1.86 &  &  \\
& Log-score & 1.78 & 3.29 & 2.33 & 2.39 &  & 1.87 &  &  \\
& CRPS & 0.861 & 1.57 & 1.36 & 1.35 &  & 0.94 &  &  \\
& Time (s) & 153 & 3609 & 3256 & 2555 &  & 3178 &  &  \\
\midrule
\multirow{2}{*}{\textbf{5}} 
& RMSE & 1.67 & 2.45 & 2.31 & 2.29 &  & 1.69 &  &  \\
& Log-score & 1.77 & 2.91 & 2.28 & 2.28 &  & 1.81 &  &  \\
& CRPS & 0.851 & 1.46 & 1.31 & 1.28 &  & 0.871 &  &  \\
& Time (s) & 452 & 5609 & 6422 & 5800 &  & 6602 &  &  \\
\bottomrule
\end{tabular}%
}
\caption{Test RMSE, log-score, CRPS and time needed for making
predictions on the test set on the MODIS 2016 data set. Every approximation makes predictions using its parameter estimates.}
\end{table}

\subsubsection{Log-likelihood evaluation}
\begin{table}[H]
\centering
\resizebox{\textwidth}{!}{%
\begin{tabular}{llcccccccc}
\toprule 
 &  & \textbf{Vecchia} & \textbf{Tapering} & \textbf{FITC} & \textbf{Full-scale}  & \textbf{FRK} & \textbf{MRA} & \textbf{SPDE} & \textbf{npspec} \\
\midrule
\multirow{2}{*}{\textbf{1}} 
& Value & 122663 & 216534 & 198554 & 203127 &  & 21483784 &  &  \\
& Time (s) & 0.0513 & 0.217 & 1.32 & 0.836 &  & 3.31 &  &  \\
\midrule
\multirow{2}{*}{\textbf{2}} 
& Value & 119805 & 201886 & 188378 & 178753 &  & 9433108 &  &  \\
& Time (s) & 0.0836 & 1.16 & 4.58 & 2.55 &  & 9.42 &  &  \\
\midrule
\multirow{2}{*}{\textbf{3}} 
& Value & 119144 & 185129 & 180404 & 168911 &  & 7585086 &  &  \\
& Time (s) & 0.164 & 4.05 & 13.5 & 4.98 &  & 16.1 &  &  \\
\midrule
\multirow{2}{*}{\textbf{4}} 
& Value & 119158 & 168029 & 174236 & 154500 &  & 6591441 &  &  \\
& Time (s) & 0.419 & 12.2 & 36.8 & 9.61 &  & 26.9 &  &  \\
\midrule
\multirow{2}{*}{\textbf{5}} 
& Value & 119144 & 159715 & 167247 & 142226 &  & 5625433 &  &  \\
& Time (s) & 1.45 & 33.3 & 94 & 20.8 &  & 65.9 &  &  \\
\bottomrule
\end{tabular}%
}
\caption{Approximate negative log-likelihood and time needed for its evaluation on the
MODIS 2016 data set. The covariance parameters estimated by a Vecchia approximation with
the largest considered number of neighbors are used.}
\label{modis_16_lik_true}
\end{table}

\begin{table}[H]
\centering
\resizebox{\textwidth}{!}{%
\begin{tabular}{llcccccccc}
\toprule 
 &  & \textbf{Vecchia} & \textbf{Tapering} & \textbf{FITC} & \textbf{Full-scale}  & \textbf{FRK} & \textbf{MRA} & \textbf{SPDE} & \textbf{npspec} \\
\midrule
\multirow{2}{*}{\textbf{1}} 
& Value & 122754 & 239799 & 200344 & 205824 &  & 3307953 &  &  \\
& Time (s) & 0.0542 & 0.217 & 1.36 & 0.867 &  & 3.32 &  &  \\
\midrule
\multirow{2}{*}{\textbf{2}} 
& Value & 119795 & 228512 & 189025 & 181639 &  & 1399985 &  &  \\
& Time (s) & 0.0794 & 1.16 & 4.47 & 2.47 &  & 9.25 &  &  \\
\midrule
\multirow{2}{*}{\textbf{3}} 
& Value & 119162 & 213135 & 180642 & 170865 &  & 1136081 &  &  \\
& Time (s) & 0.166 & 4.05 & 13.5 & 5.07 &  & 16 &  &  \\
\midrule
\multirow{2}{*}{\textbf{4}} 
& Value & 119188 & 195056 & 174364 & 155906 &  & 1011310 &  &  \\
& Time (s) & 0.426 & 12.1 & 36.9 & 10.1 &  & 26.9 &  &  \\
\midrule
\multirow{2}{*}{\textbf{5}} 
& Value & 119174 & 185366 & 167307 & 143126 &  & 886374 &  &  \\
& Time (s) & 1.44 & 33.3 & 95.1 & 21 &  & 65.9 &  &  \\
\bottomrule
\end{tabular}%
}
\caption{Approximate negative log-likelihood and time needed for its evaluation on the
MODIS 2016 data set. Two times the parameters estimated by a Vecchia approximation with
the largest considered number of neighbors are used.}
\label{modis_16_lik_fake}
\end{table}

\newpage
\subsection{MODIS 2023 data set}\label{appendix_modis_2023}
\subsubsection{Prediction}
\begin{table}[H]
\centering
\resizebox{\textwidth}{!}{%
\begin{tabular}{llcccccccc}
\toprule 
 &  & \textbf{Vecchia} & \textbf{Tapering} & \textbf{FITC} & \textbf{Full-scale}  & \textbf{FRK} & \textbf{MRA} & \textbf{SPDE} & \textbf{npspec} \\
\midrule
\multirow{2}{*}{\textbf{1}} 
& RMSE & 1.91 & 2.55 & 1.75 & 1.97 & 1.75 & 1.94 & 2.1 & 1.52 \\
& Log-score & 1.88 & 2.38 & 2.03 & 2.19 & 2.01 & 1.91 & 2.04 & 2.97 \\
& CRPS & 0.979 & 1.46 & 0.984 & 1.14 & 0.971 & 1 & 1.11 & 0.777 \\
& Time (s) & 36.6 & 541 & 562 & 628 & 459 & 9482 & 459 & 13222 \\
\midrule
\multirow{2}{*}{\textbf{2}} 
& RMSE & 1.91 & 2.47 & 2.07 & 1.89 & 4.04 &  & 2.16 &  \\
& Log-score & 1.87 & 2.31 & 2.04 & 2.12 & 3.51 &  & 2.07 &  \\
& CRPS & 0.98 & 1.39 & 1.04 & 1.08 & 1.59 &  & 1.15 &  \\
& Time (s) & 70.3 & 857 & 835 & 918 & 1607 &  & 901 &  \\
\midrule
\multirow{2}{*}{\textbf{3}} 
& RMSE & 1.92 & 2.45 & 2.81 & 1.97 &  &  & 2.2 &  \\
& Log-score & 1.87 & 2.31 & 2 & 2.08 &  &  & 2.09 &  \\
& CRPS & 0.983 & 1.38 & 1.1 & 1.06 &  &  & 1.18 &  \\
& Time (s) & 163 & 1455 & 1621 & 1520 &  &  & 1348 &  \\
\midrule
\multirow{2}{*}{\textbf{4}} 
& RMSE & 1.91 & 2.43 & 2.38 & 1.58 &  &  & 2.26 &  \\
& Log-score & 1.87 & 2.31 & 1.99 & 1.88 &  &  & 2.14 &  \\
& CRPS & 0.983 & 1.37 & 1.07 & 0.856 &  &  & 1.22 &  \\
& Time (s) & 527 & 3378 & 3525 & 2979 &  &  & 3069 &  \\
\midrule
\multirow{2}{*}{\textbf{5}} 
& RMSE & 1.91 & 2.31 & 1.47 & 1.43 &  &  & 1.87 &  \\
& Log-score & 1.87 & 2.37 & 1.87 & 1.77 &  &  & 1.85 &  \\
& CRPS & 0.982 & 1.31 & 0.83 & 0.767 &  &  & 0.955 &  \\
& Time (s) & 1393 & 13352 & 12366 & 8768 &  &  & 10990 &  \\
\bottomrule
\end{tabular}%
}
\caption{Test RMSE, log-score, CRPS and time needed for making
predictions on the test set on the MODIS 2023 data set. Every approximation makes predictions using its parameter estimates.}
\end{table}

\subsubsection{Log-likelihood evaluation}

\begin{table}[H]
\centering
\resizebox{\textwidth}{!}{%
\begin{tabular}{llcccccccc}
\toprule 
 &  & \textbf{Vecchia} & \textbf{Tapering} & \textbf{FITC} & \textbf{Full-scale}  & \textbf{FRK} & \textbf{MRA} & \textbf{SPDE} & \textbf{npspec} \\
\midrule
\multirow{2}{*}{\textbf{1}} 
& Value & 439151 & 1043380 & 1076009 & 1117564 &  & 13825680 &  &  \\
& Time (s) & 0.21 & 0.112 & 5 & 0.0511 &  & 18.8 &  &  \\
\midrule
\multirow{2}{*}{\textbf{2}} 
& Value & 430413 & 901574 & 1064069 & 1095756 &  &  &  &  \\
& Time (s) & 0.336 & 0.534 & 6.92 & 2.52 &  &  &  &  \\
\midrule
\multirow{2}{*}{\textbf{3}} 
& Value & 429066 & 878339 & 1031097 & 1007950 &  &  &  &  \\
& Time (s) & 0.673 & 0.942 & 16.5 & 6.3 &  &  &  &  \\
\midrule
\multirow{2}{*}{\textbf{4}} 
& Value & 429378 & 847272 & 963170 & 860055 &  &  &  &  \\
& Time (s) & 1.71 & 2.87 & 43.7 & 16.6 &  &  &  &  \\
\midrule
\multirow{2}{*}{\textbf{5}} 
& Value & 429420 & 726820 & 868875 & 762724 &  &  &  &  \\
& Time (s) & 5.47 & 23.2 & 228 & 49.3 &  &  &  &  \\
\bottomrule
\end{tabular}%
}
\caption{Approximate negative log-likelihood and time needed for its evaluation on the
MODIS 2023 data set. The covariance parameters estimated by a Vecchia approximation
with 240 neighbors are used.}
\label{modis_23_lik_true}
\end{table}

\begin{table}[H]
\centering
\resizebox{\textwidth}{!}{%
\begin{tabular}{llcccccccc}
\toprule 
 &  & \textbf{Vecchia} & \textbf{Tapering} & \textbf{FITC} & \textbf{Full-scale}  & \textbf{FRK} & \textbf{MRA} & \textbf{SPDE} & \textbf{npspec} \\
\midrule
\multirow{2}{*}{\textbf{1}} 
& Value & 441142 & 1016675 & 971274 & 1055451 &  & 2739914 &  &  \\
& Time (s) & 0.21 & 0.105 & 4.29 & 0.0517 &  & 20.2 &  &  \\
\midrule
\multirow{2}{*}{\textbf{2}} 
& Value & 440476 & 930184 & 950278 & 1010357 &  &  &  &  \\
& Time (s) & 0.332 & 0.53 & 6.08 & 2.28 &  &  &  &  \\
\midrule
\multirow{2}{*}{\textbf{3}} 
& Value & 442933 & 914694 & 903253 & 941800 &  &  &  &  \\
& Time (s) & 0.677 & 0.935 & 14.1 & 5.96 &  &  &  &  \\
\midrule
\multirow{2}{*}{\textbf{4}} 
& Value & 445301 & 893140 & 841040 & 841423 &  &  &  &  \\
& Time (s) & 1.7 & 2.85 & 37.5 & 15.7 &  &  &  &  \\
\midrule
\multirow{2}{*}{\textbf{5}} 
& Value & 445698 & 800784 & 778644 & 760559 &  &  &  &  \\
& Time (s) & 5.37 & 23.2 & 212 & 47 &  &  &  &  \\
\bottomrule
\end{tabular}%
}
\caption{Approximate negative log-likelihood and time needed for its evaluation on
the MODIS 2023 data set. Two times the parameters estimated by a Vecchia approximation with the largest considered number of neighbors are used.}
\label{modis_23_lik_fake}
\end{table}

\newpage    
\subsection{Simulated data with $N=100,000$ and an effective range of $0.2$, single-core setting}\label{large_sim_range_2}

\begin{table}[H]
\centering
\resizebox{\textwidth}{!}{%
\begin{tabular}{l|l|cccccc}
  & \textbf{Tuning parameter} & \textbf{} & \textbf{} & \textbf{} & \textbf{} & \textbf{} & \textbf{} \\
\midrule
\textbf{Vecchia} & nb. neighbours & 5 & 10 & 20 & 40 & 80 \\
\hline
\textbf{Tapering} & num. non-zero entries & 8 & 17 & 22 & 40 & 111\\
\hline
\textbf{FITC} & num. inducing points & 20 & 200 & 275 & 900 & 2000\\
\hline
\multirow{2}{*}{\textbf{Full-scale}} & num. inducing points & 26 & 91 & 122 & 250 & 650 \\
 & num. non-zero entries & 6 & 11 & 15 & 27 & 76 \\
\hline
\textbf{FRK} & num. resolutions & 1  & 2 & exceeds time limit &   &  &  \\
\hline
\textbf{MRA} & num. knots per partition & 1 & 2 & 3 & 4 & 5 \\
\hline
\textbf{SPDE} & max edge & 0.04 & 0.02 & 0.005 & 0.004 & 0.003   \\
\bottomrule
\end{tabular}%
}
\caption{Tuning parameters chosen for the comparison on simulated data sets with an effective range of $0.2$ and a sample size of $N=100,000$ in the single-core setting.}
\end{table}

\subsubsection{Log-likelihood evaluation}
\begin{figure}[H]

  \centering            \includegraphics[width=0.7\linewidth]{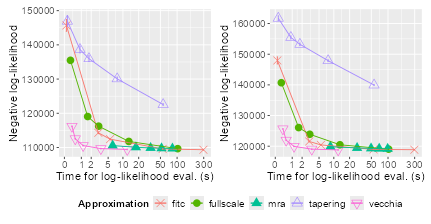} 
  \caption[$N=100,000$, range $0.2$, single-core setting. Log-likelihood evaluation.]
  {Average negative log-likelihood on
simulated data sets with an effective range $0.2$ of and a sample size N=100,000 in the single-core setting. The true data-generating parameters are used on the left, and two times these values are used on the right.}
\end{figure}

\begin{table}[H]
\centering
\resizebox{\textwidth}{!}{%
\input{tables/true_lik_02_100k_1core}
}
\caption{Mean approximate negative log-likelihood, standard error and time needed
for evaluating the approximate log-likelihood on simulated data sets with an effective range of $0.2$ of and a sample size N=100,000 in the single-core setting. Log-likelihoods are evaluated at the true data-generating parameters.}
\end{table}

\begin{table}[H]
\centering
\resizebox{\textwidth}{!}{%
\input{tables/wrong_lik_02_100k_1core}
}
\caption{Mean approximate negative log-likelihood, standard error and time needed
for evaluating the approximate log-likelihood on simulated data sets with an effective range of $0.2$ of and a sample size N=100,000 in the single-core setting. Log-likelihoods are evaluated at two-times the true data-generating parameters.}
\end{table}

\subsubsection{Parameter estimation}

\begin{figure}[H]
  \centering            \includegraphics[width=\linewidth]{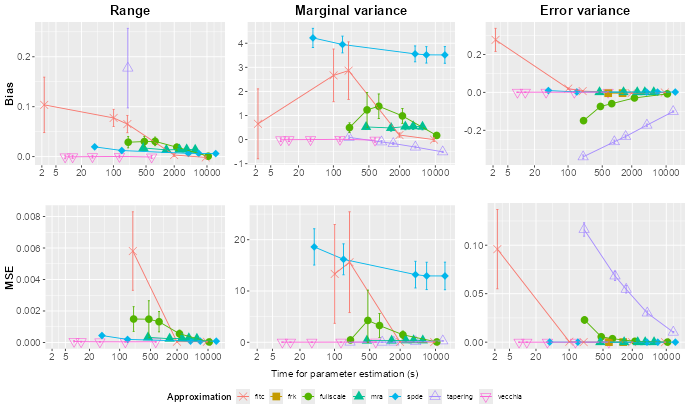} 
  \caption[$N=100,000$, range $0.2$, single-core setting. Parameter estimation accuracy.]
  {Bias and MSE of the estimates for the GP range, GP variance end error term variance on simulated
data sets with an effective range of $0.2$ and a sample size N=100,000 in the single-core setting,}

\end{figure}

\begin{table}[H]
\centering
\resizebox{\textwidth}{!}{%
\input{tables/range_02_100k_1core}
}
\caption{Bias and MSE for the GP range, standard errors and time needed for estimating covariance parameters on simulated data sets with an effective range of $0.2$ and a sample size of N=100,000 in the single-core setting.
}
\end{table}

\begin{table}[H]
\centering
\resizebox{\textwidth}{!}{%
\input{tables/variance_02_100k_1core}
}
\caption{Bias and MSE for the GP marginal variance, standard errors and time needed for estimating covariance parameters on simulated data sets with an effective range of $0.2$ and a sample size of N=100,000 in the single-core setting.
}
\end{table}

\begin{table}[H]
\centering
\resizebox{\textwidth}{!}{%
\input{tables/nugget_02_100k_1core}
}
\caption{Bias and MSE for the error term variance, standard errors and time needed for estimating covariance parameters on simulated data sets with an effective range of $0.2$ and a sample size of N=100,000 in the single-core setting.
}
\end{table}

\subsubsection{Prediction}

\begin{figure}[H]

  \centering            \includegraphics[width=\linewidth]{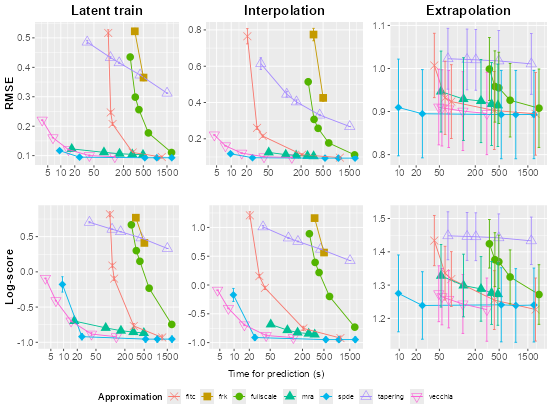} 
  \caption[$N=100,000$, range $0.2$, single-core setting. Prediction accuracy.]
  {Average RMSE and log-score on simulated data sets with
an effective range of $0.2$ and a sample size N=100,000 in the single-core setting. Predictions are done using the true data-generating parameters.}
\end{figure}

\begin{table}[H]
\centering
\resizebox{\textwidth}{!}{%
\input{tables/train_pred_02_100k_1core}
}
\caption{Average RMSE and log-score on the training set, standard errors and time
needed for making predictions on the training set on simulated data sets with an effective
range of $0.2$ and a sample size of N=100,000 in the single-core setting.
}
\end{table}

\begin{table}[H]
\centering
\resizebox{\textwidth}{!}{%
\input{tables/inter_pred_02_100k_1core}
}
\caption{Average RMSE and log-score on the test “interpolation” set, standard errors and time needed for making predictions on the test “interpolation” set on simulated
data sets with an effective range of $0.2$ and a sample size of N=100,000 in the single-core setting.
}
\end{table}

\begin{table}[H]
\centering
\resizebox{\textwidth}{!}{%
\input{tables/extra_pred_02_100k_1core}
}
\caption{Average RMSE and log-score on the test “extrapolation” set, standard errors and time needed for making predictions on the test “extrapolation” set on simulated
data sets with an effective range of $0.2$ and a sample size of N=100,000 in the single-core setting.
}
\end{table}

\newpage    
\subsection{House price data set, single-core setting}\label{house_single_core}

\begin{table}[H]
\centering
\resizebox{\textwidth}{!}{%
\begin{tabular}{l|l|ccccc}
  & \textbf{Tuning parameter} & \textbf{} & \textbf{} & \textbf{} & \textbf{} & \textbf{} \\
\midrule
\textbf{Vecchia} & nb. neighbours &  5 & 10 & 20 & 40 & 80\\
\hline
\textbf{Tapering} & avg. nb. non-zeros / row & 32 & 100 & 200 & 512 & 739  \\
\hline
\textbf{FITC} & nb. inducing points & 220 & 500 & 1000 & 2200 & 3700 \\
\hline
\multirow{2}{*}{\textbf{Full-scale}} & nb. inducing points & 106 & 252 & 444 & 1050 & 1900 \\
 & avg. nb. non-zeros / row & 17 & 36 & 71 & 256 & 420 \\
\hline
\textbf{FRK} & nb. resolutions & 1  & 2 & 3 & exceeds time limit &  \\
\hline
\textbf{MRA} & nb. knots per partition & 1 & 6 & 15 & 22 & 32 \\
\hline
\textbf{SPDE} & max edge & 3000 & 1100 & 650 & 298 & 200 \\
\bottomrule
\end{tabular}%
}
\caption{Tuning parameters chosen for the comparison on the house price data set in the single-core setting.}
\label{hyp_house_old}
\end{table}

\subsubsection{Prediction}

\begin{figure}[ht!]
      \centering  
      \includegraphics[width=\linewidth]{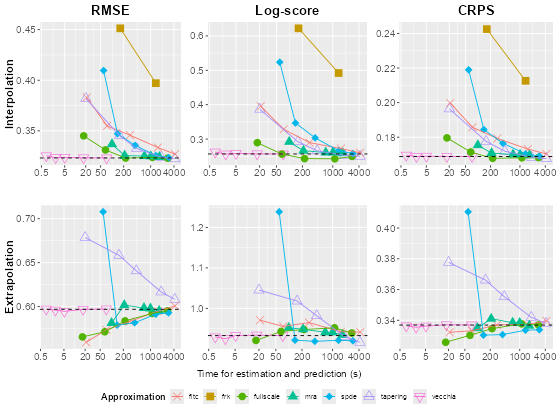} 
  \caption[House price data set, single-core setting. Prediction accuracy.]
  {RMSE, log-score and CRPS on the “interpolation” and “extrapolation” test sets of the house price data set in the single-core setting. Every approximation makes predictions using its parameter estimates. The dashed lines correspond to the results from the exact calculations.}
\end{figure}

\begin{table}[H]
\centering
\resizebox{\textwidth}{!}{%
\begin{tabular}{llcccccccc}
\toprule 
 &  & \textbf{Vecchia} & \textbf{Tapering} & \textbf{FITC} & \textbf{Full-scale} & \textbf{FRK} & \textbf{MRA} & \textbf{SPDE} \\
\midrule
\multirow{4}{*}{\textbf{1}} 
& RMSE & 0.325 & 0.382 & 0.383 & 0.345 & 0.544 & 0.337 & 0.41 \\
& Log-score & 0.262 & 0.387 & 0.395 & 0.29 & 0.81 & 0.292 & 0.524 \\
& CRPS & 0.17 & 0.196 & 0.2 & 0.18 & 0.298 & 0.175 & 0.219 \\
& Time (s) & 1.01 & 20.2 & 21.3 & 17.6 & 58.5 & 99.2 & 59.7 \\
\midrule
\multirow{4}{*}{\textbf{2}} 
& RMSE & 0.324 & 0.345 & 0.355 & 0.331 & 0.451 & 0.326 & 0.347 \\
& Log-score & 0.257 & 0.292 & 0.329 & 0.256 & 0.622 & 0.266 & 0.347 \\
& CRPS & 0.169 & 0.178 & 0.186 & 0.171 & 0.243 & 0.171 & 0.184 \\
& Time (s) & 2.5 & 157 & 73.7 & 67 & 163 & 211 & 138 \\
\midrule
\multirow{4}{*}{\textbf{3}} 
& RMSE & 0.324 & 0.332 & 0.346 & 0.323 & 0.397 & 0.325 & 0.336 \\
& Log-score & 0.258 & 0.266 & 0.29 & 0.243 & 0.492 & 0.261 & 0.304 \\
& CRPS & 0.169 & 0.172 & 0.179 & 0.168 & 0.213 & 0.17 & 0.176 \\
& Time (s) & 5.02 & 437 & 290 & 224 & 1364 & 688 & 393 \\
\midrule
\multirow{4}{*}{\textbf{4}} 
& RMSE & 0.323 & 0.324 & 0.334 & 0.324 &  & 0.325 & 0.326 \\
& Log-score & 0.257 & 0.252 & 0.272 & 0.243 &  & 0.261 & 0.266 \\
& CRPS & 0.169 & 0.168 & 0.173 & 0.168 &  & 0.17 & 0.17 \\
& Time (s) & 17.5 & 1893 & 1541 & 1115 &  & 1008 & 1364 \\
\midrule
\multirow{4}{*}{\textbf{5}} 
& RMSE & 0.323 & 0.322 & 0.327 & 0.323 &  & 0.324 & 0.324 \\
& Log-score & 0.257 & 0.25 & 0.26 & 0.25 &  & 0.259 & 0.258 \\
& CRPS & 0.169 & 0.168 & 0.17 & 0.168 &  & 0.17 & 0.169 \\
& Time (s) & 72.2 & 4229 & 4260 & 2769 &  & 1685 & 2907 \\
\bottomrule
\end{tabular}%
}
\caption{Test “interpolation” RMSE, log-score, CRPS and time needed for making predictions on the “interpolation” test set on the house price data set in the single-core setting. Every approximation makes predictions using its parameter estimates.
}
\end{table}

\begin{table}[H] 
\centering
\resizebox{\textwidth}{!}{%
\begin{tabular}{llcccccccc}
\toprule 
 &  & \textbf{Vecchia} & \textbf{Tapering} & \textbf{FITC} & \textbf{Full-scale} & \textbf{FRK} & \textbf{MRA} & \textbf{SPDE} \\
\midrule
\multirow{4}{*}{\textbf{1}} 
& RMSE & 0.597 & 0.679 & 0.559 & 0.565 & 1.45 & 0.581 & 0.708 \\
& Log-score & 0.929 & 1.05 & 0.971 & 0.921 & 3.75 & 0.952 & 1.24 \\
& CRPS & 0.336 & 0.378 & 0.332 & 0.326 & 0.976 & 0.335 & 0.411 \\
& Time (s) & 1.11 & 20 & 21.3 & 16.9 & 58.5 & 99.2 & 59.6 \\
\midrule
\multirow{4}{*}{\textbf{2}} 
& RMSE & 0.596 & 0.658 & 0.574 & 0.571 & 10.5 & 0.602 & 0.579 \\
& Log-score & 0.925 & 1.02 & 0.955 & 0.943 & 23.7 & 0.948 & 0.921 \\
& CRPS & 0.335 & 0.366 & 0.333 & 0.33 & 6.35 & 0.341 & 0.33 \\
& Time (s) & 2.6 & 155 & 73.8 & 66.2 & 163 & 212 & 138 \\
\midrule
\multirow{4}{*}{\textbf{3}} 
& RMSE & 0.595 & 0.641 & 0.583 & 0.584 & 2.76 & 0.598 & 0.582 \\
& Log-score & 0.933 & 0.982 & 0.965 & 0.948 & 2.59 & 0.941 & 0.919 \\
& CRPS & 0.336 & 0.355 & 0.337 & 0.335 & 1.5 & 0.338 & 0.331 \\
& Time (s) & 5.15 & 433 & 290 & 221 & 1364 & 681 & 393 \\
\midrule
\multirow{4}{*}{\textbf{4}} 
& RMSE & 0.597 & 0.617 & 0.595 & 0.592 &  & 0.598 & 0.591 \\
& Log-score & 0.934 & 0.931 & 0.942 & 0.952 &  & 0.941 & 0.921 \\
& CRPS & 0.337 & 0.342 & 0.337 & 0.338 &  & 0.338 & 0.334 \\
& Time (s) & 17.7 & 1876 & 1542 & 1106 &  & 998 & 1364 \\
\midrule
\multirow{4}{*}{\textbf{5}} 
& RMSE & 0.597 & 0.608 & 0.601 & 0.595 &  & 0.595 & 0.593 \\
& Log-score & 0.933 & 0.916 & 0.941 & 0.939 &  & 0.935 & 0.92 \\
& CRPS & 0.337 & 0.338 & 0.34 & 0.337 &  & 0.336 & 0.334 \\
& Time (s) & 72.6 & 4198 & 4260 & 2746 &  & 1672 & 2907 \\
\bottomrule
\end{tabular}%
}
\caption{Test “extrapolation” RMSE, log-score, CRPS and time needed for making predictions on the “extrapolation” test set on the house price data set in the single-core setting. Every approximation makes predictions using its parameter
estimates.}
\end{table}

\subsubsection{Log-likelihood evaluation}

\begin{figure}[ht!]
      \centering  
      \includegraphics[width=0.7\linewidth]{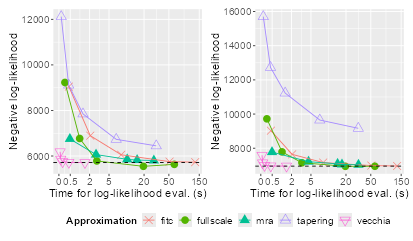} 
  \caption[House price data set, single-core setting. Log-likelihood evaluation.]
  {Negative log-likelihood on the house price data set in the single-core setting. The parameters estimated via exact calculations are used on the left and
the pointwise doubling of them on the right. The dashed lines correspond to the exact negative log-likelihood.}
\end{figure}

\begin{table}[H]
\centering
\resizebox{\textwidth}{!}{%
\begin{tabular}{llcccccccc}
\toprule 
 &  & \textbf{Vecchia} & \textbf{Tapering} & \textbf{FITC} & \textbf{Full-scale} & \textbf{FRK} & \textbf{MRA} & \textbf{SPDE} \\
\midrule
\multirow{2}{*}{\textbf{1}} 
& Value & 6184 & 12120 & 9047 & 9230 &  & 6750 &  \\
& Time (s) & 0.0437 & 0.0861 & 0.428 & 0.239 &  & 0.481 &  \\
\midrule
\multirow{2}{*}{\textbf{2}} 
& Value & 5862 & 9114 & 6893 & 6770 &  & 6063 &  \\
& Time (s) & 0.0673 & 0.418 & 2.07 & 1.1 &  & 2.77 &  \\
\midrule
\multirow{2}{*}{\textbf{3}} 
& Value & 5751 & 7847 & 6055 & 5783 &  & 5853 &  \\
& Time (s) & 0.142 & 1.37 & 8.27 & 2.88 &  & 10.5 &  \\
\midrule
\multirow{2}{*}{\textbf{4}} 
& Value & 5721 & 6732 & 5753 & 5548 &  & 5822 &  \\
& Time (s) & 0.426 & 6.76 & 51.2 & 19.5 &  & 15.4 &  \\
\midrule
\multirow{2}{*}{\textbf{5}} 
& Value & 5714 & 6446 & 5731 & 5633 &  & 5802 &  \\
& Time (s) & 1.54 & 31.5 & 130 & 61 &  & 28.8 &  \\
\bottomrule
\end{tabular}%
}
\caption{Approximate negative log-likelihood and time needed for its evaluation on the house price data set in the single-core setting. The covariance parameters estimated via exact calculations are used.
}
\end{table}

\begin{table}[H]
\centering
\resizebox{\textwidth}{!}{%
\begin{tabular}{llcccccccc}
\toprule 
 &  & \textbf{Vecchia} & \textbf{Tapering} & \textbf{FITC} & \textbf{Full-scale} & \textbf{FRK} & \textbf{MRA} & \textbf{SPDE} \\
\midrule
\multirow{2}{*}{\textbf{1}} 
& Value & 7587 & 15704 & 9027 & 9720 &  & 7773 &  \\
& Time (s) & 0.0442 & 0.0916 & 0.434 & 0.242 &  & 0.492 &  \\
\midrule
\multirow{2}{*}{\textbf{2}} 
& Value & 7142 & 12729 & 7642 & 7787 &  & 7213 &  \\
& Time (s) & 0.0667 & 0.413 & 2.06 & 1.13 &  & 4.5 &  \\
\midrule
\multirow{2}{*}{\textbf{3}} 
& Value & 6992 & 11229 & 7160 & 7150 &  & 7085 &  \\
& Time (s) & 0.141 & 1.39 & 8.29 & 3.34 &  & 14.6 &  \\
\midrule
\multirow{2}{*}{\textbf{4}} 
& Value & 6962 & 9649 & 6984 & 6941 &  & 7049 &  \\
& Time (s) & 0.428 & 7.23 & 51 & 19.6 &  & 17.2 &  \\
\midrule
\multirow{2}{*}{\textbf{5}} 
& Value & 6955 & 9166 & 6960 & 6944 &  & 7031 &  \\
& Time (s) & 1.5 & 31.6 & 130 & 58 &  & 31.8 &  \\
\bottomrule
\end{tabular}%
}
\caption{Approximate negative log-likelihood and time needed for its evaluation on the house price data set in the single-core setting. Two times the covariance parameters estimated via exact calculations are used.}
\end{table}

\subsubsection{Comparison to exact calculations}

\begin{figure}[H]
      \centering 
      \includegraphics[width=0.7\linewidth]{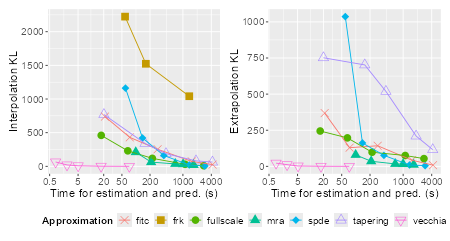} 
  \caption[House price data set, single-core setting. Comparison to exact calculations via KL divergence between predictive distributions.]
  {KL divergence between exact and approximate predictions on the “interpolation” test (left) and “extrapolation” test (right) sets on the house price data set in the single-core setting. Every approximation makes predictions using its parameter estimates.}
\end{figure}

\begin{figure}[H]
  \centering            \includegraphics[width=0.66\linewidth]{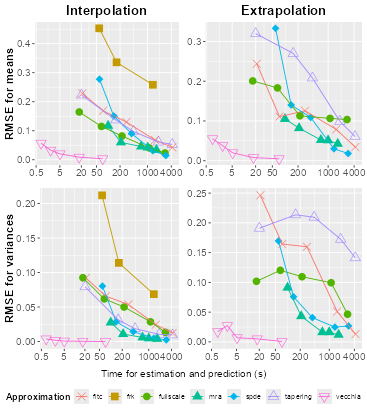} 
  \caption[House price data set, single-core setting. Comparison to exact calculations.]
  {RMSE between exact and approximate predictive means and variances on the house price data set in the single-core setting. Every approximation makes predictions using its parameter estimates.}
\end{figure}

\begin{table}[H]
\centering
\resizebox{\textwidth}{!}{%
\begin{tabular}{llcccccccc}
\toprule 
 &  & \textbf{Vecchia} & \textbf{Tapering} & \textbf{FITC} & \textbf{Full-scale} & \textbf{FRK} & \textbf{MRA} & \textbf{SPDE} \\
\midrule
\multirow{4}{*}{\textbf{1}} 
& RMSE means & 0.0571 & 0.223 & 0.229 & 0.165 & 0.453 & 0.117 & 0.278 \\
& RMSE vars & 3.78e-03 & 0.0793 & 0.0921 & 0.0925 & 0.212 & 0.0276 & 0.0804 \\
& KL-divergence & 65 & 769 & 739 & 460 & 2224 & 211 & 1162 \\
& Time (s) & 1.01 & 20.2 & 21.3 & 17.6 & 58.5 & 99.2 & 59.7 \\
\midrule
\multirow{4}{*}{\textbf{2}} 
& RMSE means & 0.033 & 0.137 & 0.169 & 0.115 & 0.336 & 0.0598 & 0.151 \\
& RMSE vars & 1.69e-03 & 0.0314 & 0.0657 & 0.0618 & 0.114 & 0.0108 & 0.0286 \\
& KL-divergence & 20.9 & 340 & 431 & 230 & 1524 & 62.4 & 421 \\
& Time (s) & 2.5 & 157 & 73.7 & 67 & 163 & 211 & 138 \\
\midrule
\multirow{4}{*}{\textbf{3}} 
& RMSE means & 0.0203 & 0.0976 & 0.13 & 0.0818 & 0.258 & 0.0455 & 0.0901 \\
& RMSE vars & 6.33e-04 & 0.0186 & 0.0537 & 0.05 & 0.0686 & 5.88e-03 & 0.0143 \\
& KL-divergence & 8.02 & 188 & 254 & 119 & 1041 & 36.4 & 161 \\
& Time (s) & 5.02 & 437 & 290 & 224 & 1364 & 688 & 393 \\
\midrule
\multirow{4}{*}{\textbf{4}} 
& RMSE means & 8.84e-03 & 0.0608 & 0.068 & 0.0406 &  & 0.0421 & 0.031 \\
& RMSE vars & 2.77e-04 & 0.0104 & 0.0245 & 0.0286 &  & 4.55e-03 & 4.69e-03 \\
& KL-divergence & 1.57 & 81.3 & 72.8 & 30.3 &  & 32 & 20.4 \\
& Time (s) & 17.5 & 1893 & 1541 & 1115 &  & 1008 & 1364 \\
\midrule
\multirow{4}{*}{\textbf{5}} 
& RMSE means & 3.29e-03 & 0.053 & 0.0425 & 0.0232 &  & 0.034 & 0.0146 \\
& RMSE vars & 4.67e-05 & 9.55e-03 & 0.0124 & 0.0133 &  & 3.85e-03 & 2.35e-03 \\
& KL-divergence & 0.205 & 63.3 & 25.7 & 9.22 &  & 20.2 & 4.66 \\
& Time (s) & 72.2 & 4229 & 4260 & 2769 &  & 1685 & 2907 \\
\bottomrule
\end{tabular}%
}
\caption{RMSE between means, RMSE between variances, KL divergence between exact and approximate predictive distributions and time needed for making predictions on the “interpolation” test set of the house price data set in the single-core setting.}
\end{table}

\begin{table}[H]
\centering
\resizebox{\textwidth}{!}{%
\begin{tabular}{llcccccccc}
\toprule 
 &  & \textbf{Vecchia} & \textbf{Tapering} & \textbf{FITC} & \textbf{Full-scale} & \textbf{FRK} & \textbf{MRA} & \textbf{SPDE} \\
\midrule
\multirow{4}{*}{\textbf{1}} 
& RMSE means & 0.0553 & 0.32 & 0.244 & 0.201 &  & 0.105 & 0.333 \\
& RMSE vars & 0.0172 & 0.191 & 0.246 & 0.102 & 0.245 & 0.0912 & 0.17 \\
& KL-divergence & 22.2 & 752 & 369 & 245 &  & 79.5 & 1036 \\
& Time (s) & 1.11 & 20 & 21.3 & 16.9 & 58.5 & 99.2 & 59.6 \\
\midrule
\multirow{4}{*}{\textbf{2}} 
& RMSE means & 0.038 & 0.27 & 0.111 & 0.183 &  & 0.082 & 0.14 \\
& RMSE vars & 0.0274 & 0.214 & 0.164 & 0.12 &  & 0.0432 & 0.0757 \\
& KL-divergence & 11.7 & 702 & 129 & 197 &  & 35.1 & 162 \\
& Time (s) & 2.6 & 155 & 73.8 & 66.2 & 163 & 212 & 138 \\
\midrule
\multirow{4}{*}{\textbf{3}} 
& RMSE means & 0.0199 & 0.208 & 0.126 & 0.113 &  & 0.0522 & 0.109 \\
& RMSE vars & 6.98e-03 & 0.21 & 0.16 & 0.11 &  & 0.0163 & 0.0408 \\
& KL-divergence & 2.38 & 519 & 142 & 97.6 &  & 16.8 & 75.9 \\
& Time (s) & 5.15 & 433 & 290 & 221 & 1364 & 681 & 393 \\
\midrule
\multirow{4}{*}{\textbf{4}} 
& RMSE means & 7.85e-03 & 0.099 & 0.0788 & 0.106 &  & 0.0514 & 0.03 \\
& RMSE vars & 5.11e-03 & 0.172 & 0.0511 & 0.0994 &  & 0.0163 & 0.0248 \\
& KL-divergence & 0.458 & 210 & 42.2 & 75 &  & 15.8 & 9.35 \\
& Time (s) & 17.7 & 1876 & 1542 & 1106 &  & 998 & 1364 \\
\midrule
\multirow{4}{*}{\textbf{5}} 
& RMSE means & 4.88e-03 & 0.0609 & 0.0342 & 0.104 &  & 0.0426 & 0.0182 \\
& RMSE vars & 7.05e-04 & 0.141 & 0.0131 & 0.0465 &  & 0.012 & 0.0267 \\
& KL-divergence & 0.114 & 115 & 8.11 & 53.4 &  & 11.3 & 4.5 \\
& Time (s) & 72.6 & 4198 & 4260 & 2746 &  & 1672 & 2907 \\
\bottomrule
\end{tabular}%
}
\caption{RMSE between means, RMSE between variances, KL divergence between exact and approximate predictive distributions and time needed for making predictions on the “extrapolation” test set of the house price data set in the single-core setting.}
\end{table}

\end{document}